\newcommand{\et}{\emph{et~al.}}
\newcommand{\eg}{\emph{e.g.}} 
\newcommand{\ie}{\emph{i.e.}}
\newcommand{\Ref}{ref.}
\newcommand{\E}[2][1]{${#1}\times10^{#2}$}
\newcommand{\cmvs}{cm$^2$V$^{-1}$s$^{-1}$} 
\newcommand{\mx}{MX$_2$}
\newcommand{\ms}{MoS$_2$}
\newcommand{\mse}{MoSe$_2$}
\newcommand{\mte}{MoTe$_2$}
\newcommand{\ws}{WS$_2$}
\newcommand{\wse}{WSe$_2$}
\newcommand{\sns}{SnS$_2$}
\newcommand{\C}{\,$^{\circ}$C}
\newcommand{\CT}{$^{\circ}$C} 
\newcommand{\npg}{Nature Publishing Group}

\newcommand{\aps}{American Physical Society}
\newcommand{\aip}{American Institute of Physics}
\newcommand{\iop}{Institute of Physics, UK}
\newcommand{\rsc}{Royal Society of Chemistry}
\newcommand{\aaas}{American Association for the Advancement of Science}
\newcommand{\Cright}[2][\acs]{Reproduced with permission from \Ref\citenum{#2}, copyright \citeyear{#2}, #1.} 
\newcommand{\cright}[2][\acs]{reproduced with permission from \Ref\citenum{#2}, copyright \citeyear{#2}, #1}  

\documentclass[twoside,twocolumn,9pt]{article} 
\pdfoutput=1
\usepackage{extsizes}
\usepackage[super,sort&compress,comma]{natbib}
\usepackage[version=3]{mhchem}
\usepackage[left=1.5cm, right=1.5cm, top=1.785cm, bottom=2.0cm]{geometry}
\usepackage{balance}
\usepackage{widetext}
\usepackage{times,mathptmx}
\usepackage{sectsty}
\usepackage{graphicx}
\usepackage{lastpage}
\usepackage[format=plain,justification=raggedright,singlelinecheck=false,font={stretch=1.125,small,sf},labelfont=bf,labelsep=space]{caption}
\usepackage{float}
\usepackage{fancyhdr}
\usepackage{fnpos}
\usepackage[english]{babel}
\usepackage{array}
\usepackage{droidsans}
\usepackage{charter}
\usepackage[T1]{fontenc}
\usepackage[usenames,dvipsnames]{xcolor}
\usepackage{setspace}
\usepackage[compact]{titlesec}
\usepackage{epstopdf}
\usepackage{multirow}
\usepackage{bm}
\usepackage{longtable}

\definecolor{cream}{RGB}{222,217,201}

\begin{document}

\pagestyle{fancy}
\thispagestyle{plain}
\fancypagestyle{plain}{
\fancyhead[C]{\includegraphics[width=18.5cm]{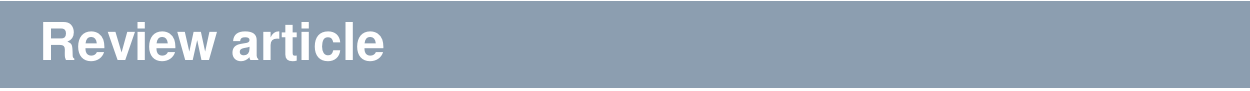}}
\fancyhead[L]{\hspace{0cm}\vspace{1.5cm}\LARGE{\makebox[5cm][l]{ \textsf{Please cite it as: \textit{Chem. Soc. Rev.}, 2016, \textbf{45}, 118}}}}
\fancyhead[R]{\hspace{0cm}\vspace{1.3cm}\includegraphics[height=55pt]{blank}}
\renewcommand{\headrulewidth}{0pt}
}

\makeFNbottom
\makeatletter
\makeatother

\renewcommand{\thefootnote}{\fnsymbol{footnote}}
\renewcommand\footnoterule{\vspace*{1pt}%
\color{cream}\hrule width 3.5in height 0.4pt \color{black}\vspace*{5pt}}
\setcounter{secnumdepth}{5}

\makeatletter
\renewcommand\@biblabel[1]{#1}
\renewcommand\@makefntext[1]%
{\noindent\makebox[0pt][r]{\@thefnmark\,}#1}
\makeatother
\renewcommand{\figurename}{\small{Fig.}~}
\sectionfont{\sffamily\Large}
\subsectionfont{\normalsize}
\subsubsectionfont{\bf}
\setstretch{1.125} 
\setlength{\skip\footins}{0.8cm}
\setlength{\footnotesep}{0.25cm}
\setlength{\jot}{10pt}
\titlespacing*{\section}{0pt}{4pt}{4pt}
\titlespacing*{\subsection}{0pt}{15pt}{1pt}

\setcounter{page}{1}  
\fancyfoot{}

\fancyfoot[RO]{\footnotesize{\sffamily{\pageref{FirstPage}--\pageref{LastPage} ~\textbar  \hspace{2pt}\thepage}}}
\fancyfoot[LE]{\footnotesize{\sffamily{\thepage~\textbar\hspace{0.01cm} \pageref{FirstPage}--\pageref{LastPage}}}}
\fancyhead{}

\renewcommand{\headrulewidth}{0pt}
\renewcommand{\footrulewidth}{0pt}
\setlength{\arrayrulewidth}{1pt}
\setlength{\columnsep}{6.5mm}
\setlength\bibsep{1pt}

\makeatletter
\newlength{\figrulesep}
\setlength{\figrulesep}{0.5\textfloatsep}

\newcommand{\topfigrule}{\vspace*{-1pt}%
\noindent{\color{cream}\rule[-\figrulesep]{\columnwidth}{1.5pt}} }

\newcommand{\botfigrule}{\vspace*{-2pt}%
\noindent{\color{cream}\rule[\figrulesep]{\columnwidth}{1.5pt}} }

\newcommand{\dblfigrule}{\vspace*{-1pt}%
\noindent{\color{cream}\rule[-\figrulesep]{\textwidth}{1.5pt}} }

\makeatother

\twocolumn[
  \begin{@twocolumnfalse}
\vspace{3cm}
\sffamily
\begin{tabular}{m{4.5cm} p{13.5cm} } \label{FirstPage}
\textsf{DOI: 10.1039/C5CS00517E} & \noindent\LARGE{\textbf{Charge transport and mobility engineering in
two-dimensional transition metal chalcogenide semiconductors}} \\
\vspace{0.3cm} & \vspace{0.3cm} \\
 & \noindent\large{Song-Lin Li,$^{\ast}$\textit{$^{a}$} Kazihito Tsukagoshi,\textit{$^{b}$} Emanuele Orgiu,$^{\ast}$\textit{$^{a}$} Paolo Samor\`i$^{\ast}
$\textit{$^{a}$}} \\

\\& \noindent\normalsize{Two-dimensional (2D) van der Waals semiconductors represent the thinnest, air stable semiconducting materials known. Their unique optical, electronic and mechanical properties hold great potential for harnessing them as key components in novel applications for electronics and optoelectronics. However, the charge transport behavior in 2D semiconductors is more susceptible to external surroundings (\eg{}\,gaseous adsorbates from air and trapped charges in substrates) and their electronic performance is generally lower than corresponding bulk materials due to the fact that surface and bulk coincide. In this article, we review recent progress on the charge transport properties and carrier mobility engineering of 2D transition metal chalcogenides, with a particular focus on the markedly high dependence of carrier mobility on thickness. We unveil the origin of this unique thickness dependence and elaborate the devised strategies to master it for carrier mobility optimization. Specifically, physical and chemical methods towards the optimization of the major factors influencing the extrinsic transport such as electrode/semiconductor contacts, interfacial Coulomb impurities and atomic defects are discussed. In particular, the use of \textit{ad-hoc} molecules makes it possible to engineer the interface with the dielectric and heal the vacancies in such materials. By casting fresh light onto the theoretical and experimental works, we provide a guide for improving the electronic performance of the 2D semiconductors, with the ultimate goal of achieving technologically viable atomically thin (opto)electronics.
} \\

\end{tabular}

 \end{@twocolumnfalse} \vspace{0.6cm}
]


\renewcommand*\rmdefault{bch}\normalfont\upshape
\rmfamily
\section*{}
\vspace{-1cm}

\footnotetext{\textit{$^{a}$\,Institut de Science et d'Ing\'{e}nierie Supramol\'{e}culaires (ISIS) and International Center for Frontier Research in Chemistry (icFRC), Universit\'{e} de Strasbourg and Centre National de la Recherche Scientifique (CNRS), Strasbourg 67083, France. \\ E-mail: songlinli@gmail.com, orgiu@unistra.fr, samori@unistra.fr}} 
\footnotetext{\textit{$^{b}$\,World Premier International Center for Materials Nanoarchitechtonics (WPI-MANA), National Institute for Materials Science (NIMS), Tsukuba, Ibaraki 305-0044, Japan.}}





\section{Introduction} \label{sect:intro}
Van der Waals (vdW) crystals represent a large family of materials that exhibit unique layered architectures, including graphite, metal oxides, chalcogenides, phosphates, cuprates, which differ on their chemical composition and crystal structure, leading to markedly different properties,  \eg{}\,their electronic characteristics can span from metallic to insulating. These materials all share a common feature---large interlayer vdW spacings and weak interlayer interactions, which results in peculiar mechanical properties like interlayer sliding and cleavability. In prehistory (the 4th millennium B.C.) graphite, the most widely known vdW material, was used as pottery paint. Nowadays it is still an unbeatable component for day-to-day applications as pencil cores, motor electrodes, and dry lubricants. The characteristic layered structure also endows vdW materials intriguing chemical and physical properties which make them the building blocks of choice for intercalation chemistry,\cite{Whitt78PiSSC_Review:chalcogenide} energy storage,\cite{Whitt04CR_Review} and superconductivity.

Remarkably, the structural anisotropy allows for mechanical exfoliation of vdW crystals down to the atomic scale.\cite{Novoselov04S,Novoselov05N,Zhang05N} The two-dimensional (2D) vdW flakes such as monolayers of metal chalcogenides represent the thinnest manifestation of stable materials that exhibit an energy bandgap. Following the success of graphene,\cite{Geim07NM,Geim09S} the research endeavor on the 2D vdW semiconductors rapidly increased. \cite{Wilson69AP,Yoffe02AP_Review,Mas-B11N_Review,Neto11RPP_Review,Hewak11NP_Review,Wang12NN,Butle13AN_Review:2D,Xu13CR,Chhowalla13NC,Pu14PCCP_Review:flexible,Kaul14JMR_Review:2D,Fiori14NN_review:2D_electronics,Pumer14JMCA_Review:chalcogenide,Jariwala14AN,Ganat14An_Review:MoS2,Schwi15N,Duan15CSR,Najma15ACR_review:defect,Lembk15ACR_Review:2Delectronics,Shi15CSR_Review:synthesis,Ji15CSR_,Wang15CSR_,Wang15N,Ferra15N_Review:2D:roadmap,Schmi15CSR,Peng15NT_Review:PV:MoS,Zeng15CSR,Zhang15CSR_Review:PhononRaman,Busce15CSR_Review:Photo} The concurrence of several unique properties, including the atomic thickness, sizable bandgap, high carrier mobility and absence of dangling bonds, and the fast-growing synthesis techniques
\cite{Al-Al77JCG_synthesis:SnS:SnSe:TaS,Feldman95S,Jen-LaPlante10JMC,Kim11L,Jin11SSS,Liu12NL,Shi12NL,Balendhran12N,Ding12N,Zhan12S,Lee12AM,Wang13JACS,Wu13APL_synthesis:CVDMSmu17,Lei13NL_,Wu13AN_synthesis:nodefect,Lee13NL_Synthesis:Seed:PTAS,Huang13JPCC_Synthesis:SnSe2,Yu13SR_synthesis:MoS2,Lu14NL,Ma14An_Synthesis:Isodoping,Mann14AM_Synthesis:Isodoping,Li14JACS_Synthesis:Isodoping,Huang14AN_Synthesis:WSe:inverter,Georg14AFM_syntesis:MoS:wafescale,Wang14An_syntesis:MoSe,Lin14AN_synthesis,Orofe14APL_synthesis,Gong14NL_synthesis,Huang14N_synthesis,Zhang14AN_synthesis:MoS,Ling14NL_Synthesis:MoS:seed,Lee14N_synthesis:MoS:waferscale,Taras14AFM_Synthesis:MoS2_3L:wafer,Zhang14AN_synthesis:MoWSe,Han15NC_,Yun15An_syntesis:MoS:1L,Pison15JPCC_synthesis:growthagent,Jeon15N_synthesis:MoS,Tao15N_synthesis:MoS:sputtering,Park15APL_synthesis:MoS:thickness,Kumar15N_synthesis:MoS:thickness,Kang15N_synthesis:MoS:waferscale} pave the way towards revolutionary applications, such as ultimate atomically-thin-body field-effect transistor (FET),\cite{britnell12s,georgiou13nn,Radisavljevic11NN,Radisavljevic13NM} stacked vdW superlattices and heterojunctions,\cite{Geim13N} valleytronics,\cite{Cao12NC,Gong13NC,Jones13NN,Mak12NN,Suzuki14NNb,Zeng12NN} and novel flexible and transparent electronics and optoelectronics.\cite{Akinw14NC,Yu13NN_Photocurernt:G/MoS2/G,Duan14NN_PV:WSe_MoS,Lee12NL_Photodetector,Lopez13NN_Photodetector,Jariw13PNAS,Britnell13S,Bernardi13NL,Xia14NP_Review:Photodetector,Koppe14NN_Photodetector:Review_2D}  Here we will focus on the role of 2D vdW crystals as electroactive  channels in FETs and, more specifically, on the factors influencing the electronic performances of the atomically-thin-body FETs and the devised strategies to improve them.

In fact, one of the prime interests in 2D crystals rests in their potential as conduction channels in digital circuits beyond silicon. The characteristic FET scaling length is derived as $\lambda =\sqrt{{{t}_{\text{s}}}{{t}_{\text{ox}}}{{\varepsilon }_{\text{s}}}/{{\varepsilon }_{\text{ox}}}}$,\cite{Yan92EDITo_Scalingrule} where $\epsilon$ and $t$ are electrical permittivity and thickness, and the subscripts s and ox denote semiconductor and oxide dielectric. The thinner the FET channels, the smaller and faster the FETs will be.\cite{Chang03PI,Ieong04S,Vogel07NN} Given the material physical limitation (such as surface roughness control) and production yield, the thickness of silicon channels can hardly be less than 5 nm,\cite{ITRS2013}  being much larger than the atomic scale. Exploiting 2D vdW semiconductors as FET channels would enable further device miniaturization after silicon.\cite{schwierz10nn,Wang12NL_Logic:MoS2}

It is noteworthy that the 2D planar structure also offers full compatibility to conventional semiconductor processing such that they can be perfectly carved for making highly ordered FET arrays, being a critical factor rivaling the 1D nanostructures. The third figure of merit of the vdW semiconductors is the self-saturated nature of the surfaces which, in principle, contain no dangling bonds and are free of the composition fluctuation at the channel/dielectric interfaces, making them immune to the notorious `sixth-power law' mobility degradation\cite{Gomez07IEDL} due to surface roughness (\ie{}\,interface asperity) that occurs in non-vdW superlattices and silicon.\cite{Sakaki87APL,Gold87PRB} The chemical stability is the fourth advantage which makes them stand out over other semiconductor membranes carved from 3D materials (\eg{}\,silicene\cite{Tao15NN_FET:2Dsilicene} and germanene), which degrade rapidly in ambient conditions. In contrast, most vdW crystals are stable in air; some of them like graphite and molybdenite exist as minerals in nature.

In the framework of post-silicon microelectronics, a great attention was initially devoted to the metallic graphene for its ultrahigh carrier mobility\cite{Novoselov04S,Novoselov05N,Zhang05N} rather than the 2D vdW semiconductors.\cite{Novos05PNAS_,Ayari07JAP_} It was then realized that it would be extremely difficult to use graphene for any digital application due to the absence of a bandgap, despite sustained efforts on bandgap and device engineering.\cite{sordan09apl,traversi09apl,li10nl,harada10apl,li11an,rizzi12nl,guerriero13an,li11s,Miyazaki10SST,Nakaharai13AN,Aparecido-Ferreira12N,Miyazaki12APL,Nakah14ITN_,Nakaharai12APE} Renewed interest on 2D vdW semiconductors arose in 2011 when Kis \et{}\,reported high carrier mobility in monolayer \ms{} FETs.\cite{Radisavljevic11NN,Fuhrer13NN}

As far as the FET performance is concerned, one of the essential figures of merit is the field-effect mobility ($\mu_\text{FE}$), which determines how fast a charge can move through a semiconductor or a metal under the effect of an external electric field. For 2D materials, where surface and bulk structurally coincide, a major yet not fully unanswered question is why in such atomically thin semiconductors carrier mobility undergoes degradation\cite{Novoselov05N,Ayari07JAP_} unlike in the corresponding bulk systems,\cite{Fivaz67PR} in spite of the immunity to the surface roughness scattering. It appears obvious that the full exposure of the lattice atoms to the environment can lead to strong carrier scattering and lower carrier mobility. In order to find out new strategies for improving carrier mobility, in-depth and quantitative answers to the thickness dependence of electronic performances are highly desirable.

Several theoretical studies were performed to cast light onto the charge transport behavior of the 2D vdW semiconductors.  Kaasbjerg \et{}\,extensively investigated the role of lattice phonons in \ms{} monolayers and predicted an intrinsic transport mobility of $\sim$410\,\cmvs{} at room temperature.\cite{Kaasbjerg12PRB,Kaasb13PRB_Transport:Phonon} Jena \et{}\,first considered the scattering generated by long-range Coulomb impurities in multilayer \ms{}.\cite{Kim12NC} Li \et{}\,addressed the role of the channel thickness in carrier scattering by considering various scattering mechanisms and ascribed the interfacial impurity scattering as the origin of the strong thickness dependence of mobility.\cite{Li13NL} Alongside phonons and Coulomb impurities, Ma \et{}\,were the first to consider the role of remote interface phonons, located in the dielectric, on the electronic behavior and identified the implications of using high-$\kappa$ dielectric in the atomically-thin-body \ms{} FET.\cite{Ma14PRX} The above works represent the theoretical framework of this review. On the other hand, a notable experimental effort was devoted to improving the mobility of the 2D vdW flakes by i) eliminating adverse extrinsic factors to attain material characteristics close to their intrinsic behavior, and ii) upon strain engineering to gain extra performance enhancement. To date, dramatic progress has been achieved on the first route in particular through contact optimization and carrier scattering suppression.

The review will discuss the origin of the high thickness dependence of electronic performance exhibited by 2D vdW semiconductors, providing a theoretical insight and summarizing the devised strategies to minimize its effect. A brief introduction is first given in section \ref{sect:intro} to illustrate the advantages and current hurdles in using the 2D vdW semiconductors as the active layer in FET devices. Section \ref{sect:structure}  outlines the material parameters regarding the electronic behavior, including band structure, carrier effective mass, and lattice phonons. In order to provide the reader with information on the typical electronic properties of \mx{} flakes, section \ref{mu_before} gives an exhaustive list of the values of carrier mobility measured so far, together with fabrication and measurement details. The extrinsic and intrinsic factors responsible for the charge transport behavior are outlined in section \ref{sect_factor}, shining light onto the origin of the dependence of the electronic performances on thickness. Section \ref{sect:engineering} describes various physical and chemical strategies on mobility engineering developed in recent years, followed by the state-of-the-art performance achieved after the mobility engineering. Section \ref{sect:trap} presents the experimental standards one should follow to avoid experimental traps and unintentional errors, which are neglected in some literature. Finally, a summary and outlook on the above-mentioned research field are given that are meant to suggest new avenues to minimize the charge scattering while paving the way towards chemical strategies to be adopted.

\section{Basic material properties} \label{sect:structure}

The term \textit{chalcogen} was proposed around 1930 by Werner Fischer to denote the elements of Group 16. The use of such a term was approved in 1938 by the Committee of the International Union of Chemistry (later IUPAC).\cite{bouro10_book:chalcogenide} It was then widely accepted that the elements sulfur, selenium, and tellurium are named chalcogens whereas their compounds chalcogenides. A large number of chalcogenides exhibit a layered structure and lend themselves to the application as the conduction channels in FETs.

In this section we outline the material parameters pertinent to electronic transport behavior in 2D vdW semiconductors, including crystal structure, phonon vibration mode, band structure, carrier effective mass, and electrical permittivity. Special attention is paid to the variation of these parameters with reducing material thickness, which may lead to mobility change.

\subsection{Atomic structure}
\begin{figure}[h!bt]
 \centering
 \includegraphics[width=0.5\textwidth]{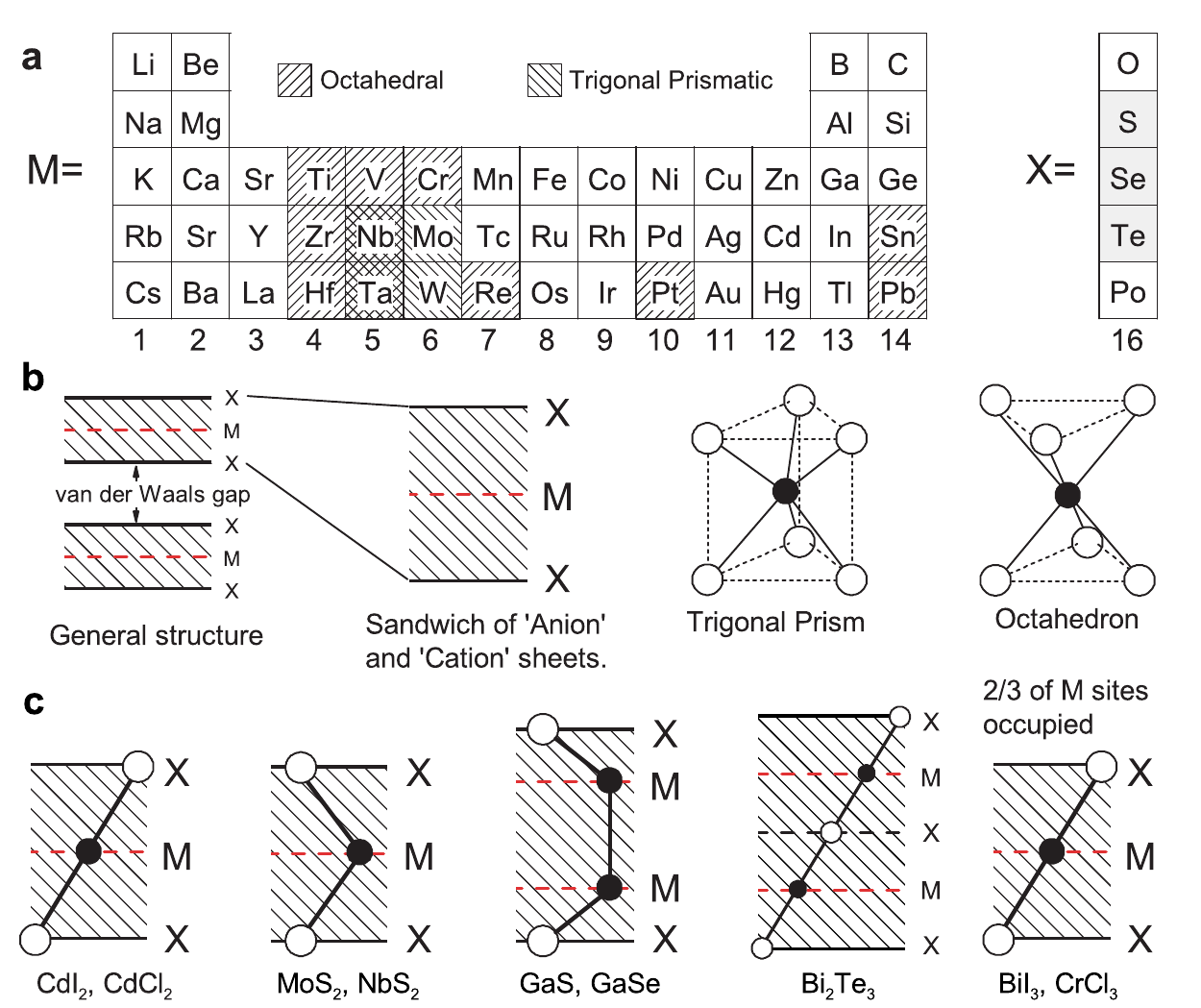}\\ 
 \includegraphics[width=0.5\textwidth]{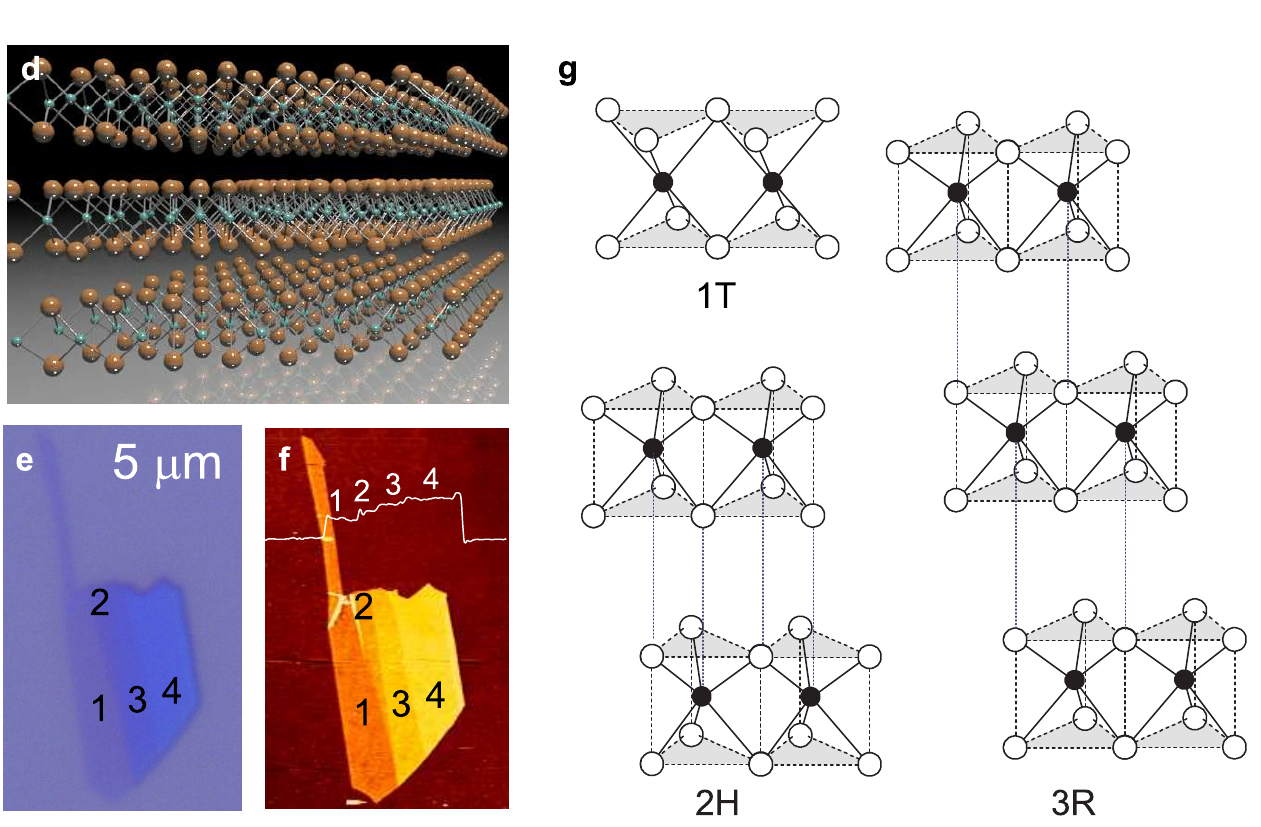}
  \caption{(a) Element periodic table showing the metal and chalcogen elements that form the \mx{} type van de Waals crystals. The shadows indicate the structure coordination of the crystals (octahedral or trigonal prismatic). \Cright[Elsevier Ltd]{Whitt78PiSSC_Review:chalcogenide} (b) The two basic trigonal prismatic and octahedral coordination units for the \mx{} crystals. (c) Cross-sectional (along 11\={2}0 plane) atomic coordination for the \mx{} chalcogenides and other typical van de Waals crystals. Panels (b) and (c) are \cright[Taylor \& Francis Ltd]{Wilson69AP}. (d) Three-dimensional schematic representation of a typical 2H-\mx{} structure with the chalcogen atom X in yellow and the metal atom M in cyan. (e) and (f) show optical and corresponding AFM images for a \ms{} flake with consecutive thickness values from 1 to 4 layers. Panels (e) and (f) are \cright{Li12AN}. (g) Schematic arrangements of sandwich units for the typical three phases of \ms{} crystals: 1T (tetragonal symmetry, one layer per repeat unit, octahedral coordination), 2H (hexagonal symmetry, two layers per repeat unit, trigonal prismatic coordination), and 3R (rhombohedral symmetry, three layers per repeat unit, trigonal prismatic coordination). The dotted vertical lines indicate the alignments of interlayer M and X atoms.}
 \label{fgr:periodictable_unit}
\end{figure}

The metal chalcogenides chemical composition discussed in the review can be described by the formula \mx{} (M=Ti, Zr, Hf, V, Nb, Ta, Cr, Mo, W, Pt and X=S, Se, Te). Figure \ref{fgr:periodictable_unit}a displays the location of these elements in the element periodic table. The layered structure originates from the stacking of hexagonally packed X--M--X trilayer sandwich units. The metal and chalcogen atoms are covalently bonded as individual `tricomponent' (trilayered) sandwich units while the different sandwich units are held together by weak vdW force, resulting in a remarkably easy mechanical cleavage. In compliance with the terminology used in literature, we call an individual `tricomponent' sandwich unit as a \mx{} monolayer. Within each layer, the atoms are arranged in configuration of either trigonal prism or octahedron (Fig.\,\ref{fgr:periodictable_unit}b), resulting in different lattice symmetries. It is worth noting that the layered structure is not only limited to the dichalcogenides composed of transition metal elements; some non-transition metal (\eg{}\,Ga, In, Bi, Sn, Pb) chalcogenides and halides also show layered structures. Figure \ref{fgr:periodictable_unit}c illustrates some examples of monochalcogenides, trichalcogenides and halides that also possess a layered structure. Notably, Bi$_2$Se$_3$ and Bi$_2$Te$_3$ are well-known topological insulators.

Due to compositional variation, the \mx{} family covers a wide range of electronic properties, spanning from those of an insulator like HfS$_2$, to semiconductors like \ms{} and semi-metals like WTe$_2$ and TeS$_2$, way down to true metals like NbS$_2$ and VSe$_2$.\cite{Wilson69AP} In this review article we focus our attention on semiconductors with bandgap around 1--2\,eV. As a prototype \mx{} semiconductor, we will especially concentrate on the structure and properties of \ms{} layers. Figure\,\ref{fgr:periodictable_unit}d shows the atomic structure for typical 2H-phased \ms{}. It exists in nature as the mineral \textit{molybdenite} and can be easily mechanically exfoliated into few-layer flakes. Figures\,\ref{fgr:periodictable_unit}e and \ref{fgr:periodictable_unit}f illustrate the optical and atomic force images for an exfoliated \ms{} flake with consecutive numbers of layers (NL, N is an integer) from 1 to 4. In fact, \ms{} has three different structural phases: 1T (tetragonal symmetry), 2H (hexagonal symmetry) and 3R (rhombohedral symmetry), as illustrated in Fig.\,\ref{fgr:periodictable_unit}g. Among them, the 2H and 3R phases are semiconducting while the 1T phase is metallic. Phase change can occur under external stimuli\cite{Lin14NN_1Tphase:TEM,Cho15S} or chemical treatment,\cite{Kappera14NM,Kappe14AM_FET:MoS2:1Tcontact,Voiry15CSR} for example, by soaking in n-butyl lithium \ms{} can undergo phase change from semiconducting 2H to  metallic 1T phase. The phase change induced property change has been employed to reduce the contact resistance, as will be discussed in section \ref{sect:engineering}.

\subsection{Lattice phonon modes}

In elastic materials the lattice phonon is a collective atom displacement with atoms vibrating around their equilibrium positions. Such a displacement can modify the carrier pathway in two ways: 1) deformation of the local lattice potential, and 2) formation of electric fields due to polarizability and piezoelectricity of lattices. These two scattering mechanisms will be discussed in section \ref{sect_mechanism}. Lattice phonons have a thermal origin and exist at non-zero temperatures; hence, unlike other scattering centers phonon is an intrinsic scattering factor.
\begin{figure*}[h!bt]
 \centering
  \includegraphics[width=0.98\textwidth]{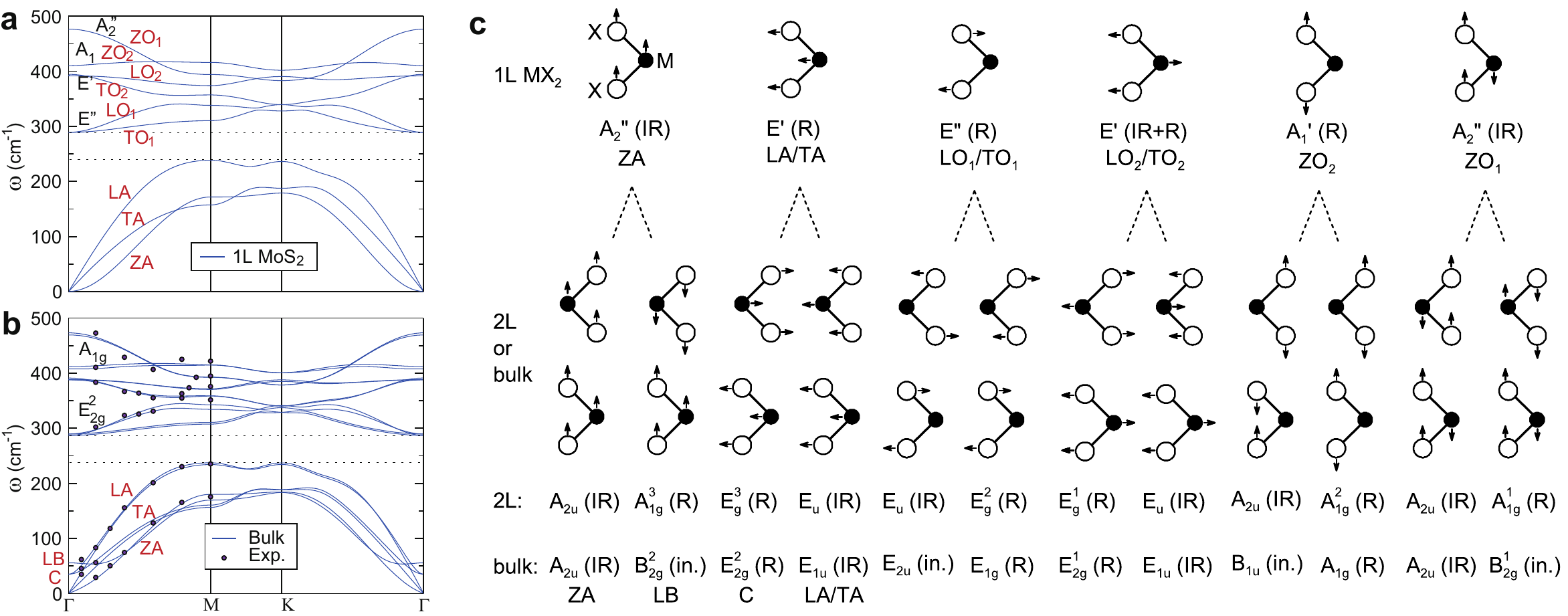}
  \caption{Calculated phonon dispersion curves for (a) 1L and (b) bulk \mx{}. The dots in (b) are the data from inelastic neutron scattering experiments. Panels (a) and (b) are \cright[\aps]{Molina-Sanchez11PRB}. (c) Vibration modes, symmetry representation, and optical activities (Raman: R; Infrared: IR; inactive: in.) of the lattice phonons for 1L, 2L and bulk \mx{}. There are 6 optical and 3 acoustic branches in the 1L flake while the numbers of optical branches increase to 3$\times$6$-$3$=$15 for 2L and bulk samples, due to the doubling of the numbers of atoms in unit cells. Note that the low-frequency LB and C modes are of optical characteristics although they share quite close dispersion behavior to the 3 acoustic modes at high wavenumbers. \Cright[\rsc]{Zhang15CSR_Review:PhononRaman}  }
 \label{fgr:phonon_mode}
\end{figure*}

\begin{table}[h!bt]
\small
\caption{\ Relevant phonon symmetry representation and optical activity (Raman: R, Infrared: IR, and inactive: in.)
of single-layer (point group $D_{3h}$), bilayer (point group $D_{3d}$), and
bulk (point group $D_{6h}$) MoS$_2$. Vibration direction is along the azimuth axis of the unit
cell. Phonon frequencies ($\omega_{MoS_{2}}$) are calculated values. \Cright[\aps]{Molina-Sanchez11PRB}}
  \label{tbl:phonon_mode}
  \begin{tabular*}{0.5\textwidth}{@{\extracolsep{\fill}}llllllll}
    \hline
$D_{3h}$                    	&$D_{3d}$/$D_{6h}$   	&Activity 	& Direction 	& Atoms   	& \multicolumn{2}{c}{$\omega_{MoS_{2}}$ (cm$^{-1}$)} \\	
\hline							
\multirow{2}{*}{$A_2^{''}$ (IR)} 	&$A_{2u}$/$A_{2u}$   	& IR 	&\multirow{2}{*}{ z axis} 	&\multirow{2}{*}{Mo+S} 	& \multirow{2}{*}{0.0} 	& 0.0    \\
                            	&$A^3_{1g}$/$B^2_{2g}$ 	& R/in.  	&    	&    	&    	& 55.7    \\
\hline						
\multirow{2}{*}{$E^{'}$ (R)} 	&$E_{g}^3$/$E_{2g}^2$ 	& R	&\multirow{2}{*}{ xy plane} 	&\multirow{2}{*}{Mo+S} 	& \multirow{2}{*}{-} 	& 35.2  \\
                            	&$E_{u}$/$E_{1u}$ 	& IR  	&    	&    	&  	& -  \\
\hline						
\multirow{2}{*}{$E^{''}$ (R)}      	&$E_{u}$/$E_{1u}$    	& IR/in.  	&\multirow{2}{*}{ xy plane} 	&\multirow{2}{*}{S} 	& \multirow{2}{*}{289.2}     	& 287.1 \\
                            	&$E_{g}^2$/$E_{1g}$   	& R     	&    	&    	&                       	& 288.7 \\
    \hline						
\multirow{2}{*}{$E^{'}$ (IR+R)}       	&$E_{g}^1$/$E_{2g}^1$ 	&R  	&\multirow{2}{*}{ xy plane} 	&\multirow{2}{*}{Mo+S} 	& \multirow{2}{*}{391.7} 	& 387.8 \\
                            	&$E_{u}$/$E_{1u}$   	&IR ($\bm{E}\bot$z)  	&    	&    	&                 	& 391.2 \\
\hline						
\multirow{2}{*}{$A_1$ (R)}      	&$A_{2u}$/$B_{1u}$   	& IR/in.  	&\multirow{2}{*}{ z axis} 	&\multirow{2}{*}{S} 	& \multirow{2}{*}{410.3} 	& 407.8 \\
                            	&$A_{1g}^2$/$A_{1g}$   	& R     	&    	&    	&                      	& 412.0 \\
\hline						
\multirow{2}{*}{$A_2^{''}$ (IR)} 	&$A_{2u}$/$A_{2u}$   	& IR ($\bm{E}||$z)  	&\multirow{2}{*}{ z axis} 	&\multirow{2}{*}{Mo+S} 	& \multirow{2}{*}{476.0} 	& 469.4 \\
                            	&$A^1_{1g}$/$B^1_{2g}$ 	& R/in.  	&    	&    	&                        	& 473.2 \\

\hline

  \end{tabular*}
\end{table}

The phonon modes of bulk \mx{} vdW crystals have been fully investigated in the 1970--80s. Related information such as symmetry representation, vibration mode, and optical activity are well documented in literature.\cite{Verble70PRL,Wieting71PRB,Agnihotr72PM,Agnihotr73SSC,Sekine80JPSJ,Sugai82PRB,Sekine84JPSJ,Wieting80PB,SEKINE80SSC,SOURISSEAU91CP,Frey98JMR} Taking advantage of the capacity to reduce the thickness of the crystal, new information such as frequency shift\cite{Lee10AN_Raman:MoS2} and excitation of new phonon modes\cite{Terro14SR_Raman:MX2:newMode,Yamam14AN_Raman:MoTe:NewMode} have been acquired in recent years.

According to the energy-momentum ($\omega$-$k$) dispersion relations, phonons are categorized into two types: acoustic ($\omega \propto k$ at $k\approx0$) and optical ($\omega\approx$\,constant), which represent the relative motion phase for adjacent atoms. A simple rule to discern the phonon feature, for instance in a 1D diatomic chain, is that the optical modes are produced when two adjacent atoms move against each other (out-of-phase), while the acoustic modes are produced when they move together (in-phase). Figures\,\ref{fgr:phonon_mode}a and \ref{fgr:phonon_mode}b show the calculated dispersion relations for \mx{} monolayer and bulk, respectively. Specifically, for monolayer \mx{} one unit cell comprises one X--M--X sandwich with 3 atoms and thus there are 9 phonon modes (3 acoustic and 6 optical modes). The numbers of atoms in the unit cell increases to 6 for bulk and, accordingly, the numbers of optical modes increases to 15.

Figure\,\ref{fgr:phonon_mode}c illustrates the schematic atomic vibration modes, optical activities (Raman, infrared, or inactive, abbreviated here as R, IR and in., respectively), and acoustic/optical features for the monolayer (1L), bilayer (2L) and bulk \ms{}. Lattice vibration modes are normally classified according to the irreducible representation of the crystal symmetry. For few-layer flakes, the symmetries differ if the flakes have an odd or even number of layers. The odd numbered flakes have a point group symmetry of D$_{3h}$ owing to the presence of the horizontal reflection plane ($\sigma_h$) that passes through the transition metal atom (M). The corresponding representation is
$\Gamma$= $2A''_2$+$A'_1$+$2E'$+$E''$,\cite{Jim91PRB_Phonon:mode:1LMoS2,Ataca11JPCCb}
where one $A''_2$ and one $E'$ are acoustic modes, another $A''_2$ is IR active, $A'_1$ and $E''$ are R active, and another $E{'}$ is both R and IR active, as shown in Fig.\,\ref{fgr:phonon_mode}a.
In contrast, due to the presence of the inversion symmetry, the symmetry of the even numbered flakes is D$_{3d}$ with the representation:
$\Gamma$= $3A_{1g}$+$3A_{2u}$+$3E_{g}$+$3E_{u}$,\cite{Ribei14PRB_Phonon:mode,Zhao13NLb}
where one $A_{2u}$ and one $E_{u}$ are acoustic modes, the other $A_{2u}$ and $E_{u}$ are IR active, and $A_{1g}$ and $E_{g}$ are R active.

For bulk \mx{}, the point group symmetry is enhanced to D$_{6h}$ due to the gain of translational symmetry along the $z$ axis.\cite{Ribei14PRB_Phonon:mode} The lattice vibrations at $\Gamma$ point is:
 $\Gamma$=$A_{1g}+2A_{2u}+2B_{2g}+B_{1u}+E_{1g}+2E_{1u}+2E_{2g}+E_{2u}$,\cite{Verble70PRL,Ataca11JPCCb}
 where one $A_{2u}$ and one $E_{1u}$  are acoustic modes, $A_{1g}$, $E_{1g}$, and $E_{2g}$ are R active, another $A_{2u}$ and $E_{1u}$ are IR active, and $B_{2g}$, $B_{1u}$, and $E_{2u}$ are optically inactive. Here the modes denoted by the letter ``$E$'' are doubly degenerate in the $xy$ plane. For the sake of clarity, Table\,\ref{tbl:phonon_mode} also lists the crystal symmetry, vibration mode, and Raman frequency for the 1L, 2L and bulk \ms{}.

\subsection{Band structure and electrical permittivity}
\begin{figure}[h!bt]
 \centering
 \includegraphics[width=0.5\textwidth]{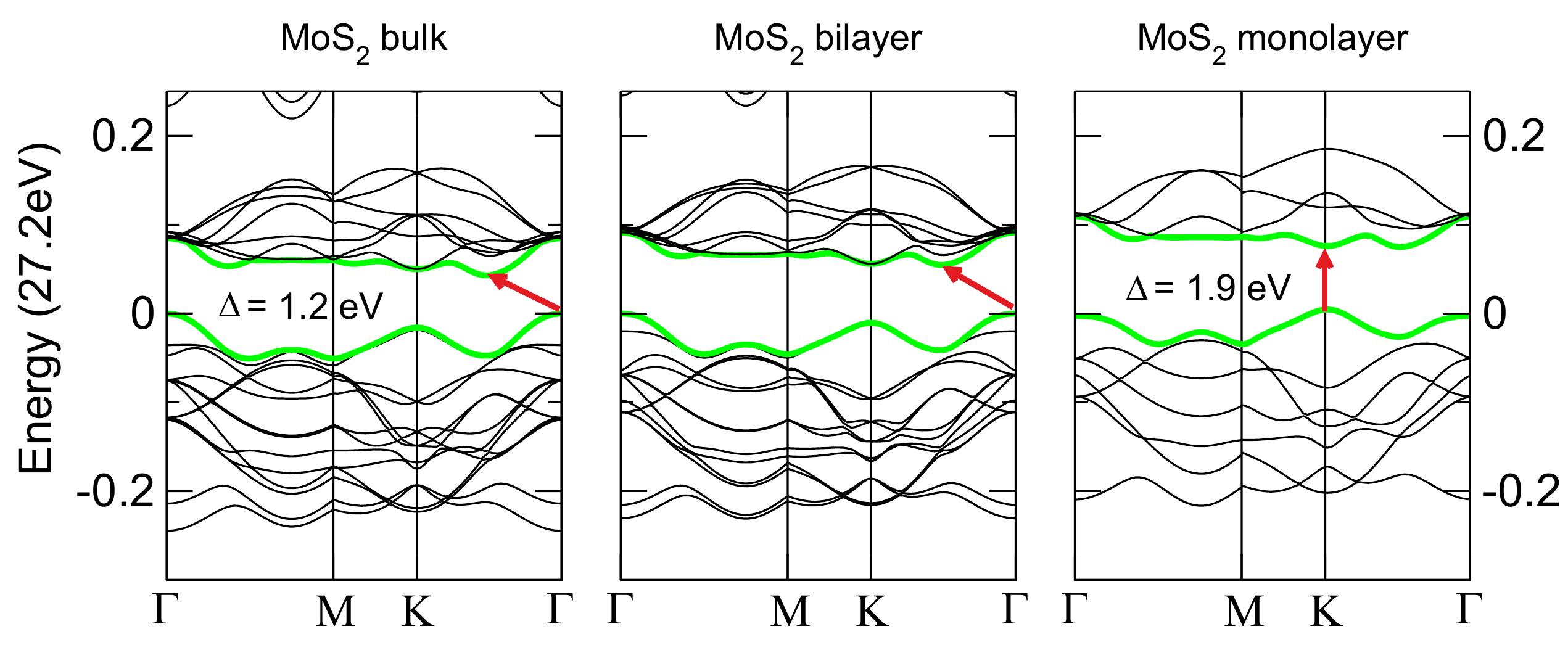}
  \caption{Band structures of bulk, bilayer, and monolayer \ms{}. The top of valence band and bottom of conduction band are highlighted in green. The red arrows indicate the smallest value of the bandgap (direct or indirect) for a given thickness. Adapted with permission from \Ref \citenum{Kuc11PRB}, copyright 2011, American Physical Society.}
 \label{fgr:band_structure}
\end{figure}
When compared to bulk materials, the band structures of 2D materials are considerably modified owing to the quantum confinement effect. The energy-momentum relation (of electrons) and even the positions of band edges can be altered that can lead to a fundamental change of physical properties such as carrier effective mass and dielectric constant, which requires a special attention when comparing the performances of the same material measured at different thicknesses.
\begin{table}
\small
  \caption{\ Calculated the hole and electron effective masses (in unit of electron mass $m_\text{e}$) at the high symmetry points. Reproduced with permission from \Ref \citenum{Yun12PRB}. Copyright 2012, American Physical Society.}
  \label{tbl:effective_mass}
  \begin{tabular*}{0.45\textwidth}{@{\extracolsep{\fill}}lllll}
    \hline
  Type & Symmetry point & Bulk & 2L & 1L \\
  \hline
  hole & $\Gamma$ & 0.711 & 1.168 & 3.524 \\
   &K & 0.625 & 0.628 & 0.637\\
  electron &midpoint of $\Gamma$--K & 0.551 & 0.579 & 0.569 \\
   &K & 0.821 & 0.542 & 0.483 \\
    \hline
  \end{tabular*}
\end{table}

In particular, two aspects are closely related to the charge transport. First, the bandgap magnitude determines the height of the Schottky barrier at the semiconductor/electrode interface. Carrier injection into monolayer is more difficult than into bulk owing to a broader bandgap. Second, the effective mass $m^*$ directly reflects the intrinsic mobility following the equation $\mu \propto 1/{m^*}$. Kuc \et{}\,calculated the band structures for \ms{} with different thickness values from 1L to bulk and systematically revealed the influence of thickness on the position of band edges as well as the size of bandgap.\cite{Kuc11PRB}  Figure\,\ref{fgr:band_structure} shows the band structures of bulk, 2L and 1L \ms{}, with the band edges of the valence and conduction bands indicated by arrows. For 2L and bulk \ms{} the conduction band minimum and the valence band maximum are located at the $\Gamma$ point and a midpoint between $\text{K}$ and $\Gamma$, respectively. Both of them shift to the $\text{K}$ point for the 1L \ms{}. The energy-momentum relations at different values of the momentum are not necessarily the same and thus the shift in the conduction band minimum may change the carrier effective mass and the intrinsic mobility. Table\,\ref{tbl:effective_mass} lists the thickness modulated carrier effective mass values in \ms{} calculated by Yun \et{}\,\cite{Yun12PRB} Evidently, the electron effective mass is reduced from 0.551 to 0.483\,$m_\text{e}$ where $m_\text{e}$ is the electron mass. The slight reduction of carrier effective mass is favorable to achieving high mobility.
\begin{figure}[h!bt]
 \centering
 \includegraphics[width=0.35\textwidth]{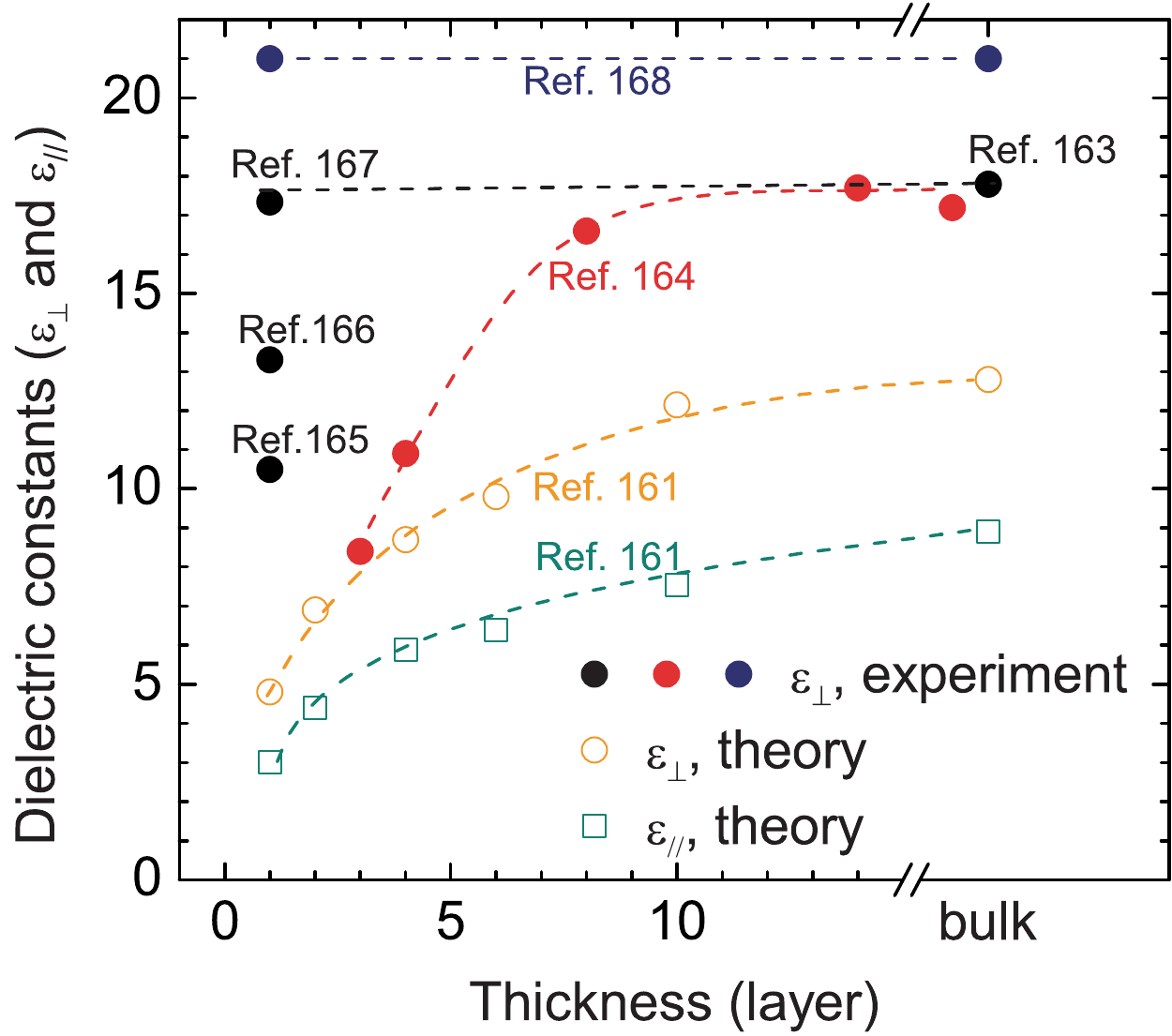}
  \caption{Theoretical and experimental values of dielectric constants for \ms{} at different thicknesses. While most theoretical calculations favor the increasing trend of dielectric constant with increasing thickness, there are controversies in experiments on the trend of dielectric constant with thickness.}
 \label{fgr:dielectric_vs_thickness}
\end{figure}

Also, reducing thickness by exfoliating the top layers changes the electrostatic surroundings of the remaining low lying layers and, consequently, may alter the carrier screening and electrical permittivity (dielectric constant, $\epsilon$), forming a third way to modify charge transport. Figure\,\ref{fgr:dielectric_vs_thickness} summarizes theoretical and experimental values of dielectric constants for \ms{} at different thicknesses. Monotonic thickness dependence of $\epsilon$ can be traced. Using first-principle calculations, Kumar \et{}\,theoretically studied the influence of thickness on the dielectric properties (in-plane $\epsilon_{\parallel}$ and out-of-plane dielectric constant, $\epsilon_{\perp}$) of Mo and W based chalcogenides.\cite{Kumar12PB_Dielectric:MX2} In their calculation, both $\epsilon_{\parallel}$  and $\epsilon_{\perp}$ decrease as thickness reduces. For instance, $\epsilon_{\perp}$ of \ms{} is reduced from 12.8 to 4.8 when thinned from bulk to monolayer. However, it should be noted that even for a specific sample large discrepancies still exist among different theoretical research groups, which results in tremendous variation in adopting the $\epsilon$ value when calculating the field-effect mobility. Taking the 1L \ms{} as an example, Yoon \et{}\,use 3.3 in their non-equilibrium Green's function calculation,\cite{Yoon11NL_} while Li \et{}\,adopt 17.8 following the value of bulk,\cite{Li13NL} and Ma \et{}\,employ 7.6.\cite{Ma14PRX} Therefore, more accurate measurements or techniques that can lead to more reliable permittivity information need to be developed.

The electrical permittivity can be determined experimentally by optical absorption and reflection techniques.\cite{Beal79JPCSSP,Yim14APL_Dielectric:MoS2:thickness,Zhang15SR_Dielectric:MoS2,Mukhe15OME_Dielectric:MoS2_1L,Liu14APL_Dielectric:MX2,Li14PRB_Dielectric:MX2}  Several groups have measured the $\epsilon$ in thick \ms{} with different thicknesses. The data from Yim \et{}\,\cite{Yim14APL_Dielectric:MoS2:thickness} seems to support the reducing trend of $\epsilon$ upon reducing thickness as predicted theoretically by Kumar \et{}, but the magnitudes are generally higher. On the other hand, owing to the influence of surface adsorbate layers on the ultrathin samples (\eg{}\,water and chemical residues on substrates), inconsistent experimental results were reported for the monolayer \ms{}, with the real part of static $\epsilon$ varying from 10.5 to 21.\cite{Zhang15SR_Dielectric:MoS2,Mukhe15OME_Dielectric:MoS2_1L,Liu14APL_Dielectric:MX2,Yim14APL_Dielectric:MoS2:thickness,Li14PRB_Dielectric:MX2} Among them, Li \et{}\,measured a $\epsilon_{\perp}$  value of 17.3 in monolayer \ms{},\cite{Liu14APL_Dielectric:MX2} being quite close to the bulk value of 17.8 reported by Hughes \et{}\cite{Beal79JPCSSP} If these values are reliable, it would imply no variation of the dielectric permittivity with reducing the thickness. This conclusion is further supported by the optical reflectance measurements from Heinze \et{}\,where they observed nearly similar $\epsilon_{\perp}$ between bulk and monolayer in all the visible regime  for four types of \mx{} (\ms{}, MoSe$_2$, WS$_2$, WSe$_2$).\cite{Li14PRB_Dielectric:MX2} As it will be discussed in section\,\ref{sect_factor}, $\epsilon$ determines the polarization function as well as the frequency and the coupling intensity of the surface polar phonon, which is an essential parameter for studying the carrier scattering mechanisms. Reliable information on electrical permittivity is instrumental to gain more accurate understanding on the electronic transport behavior.

\section{Electronic performance at early time (with slight or without mobility engineering)} \label{mu_before}

For 2D chalcogenides, their charge transport behavior is more susceptible to lattice defects and external surroundings (\eg{}\,gaseous adsorbates from air and trapped charges in substrates) due to the fact that surface and bulk coincide. In experiment, a wide distribution of carrier mobility exists as a result of varied sample quality and measurement conditions. Table\,\ref{tbl:mobility_typical} lists typical carrier mobility values of \mx{} chalcogenides reported in recent years. In order to find out the relationship between carrier mobility and extrinsic factors (contact quality, densities of charged impurities and structural defects), the detailed device information, when available, are all reported, including preparation methods (exfoliated or synthesized), channel thickness, contact metals, thermal annealing condition (\textit{in situ} or \textit{ex situ}, gas environment, temperature, and duration), interface surroundings, and measurement environment.

\vspace*{-10pt}
In spite of the mobility variation, some tendencies can still be singled out. First, the quality of the contact plays a crucial role. High mobility is often seen in samples with appropriate annealing and/or work function matching by suitable electrodes. It is worth noting that devices operated by high-capacitive ionic liquid/gel normally exhibit higher mobility than those gated by common oxide dielectrics. This behavior can be attributed to the improved carrier injection at high carrier density. Second, thicker samples (below $\sim$10\,nm) normally show higher mobility, as a result of protection effect of the outer layers to external scattering centers. Third, the electronic characteristics are temperature dependent, indicative of the important role played by lattice phonons and/or other thermally related scattering factors.

\subsection{Thickness dependence}
\begin{figure}[h!bt]
 \centering
\includegraphics[width=0.5\textwidth]{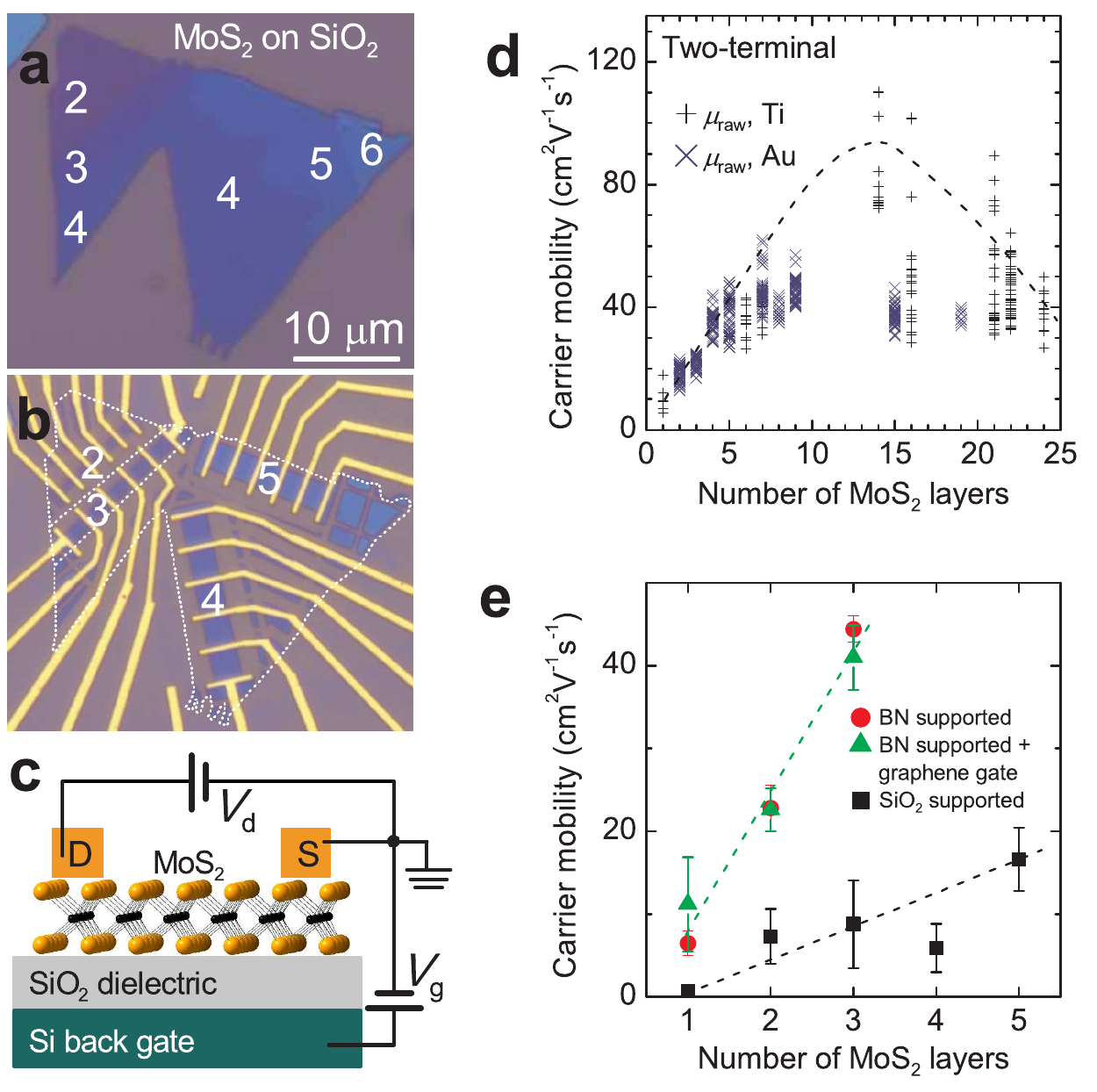}
 \caption{(a) and (b) Optical images for as-transferred \ms{} flakes with consecutive numbers of layers from 2 to 6 and corresponding FETs with bottom SiO$_2$ as gate dielectric. (c) Schematic diagram for the back-gated \ms{} FETs.  (d) and (e) Dependence of carrier mobility on thickness in \ms{} FETs.  Panels (a)--(d) are \cright{Li13NL}. Panel (e) is \cright{Lee13AN_BN:u45_3LMoS_thickness}.}
 \label{fgr:thickness}
\end{figure}
The trace of dependence of field-effect mobility on thickness was observed already in 2005 when the \mx{} monolayers were first prepared by mechanical exfoliation.\cite{Novos05PNAS_} A low mobility range from 0.5 to 3.0\,\cmvs{} was reported for \ms{} and NbSe$_2$, showing about two orders of magnitude lower than the corresponding bulk materials. By collecting more data on a broad thickness range, a deeper understanding of the charge transport mechanism was put forward. Figure\,\ref{fgr:thickness} shows a schematic measurement configuration in a bottom-contact FET and the values of carrier mobility at different semiconductor thicknesses.\cite{Li13NL,Lee13AN_BN:u45_3LMoS_thickness} In scandium (Sc) contacted \ms{} samples, Appenzeller \et{}\,first reported a parabolic-like mobility dependence on thickness, with the highest mobility of 180\,\cmvs{} obtained at around 10 nm.\cite{Das13NLb}  Later, Li \et{}\,focused the attention to the few-layer regime (Figs.\,\ref{fgr:thickness}a--\ref{fgr:thickness}b) and reported mobility varying from 10 to 50\,\cmvs{} in Au or Ti contacted samples while the thickness changed from 1 to 5 layers (Fig.\,\ref{fgr:thickness}d).\cite{Li13NL}  Similar monotonic decrease of mobility with reducing thickness was also confirmed by Hone \et{}\,on samples supported by SiO$_2$ and hexagonal boron nitride (hBN) dielectrics (Fig.\ref{fgr:thickness}e).\cite{Lee13AN_BN:u45_3LMoS_thickness} The origin of the thickness dependence is rather complex. As will be discussed in section\,\ref{sect_CI}, one of the main reasons is the interfacial Coulomb impurity,\cite{Li13NL} which is also an extrinsic scattering factor to be suppressed for achieving high channel mobility.

\subsection{Temperature dependence}
Alongside the thickness dependence, temperature ($T$) is a powerful parameter whose tuning allows for exploring the intrinsic carrier scattering mechanisms due to lattice phonons, because the phonon number depends highly on temperature. In contrast, other scattering mechanisms such as Coulomb impurity and atomic vacancy possess a moderate or weak dependence on temperature where temperature mainly embodies its effect through tuning the carrier screening ability and the carrier distribution near the Fermi level.

Several theoretical studies have been carried out to investigate the intrinsically phonon-dominated mobility dependence on temperature, which predict a power-law relation between mobility and temperature $\mu\propto T^{\gamma}$ with $\gamma$ being a parameter dependent on the phonon type. In high-quality bulk, Fivaz and Mooser predicted $\gamma$ values of $\sim$ 2.6, $\sim$ 1.6, and $1$ for homopolar optical, polar optical and acoustic phonons, respectively.\cite{Fivaz67PR} They determined that the homopolar optical phonons are the primary scattering centers in most \mx{} chalcogenide crystals. For \ms{} monolayer, in contrast, Kaasbjerg \et{}\,showed that the scattering around room temperature is co-dominated by the deformation potential of optical phonons (see LO$_2$/TO$_2$ and LO$_1$/TO$_1$ modes in Fig.\,\ref{fgr:phonon_mode}a) and the Fr\"{o}hlich interaction (polar optical phonons, see LO$_2$/TO$_2$ modes in Fig.\,\ref{fgr:phonon_mode}a), which gives rise to $\gamma \sim 1.69$.\cite{Kaasbjerg12PRB} The acoustic phonons ($\gamma=1$) become leading at temperatures lower than 100\,K.

\begin{figure}[h!bt]
 \centering
\includegraphics[width=0.5\textwidth]{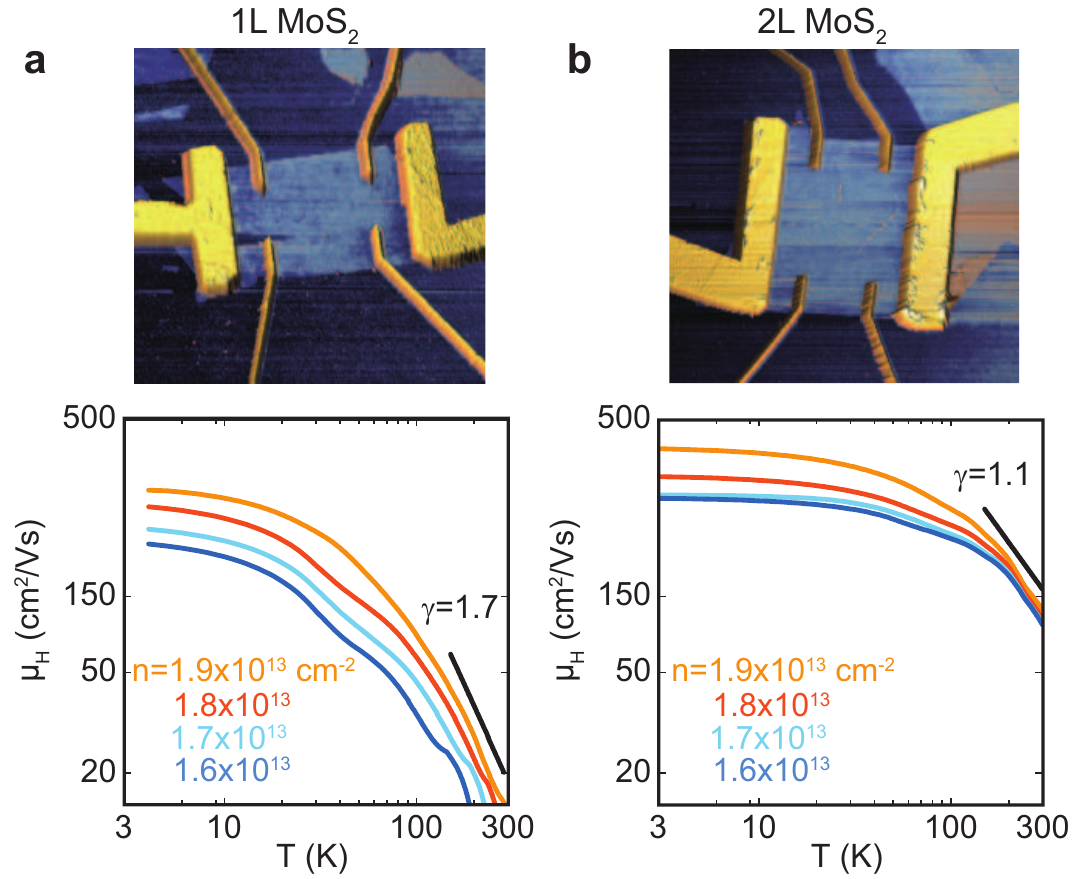}
 \caption{Dependence of Hall mobility ($\mu_\text{H}$) on temperature for (a) monolayer and (b) bilayer \ms{} FETs after long-time thermal annealing at different carrier densities. Top panels are the color-enhanced AFM height images for the corresponding devices. Bottom panels are the mobility data and the power-law temperature fitting near room temperature. Reproduced with permission from \Ref\citenum{Baugher13NL}, copyright 2013, American Chemical Society.}
 \label{fgr:temperature}
\end{figure}
Experimentally, however, there have been no reports showing 2D vdW samples that can reach the intrinsic phonon-limited transport regime and exhibit the predicted $\gamma$ values. As can been seen in Table\,\ref{tbl:mobility_typical}, the measured $\gamma$ values range broadly from 0.56 to 2.85 near room temperature, implying the existence of other scattering mechanisms that deviate the anticipated $\gamma$ values from pure phonon scattering. In high-quality 1L \ms{} samples with mobility of 60--70\,\cmvs{}, Hersam \et{}\,extracted a low $\gamma$ value $\sim$ 0.62,\cite{Jariwala13APL} being much smaller than 1.69. They attributed the deviation to the presence of remote phonons from underlying oxide substrates and the effect of contact resistance. A close result ($\gamma = 0.72$) was observed by Wang \et{}\,in their high-quality vacancy healed 1L \ms{} samples.\cite{Yu14NC_MoS2_mu80} In another study on superclean hBN encapsulated \ms{}, Hone \et{}\,reported high $\gamma$ values of 1.9 for 1L and 2.5 for 2L samples.\cite{Cui15NN_FET:G/MoS/BN_mu34k}
All the experimental data indicate that at room temperature the charge transport in 2D vdW semiconductors is not dominated by lattice phonons.

This conclusion is further corroborated by the absolute magnitude of mobility. The theoretically predicted room-temperature mobility limited by phonons amounts to 410\,\cmvs{} for 1L \ms{}. Figure\,\ref{fgr:temperature} shows the Hall mobility versus temperature for 1L ($\gamma = 1.7$) and 2L ($\gamma = 1.1$) \ms{} measured by Jarillo-Herrero \et{} Although they observed $\gamma= 1.7 $ in long-time \textit{in situ} annealed 1L \ms{},\cite{Baugher13NL} appearing to match the theoretical prediction, the absolute value of mobility is only 20\,\cmvs{}, \ie{}\,it is much lower than the predicted value. Hitherto, this high theoretical value has never been reached experimentally (See Table\,\ref{tbl:mobility_typical}), indicating that there is still large room for mobility improvement if the extrinsic scattering centers can be effectively suppressed or minimized.

\subsection{Dependence of electronic phase on carrier density}
\begin{figure*}[h!bt]
 \centering
\includegraphics[width=0.9\textwidth]{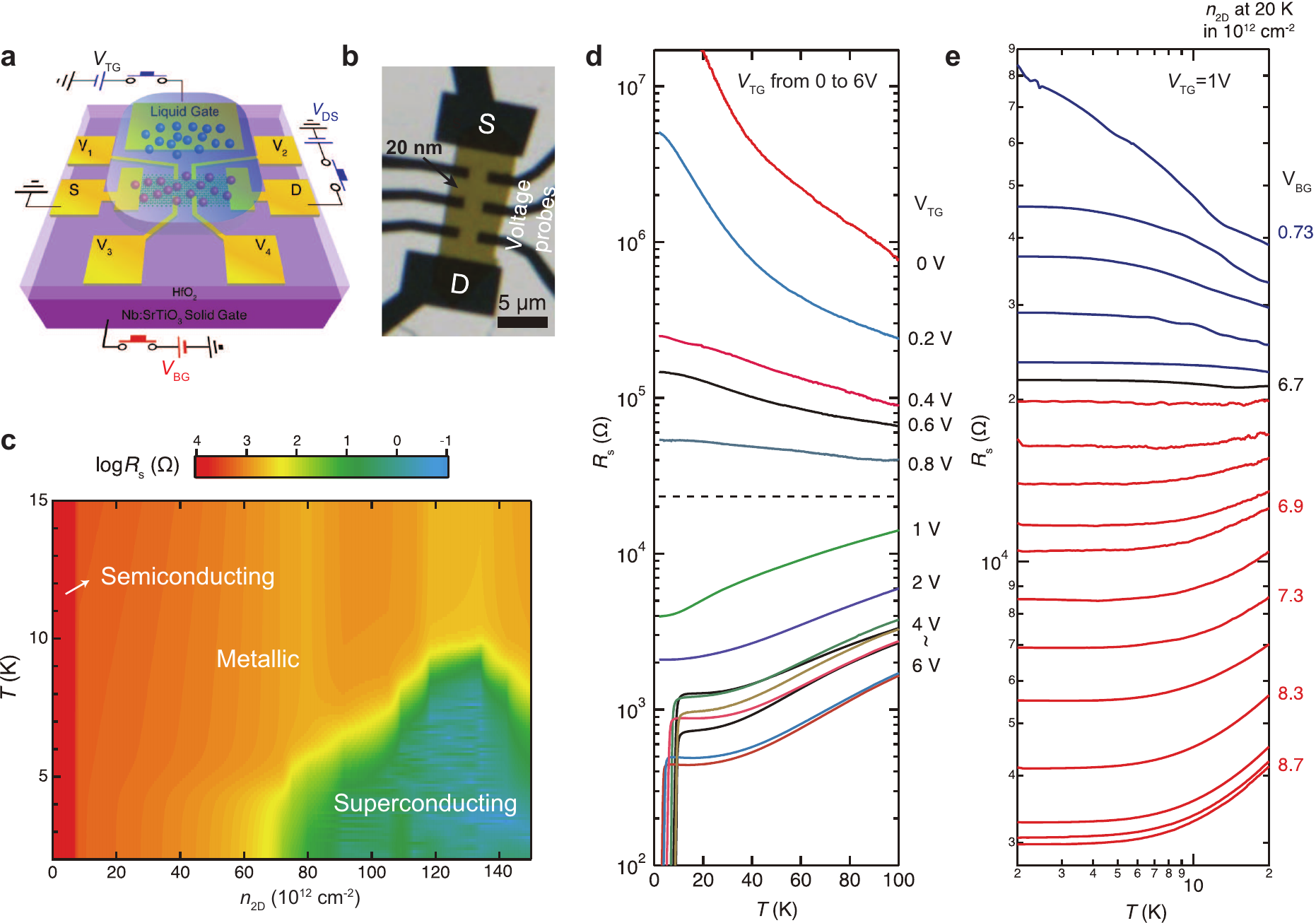}
 \caption{ (a) Schematic and (b) optical images for an ionic liquid gated (TG) \ms{} FET. (c) Phase diagram showing the evolution of electronic phases as a function of carrier density $n_\text{2D}$. (d) Temperature
dependence of the channel sheet resistance $R_\text{s}$ at different $V_\text{TG}$ gate biases
ranging from 0 to 6 V (indicated on the right). (e) Temperature dependence
of $R_\text{s}$ at $V_\text{TG} =$ 1 V and different $V_\text{BG}$ showing a
metal-insulator transition at $n_\text{2D}$ = \E[6.7]{12}\,cm$^{-2}$. For each $V_\text{BG}$, the corresponding $n_\text{2D}$ is determined by Hall measurement at 20 K. \Cright[\aaas]{Ye12S_IonicLiquid:SC}}
 \label{fgr:mit}
\end{figure*}

The layered \mx{} chalcogenides exhibits a rich phase diagram depending on carrier density ($n_\text{2D}$) as a result of complicated electron-electron interaction. Figure\,\ref{fgr:mit} shows the results on carrier density by Iwasa \et{}\,who first investigated the electronic behavior of 20 nm thick \ms{} over a wide range of $n_\text{2D}$ through ionic liquid gating. With this, an extremely high $n_\text{2D}$ value up to $10^{14}$\,cm$^{-2}$ is reached such that the superconducting phase can be accessed. As shown in the phase diagram (Fig.\,\ref{fgr:mit}b), the thick \ms{} flake exhibits a semiconducting (insulating) phase as $n_\text{2D}<$\,\E[6.7]{12}\,cm$^{-2}$, a metallic phase from \E[6.7]{12} to \E[6.8]{13}\,cm$^{-2}$, and a domelike superconducting phase as $n_\text{2D}>$\,\E[6.8]{13}\,cm$^{-2}$. The critical density for the insulator-metal transition $n_\text{2D}$\,=\,\E[6.7]{12}\,cm$^{-2}$ corresponds to a sheet resistance $R_\text{s} = 21.7$\,k$\Omega$ per square (Fig.\,\ref{fgr:mit}d), which is close to the quantum resistance $h/e^2$ and is consistent with metal-insulator transition found in other 2D systems.

The metal-insulator transition was soon confirmed in the monolayer \ms{} samples.\cite{Baugher13NL,Radisavljevic13NM,Schmidt14NL,Chu14SR_FET:MoS2:IonicGated,Yu14NC_MoS2_mu80}  However, the superconducting phase is absent,\cite{Chu14SR_FET:MoS2:IonicGated,Schmidt14NL} indicating the detrimental role played by the interfacial impurities that are strong enough to destroy the electrostatic surrounding required to form the superconducting cooper pairs.

The semiconducting regime with medium $n_\text{2D}$ is critical for FET applications. In this regime, the variation of carrier density has two opposite effects to the channel mobility. On the one hand, high $n_\text{2D}$ is beneficial for screening the interfacial impurity potential that increases mobility. On the other hand, high $n_\text{2D}$ also increases the carrier energy such that they are interacted with high-energy scattering centers, which may reduce mobility.\cite{Li13NL} Hence, there is normally an optimized carrier density for achieving the best carrier mobility.

In very low $n_\text{2D}$ regime, some groups reported a hopping-like transport behavior,\cite{Ghatak11AN,Jariwala13APL,Qiu13NC} which was interpreted that most carriers are filled into the band tails where carriers are highly localized. Another possible explanation to this behavior is the presence of large contribution of the contact resistance, which becomes increasingly important at low temperature and produces the artificial behavior of $\text{ln}{\sigma}\sim T^{-1/3}$ since most reports of hopping transport behavior are observed in two-terminal devices.\cite{Ghatak11AN,Jariwala13APL,Yu14NC_MoS2_mu80} No trace of hopping behavior was seen in the devices with a four-terminal measurement where the contact contribution is eliminated.\cite{Radisavljevic13NM} Instead, only thermal activation behavior was observed in this regime.

\section{Factors related to electronic transport} \label{sect_factor}

The carrier mobility of bulk \mx{} chalcogenides can reach $\sim$200\,\cmvs{} at room temperature.\cite{Fivaz67PR} Last section showed that most 2D \mx{} chalcogenides exhibit reduced mobility in comparison with their bulk phase. To develop technologically viable 2D semiconductors, especially for the atomically-thin-body FETs, it is highly desired to unravel the origin of such adverse thickness dependence.

It is well known that silicon shows thickness dependence below $\sim$4 nm with a power-law thickness scaling behavior ($\mu\sim t^{-6}$),\cite{Gomez07IEDL,Gold87PRB} as a result of the inevitable compositional transition from SiO$_2$ dielectric to Si channel, \ie{}\,the issue of surface roughness (SR). However, this factor can be confidently ruled out in case of 2D vdW semiconductors because of the atomically well-defined interlayer interfaces. Studies indicate that there are two aspects responsible for the thickness dependence in the 2D vdW semiconductor:\cite{Li15JPSJ} 1) carrier injection at the electrode/channel contacts, and 2) carrier scattering mechanism within the conduction channels.

\subsection{Electrode/semiconductor contacts}
\begin{figure}[h!bt]
 \centering
 \includegraphics[width=0.48\textwidth]{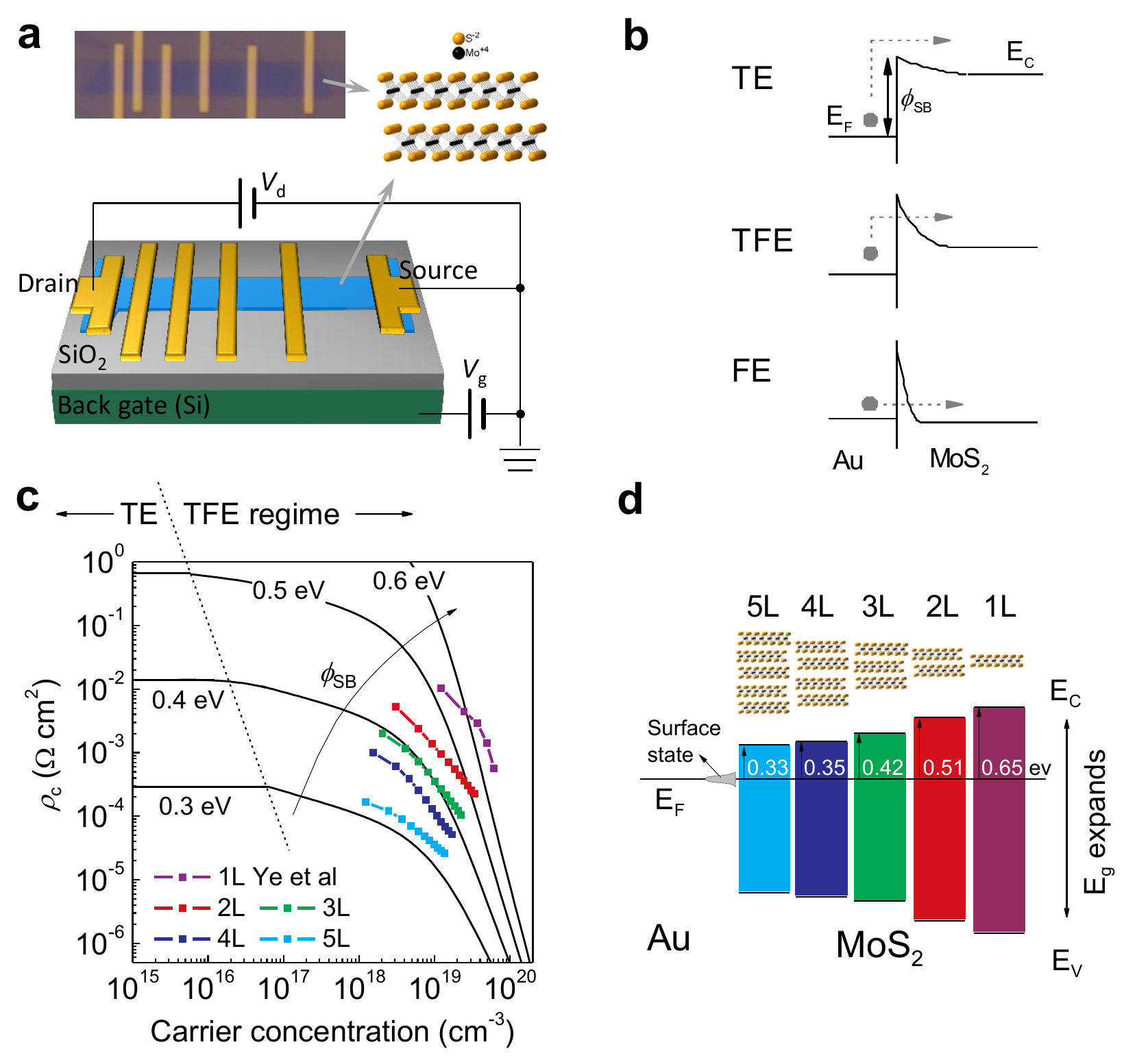}
  \caption{(a) Schematic diagram and real optical image for the geometry of transfer line measurement. The inset shows the atomic structure of \ms{}.(b) Schematic band alignments between Au electrode and channel \ms{} and the evolution of the three carrier injection mechanisms from low to extremely high carrier densities: thermal emission (TE), thermal field emission (TFE), and field emission (FE). The difference among the three injection mechanisms lies in the width of interfacial barrier which changes with the carrier density in channels. (c) Comparison of the room temperature contact resistivity (solid dots) with theoretical calculation of TE and TFE conduction mechanisms to extract injection barrier height (${\phi}_{\text{SB}}$). (d) Evolution of energy level alignment at the Au/\ms{} interfaces as \ms{} thickness reduces from 5 to 1 layer. Adapted with permission from \Ref\citenum{Li14AN_MoS2:Contact}, copyright 2014, American Chemical Society.}
 \label{fgr:barrier}
\end{figure}
Contact issue is ubiquitous at all metal/semiconductor interfaces due to the formation of interfacial Schottky barriers which blocks efficient carrier transfer. Figure\,\ref{fgr:barrier} shows related studies on the contact resistance for \ms{}. Strictly speaking, the contact does not affect the intrinsic carrier mobility in semiconductors; this unfavorable effect can be ruled out by employing a transfer-line method or four-terminal measurement (Fig.\,\ref{fgr:barrier}a). However, the practical FETs collect current with two electrodes (source and drain). In most cases, the field-effect mobility $\mu_{\mathrm{FE}}$, defined as the derivative of drain current to gate bias, is used to evaluate the channel performance. Normally the contact resistance does not depend on gate bias as strongly as channel resistance, the presence of contact resistance thus leads to underestimation of semiconductor performance when adopting $\mu_{\mathrm{FE}}$ as the criterion.

\subsubsection{Schottky barrier and Fermi pinning}
A Schottky barrier (${{\phi }_{\text{SB}}}$) often forms at the metal/semiconductor interfaces due to the difference of chemical potentials and mismatch of energy levels of the two contacting components. This understanding leads to an empirical rule by reducing the barrier through matching the energy levels of contacting materials. In solid-state physics, low work function metals are expected to inject effectively electrons into n-type semiconductors while metals which possess a high work function are normally employed for p-type semiconductors. Note that the barrier heights for electron and hole injection are normally different, depending on the exact alignment of the energy levels. The carrier conduction type (electron: n-type or hole: p-type) in semiconductor and the practical magnitude of interfacial barrier are determined by the smaller barrier.

However, the actual injection barrier height is also governed by the effect of Fermi level pinning due to the presence of interface states of semiconductors. The pinning effect is more pronounced when the channels becomes atomically thick because the density of interface states enhances considerably. The barrier height $\phi_\text{SB}$ is proportional to the potential difference of the energy levels $\Delta u$, expressed as $\phi_\text{SB}=\beta\Delta u$ with $\beta$ a coefficient between 0 and 1, representing the strong and weak pinning limits, respectively.\cite{Sze07_textbook}

It is widely accepted that the electrode/semiconductor contacts play a crucial role in the overall device performance. In early time, the variation of barrier width upon applying gate bias has ever been suggested as the current switching mechanism in FETs with nanostructured channels (\eg{}\,carbon nanotube FETs).\cite{Heinze02PRL}  Transistors operated under this switching mechanism are termed as `Schottky barrier transistors'. At low carrier density the injection is dominated by a thermal emission process, while the injection becomes thermal-field emission (thermally assisted tunneling) or even field-emission (direct tunneling) at high carrier density, as shown in Fig.\,\ref{fgr:barrier}b.

Such a switching mechanism has also been proposed on monolayer \ms{} FETs where a faster variation is observed in contact resistance than channel resistance in the sub-linear conduction regime.\cite{Liu14AN} It is found later, however, that the influence of contact resistance also depends highly on channel thickness and channel length. The contact may not dominate in FETs with a long channel length. In another report on mechanically exfoliated bilayer and hexalayer \ms{} samples, Chen \et{}\,reported that the contact resistance comprises only 5--20\% of the total channel resistance,\cite{Guo14AN} indicating that the heights of contact barriers are smaller in thick flakes.

Furthermore, for the few-layer thick 2D semiconductors, the contact barrier height can be modified by the quantum confinement effect through the change of semiconductor bandgap.\cite{chen05nl} To address this issue, Li \et{}\,performed a systematic thickness scaling study on the Au/2D \ms{} contacts using the transfer line method to extract the area normalized contact resistivity ($\rho_c$) for each \ms{} layer.\cite{Li14AN_MoS2:Contact} For \ms{} thinner than 5 layers, the contact resistivity sharply increases with reducing \ms{} thickness, as a consequence of bandgap expansion (Fig.\,\ref{fgr:barrier}c). Figure\,\ref{fgr:barrier}d plots a full evolution diagram of energy level alignment to elucidate the thickness scaling effect. The interfacial potential barrier is varied from 0.3 to 0.6 eV with merely reducing \ms{} thickness. The thickness-dependent barrier height for charge injection is one of the reasons responsible for the field-effect mobility degradation in the ultrathin flakes. Hence, optimizing the contact quality is crucial for improving the mobility of the two-terminal FETs with 2D semiconductor channels.

\subsubsection{Current crowding effect} \label{crowding}
\begin{figure}[!bth]
 \centering
 \includegraphics[width=0.48\textwidth]{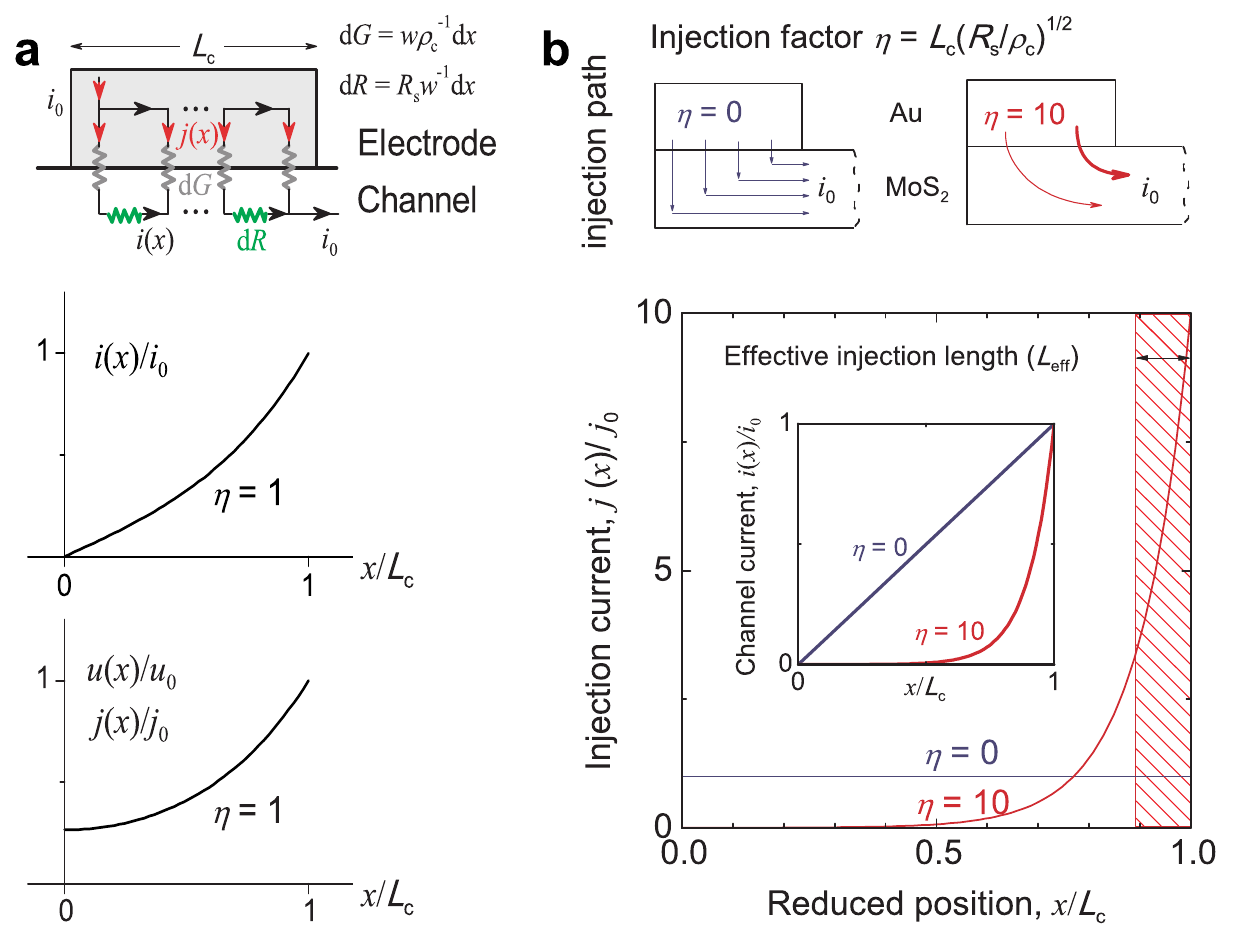}
  \caption{(a) Resistor network model for analysing the distributions of injection current along a probe/channel stack. Top: Current injection path and distribution of impedance elements at the probe/channel interface (modeled as $\text{d}G$) and in the channel (modeled as $\text{d}R$). Bottom: Plots for lateral channel current $i(x)$, vertical interfacial potential $u(x)$ and injection current density $j(x)$ at $\eta=1$. (b) Comparison for two extreme cases of $\eta=0$ and $10$. Reproduced with permission from \Ref\citenum{Li14AN_MoS2:Contact}, copyright 2014, American Chemical Society.}
 \label{fgr:crowding}
\end{figure}
At electrode/semiconductor contacts, current is not injected uniformly along the entire contact length ($L_\text{c}$) but rather concentrates at the two near-end edges of the two injection contacts, which is called `current crowding effect'. It is important to understand this effect if one needs to compromise the contact length and area occupancy of electrodes in a limited chip area during device design. We discuss it here as basic knowledge aiming to emphasize some experimental standards for four-terminal measurements, as will be seen in section\,\ref{trap:4w}.

For 2D channels, the vertical current injection path at the electrode/channel interface and the lateral current distribution in the channels can be analytically solved via a resistor network model,\cite{Murrm69ITED_Rc:CurrentCrowding} in which the electrode/channel stack is divided into infinite resistor and conductor elements. Figure\,\ref{fgr:crowding}a shows the schematic distribution of the impedance elements at the contact interface $\text{d}G={\rho_\text{c}^{-1}}w\text{d}x$ and in the channel $\text{d}R={R_\text{s}}{{w}^{-1}}\text{d}x$, where ${{R}_\text{s}}$, $\rho_\text{c}$, $w$, and $x$ denote the channel square resistivity, area contact resistivity, channel width, and channel coordinate, respectively. It has been derived that the lateral channel current $i(x)$, vertical interface potential $u(x)$, and vertical injection density $j(x)$ satisfy the relations below\cite{Murrm69ITED_Rc:CurrentCrowding}
\begin{equation}
 i(x) ={i_0}\frac{\text{sinh}(\eta x/ L_\text{c})}{\text{sinh}(\eta)}
\end{equation}
\begin{equation}
u(x)={i_0}\frac{\sqrt{R_\text{s} \rho_\text{c}}}{w}\frac{\text{cosh}(\eta  /L_\text{c} )}{\text{sinh}(\eta)}
\end{equation}
\begin{equation}
  j(x)=u(x)/\rho_\text{c}
\end{equation}
where $\eta  = L_\text{c}/L_\text{T}$ is the injection factor, and $L_\text{T}=\sqrt {{\rho_\text{c}}/{R_\text{s}}}$ is the transfer length, which is a characteristic length for current injection phenomena. In particular, $R_\text{c}$, $R_\text{s}$, and $\rho_\text{c}$ are linked by
\begin{equation}
{{R}_\text{c}}w =\sqrt{{{R}_\text{s}}{{\rho }_\text{c}}}\text{ coth}\eta .
\end{equation}
According to Eqs. (1)--(3), the carrier injection at contact is governed by the $\eta$ factor which is a function of $L_\text{c}$, $\rho_\text{c}$, and ${R_\text{s}}$. Figure\,\ref{fgr:crowding}a depicts an illustrative distribution of $i(x)$, $u(x)$, and  $j(x)$ at $\eta=1$. Apparently, the injection is rather asymmetric along the $x$ axis. To deepen our understanding, Fig.\,\ref{fgr:crowding}b plots two extreme cases for $\eta=0$ and $10$. As $\eta=0$ (\eg{}\,in presence of a superconducting channel with $R_\text{s}=0$ or a bad contact with $\rho_\text{c}\rightarrow\infty$), the current is uniformly injected along the entire channel. At $\eta=10$ (\eg{}\,in case of a wide electrode with large $L_\text{c}$ or small $\rho_\text{c}$), the crowding behavior is aggravated and half of the current is injected from $\sim$10\% portion from the side. For typical \ms{} devices $\eta$ is in the range of 2--5 depending on gate bias.

In device physics, one can use the transfer length ${{L}_{\text{T}}}=\sqrt{{{\rho }_{\text{c}}}/{{R}_{\text{s}}}}$ to estimate the minimum electrode length (few times of $L_\text{T}$) that enables efficient current injection. For Ti contacted 1L \ms{}, Ye \et{}\,estimated $L_\text{T}$ is 1.26 $\mu$m at 0 V gate bias and drops to 0.63 $\mu$m at high gate biases.\cite{Liu14AN} They suggested that the contact length should be at least 1 $\mu$m (1.5 $L_\text{T}$) to guarantee good contact at device on state. In contrast to the shrinking tendency with increasing gate bias, Chen \et{}\,observed an $L_\text{T}$ behavior increasing with elevating gate bias in their few-layer samples.\cite{Guo14AN} It increases from 20 to 80 nm for Ti/2L \ms{} contact, from 50 to 180 nm for Ti/6L \ms{}, and from 30 to 200 nm for Au/6L \ms{}. Independently, Li \et{}\,extracted $L_\text{T}$ from 200--400 nm for thermally annealed Au contacted few-layer \ms{} (2--9 L).\cite{Li14AN_MoS2:Contact} It seems that the contact length depends highly on the semiconductor thickness and whether the samples are undergone annealing treatment. For most appropriate annealed devices, contact length would not be a limit to \ms{} performance since most devices employ electrode longer than 500 nm (limited by lithographic resolution).

\subsection{Carrier scattering mechanisms} \label{sect_mechanism}
\begin{figure*}[h!bt]
 \centering
 \includegraphics[width=0.9\textwidth]{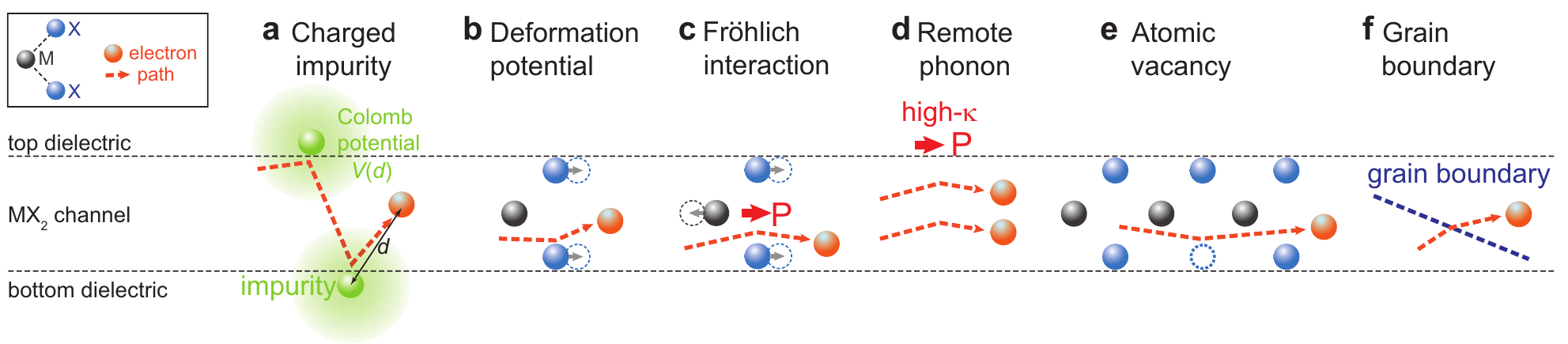}
  \caption{Schematic diagrams of leading carrier scattering mechanisms in 2D \mx{} channels. The black and blue balls denote the M and X atoms, respectively. The orange balls and corresponding arrows denote the electrons and their paths in the channels.  (a) Charged interfacial impurities at both the top and bottom channel surfaces. The green balls and the outer surroundings denote the impurities and corresponding scattering potential $V(d)$. (b) Deformation potential of phonons due to atom vibration. (c) Fr\"{o}hlich interaction due to polar optical phonons. The red arrow denotes the oscillating polar electric field induced by the polar optical phonon. (d) Remote interface phonon from dielectric. The red arrow denotes the polar phonon in the top dielectric. (e) Atomic vacancies (dashed circle) which tend to form in both natural and synthetic chalcogenides. (f) Grain boundaries (dashed line) which are typically present in synthetic chalcogenides.}
 \label{fgr:mechanism}
\end{figure*}
In this subsection, we give a fundamental introduction to the main scattering mechanisms in semiconductors, including Coulomb impurities, lattice phonons and remote interfacial phonons. Figure\,\ref{fgr:mechanism} schematically depicts the carrier scattering processes where the orange balls and dashed arrows represent the carriers and their transport paths, respectively. The change in the direction of carrier path results in a scattering event. The common method to calculate the scattering rates for each mechanism is to solve the Boltzmann transport equation within the relaxation-time approximation. To this end, one has to derive the scattering matrix elements $M_{2D}$, electron polarization function for the 2D channels ${{\varepsilon }_{2D}}$, and configurative form factor $\Phi(q, t)$ according to the device configuration. Here, we will summarize the main theoretical results and show the derived calculating formulae without dealing with the derivation. For the detailed theoretical analyses, we recommend the readers refer to related literature.\cite{Li06_transport:introductory,ridley1997electrons,Lundstrom00,Ando82RMP}

\subsubsection{Interfacial impurities} \label{sect_CI}
\begin{figure}[h!bt]
 \centering
 \includegraphics[width=0.4\textwidth]{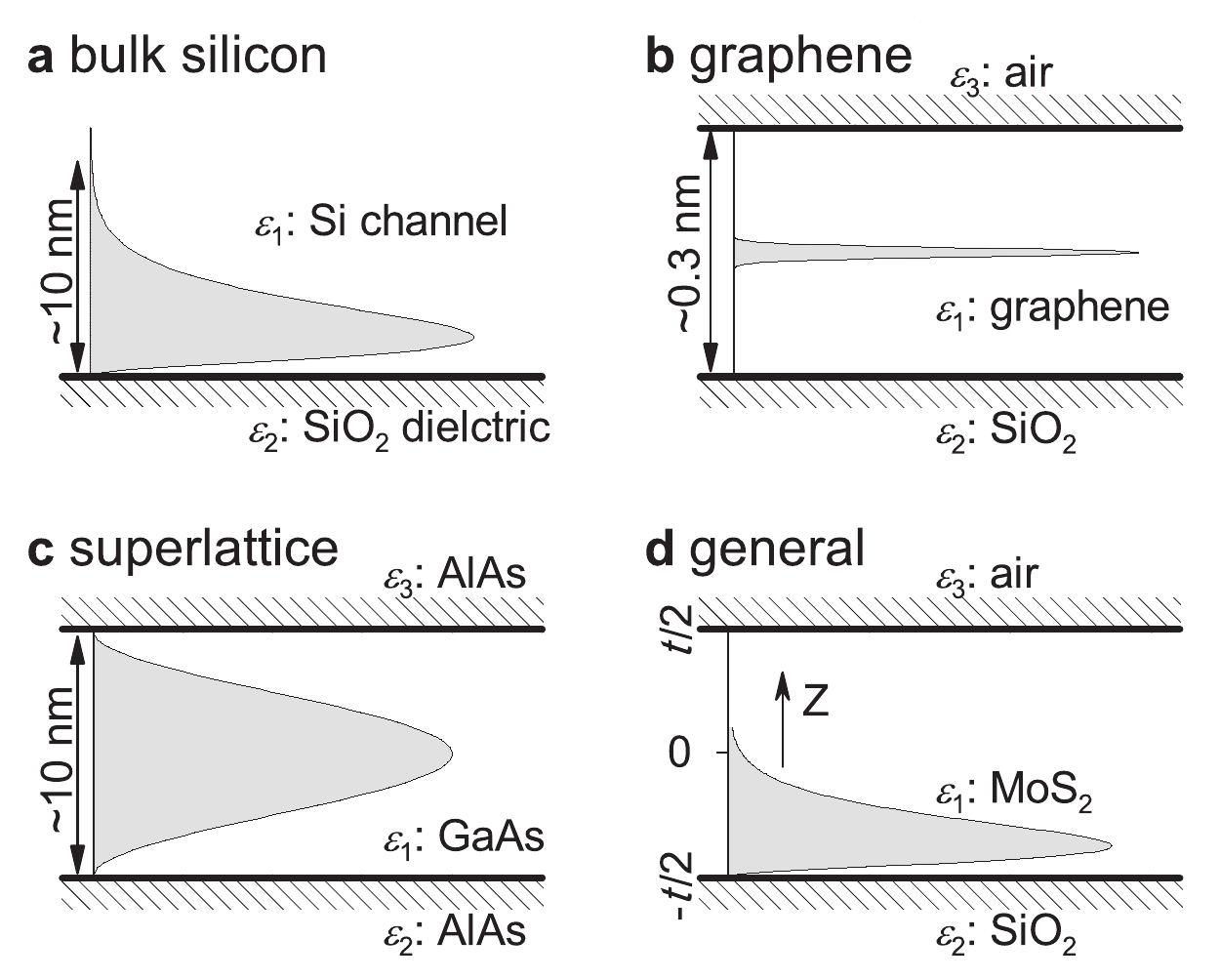}
  \caption{Schematic diagrams of dielectric environments and carrier distributions for different FET configurations. (a) Bulk silicon: one boundary which produces only one image charge.\cite{Ando82RMP} (b) Graphene: negligible thickness in the middle channel ($t\sim0.3$\,nm) that enables the approximation of a simple pulse-like carrier distribution.\cite{Ando06JPSJ,DasSarma11RMP} (c) Superlattice: symmetric dielectrics and trigonometric carrier wavefunction.\cite{Gold87PRB,jena07prl} (d) A general FET with complicated device configuration: 1) two channel boundaries which can produce infinite image charges when considering charge-charge interaction; 2) a lopsided carrier distribution which leads to complicated configurative form factors  $\Phi(q, t)$ in scattering matrix elements $M _{2D}$ and  electron polarization function $\epsilon _{2D}$. \Cright{Li13NL}.}
 \label{fgr:modelDiagram}
\end{figure}
\begin{figure}[h!bt]
 \centering
 \includegraphics[width=0.48\textwidth]{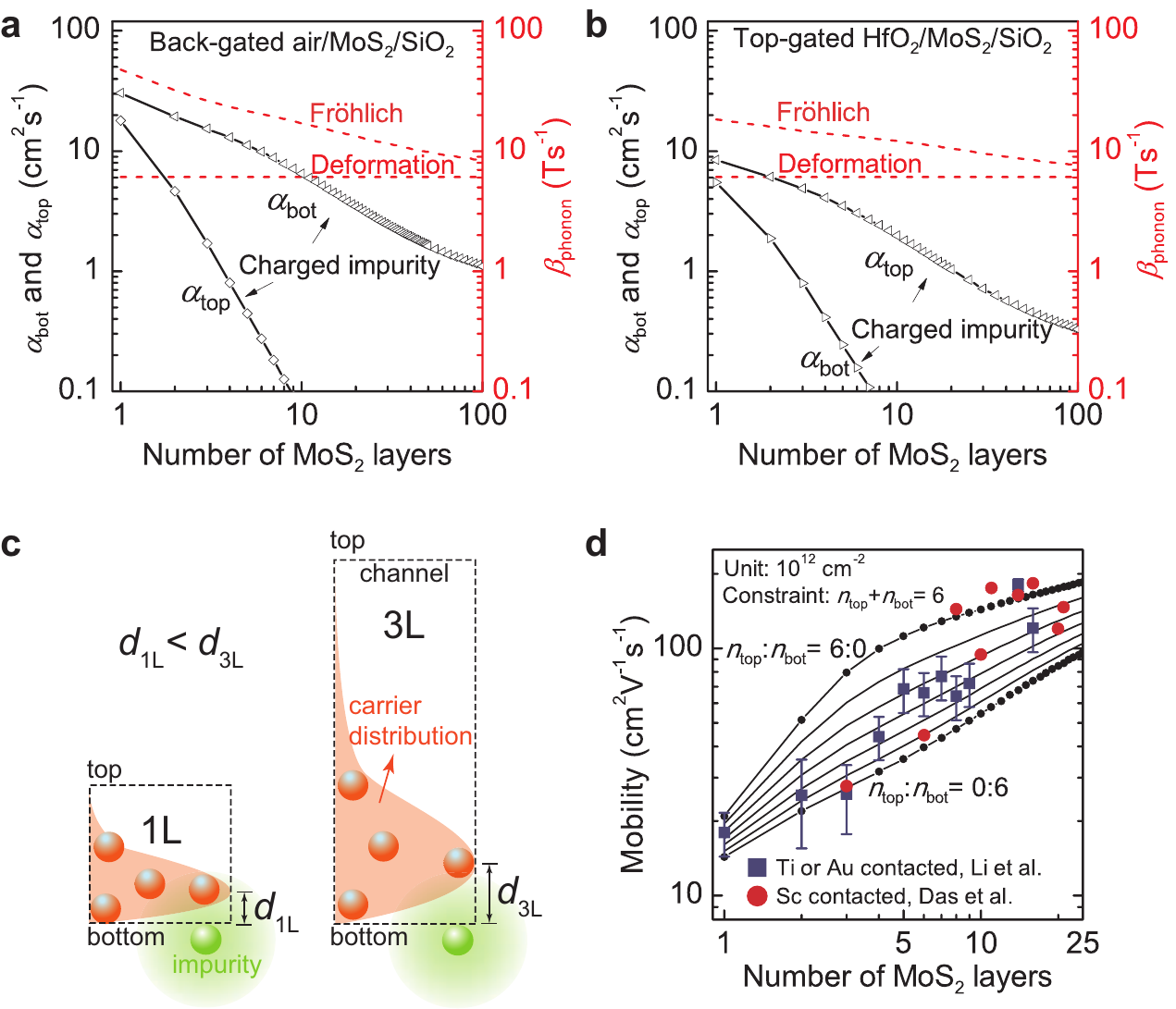}
  \caption{ (a) and (b) Calculated Coulomb impurity scattering coefficients at the top and bottom channel surfaces ($\alpha _\text{top}$ and $\alpha _\text{bot}$) for different channel thickness. For comparison, the scattering rates of phonon deformation and Fr\"{o}hlich interaction ($\beta_\text{phonon}$) are also plotted on the right longitudinal axis in red. (a) is for back-gated air/\ms{}/SiO$_2$ and (b) for top-gated HfO$_2$/\ms{}/SiO$_2$ FETs. (c) Schematic diagram for the origin of dependence of impurity scattering intensity on channel thickness. The underlying nature is the variation of interaction distance ($d_\text{NL}$) between the impurities and the channel carriers as channel thickness changes. For instance, for the 1L and 3L \ms{} samples $d_\text{3L}>d_\text{1L}$ and hence the scattering rate is lower in the 3L sample (\ie{}, carrier mobility is higher). (d) Comparison between the experiment and calculation based on impurity and phonon scattering. \Cright{Li13NL}}
 \label{fgr:s3_impurity}
\end{figure}
In 2D systems, carriers are within a confined space and the Coulomb/charged impurities (CIs) are randomly distributed at the channel/dielectric interface. Each point impurity imposes its scattering potential $V(d)$ to channel carriers via the long-range Coulomb interaction. Figure\,\ref{fgr:mechanism}a schematically shows two point impurity located at the top and bottom channel surfaces and the evanescent scattering potential around them. The interaction distance between a point impurity and a conduction electron is $d$. With the advent of silicon FETs,\cite{Ando82RMP} superlattice,\cite{Gold87PRB,jena07prl} and graphene,\cite{Ando06JPSJ,DasSarma11RMP} it is proved that the randomly distributed Coulomb impurities are one type of leading scattering mechanisms. In comparison with silicon FETs, where impurities come mainly from the residual metals ions and unsaturated silicon bonds at the channel/dielectric interfaces, the sources of the Coulomb impurities in 2D semiconductors are much richer. Additional impurities such as gaseous adsorbates and chemical residues can also be introduced during device fabrication. The  `dirty' surface condition is one of the main reasons for the low mobility in 2D vdW crystals.

While the theoretical calculations on various symmetric systems (\ie{}\,superlattices, graphene, and silicon FETs) have been performed, the calculation on a generalized asymmetric system remains invalid. As mentioned, the challenging parts are the derivations of scattering matrix element $M _{2D}$, configurative form factor $\Phi(q, t)$, and carrier polarization function ${{\varepsilon }_{2D}}$ according to the exact device configurative parameters (\ie{}, carrier distribution, symmetry of dielectric surroundings, and channel thickness). In most symmetric systems, the three terms can be analytically expressed,\cite{Ando82RMP,Gold87PRB,jena07prl,Ando06JPSJ,DasSarma11RMP} while they are only numerically solvable for complex systems.\cite{Kim12NC}

Figure\,\ref{fgr:modelDiagram} depicts the difference for four device configurations. Here we use $\epsilon_\text{i}$ (i= 1, 2, 3) to denote the dielectric constants for different layers of channel and dielectrics. For superlattice with a periodic structure of two semiconductor layers (Fig.\,\ref{fgr:modelDiagram}c), one often adopts a symmetric carrier distribution (trigonometric) and a dielectric environment (${{\epsilon }_{2}}={{\epsilon }_{3}}$). For graphene (Fig.\,\ref{fgr:modelDiagram}b), the thickness of channel (0.35 nm) is negligible which enables the use of a pulse-like ($\delta$ function) to represent the carrier distribution. In general, the high symmetries exhibited enable analytic expression for the three factors of $M_{2D}$, ${{\epsilon}_{2D}}$, and $\Phi(q, t)$. The situation is difficult in case of silicon FETs (Fig.\,\ref{fgr:modelDiagram}a) because it involves many configurative asymmetries.  Upon adopting an empirical form of carrier distribution, analytic forms for the three items can still be reached.\cite{Ando82RMP} Nevertheless, none of the three device configurations is suitable for the generalized \ms{} FET since it possesses a finite channel thickness (non-negligible thickness), asymmetric dielectric environments (\ie{}, ${\epsilon _2}\ne {\epsilon _3}$), and lopsided carrier distribution (close to the gated dielectric). Compared to silicon, the \ms{} FET has one more top dielectric needed to be considered.

Jena \et{}\,are the first to calculate the scattering of charged impurities in 2D \ms{} FETs.\cite{Kim12NC} They achieved the carrier distribution in multilayer \ms{} by numerically solving the Schr\"{o}dinger-Poisson equation. Such a treatment, albeit accurate, requires a case-by-case calculation for each channel thickness or gate bias because any variation of them would change the carrier distribution. Later on, they did an analogue simplification to graphene for calculating monolayer \ms{} with adopting zero channel thickness and symmetric carrier distribution.\cite{Ma14PRX}

To combine accuracy and convenience, Li \et{}\,employed a phenomenological method by adopting the carrier distribution from silicon FET.\cite{Li13NL} In their calculation they strictly considered the configurative parameters of devices, including non-zero channel thickness, asymmetric surroundings, positions of interfacial impurities. This method allowed them, for the first time, to shed light onto the dependence of mobility over the full range thickness. In their calculation, the impurity scattering rate can be expressed as a linear combination of the contributions from the top and bottom channel surfaces as
\begin{equation}\label{rate_CI}
  {{\tau }_\text{CI}}^{-1}={{\alpha }_{\text{bot}}}(t){{n}_{\text{bot}}}+{{\alpha }_{\text{top}}}(t){{n}_{\text{top}}}
\end{equation}
where the subscript bot/top denotes the bottom/top interface, $t$, and $n$ are the channel thickness and density of impurity, and $\alpha$ is the scattering coefficient calculated. They compared the scattering rates for impurities located at bottom and top surfaces, as well as the dependence of the scattering rates for different scattering centers (including lattice phonons) on channel thickness. The scattering coefficients $\alpha_\text{bot}$ and $\alpha_\text{top}$ versus channel thickness for back-SiO$_2$-gated and top-HfO$_2$-gated FET configurations are given in Figs.\,\ref{fgr:s3_impurity}a and \ref{fgr:s3_impurity}b, respectively.

Two major pieces of information can be understood. First, the impurity scattering from the interface of gated dielectric is stronger than the ungated interface as a result of distribution unbalance of carriers upon applying gate bias. For the monolayer \ms{}, the scattering from charged impurity would outperform that from phonons if the gated interface had impurity density higher than $\sim2\times10^{12}$\,cm$^{-1}$. Second, the impurity scattering is considerably enhanced in extremely thinned channels, resulting from the reduction of interaction distance ($d_\text{NL}$) between impurities and carriers. As an example, Fig.\,\ref{fgr:s3_impurity}c shows the carrier distribution and the interaction distance $d_\text{1L}$ and $d_\text{3L}$ for the back-gated 1L and 3L FET channels, respectively. The carriers in the thinner 1L channel are located closer to the gated bottom dielectric due to electrostatic equilibrium, resulting in a smaller  $d_\text{1L}$ than $d_\text{3L}$. Since scattering potential $V(d)\propto d^{-1}$, the scattering intensity on the thin channels is stronger than on the thick ones. Therefore, the variation of interaction distance with channel thickness is the direct origin for the dependence of carrier mobility on thickness.\cite{Li13NL} Figure\,\ref{fgr:s3_impurity}d compares the experiment and calculation of the mobility values at different thicknesses. Reasonable agreement is reached in terms of the thickness dependence.

Based on the above results, a general conclusion can be drawn that it is crucial to achieve clean channel interfaces in order to realize high mobility in the extremely thinned 2D semiconductors. Specific strategies will be discussed in section\,\ref{engineering:impurity}.

\subsubsection{Deformation potential}

\begin{table}
\small
\caption{\ Physical parameters of lattice phonons for monolayer \ms{}. The $\Gamma$/K subscript
on the optical deformation potentials indicate couplings to the intra/intervalley phonons. Reproduced with permission from \Ref \citenum{Kaasbjerg12PRB}, copyright 2012, American Physical Society.}
  \label{tbl:phonon_parameter}
  \begin{tabular*}{0.45\textwidth}{@{\extracolsep{\fill}}lll}
    \hline
Parameter &  Symbol  & Value  \\
\hline
Lattice constant               &   $a$                     &   3.14 \AA            \\
Ion mass density               &   $\rho$                  &   $3.1\times 10^{-7}$ g/cm$^2$ \\
Effective electron mass        &   $m^*$                   &   0.48 $m_\text{e}$          \\
Valley degeneracy              &   $g_v$                   &   2           \\
Effective layer thickness      &   $\sigma$                &   5.41 \AA            \\
Piezoelectric constant         &   $e_{11}$                &   $3.0 \times 10^{-11}$~C/m \\
Transverse sound velocity      &   $c_\text{TA}$           &   $4.2 \times 10^3$ m/s  \\
Longitudinal sound velocity    &   $c_\text{LA}$           &   $6.7 \times 10^3$ m/s  \\
Acoustic deformation potentials\\
TA                             &   $\Xi_\text{TA}$         &   $1.5$ eV       \\
LA                             &   $\Xi_\text{LA}$         &   $2.4$ eV       \\
Optical deformation potentials\\ 
TA                             &   $D_{\mathbf{K},\text{TA}}^1$       &   $5.9$ eV       \\
LA                             &   $D_{\mathbf{K},\text{LA}}^1$       &   $3.9$ eV       \\
TO                             &  $D_{\Gamma,\text{TO}}^1$   &   $4.0$ eV       \\
TO                             &  $D_{\mathbf{K},\text{TO}}^1$        &   $1.9$ eV       \\
LO                             &  $D_{\mathbf{K},\text{LO}}^0$        &   $2.6 \times 10^8$ eV/cm  \\
Homopolar                      &  $D_{\Gamma,\text{HP}}^0$   &   $4.1 \times 10^8$ eV/cm  \\
Phonon energies                \\ 
TA                             &  $\hbar\omega_{\mathbf{K},\text{TA}}$       & 23 meV \\
LA                             &  $\hbar\omega_{\mathbf{K},\text{LA}}$       & 29 meV \\
TO                             &  $\hbar\omega_{\Gamma,\text{TO}}$   & 48 meV    \\
                               &  $\hbar\omega_{\mathbf{K},\text{TO}}$       & 47 meV    \\
LO                             &  $\hbar\omega_{\Gamma,\text{LO}}$  & 48 meV    \\
                               &  $\hbar\omega_{\mathbf{K},\text{LO}}$       & 41 meV    \\
Homopolar                      &  $\hbar\omega_\text{HP}$  &   50 meV\\
    \hline
  \end{tabular*}
\end{table}

In semiconductors the lattice potential determines the band structure. The atomic displacement ($a$) due to lattice phonon forms a perturbation to band edges. Lattice phonons can scatter off the electron waves through the potential deformation, as shown in Fig.\,\ref{fgr:mechanism}b. For acoustic phonons, adjacent atoms move in the same direction and the modification of interatomic distance $\delta A\propto \delta a$, with $A$ the interatomic distance in equilibrium. Hence the shift of band edges can be written as 	$\delta E=\Xi \cdot \delta a/A$ in a linear approximation, where $\Xi$ is the deformation potential of acoustic phonons. For optical phonons, adjacent atoms move in the opposite direction and $\delta A\propto a$. Thus energy shift can be expanded as $\delta E={{D}_{0}}a+{{D}_{1}}\cdot \delta a/A$ with ${{D}_{i}}$ the $i$-th order deformation potential.

Phonon scattering depends highly on temperature because the number of phonons follows the Bose-Einstein distribution ${{N}_{q}}=1/[\exp (\hbar \omega /{{k}_{B}}T)-1]$, where $\hbar \omega$ is the phonon energy. The expressions of phonon scattering rates and related physical parameters for monolayer \ms{} have been derived by Kaasbjerg \et{}\,\cite{Kaasbjerg12PRB,Kaasb13PRB_Transport:Phonon} Table\,\ref{tbl:phonon_parameter} summarizes the phonon parameters required in the calculation. To account for the screening effect, Jena \et{}\,included the electron polarization function $\epsilon _{2D}$  in the calculation. Hence the scattering rate due to deformation potential of the acoustic phonon is written as\cite{Ma14PRX}
\begin{equation}\label{rate_ac}
  \frac{1}{\tau{}_{\text{ac}}}=\frac{\Xi _{\text{ac}}^{2}{{k}_{B}}Tm^*}{2\pi {{\hbar }^{3}}{{\rho }_{s}}v_{s}^{2}}\int\limits_{-\pi }^{\pi }{\frac{(1-\cos \theta )}{\epsilon _{2D}^{2}}},
\end{equation}
where $k_B$, $T$, and $\hbar$ are the Boltzmann constant, temperature, and Planck constant, respectively, and $\theta$ is the elastic scattering angle from the initial momentum $k$ to the final momentumto $k'$, and $\epsilon _{2D}$  is the electron polarization function in the 2D semiconductor. The meanings and values of other parameters related to \ms{} are listed in Table\,\ref{tbl:phonon_parameter}. Similarly, the scattering rate of the polar phonons is given by\cite{Ma14PRX}
\begin{equation}\label{rate_op1}
  \frac{1}{\tau{}_{\text{op}}}=\frac{\Theta [{{E}_{k}}-\hbar \omega _{\text{op}}^{\nu }]}{\tau _{\text{op}}^{\text{+}}}+\frac{1}{\tau _{\text{op}}^{-}},
\end{equation}
\begin{equation}\label{rate_op2}
  \frac{1}{\tau _{0\text{-ODP}}^{\pm }}=\frac{D_{\text{0}}^{2}m^*({{N}_{q}}+\tfrac{1}{2}\pm \tfrac{1}{2})}{4\pi {{\hbar }^{2}}{{\rho }_{s}}\omega }\int\limits_{-\pi }^{\pi }{\frac{(1-({k}'/k)\cos \theta )\text{d}\theta }{\epsilon _{2D}^{2}}},
\end{equation}
\begin{equation}\label{rate_op3}
  \frac{1}{\tau _{\text{1-ODP}}^{\pm }}=\frac{D_{\text{1}}^{2}m^*({{N}_{q}}+\tfrac{1}{2}\pm \tfrac{1}{2})}{4\pi {{\hbar }^{2}}{{\rho }_{s}}\omega }\int\limits_{-\pi }^{\pi }{\frac{{{q}^{2}}(1-({k}'/k)\cos \theta )\text{d}\theta }{\epsilon _{2D}^{2}}}.
\end{equation}
where $\Theta[ ]$ is the step function, $E_k$  is the carrier energy at momentum $k$, $\omega$ is the phonon frequency, and $q=2k\cdot \text{sin}(\theta /2)$ is the scattering vector. The superscript $+$ and $-$ represent scattering processes with absorbing and releasing a phonon, respectively. The subscripts 0 and 1 denote the zeroth- and first-order optical deformation potential, respectively.

\subsubsection{Fr\"{o}hlich and piezoelectric interactions}
In compound semiconductors like GaAs and all \mx{} chalcogenides, dipole moments forms between adjacent cation and anion due to the polar nature of the chemical bonds. Deformation of the lattice by polar phonons perturbs the dipole moment between atoms which results in an electric field that is coupled to carriers, as shown in Fig.\,\ref{fgr:mechanism}c. Polar optical phonon scattering, known as Fr\"{o}hlich interaction, is normally strong near room temperature. In contrast, the polar acoustic phonon scattering, known as piezoelectric interaction, is typically weak.\cite{Kaasb13PRB_Transport:Phonon}

The scattering rate for Fr\"{o}hlich interaction is given by\cite{Gelmont95JAP,Ma14PRX}
\begin{equation}\label{rate_FL}
\begin{aligned}
   \frac{1}{\tau _{\text{LO}}^{\pm }}= & \frac{{{e}^{2}}\omega m^*}{8\pi {{\hbar }^{2}}}\frac{1}{{{\varepsilon }_{0}}}(\frac{1}{{{\varepsilon }_1^{\infty }}}-\frac{1}{{{\varepsilon }_1^0}})({{N}_{q}}+\tfrac{1}{2}\pm \tfrac{1}{2})  \\
 & \int\limits_{-\pi }^{\pi }{\frac{1}{q} \Phi(q, t) \frac{(1-({k}'/k)\cos \theta )\text{d}\theta }{\varepsilon _{2D}^{2}}}
 \end{aligned}
\end{equation}
where $e$ and  $\epsilon_0$  are the elementary charge and vacuum permittivity, respectively, the superscripts of 0 and $\infty$ in the permittivity denote the static and high-frequency values, respectively, and $\Phi(q, t)$ is the thickness-dependent configurative form factor. A simplified version of $\Phi(q, t)$ for monolayer channels (within zero-thickness approximation) is available in \Ref\citenum{DasSarma11RMP}. A strict solution for a generalized FET configuration can be found in refs.\,\citenum{Li13NL} and \citenum{Li15JPSJ}.

Kaasbjerg \et{}\,derived the scattering matrix element for the piezoelectric interaction in 2D \ms{},\cite{Kaasb13PRB_Transport:Phonon}
\begin{equation}\label{matrix_PE}
  \left| M^{\mathbf{q}\lambda }_{\text{PE}} \right|={{e}_{11}}eq/{{\epsilon }_{0}}\times \text{erfc}(q\sigma /2)\left| {{A}_{\lambda }}(\mathbf{q}) \right|
\end{equation}
where $M^{\mathbf{q}\lambda }_{\text{PE}}$ is the independent component of the piezoelectric tensor of the 2D hexagonal lattice, $e_{11}$ is the piezoelectric constant, erfc() is the complementary error function, $\sigma$  is an effective width of the electron wave functions, and $A_{\lambda}(\mathbf{q})$  is the anisotropy factor that accounts for the anisotropic angular dependence of the piezoelectric interaction. Within the long-wavelength approximation, the high-temperature relaxation time for piezoelectric scattering can be calculated together with the acoustic phonon scattering as\cite{Kaasb13PRB_Transport:Phonon}
\begin{equation}\label{rate_op3}
 \frac{1}{\tau{}_\text{PE}} =  \frac{1}{\tau{}_\text{ac}} \times  {{{{\left( {{e_{11}}e/{\varepsilon _0}} \right)}^2}} \over {2\Xi _{{\rm{ac}}}^2}}.
\end{equation}

\subsubsection{Remote interfacial phonons}
\begin{table}[h]
\small
  \caption{\ RIP modes for commonly used dielectrics.}
  \label{tbl:RIP_parameter}
  \begin{tabular*}{0.45\textwidth}{@{\extracolsep{\fill}}lllll}
    \hline
 Dielectric &  $\epsilon_\text{ox}^0$  & $\epsilon_\text{ox}^\infty$ &$\omega_\text{RIP}^1$ & $\omega_\text{RIP}^2$ \\
\hline
  SiO$_2$   & 3.9 & 2.5 & 55.6 & 138 \\
  BN        & 5.1 & 4.1 & 93.1 & 179 \\
  AlN       & 9.1 & 4.8 & 81.4 & 88.5 \\
  Al$_2$O$_3$ & 12.5 & 3.2 & 48.2 & 71.4 \\
  HfO$_2$   & 22 & 5.0 & 12.4 & 48.4 \\
  ZrO$_2$   & 24 & 4.0 & 16.7 & 57.7 \\
    \hline
  \end{tabular*}
\end{table}

Electrons in semiconductors, especially in the inversion layer of electrically gated FET channels, can excite phonons in the surrounding dielectrics via long-range Coulomb interactions, if the dielectrics support polar vibrational modes, as shown in Fig\,\ref{fgr:mechanism}d. They are long recognized as `remote interface phonons' (RIP) or `surface optical phonons' (SOP) and exist in dielectrics near the inversion layers in silicon\cite{moore80jap,fischetti01jap}, organic FETs,\cite{stass04apl_,veres03afm_,hulea06nm_rip} and graphene.\cite{chen08nn,fratini08prb,sabio08prb,konar10prb,zou10prl,dasilva10prl} The RIP scattering may not be a significant scattering mechanism in low-field transport or in FETs using low-$\kappa$ dielectrics, but it can become important at high fields,\cite{dasilva10prl}  large inversion densities or high-$\kappa$ dielectric surroundings, as pointed out by Moore \et{}\,\cite{moore80jap,fischetti01jap} Experimental studies on organic FETs indicate that the use of high $\kappa$ dielectric degrades FET carrier mobility.\cite{stass04apl_,veres03afm_,hulea06nm_rip} Table\,\ref{tbl:RIP_parameter} lists the RIP modes for commonly used dielectrics, which are useful for theoretical calculation and device design.

On the assumptions of zero thickness for 2D channels and semi-infinite for dielectrics, the electron-RIP coupling parameter is
\begin{equation}\label{matrix_rip}
  F_{\upsilon }^{2}=\frac{\hbar \omega _{\text{RIP}}^{\nu }}{2S{{\epsilon }_{0}}}\left( \frac{1}{\epsilon _{2}^{\infty }+\epsilon _{3}^{\infty }}-\frac{1}{\epsilon _{2}^{0}+\epsilon _{3}^{\infty}} \right)\end{equation}
where $\epsilon_{\textrm{i}}$ (i$= 1, 2, 3$) denotes the dielectric constants for different layers of channel and dielectrics as shown in Fig.\,\ref{fgr:modelDiagram}, and
\begin{equation}\label{freq_rip}
  \omega _{\text{RIP}}^{\nu }=\omega _{\text{TO}}^{\nu }{{\left( \frac{\epsilon _{2}^{0}+\epsilon _{3}^{\infty }}{\epsilon _{2}^{\infty }+\epsilon _{3}^{\infty }} \right)}^{1/2}}.
\end{equation}
The scattering rate due to RIP can be written as\cite{Ma14PRX}
\begin{equation}\label{rate_rip}
\begin{aligned}
  \frac{1}{\tau _{\text{RIP}}^{\pm }}=&\frac{32{{\pi }^{3}}{{e}^{2}}F_{\upsilon }^{2}m^*S}{{{\hbar }^{3}}{{a}^{2}}}\left( {{N}_{q}}+\tfrac{1}{2}\pm \tfrac{1}{2} \right) \\ &\int\limits_{-\pi }^{\pi }{\frac{1}{q} {\left( {{{{\rm{sinh (}}aq/2{\rm{)}}} \over {4{\pi ^2}q + {a^2}{q^3}}}} \right)^2} \frac{(1-({k}'/k)\cos \theta )\text{d}\theta }{\epsilon _{2D}^{2}}}.
\end{aligned}
\end{equation}

The significant room-temperature RIP scattering rate for high-$\kappa$ dielectric in monolayer \ms{} FETs poses a challenge to the dielectric screening engineering advocated,\cite{jena07prl,Radisavljevic11NN} which is oriented to enhance carrier mobility. We will elaborate this issue in section\,\ref{RIP}.

\subsubsection{Atomic and structural defects}
Atomic and structural defects can create midgap energy states or highly localized band tails and considerable affect charge transport in semiconductors. There are many different types of defects, ranging from spatially extended structures (\eg{}, grain boundaries, dislocations, and precipitates), to pairs and complexes, to isolated vacancies or impurities. The theoretical and experimental studies on the defects are very active in the field of electronic band structure calculation\cite{Caulf97JPCM_defect:vacancy,Feng14MCP_defect:vacancy,Feng14JAC_defect:vacancy,Gan14PLA_defect:vacancy,Heine15ACR_defect:e-structure,Santo14N_,Spirk04SS_,Tianm15RMMaE_defect:vacancy,Komsa12PRL_defect:e-structureTEM}   because defect chemistry brings about industrial applications. For instance, the sulfur vacancies in \ms{} enable its use as chemical catalysis for desulfurization in petrochemistry\cite{Moses07JC_defect:vacancy,Paul03JPCB_defect:vacancy} and water splitting.\cite{Moral14ACR_Energy:H2,Kibsg12NM_Energy:H2,Voiry13NM_Energy:H2,Karun12S_Energy:H2}
\begin{figure*}
 \centering
 \includegraphics[width=0.75\textwidth]{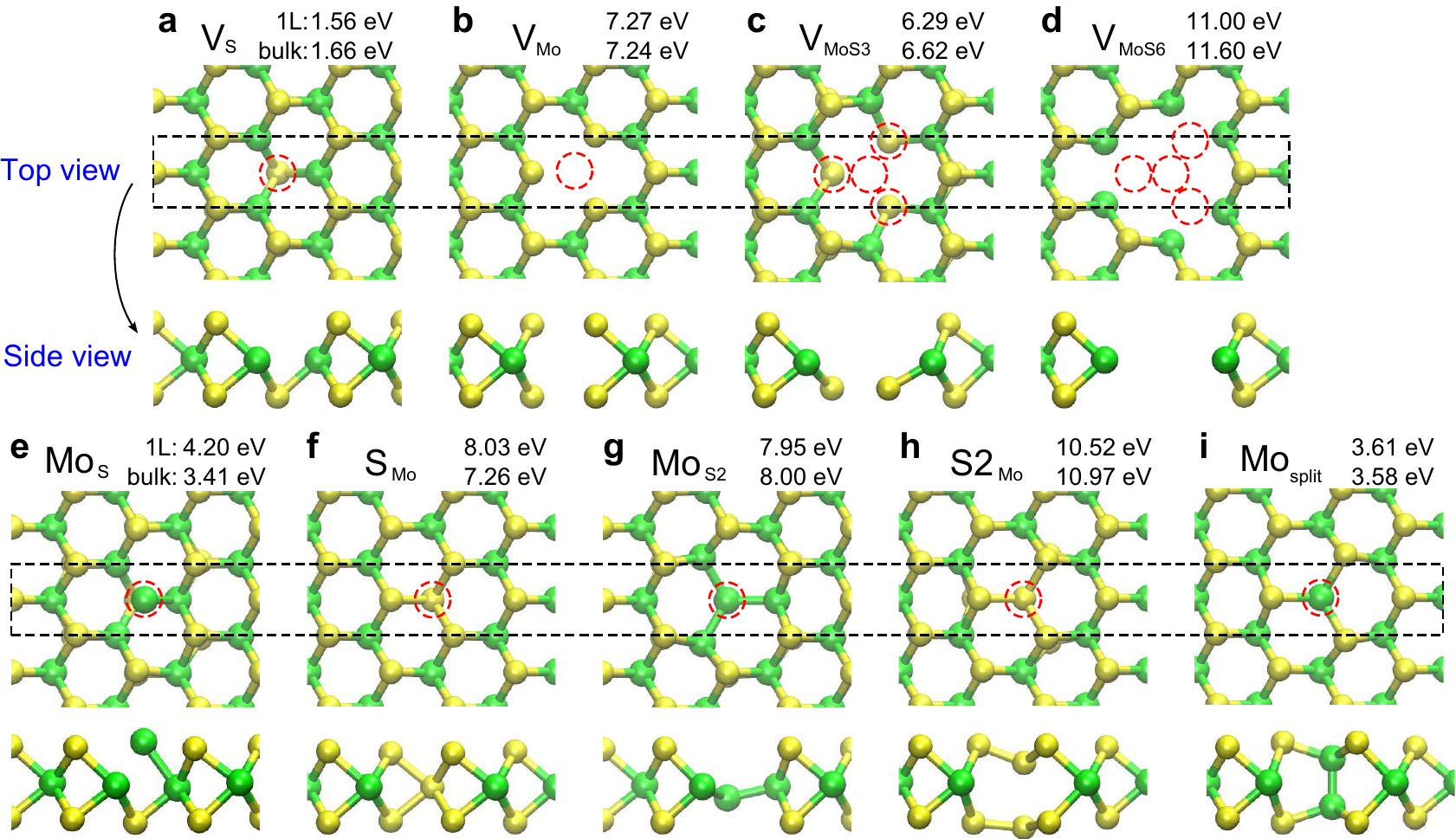}
  \caption{Optimized atomic structures and formation energies for different defects in \ms{} monolayer (1L) and bulk. (a)--(d): vacancies, (e)--(h):antisites, (i) Mo-Mo split interstitial defect. Dashed circles (red) denote the position of the defect. Owing to a rather low formation energy of $\sim$1.6\,eV, sulfur vacancies tend to form in \ms{}. \Cright[\aps]{Komsa15PRB_Vacancy:MoS}}
 \label{fgr:defect_type}
\end{figure*}

The defect scattering is normally not considered in high-quality superlattices and silicon FETs,\cite{Ando82RMP}  because the very low density causes negligible scattering rates relative to other mechanisms. However, the defect density is by no means low in the 2D vdW crystals. Figure\,\ref{fgr:defect_type} show the structures and corresponding formation energies of different atomic defects in \ms{}. The sulfur vacancies have a rather low formation energy of $\sim$1.6 eV, hence the anion vacancies tend to form in chalcogenides, just like in oxides (\eg{}\,ZnO), which is presumably a strong scattering source when the sample quality is not sufficiently high.

High-resolution TEM experiments revealed the presence of large amounts of point defects and grain boundaries in natural and synthesized \ms{}.\cite{Komsa12PRL_defect:e-structureTEM,Qiu13NC,Zande13NM,Hong15NC_defect:vacancy} It is found that the structures of dominant defects correlate closely with material growth methods. For samples prepared by mechanical exfoliation and chemical vapor deposition (CVD), sulfur vacancy defects with one (denoted as V$_\text{S}$, Fig.\,\ref{fgr:defect_type}a) or two (V$_\text{S2}$) S atoms absent are frequently observed, while the dominant defects for the physical vapor deposited (PVD) samples are antisite defects with one Mo atom replacing one (Mo$_\text{S}$, Fig.\,\ref{fgr:defect_type}e) or two (Mo$_\text{S2}$, Fig.\,\ref{fgr:defect_type}g) S atoms.\cite{Hong15NC_defect:vacancy}  The density of sulfur vacancy can reach \E[(1.2 \pm 0.4)]{13}\,cm$^{-2}$ in exfoliated and CVD samples, corresponding to a surprisingly high atomic percentages of 0.4\%. Undoubtedly, atomic defects would play an important role in carrier scattering, if such high-level defects are present in device channels.

Wang \et{}\,attributed the presence of sulfur vacancies as the reason for the hopping transport behavior observed in low carrier density regime.\cite{Qiu13NC} The short-range vacancy scattering is also proposed as one of the scattering mechanisms in the CVD \ms{} flakes by Eda \et{}\,\cite{Schmidt14NL} With long-time \textit{in situ} thermal annealing to minimize contact resistance, they estimated a high room-temperature mobility of 45\,\cmvs{} and an intrinsic mobility of 58\,\cmvs{} after deducting the effect of short-range vacancy scattering. Since the anion vacancies could be the leading scattering centers, vacancy repair is expected to improve device performance.

Hitherto, there has been no theoretical work on the vacancy scattering in 2D vdW semiconductors. However, one can quickly grasp its basic characteristics by looking through previous works on their bulk \cite{Ravic71pssb_defect:PbTe,Zayac97S_defect:PbTe} and 2D analogue graphene.\cite{Hwang07PRLb,Hwang08PRB} Unlike Coulomb impurity scattering, vacancy scattering is a kind of short-range interaction with interaction range comparable to the lattice spacing (Fig.\,\ref{fgr:mechanism}e). In this case, the scattering potential can be treated as a $\delta$ function.\cite{Hwang08PRB} The scattering matrix in low-energy regime becomes dispersionless on energy for 2D semiconductors because the Fourier transformation of the real space $\delta$ potential is a constant in the reciprocal momentum space, which results in a constant scattering matrix. Hence, the scattering rate of atomic vacancies is independent on carrier density, manifested itself as a resistivity background as in graphene.\cite{Hwang07PRLb}   In addition the vacancy scattering in 2D systems should be weakly dependent on temperature or channel thickness because the defect density is independent on these two parameters. It would reduce the temperature dependence of mobility and the temperature exponent $\gamma$, once vacancy scattering becomes important. In this sense, the low $\gamma$ values ($\sim$0.7) observed in clean samples\cite{Jariwala13APL,Yu14NC_MoS2_mu80} is likely indicative of the emergent dominance of vacancy scattering. Overall, it is expected that defect scattering would become paramount in very clean samples or at low temperature when Coulomb impurity and phonon scattering rates are low.\cite{Cui15NN_FET:G/MoS/BN_mu34k}

\begin{figure}
 \centering
 \includegraphics[width=0.5\textwidth]{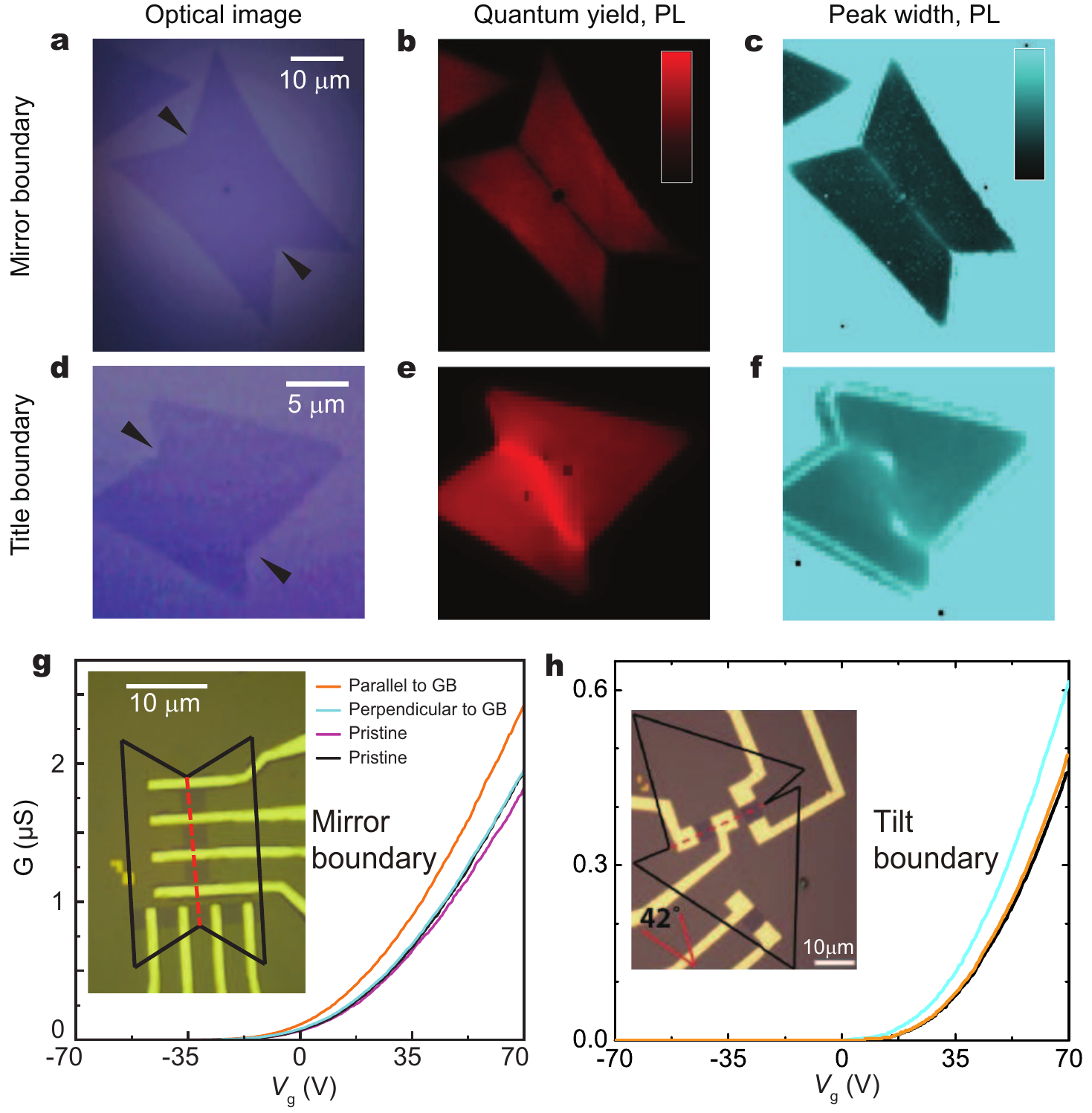}
 \caption{Optical and electronic properties of mirror and tilt boundaries in synthesized \ms{} flakes. (a)--(c) Optical measurements of a \ms{} flake containing a mirror twin boundary. (d)--(f) Corresponding measurements for a \ms{} flake containing a tilt boundary. Panels (a) and (d) are optical images; Panel (b), (c), (e), and (f) are colour plots of photoluminescence. Panels (g) and (h) show the transfer curves of four FETs fabricated from the \ms{} flakes with mirror and tilt boundaries in the insets. \Cright[\npg]{Zande13NM}}
 \label{fgr:defect_GB}
\end{figure}
Grain boundary is another typical form of defect commonly present in synthesized samples. Figures\,\ref{fgr:defect_GB}a--\ref{fgr:defect_GB}f show two types of grain boundaries (the mirror and title type), as seen by optical microscopy and photoluminescent mapping.\cite{Zande13NM,Najmaei13NM,Zhang13AN_defect:GB:WS}  An important question for device applications is whether these grain boundaries disrupt or modify electronic transport. Hone \et{}\,compared the electronic transport properties of exfoliated and synthesized \ms{} flakes with room temperature mobilities from 1 to 8\,\cmvs{}.\cite{Zande13NM} They found that device performance strongly depends on the boundary type as well as current flow direction. The mirror twin boundary has little effect on channel conductivity when current is perpendicular to boundary and, surprisingly, it slightly increases the on-state conductivity by when current flows in parallel (Fig.\,\ref{fgr:defect_GB}g). This observation suggests that the few-atom-wide twin boundaries, although still semiconducting, have similar conductivity of pristine \ms{}. In contrast, the tilt boundary generally degrades the device performances to large extent in any current directions (Fig.\,\ref{fgr:defect_GB}h). A wide variation of the conductance (5--80\%) is observed among devices, implying a complicated dependence of electronic structure at boundaries on the tilt angle and atomic structure.\cite{Zou13NL}

\subsubsection{Other scattering mechanisms}
Besides above mechanisms, other scattering factors such as electron-electron collision and surface corrugation could also be a correction the overall electronic performance. It is noteworthy that all the above mechanisms are discussed within the assumption of low-energy, intraband approximation and they are suitable only for the transport behavior at low fields. More complicated scattering processes including interband scattering under high electrical and magnetic fields are beyond the scope of this review and will not be discussed here.

\section{Mobility engineering strategies and state-of-the-art performance} \label{sect:engineering}
\subsection{Contact optimization}
It is well known that the presence of contact resistance strongly limits the current carrying capacity in FETs. The contact resistance for 2D chalcogenides ranges from 10--100\,k$\Omega\cdot\mu$m at Au/\ms{} contacts\cite{Li14AN_MoS2:Contact} to $\sim1\,$k$\Omega\cdot\mu$m with graphene/2L \wse{} contacts.\cite{Das14NL} These values are at least 1--2 orders higher than the requirement of the deep-nanometer technological node in semiconductor industry. Hence, optimizing the electrode/channel contacts is highly desired.
\begin{figure*}[h!bt]
 \centering
 \includegraphics[width=0.98\textwidth]{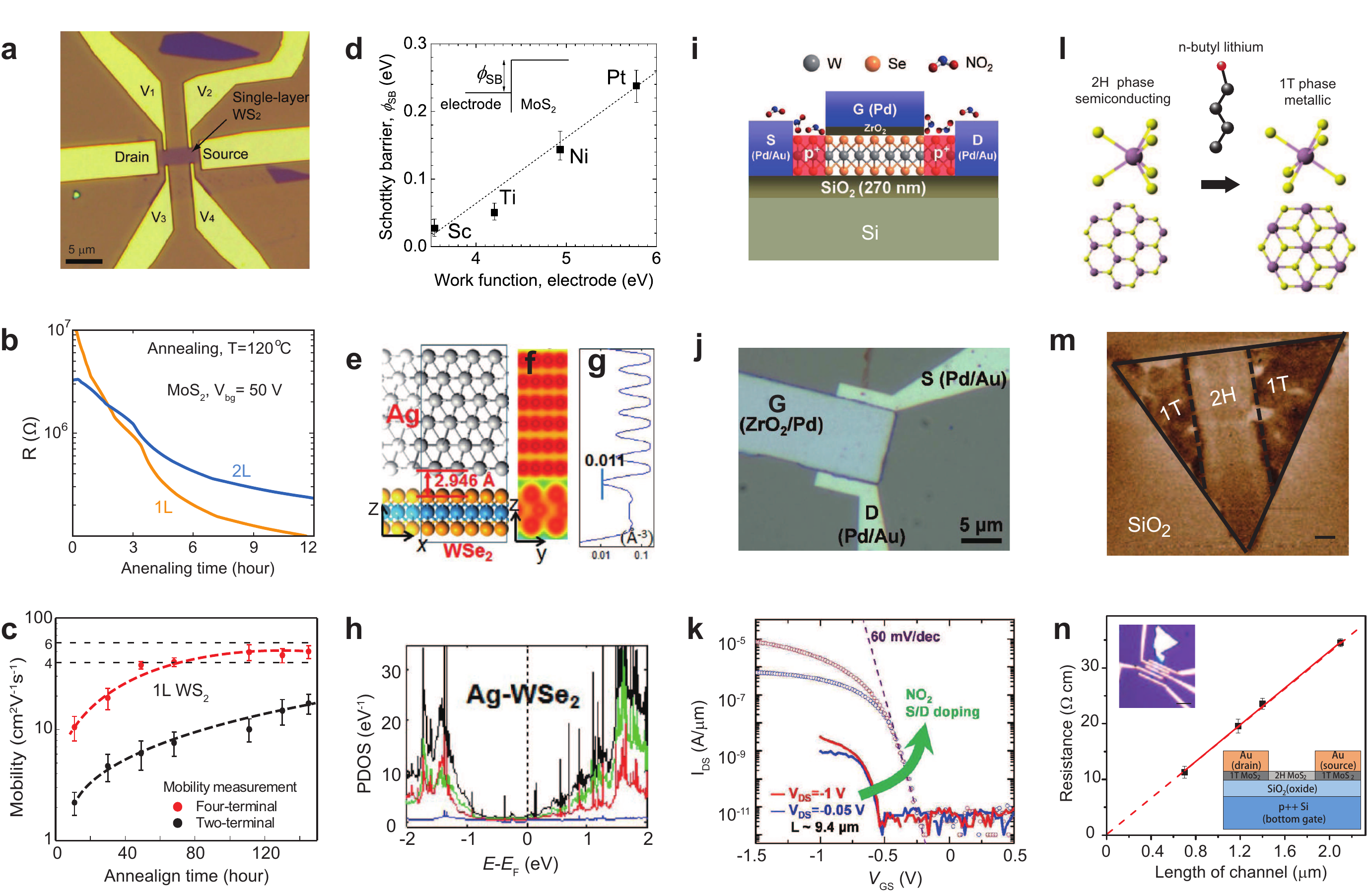}
 \caption{Mobility engineering for contact optimization. (a)--(c) Thermal annealing. (a) Optical image of 1L \ws{} four-terminal measurement geometry which can be used to estimate the contact quality. (b) Evolution of contact resistance at Ti/\ms{} interface with time under 120\C{} thermal annealing up to 12 h. (c) Mobility comparison for 1L \ws{} under different wiring methods (two-terminal and four-terminal), showing that the contact resistance is difficult to be eliminated since the difference of mobility values under the two wiring methods exists all the time (up to 150 h annealing). (d) Energy level matching, which shows that electrodes with low work functions in general lead to low contact barrier to n-type semiconductors. Adapted with permission from \Ref \citenum{Das13NLb}, copyright 2013, American Chemical Society.  (e)--(h) Orbit hybridizing engineering. (e) Side view of the relaxed contact regions at the \ws{}/ Ag (111) interface. (f) and (g) show the contour plot of the spatial electron density and the average line electron density, respectively. (h) Partial density  of states (PDOS) the Ag-\ws{} system. Hybridization states occur at the contacts which are beneficial for carrier transfer. (i)--(k) Contact doping engineering. (i) Schematic diagram for using NO$_2$ gas to dope the channel edges in vicinity of contacts. (j) Optical device image with top ZrO$_2$/Pd gate stacks. (k) Comparison of channel current before and after NO$_2$ doping. (l)--(n) Phase change engineering. (l) Atomic phase change of \ms{} from 2H to 1T structure after soaking in n-butyl lithium. (m) Electrostatic force microscopy phase image of a 1L \ms{} nanosheet showing the difference between locally patterned 2H (bright colour) and 1T (dark colour) phases. The scale bar is 1 $\mu$m. (n) Transfer line plot for the 1T phase contacted \ms{} channels. The intercept is close to zero, indicating the contact resistance is very small.  Panels (a) and (c) are \cright{Ovchi14AN_FET:WS}. Panel (b) is \cright{Baugher13NL}. Panels (e)--(h) are \cright{Liu13NL}. Panels (i)--(k) are \cright{Fang12NL}. Panels (l)--(n) are \cright[\npg]{Kappera14NM}.}
 \label{fgr:contact}
\end{figure*}

Figure\,\ref{fgr:contact} summarizes the mobility engineering strategies devised for contact optimization. Thermal annealing is traditionally used to reduce contact resistance. This technique is particularly necessary and effective for the 2D chalcogenide FETs. Jarillo-Herrero \et{}\,found that large contact resistance is the main reason for the low field-effect mobility in thermally untreated 2D FETs.\cite{Baugher13NL}  A huge drop in contact resistance, with 2 orders of magnitude in monolayer and more than 1 order of magnitude in bilayer \ms{} devices, was observed after long-time (12 h), low-temperature (120\C{}) \textit{in situ} thermal annealing (Fig.\,\ref{fgr:contact}b). Remarkably high room temperature mobility values of 20--50 and 80--150\,\cmvs{} were observed in monolayer and bilayer \ms{}, respectively. In their experiment, low annealing temperature was used likely because of the safety limitation of test cables and cryostat, which, however, favorably prevents the decomposition of \ms{} during annealing and the creation of sulfur vacancies, a kind of short-range carrier scattering center. The similar behavior has been observed in CVD \ms{}  samples by Eda \et{}\,\cite{Schmidt14NL}  and 1L \ws{} samples by Kis \et{}\,\cite{Ovchi14AN_FET:WS} (Fig.\,\ref{fgr:contact}a). Figure\,\ref{fgr:contact}c shows the real-time monitor of the mobility evolution with annealing time for the 1L \ws{} sample under both two-terminal and four-terminal measurement. After a long annealing time of 150 h, the mobility extracted by four-terminal method remains 2-fold higher than that extracted by two-terminal method, suggesting that the contact issue, though largely reduced, cannot be completely eliminated by thermal annealing.

Energy level matching between electrodes and semiconductors was also employed to improve the field-effect mobility. The height of the contact barrier is highly related to the work function of metal electrodes, as shown in inset of Fig.\,\ref{fgr:contact}d. The metal Sc, with a low work function of 3.5 eV which is close to the electron affinity of \ms{} $\sim$4.0 eV, is found to form electrically transparent interface with thick \ms{}.\cite{Das13NLb} The mobility reaches 180\,\cmvs{} in $\sim$10 nm \ms{} devices. The pinning coefficient is estimated to be 0.1, as compared with $\sim$0.3 in bulk silicon, suggesting the presence of extremely strong pining effect in the 2D semiconductors due to the full exposure of channel surfaces. Nevertheless, the energy level matching with Sc electrode seems an effective way to improve mobility for thick channels but to be much less influential for ultrathin samples. The carrier mobility was found to sharply drop to 25\,\cmvs{} in $\sim$2 nm \ms{}, indicating a more complicated behavior in the ultrathin regime. This can be ascribed to the expansions of bandgap and contact barrier due to quantum confinement (Fig.\,\ref{fgr:barrier}d). Owing to the upshift of conduction band in the few-layer \ms{}, electrodes with lower work functions are required to match the high-lying conduction band.

Graphene is also explored as a contact material to chalcogenides because it is a unique Dirac metal whose work function can be broadly tuned upon applying gate voltage.\cite{Kwak14NL_FET:MoS:Gcontact,Liu15NL_FET:G/MoS/BN,Cui15NN_FET:G/MoS/BN_mu34k,Chuan14NL_Rc:WSe2:IonicLiquid} In particular, graphene contacted \ms{} shows extremely high low-temperature mobility when encapsulated by clean BN.\cite{Liu15NL_FET:G/MoS/BN,Cui15NN_FET:G/MoS/BN_mu34k}  The devices with graphene contacts will be discussed in section\,\ref{state-of-the-art}.

Besides the empirical strategy, band structure calculation is also used to understand the interface physics to facilitate contact design between metal and monolayer \wse{} (Figs.\,\ref{fgr:barrier}e--\ref{fgr:barrier}g). Calculations based on \textit{ab initio} density functional theory (DFT) indicate that using d-orbital electron abundant transition metals as contacts like Ag and Ti are beneficial to form better electron injection into \wse{}  because the d-orbitals in these metals can hybridize with the d-orbitals in the Se and W atoms (Fig.\,\ref{fgr:barrier}h).\cite{Liu13NL}  A considerably high carrier mobility of 142\,\cmvs{} is measured in thermally annealed Ag contacted monolayer \wse{} samples in vacuum environment.

Apart from the rational choice of contact metals, contact doping by gas molecules or alkali metals is also developed as a chemical strategy to optimize contact quality.\cite{Fang13NL,Fang12NL} By introducing highly active electron donor NO$_2$ to form heavily doped contact areas (Fig.\,\ref{fgr:contact}i), Javey \et{}\,observed a high mobility of 250\,\cmvs{} in top ZrO$_2$ gated, Pd contacted p-type monolayer \wse{} FETs (Fig.\,\ref{fgr:contact}j).\cite{Fang12NL}  They carefully compared the device current before and after NO$_2$ doping and about 3 orders of magnitude enhancement in device current was observed (Fig.\,\ref{fgr:contact}k), indicating the critical role of contact in achieving high device performance. Additionally, they observed a high mobility of >110\,\cmvs{} in a potassium doped 3L \wse{} FET.\cite{Fang13NL}

For the same reason, ionic liquid gating is used for contact engineering.\cite{Chuan14NL_Rc:WSe2:IonicLiquid,Wang15NL_FET:WSe_BN} This is because the injection of charges is assisted by high carrier density in channels near the metal/semiconductor contacts, which largely decreases the width of the injection barrier. As a result, carrier injection from the source/drain contacts is controlled by tunneling instead of by the over-the-barrier thermal activation process (Fig.\,\ref{fgr:barrier}b). Zhou \et{}\,reported a high electron and hole mobility of $\sim$200\,\cmvs{} at 160 K and at $\sim$300\,\cmvs{} at 77 K in 6 nm \wse{}.\cite{Chuan14NL_Rc:WSe2:IonicLiquid} Similarly, Jarillo-Herrero \et{}\,achieved a low contact resistance of <330$\,\Omega \cdot \mu$m and a high mobility of $\sim$600\,\cmvs{} at 220K in 3L \wse{}.\cite{Wang15NL_FET:WSe_BN}

Artificial engineered phase change has also been developed to increase contact quality.\cite{Kappera14NM,Kappe14AM_FET:MoS2:1Tcontact,Cho15S} For chalcogenides, different atomic structures and phases may exhibit distinct electronic properties. Typically, the 2H phase of \ms{} is metallic while the 1T phase is semiconducting. After replacing the 2H phase with the 1T phase, Chhowalla \et{}\,observed a $\sim$5-fold decrease of contact resistance from 1.1 to 0.2\,k$\Omega\cdot\mu$m at zero gate bias,\cite{Kappera14NM} a value comparable to the source/drain parasitic resistance in silicon FETs (290 $\Omega\cdot\mu$m and $\Omega\cdot\mu$m for planar and SOI low standby power CMOS, respectively). As a result, the carrier mobility increases from 19 to 46\,\cmvs{}. Direct doping and metallization at contacts is a promising way to achieve technologically viable 2D FETs.

\subsection{Reduction of impurity scattering}\label{engineering:impurity}

\begin{figure*}[h!bt]
 \centering
 \includegraphics[width=0.95\textwidth]{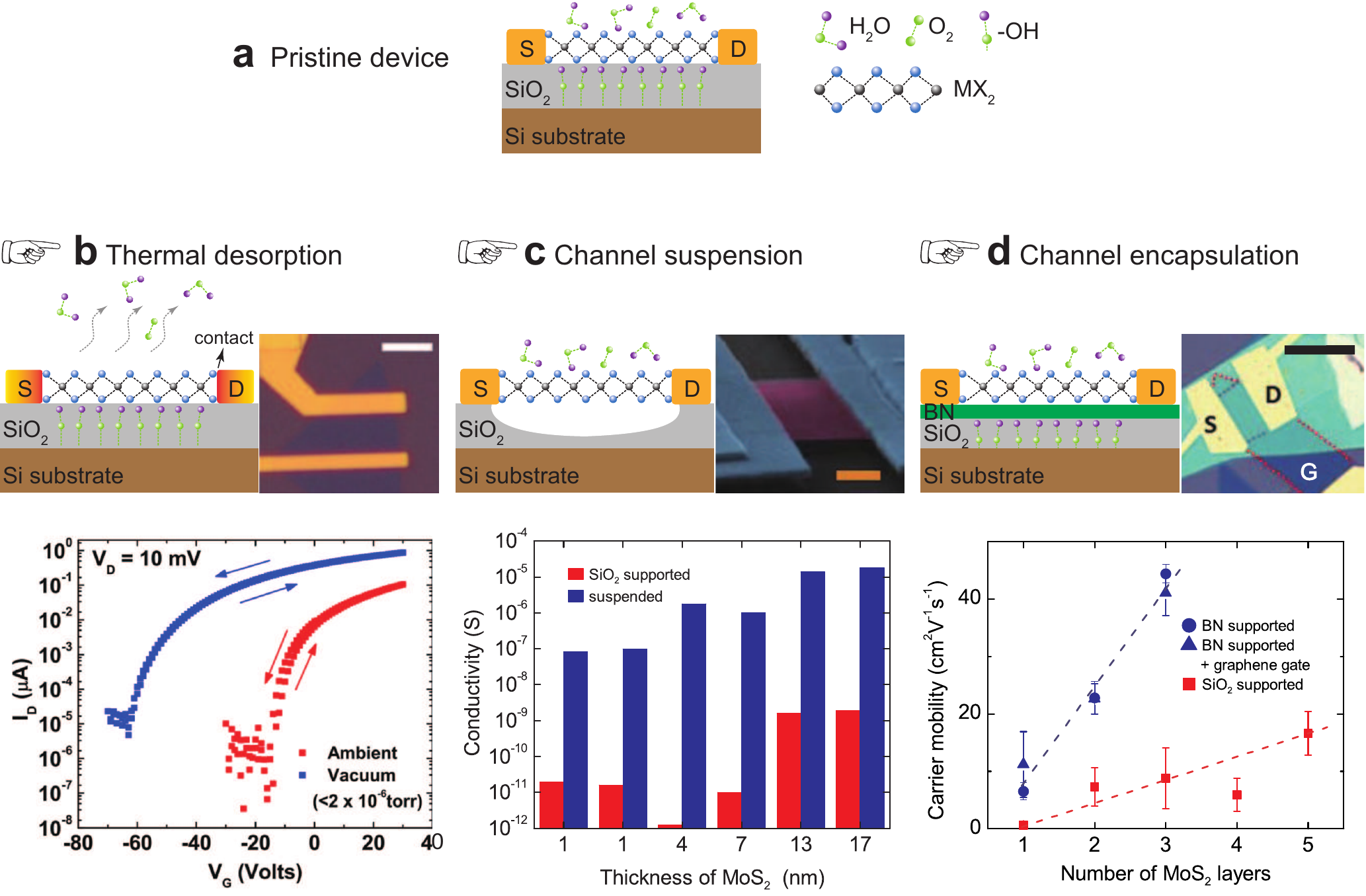}
 \caption{Mobility engineering for interfacial impurity suppression. (a) Schematic diagram for a pristine SiO$_2$ back-gated FET consisted of a 2D \mx{} channel. The main Coulomb impurities arise from the molecular absorbates on the top channel surface and the dangling bonds and hydroxyl group (--OH) on the bottom channel surface (\ie{}\,channel/dielctric interface). Three techniques are developed to suppress the impurity sources: (b) thermal desorption at ultrahigh vacuum, (c) channel suspension to remove the bottom impurities, (d) channel encapsulation by clean dielectrics such as boron nitride or polymers (\eg{}\,PMMA).  In each panel from (b) to (d), the schematic techniques, typical device images, and corresponding effects on performance are illustrated. Panel (b) is \cright[\aip]{Jariwala13APL}. Panel (c) is \cright[\iop]{Wang15N_FET:MoS:suspended}. Panel (d) is \cright{Lee13AN_BN:u45_3LMoS_thickness}.} \label{fgr:s5_impurity}
\end{figure*}

Before discussing the technical routes, we first remark on the origins of the Coulomb impurities and corresponding experimental evidence, which may help to understand the nature behind. Figure\,\ref{fgr:s5_impurity}a schematically displays the sources of Coulomb impurities at the both interfaces of the FET channels, including gaseous adsorbates (\eg{}\,humidity and oxygen molecules), unsaturated chemical bonds/groups at dielectric surface, and chemical residues during device fabrication. It has been pointed out that the device current is higher in vacuum than in ambient measurement environment (Fig.\,\ref{fgr:s5_impurity}b).\cite{Qiu12APL_Transport:2L,Jariwala13APL}  The gaseous adsorbate such as humidity and oxygen molecules\cite{Qiu12APL_Transport:2L,Late12AN_FET:absorbate:MoS2,Tongay13NL} which results in charge transfer to channel are believed to be an important Coulomb impurity source. The second impurity source can be attributed to chemical residues introduced during device fabrication, as evidenced by the increase in channel current in both exfoliated and CVD devices after performing long-time thermal annealing under ultrahigh vacuum.\cite{Baugher13NL,Schmidt14NL} The dangling bonds, hydroxyl groups and charged ions in the surface of dielectric\cite{Dolui13PRB_CIsourcce:danglingbondandNa}  are believed to be the third type of impurity source since a remarkable current increase is observed in suspended \ms{} channels by etching out the underlying SiO$_2$ dielectric (Fig.\,\ref{fgr:s5_impurity}c).\cite{Jin13JAP_SuspendedMoS2_mu1,Klots14SR_SuspendedMoS2:Photocurrent,Wang15N_FET:MoS:suspended}

Figure\,\ref{fgr:s5_impurity}b--\ref{fgr:s5_impurity}d show the main three schemes developed to suppress the Coulomb impurity sources including 1) thermal desorption at ultrahigh vacuum (Fig.\,\ref{fgr:s5_impurity}b), 2) channel suspension to remove the bottom impurities (Fig.\,\ref{fgr:s5_impurity}c), and 3) channel encapsulation by clean dielectrics such as the vdW dielectric hexagonal boron nitride (hBN) and hydrophobic polymers (\eg{}\,PMMA) (Fig.\,\ref{fgr:s5_impurity}d, only encapsulation at the bottom side is shown).

Thermal desorption is proven one of the most effective methods to improve mobility in chalcogenide. In the high vacuum measurement (2$\times$10$^{-6}$\,Torr), Hersam \et{}\,observed the highest room-temperature mobility of 60--70\,\cmvs{} for pristine (neither encapsulated nor suspended), two-terminal (mobility underestimation due to contact resistance) 1L \ms{},\cite{Jariwala13APL} demonstrating that the gaseous adsorbates are one of the leading scattering centers. Here, it deserves to note the different roles played by the annealing surroundings. Annealing \textit{ex situ} can only influence the contact quality while \textit{in situ} vacuum annealing can further remove gaseous adsorbates. In addition, an ultrahigh vacuum and long-time treatment is particularly critical to complete the adsorbate desorption. \cite{Ovchi14AN_FET:WS,Baugher13NL}

Among the three schemes to suppress interfacial impurities, the first and third are technological viable but the second one is only of scientific interest since a suspended channel would cause severe issues in mechanical stability. Besides, the channel suspension scheme on 2D chalcogenides seems not as successful as graphene. Naively, one expects suspended devices to have higher mobility, due to removal of substrates that introduce trapped charges and other scattering centers. However, the electronic performance of most suspended 1L \mx{} flakes fabricated to date remains much lower than SiO$_2$ supported samples.\cite{Wang15N_FET:MoS:suspended,Klots14SR_SuspendedMoS2:Photocurrent,Jin13JAP_SuspendedMoS2_mu1}   One possible explanation is the higher chemical activity of \mx{} chalcogenides than graphene in the substrate etchant which produces new scattering sites on channels after dielectric etching treatment.\cite{Wang15N_FET:MoS:suspended}  Additionally, the general worse contact in chalcogenides devices could be another factor in limiting the performance of suspended \mx{} flakes. Recently, Lau \et{}\,investigated the effect of thermal annealing condition on the suspended \ms{} samples and they showed remarkable increase in mobility from 46 to 105\,\cmvs{} for a 17 nm thick sample.\cite{Wang15N_FET:MoS:suspended}

Channel encapsulation is always a challenging task due to either the incompatibility of encapsulation materials to processing (\eg{}\,PMMA) or the complexity in fabrication (\eg{}\,hBN). Nonetheless, interface engineering via channel encapsulation has been applied to fabricate high-performance \ms{} FETs. By placing multiple \ms{} flakes on PMMA dielectric and using four-terminal measurements, Fuhrer \et{}\,demonstrated a high mobility value of $\sim$500\,\cmvs{} in $\sim$50 nm \ms{}.\cite{Bao13APL} With a similar dielectric material, Hu \et{}\,observed a room-temperature mobility of 1055\,\cmvs{} in back-gated multilayer InSe,\cite{Feng14AM} being an extremely high value comparable to silicon.

The atomically flat vdW dielectric hBN, which is free of dangling bond and thus is in principle free of trapped charges, has proven to be beneficial for graphene electronics.\cite{dean10nn,dean11np,xue11nm}  It is naturally introduced into the chalcogenide devices. Kim \et{}\,first adopted hBN as dielectric layer in \ms{} FETs.\cite{Lee13AN_BN:u45_3LMoS_thickness} An order of magnitude enhancement in field-effect mobility, from 0.5 to 7.6\,\cmvs{}, is observed in the hBN supported monolayer \ms{} FETs, as compared with the conventional SiO$_2$ supported devices. A high mobility of $\sim$45\,\cmvs{} is measured in a trilayer device. By comparing the mobility trend versus channel thickness, the authors pointed out that there is likely a remarkable contribution from the electrode/channel contacts that limit the mobility because contact resistance decreases with increasing channel thickness. Independently, Tsukagoshi \et{}\,also fabricated hBN supported \ms{} FETs and observed similar behavior of mobility enhancement and thickness dependence.\cite{Chan13N} Moreover, they studied the temperature dependence and revealed that the clean hBN dielectric can suppress the carrier traps at channel/dielectric interface, which converts the carrier transport in monolayer \ms{} from hopping to thermal activation mode. Besides \ms{}, other \mx{} chalcogenides were also explored. The mobility enhancement from 17 to 80\,\cmvs{} is observed in 4L \ws{} after replacing the underlying dielectric from SiO$_2$ to clean hBN.\cite{Withe14SR_FET:WS:BN}

\subsection{Dielectric screening versus RIP scattering} \label{RIP}
Since Coulomb impurity is one of the main scattering mechanisms in FETs, a dielectric engineering, \ie{}, employing high $\kappa$ dielectrics, was proposed to be a strategy to reduce the Coulomb impurity scattering to carriers for the reason that the elevated permittivity of dielectrics is expected to enhance the screening ability to the adverse scattering potentials from internal and external Coulomb impurities.\cite{jena07prl} This strategy would be justifiable, if the adopted high $\kappa$ dielectrics do not bring remote interfacial phonons, and appears working on 2D chalcogenides with low initial mobility. A noticeable performance enhancement was observed in monolayer \ms{} and SnS2  flakes,\cite{Radisavljevic11NN,Song13N} with carrier mobility increasing from few to $\sim$50\,\cmvs{}.

In fact, there is a growing controversy on the exact role of the dielectric environment in suppressing carrier scattering. Early on, it was found that high-$\kappa$ dielectrics lead to low carrier mobility in FETs made of silicon and organic materials because of the carrier scattering of polar phonon at the channel/dielectric interfaces.\cite{moore80jap,fischetti01jap,stass04apl_,veres03afm_,hulea06nm_rip} In high-quality graphene, Geim \et{}\,also pointed out that screening of Coulomb impurities by high-$\kappa$ dielectrics has little effect on mobility.\cite{Ponomarenko09PRL} The same is also true in chalcogenides. Several groups confirmed that the introduction of high-$\kappa$ dielectrics deteriorates, rather than improves, the mobility of high-quality chalcogenide devices.\cite{Kappera14NM,Chuan14NL_Rc:WSe2:IonicLiquid} In \ms{} FETs with high initial mobility, Liao \et{}\,recently observed a mobility degradation of 30-50\% in top HfO$_2$ gated \ms{}, as compared with back SiO$_2$ gated devices.\cite{Wang15S} Conversely, when dielectrics without available RIP modes are used such as polymer parylene, 2--3 fold enhancement in mobility, from $\sim$50 to 100--150\,\cmvs{}, was observed in 10 nm \mse{} samples.\cite{Chamlagain14AN}  All the experimental observations can be well explained by taking into account the adverse RIP scattering accompanied by the use of high-$\kappa$ dielectrics. Ma and Jena \et{}\,found that a FET can gain additional mobility enhancement from dielectric screening when the average permittivity of the top and bottom dielectrics is smaller than a critical value of 10.\cite{Ma14PRX} The adverse RIP scattering would outperform and degrade mobility once the average permittivity is higher than the critical value.

In this sense, the early observation of mobility enhancement after high-$\kappa$ dielectric deposition in devices with low initial mobility is likely due to the facts below: 1) improvement of device contact by thermal annealing during the time-consuming atomic layer deposition (ALD) process for the high $\kappa$ dielectrics; 2) reaction and cleaning effect on the surface gaseous adsorbates (humidity and oxygen, which are one of the main Coulomb impurity source) with the ALD precursors; 3) encapsulation of the conduction channels and consequent insulation from external surroundings.

\subsection{Atomic vacancy healing}
Apart from the long-range charged impurities, short-range scattering factors such as anion vacancy in chalcogenides is also likely one of the leading carrier scattering mechanisms. Hence, it is desired to repair them with a chemical approach based on interface and molecular engineering when contact optimization and impurity scattering are already well addressed.
\begin{figure}[h!bt]
 \centering
 \includegraphics[width=0.5\textwidth]{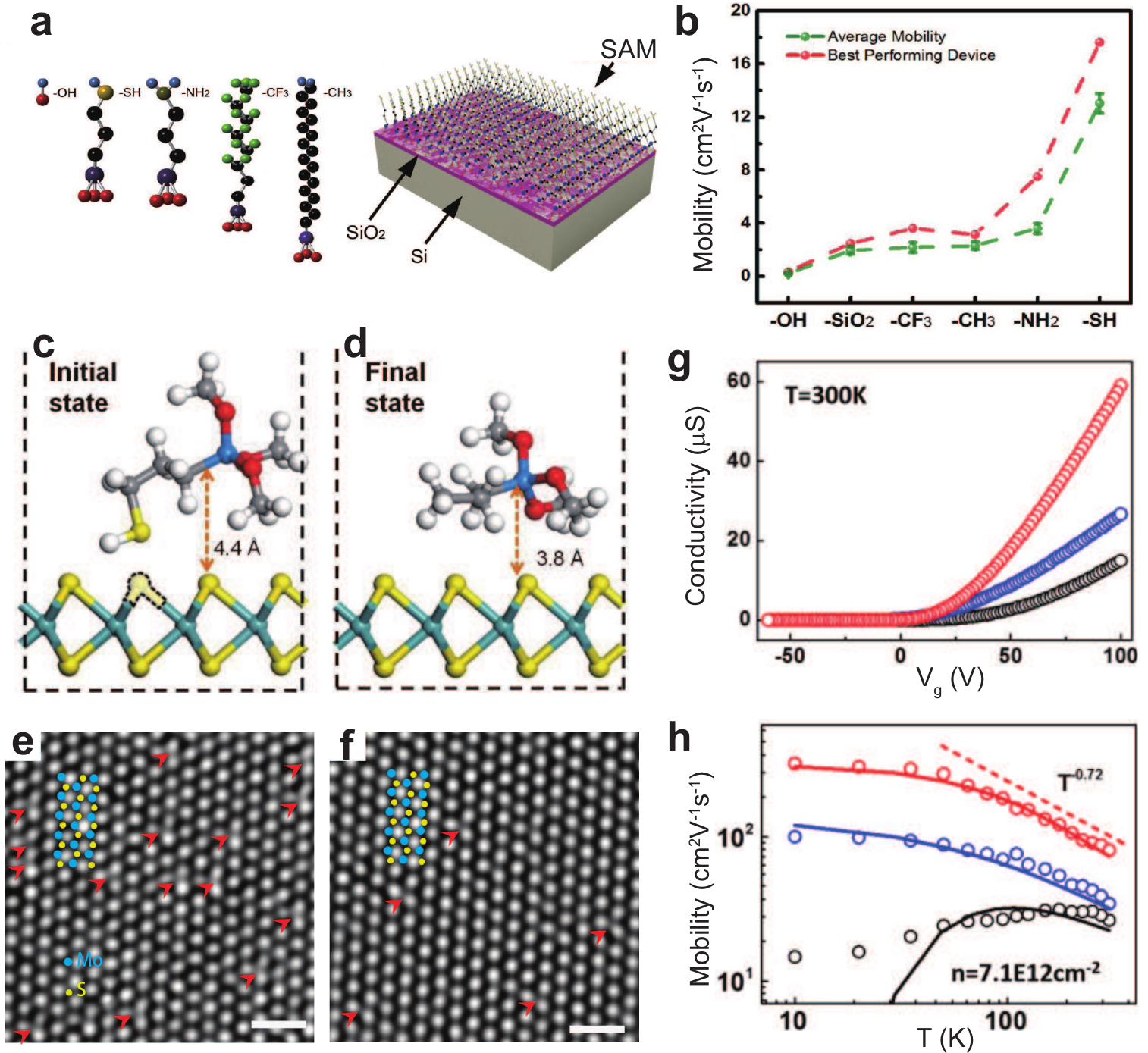}
 \caption{Mobility engineering for atomic vacancy healing. (a) Schematic molecules which can form self assembly monolayer (SAM) onto SiO$_2$ substrates. (b) Comparison on the \ms{} mobility modified by different surfacial chemical groups present at the SiO$_2$ dielectric after SAM modification. Panels (a) and (b) are \cright{Najma14NL_Interface:SAM_mu13}. (c) and (d) show the calculated molecular reaction and layout before and after sulfur vacancy healing. (e) and (f) are the corresponding high-resolution TEM images for (c) and (d) to evaluate the effect of vacancy healing.  (g) Comparison of room-temperature conductivity of samples under different heling conditions. Black: pristine, blue: one-sided vacancy healing, red: double-sided vacancy healing. (h) Temperature dependence of mobility for corresponding samples shown in (g). Panels (c)--(h) are \cright[\npg]{Yu14NC_MoS2_mu80}.}
 \label{fgr:vacancy}
\end{figure}

\begin{figure*}[!bt]
 \centering
 \includegraphics[width=0.9\textwidth]{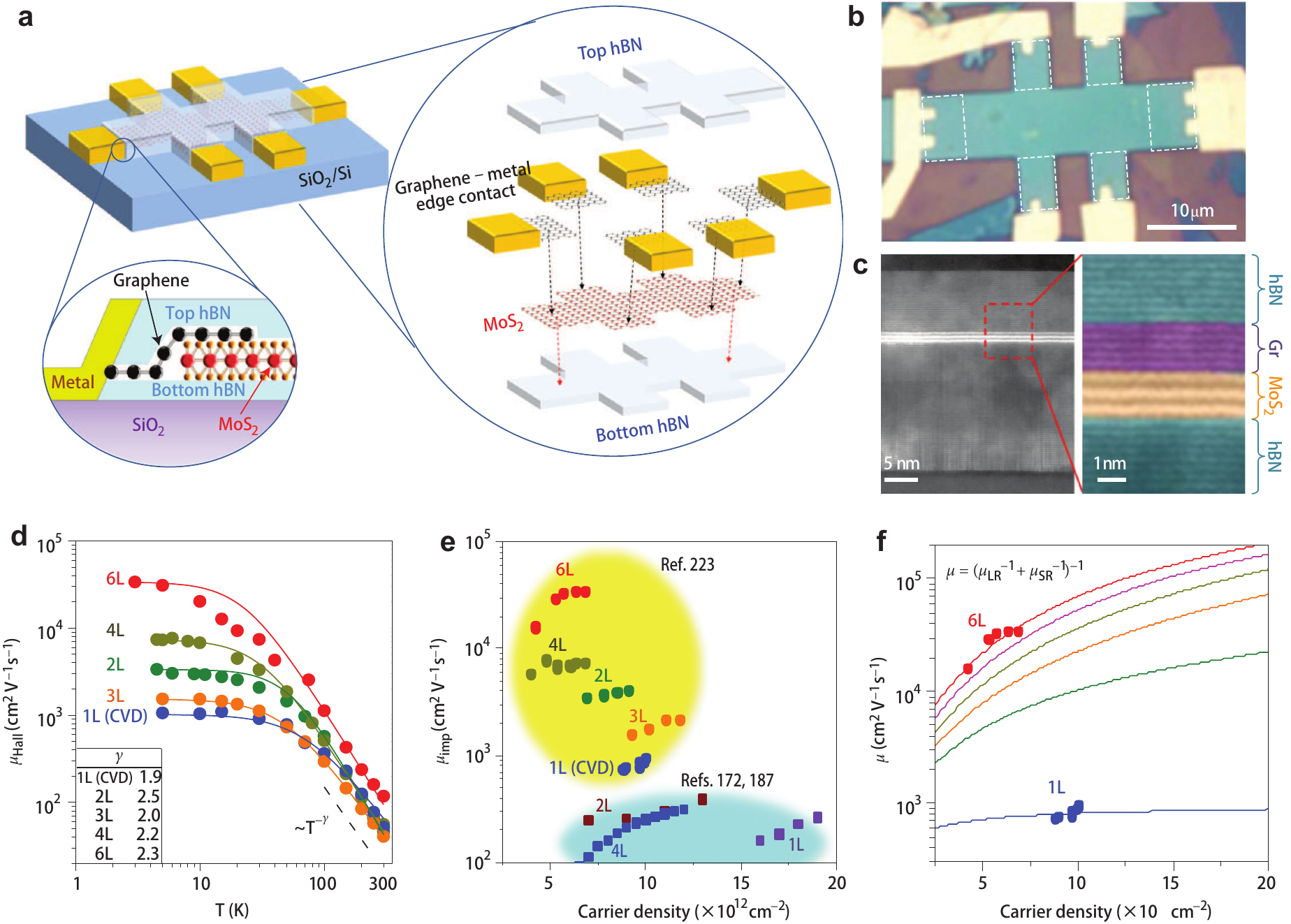}
 \caption{A typical case with combined schemes of mobility engineering. (a) Schematic of the hBN-encapsulated \ms{} multi-terminal device. The exploded view shows the individual components that constitute the heterostructure stack. Bottom: Zoom-in cross-sectional schematic of the metal-graphene-\ms{} contact
region. (b) Optical microscope image of a fabricated device. Graphene contact regions are outlined by dashed lines. (c) Cross-sectional TEM image of the fabricated device. The zoom-in false-colour image clearly shows the ultra-sharp interfaces between different layers (graphene, 5L; \ms{}, 3L; top hBN, 8 nm; bottom hBN, 19 nm). (d) Hall mobility of hBN-encapsulated \ms{} devices (with
different numbers of layers of \ms{}) as a function of temperature. The solid fitting lines are drawn by a combined phonon and impurity scattering model. As a visual guide, the dashed line shows the power law $\mu\sim T^{-\gamma}$, and fitted values of $\gamma$ for each device are listed in the inset table. (e) Impurity-limited mobility ($\mu_\text{imp}$) as a function of the \ms{} carrier density. For comparison, previously reported values from \ms{} on SiO$_2$ substrates (refs. \citenum{Baugher13NL} and \citenum{Neal13AN_FET:MoS2}) are plotted. (f) The solid lines show the theoretically calculated mobility including both long-range (LR) impurity scattering and short-range (SR) defect scattering based on Matthiessen's rule, $\mu^{-1}=\mu_\text{LR}^{-1}+\mu_\text{SR}^{-1}$, as a function of carrier density for 1L to 6L \ms{}. Experimental data are shown as solid circles.
\Cright[\npg]{Cui15NN_FET:G/MoS/BN_mu34k}}
 \label{fgr:MoS_BN}
\end{figure*}

Figure\,\ref{fgr:vacancy} show a typical vacancy healing scheme assisted by interface engineering on dielectric and/or semiconductor with designed functional molecules. Due to the full exposure of sulfur planes, a post molecular surface modification and healing becomes realistic. By introducing a long-chain molecule with chemical group at one end attaching to SiO$_2$ and the other end modifying \ms{} channel, one can form self-assembly monolayer (SAM) at the surface of `dirty' SiO$_2$ dielectric (Fig.\,\ref{fgr:vacancy}a). Lou \et{}\,intentionally modified the chemical surroundings of the conduction channels in FETs with a series of designed SAMs.\cite{Najma14NL_Interface:SAM_mu13}  The higher mobility of $\sim$18\,\cmvs{} was observed in the thiol group (--SH) contacted \ms{} channels than those contacted by other groups of --OH, --SiO$_2$, --CF$_3$, --CH$_3$  and --NH$_2$  (Fig.\,\ref{fgr:vacancy}b), due partially to the vacancy repair effect. They concluded that the mobility improvement is a combined effect of interface-related effects of charge transfer, built-in molecular polarities, varied densities of defects, and remote interfacial phonons.

Later on, with a double-side encapsulation of \ms{} channels by thiol group ended (3-mercaptopropy)trimethoxysilane molecules and appropriate thermal healing, Wang \et{}\,achieved a record high room-temperature mobility of 80\,\cmvs{} in monolayer \ms{}.\cite{Yu14NC_MoS2_mu80} By comparing the high-resolution TEM images for the healed and untreated samples (Figs.\,\ref{fgr:vacancy}e and \ref{fgr:vacancy}f), they revealed that that the density of sulfur vacancies is largely reduced in the healed samples. Meanwhile, a remarkable enhancement in the room-temperature and low-temperature mobility is achieved in the double-side treated samples when compared with the single-side treated or untreated samples. Based on the observation and theoretical calculation, they attributed the large performance improvement as the healing effect on sulfur vacancy by the thiol group during thermal treatment. It worth noting that the above result is not merely a simple vacancy healing effect for the reasons below. Large suppression of Coulomb impurities and trap densities are simultaneously seen in their analysis, indicating the existence of cleaning effect on the channel interfaces after applying the SAM. In other word, the SAM layers can act the same role as hBN encapsulation to isolate the external Coulomb impurity scattering. It would be more reasonable to ascribe the mobility enhancement to the synergetic effects of the interface cleaning and the vacancy healing. Anyway, the proposed molecular healing strategy is a very promising technique to propel the 2D vdW semiconductors to their performance limit.

Apart from the iso-elemental repair technique, Soe \et{}\,also proposed a novel healing technique of surface laser passivation by using a hetero-element of oxygen.\cite{Lu15NL_defect:healing} The basic concept is to passivate the chalcogen vacancies by adsorbed oxygen atoms which can, meanwhile, suppress the midgap states and repair the material electronic structure. They demonstrated that this technique can enhance the conductivity of monolayer \ws{} by 400 times and the photoconductivity by 150 times.

\subsection{state-of-the-art performance} \label{state-of-the-art}
\begin{table*}[h!bt]
\small
  \caption{The state-of-the-art carrier mobility values in 2D chalcogenide \ms{} and \wse{}.}
  \label{tbl:state-mobility}
  \begin{tabular*}{\textwidth}{@{\extracolsep{\fill}}llllllllll}
  \hline
Channel & Channel  & Contact \& & Thermal    & Dielectric \&   & Measurement  & $\mu$@RT     &$\mu$@LT   & $\gamma$ value  & Ref.\\
material& thickness  & doping   & annealing  &encapsulation    & pressure     &\,\cmvs{}        &\,\cmvs{}   & near RT         & no.\\
\hline
\ms{} &	 1L &	Gr. &	 No &	DE:BN&	 vac. &	 $\sim$100 &	 328 (1.9 K) &	 1.2 &	 \citenum{Liu15NL_FET:G/MoS/BN}\\
\ms{} &	 1L &	 Ti/Pd &	 ex. 350\CT{}  &	DE:SAM&	 $\sim$1.3 mPa &	 81 &	 >300 (10 K) &	 0.72 &	\citenum{Yu14NC_MoS2_mu80}\\
\ms{} &	 1L &	 Ti/Pd &	 in. 77\CT{}  &	DE:SAM&	 $\sim$1.3 mPa &	 81 &	 >300 (10 K) &	 0.72 &	\citenum{Yu14NC_MoS2_mu80}\\
\ms{} &	 1L &	 Au &	 No &	BG:SiO$_2$ &	 <0.3 mPa &	 60-70 &	 $\sim$110 ($\sim$5 K) &	 0.62 &	 \citenum{Jariwala13APL} \\
\ms{}-C&	 1L&	Gr.&	 &	DE:BN&	 vac.&	 $\sim$50&	 1020 ($\sim$4K)&	 1.9&	 \citenum{Cui15NN_FET:G/MoS/BN_mu34k} \\
\ms{}-C &	 1L &	 Au &	 in. 120\CT{}  &	BG:SiO$_2$ &	vac. (PPMS)&	 45 &	 $\sim$500 (10 K) &	 0.62 &	 \citenum{Schmidt14NL} \\
\ms{}&	 2L&	Gr.&	 &	DE:BN&	 vac.&	 $\sim$40&	 $\sim$4000 ($\sim$4K)&	 2.5&	 \citenum{Cui15NN_FET:G/MoS/BN_mu34k} \\
\ms{}&	 3L&	Gr.&	 &	DE:BN&	 vac.&	 $\sim$40&	 $\sim$2000 ($\sim$4K)&	 2&	 \citenum{Cui15NN_FET:G/MoS/BN_mu34k} \\
\ms{}&	 4L&	Gr.&	 &	DE:BN&	 vac.&	 $\sim$55&	 $\sim$7000 ($\sim$4K)&	 2.2&	 \citenum{Cui15NN_FET:G/MoS/BN_mu34k} \\
\ms{} &	 5L &	Gr. &	 No &	DE:BN&	 vac. &	 <100 &	 1300($\sim$10 K) &	 N.A. &	 \citenum{Liu15NL_FET:G/MoS/BN}\\
\ms{}&	 6L&	Gr.&	 &	DE:BN&	 vac.&	 $\sim$120&	 34000 ($\sim$4K)&	 2.3&	 \citenum{Cui15NN_FET:G/MoS/BN_mu34k} \\
\wse{} &	 1L &	 Ag &	 in. 170\CT{}  &	BG:SiO$_2$ &	 <0.1 mPa &	 140 &	 N.A. &	 N.A. &	 \citenum{Liu13NL}\\
\wse{}:p{} &	 1L &	 Pd+NO$_2$ &	 N.A. &	 TG: ZrO$_2$ &	 <0.1 mPa &	 250 &	 N.A. &	 N.A. &	 \citenum{Fang12NL}\\
  \hline
  \multicolumn{10}{l}{Abbreviations and notes are same as Table\,\ref{tbl:mobility_typical}.}\\ \hline
\end{tabular*}
\end{table*}

We have showed that the optimization of an individual aspect leads to improved mobility. The highest device performance is attained when multiple schemes are employed. Very recently, Hone \et{}\,carefully tackled both the issues of interfacial impurity and channel/electrode contact with a combined `hBN encapsulation + graphene contact' scheme, as shown in Fig.\,\ref{fgr:MoS_BN}. A superclean interfacial condition is achieved by using a dry transfer method where the \ms{} channel is encapsulated by high-quality vdW hBN dielectrics that act as excellent channel isolator to external Coulomb impurity sources. Furthermore, graphene is used as the contact to channels, which reduces contact considerably. Hall geometry is adopted in the characterization to further eliminate contact issue. With the delicate experimental design, they observed a record high low temperature (<5 K) mobility from 1000 to 34000\,\cmvs{} for 1L--6L \ms{}.\cite{Cui15NN_FET:G/MoS/BN_mu34k} A high room temperature mobility from 45 to 120\,\cmvs{} is also achieved. They conclude that the interfacial scattering centers, including both long-range Coulomb impurities and short-range defects are the limiting scattering mechanisms in the high-quality 2D samples, rather than the scattering centers within the bulk. A ultralow interfacial impurity density of  6$\times10^9$\,cm$^{-2}$ is fitted for their superclean samples.

While the above results are extracted from the four-terminal measurement, Duan \et{}\,recently confirmed the presence of high performance in the two-terminal measurement, \ie{}\,in the practical FET device configuration, with the similar `hBN encapsulation + graphene contact' scheme. They reported a record high RT mobility of $\sim$100\,\cmvs{} and low-T mobility of >300\,\cmvs{} in hBN sandwiched monolayer \ms{}.\cite{Liu15NL_FET:G/MoS/BN} Impressively, extremely high low-temperature two-terminal field-effect mobility of 1300\,\cmvs{} was observed in multilayer samples.

So far, the highest room temperature mobility has been achieved in 1L \mx{} chalcogenides is in \wse{} with a value of 250\,\cmvs{}. In this case, synergetic engineering on contact optimization by NO$_2$ doping and channel encapsulation by ZrO$_2$ top dielectric was employed. To gain a quick on the state-of-the-art carrier mobility results achieved to date, Table\,\ref{tbl:state-mobility} summarizes the works with notably high mobility. Apparently, multiple schemes of mobility engineering are used for most of them, pointing out a clear direction for performance optimization and device design.

\section{Experimental traps and standards} \label{sect:trap}
Modern sciences are highly developed and involve interdisciplinary research activities. As for the research field of 2D vdW crystals, numerous chemists, physicists and materials scientists participated into this extremely active subject and stimulated inspirations. However, the lack of solid training on a specific field tends to cause low-level mistakes. Here we would like to point out some apparent `traps' and emphasize some basic experimental standards one should follow.

\begin{figure}[h!bt]
 \centering
 \includegraphics[width=0.28\textwidth]{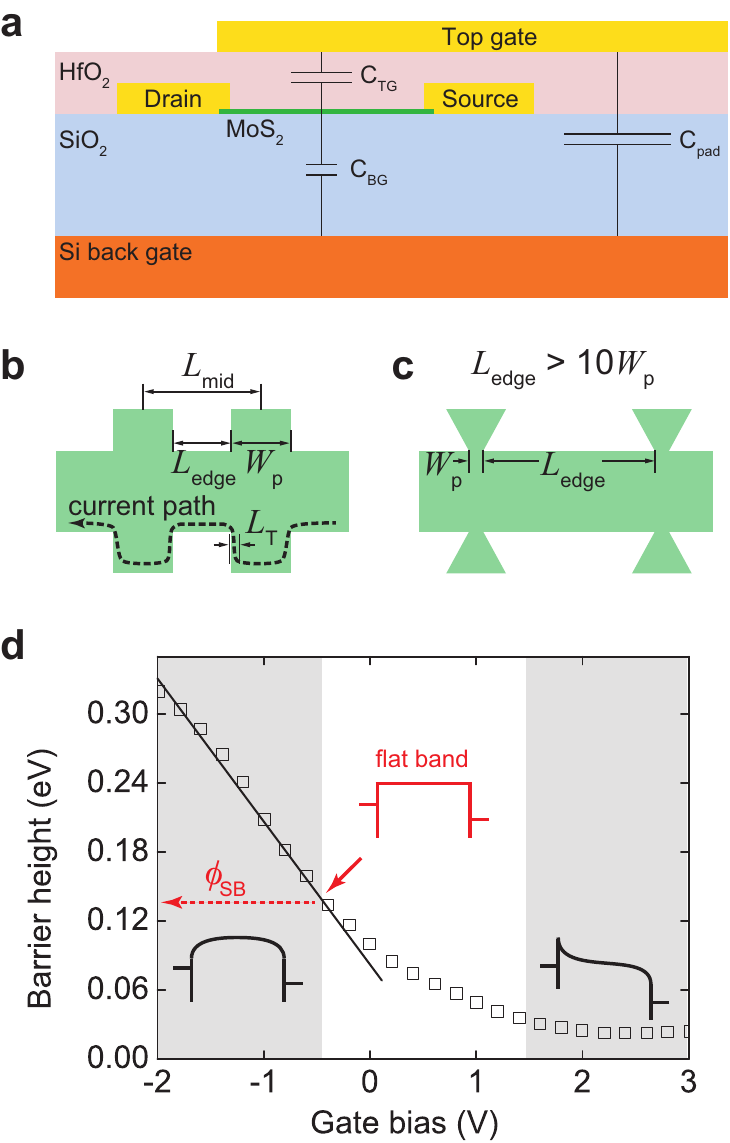}
 \caption{(a) Layout of capacitance distribution in a dual-gate FET. \Cright[\npg]{Fuhrer13NN} (b) A typical four-terminal geometry with a comparable edge distance ($L_\text{edge}$) to the width of voltage probe ($W_\text{p}$), which may cause large overestimation in mobility due to the current crowding effect in the inner voltage probes. (c) A standard four-wire geometry in which $L_\text{edge}$>10\,$W_\text{p}$ such that measurement error can be controlled less than 10\%. (d) Extraction barrier height $\phi_\text{SB}$ at the flat band condition. \Cright{Das13NLb}}
 \label{fgr:trap}
\end{figure}

\subsection{Mobility overestimation in dual-gated structure}
A common trap in early time is the mobility overestimation in a dual-gated FET structure, pointed out by Fuhrer and Hone.\cite{Fuhrer13NN} Figure\,\ref{fgr:trap}a illustrates the layout of capacitance for a dual-gated FET geometry. The capacitive coupling between the back and top gates through the large measurement pads of the top gate can lead to a significant deviation of the back-gate capacitance from its nominal parallel-plate value. In a typical device with a thin (few tens nm) high-$\kappa$ top-gate dielectric, 300 nm bottom gate dielectric, and $\sim$(100\,$\mu$m)$^2$ measurement pads, $C_\text{pad} \gg C_\text{TG} \gg C_\text{BG}$.  If the device is characterized by sweeping the back gate with floating the top-gate electrode, then the effective back-gate capacitance $C=C_\text{BG}+ (C_\text{TG}^{-1}+ C_\text{pad}^{-1})^{-1}\approx C_\text{TG}$. Neglecting this coupling would lead to a mobility overestimation by a factor around $C_\text{TG}/C_\text{BG}$. This hidden trap has led to mobility overestimation in several early reports.\cite{Radisavljevic11NN,Liu12IEDL_b,Wang12NL_Logic:MoS2,Das13NLb} As suggested by Fuhrer and Hone,\cite{Fuhrer13NN}  the correct measurement method for the dual-gated FET structure is to sweep one gate while grounding the other one such that removing the capacitive coupling effect.

\subsection{Four-terminal measurement} \label{trap:4w}
It is often to adopt the four-terminal geometry in electronic characterization to rule out the contact issue, which may also produce experimental artifacts in some cases. The size of the exfoliated 2D vdW flakes is typically small, which often restricts one to define a long enough distance between the inner voltage collection probes. Figure\,\ref{fgr:trap}b shows a typical four-terminal layout used in experiment with comparable probe width and probe distance, which would cause large experimental error. Normally one defines the channel length by calculating the distance between the midpoints ($L_\text{mid}$) rather than between the near edges ($L_\text{edge}$) of the voltage probes. As will be seen, ($L_\text{edge}$) is slightly shorter but more close to the realistic channel length in most cases. Large mobility overestimation ($\sim$2 folds) would arise if the probe width ($W_\text{p}$) is comparable to $L_\text{edge}$, as shown in Fig.\,\ref{fgr:MoS_BN}b. \cite{Cui15NN_FET:G/MoS/BN_mu34k,Pradhan13APL}

Taking into account the current crowding effect discussed in section\,\ref{crowding}, the real channel length should be $L_\text{edge}+2 L_T$, where $L_\text{T}$ is the transfer length. $L_\text{T}$ is typically in range of about 20--600 nm (dependent on channel thickness and gate bias) for the Au or Ti contacted \ms{} \ms{}\cite{Liu14AN,Guo14AN} and it is expected to be much smaller if considering the smaller $\rho_c$ for the graphene/\ms{} contacts. \cite{Cui15NN_FET:G/MoS/BN_mu34k}  Since $L_\text{T}  < L_\text{edge}$, a more accurate channel length should be $L_\text{edge}$  rather than the normally used $L_\text{mid}$ ($\sim$2 $L_\text{edge}$ if $W_\text{p} \sim L_\text{edge}$). In a standard four-terminal measurement, the ratio of $L_\text{edge}$ to $W_\text{p}$ should be greater than 10 to guarantee an experimental error within 10\%, as shown in Figs.\,\ref{fgr:trap}c and \ref{fgr:temperature}a.

\subsection{Barrier height extraction by thermionic emission}
It is frequent to estimate the contact barrier height by using the thermionic emission theory
\begin{equation}
{{I}_{\text{DS}}}=A{{T}^{2}}\exp \left( \frac{q{{\phi }_{\text{SB}}}}{{{k}_{\text{B}}}T} \right)\left( 1-\exp \left( \frac{q{{V}_{\text{DS}}}}{{{k}_{\text{B}}}T} \right) \right),
\end{equation}
which enables an Arrhenius plot  $\ln {{I}_{\text{DS}}}=\ln (A{{T}^{2}})+q({{\phi }_{\text{SB}}}-{{V}_{\text{DS}}})/({{k}_{\text{B}}}T)$ to extract the barrier height at large $V_\text{DS}$. For the two-terminal FETs, the basic assumptions which validate the use of the Arrhenius plot to extract barrier height are: 1) contact dominates the overall device current; and 2) Thermally assisted tunneling current is negligible. The verification of the two assumptions has been neglected in a large number of reports. For instance, the derivation of a negative barrier height of $-$5.7 meV at the permalloy/\ms{} contacts contradicts the first assumption and thus is unreliable.\cite{Wang14SR_Rc:MoS2:Permalloy} It is well known that the barrier is effective for carrier blocking only when it is higher than $3k_BT$. If the extracted barrier height is smaller than $3k_BT$, $\sim$ 80 meV at RT, one has to consider the validity of the original assumptions.

It has been shown that for general long-channel \mx{} devices the contact resistance comprises 10--20\% of the total resistance,\cite{Guo14AN} indicating the inapplicability to apply this method in most devices. To make the first assumption valid, one has to use an ultrashort channel (\eg{}\,50\,nm long) such that the contact resistance exceeds the channel resistance. Another way to avoid the first assumption is to directly use the net contact results extracted from the four-terminal\cite{Radis13NN_CorrectionCg,Cui15NN_FET:G/MoS/BN_mu34k} or transfer-line measurement\cite{Li14AN_MoS2:Contact} (Fig.\,\ref{fgr:barrier}a).

In addition, the current injection is controlled by two components: thermionic and tunneling. The ratio between them varies with the carrier density (\ie{}\,gate bias). At high carrier density, the channel current actually comprises a high ratio of the tunneling current, which would lead to underestimation of barrier height by using the Arrhenius plot. One may also notice that the extracted barrier value highly depends on gate bias, a signature of the involvement of tunneling current (since it is gate-bias dependent). In this sense, the extracted barrier value at the high current regime can at most be viewed as an effective parameter.

In order to suppress the tunneling component, one has to tune the device to the flat band condition. Figure\,\ref{fgr:trap}d shows the extracted harrier height versus gate bias. Apparently, the presence of tunneling current results in underestimation of the barrier height. As suggested by Appenzeller \et{},\cite{Das13NLb} a more accurate way is to plot the effective barrier height versus gate voltage and extract the turning point between the sublinear and linear regime. The value at the turning point can be more reliably adopted as the barrier height than those extracted in the conventional way.

\section{Summary and outlook} \label{sect:summary}

The last few years have witnessed the significant progress on the electronic performance of 2D vdW semiconductors. With fundamental understanding in the transport physics and developing delicate mobility engineering strategies, their room temperature mobility increases rapidly from few to few hundred\,\cmvs{}. While most mobility engineering efforts are devoted to device fabrication that rely on physical methods (\ie{}, thermal annealing, energy level matching, hBN encapsulation, graphene contact), the chemical strategies concerning interfacial and molecular engineering remains highly unexplored.

Interface modification by growing self-assembled monolayers (SAMs) on dielectric supports\cite{Yu14NC_MoS2_mu80,Najma14NL_Interface:SAM_mu13} is a very promising method to create structurally well-defined surfaces with controlled chemical and physical properties like composition, surface energy, hydrophobicity, etc. The use of silane chemistry allows to chemisorb alkyl substituted molecules on SiO$_2$ or other oxide dielectrics and to have an atomically smooth SAMs surfaces as \textit{ad-hoc} support for the 2D channels in the interelectrodic channel, which is expected to minimize the density of interfacial impurities and thus the extrinsic impurity scattering. Besides, by virtue of the molecular design and engineering, more advanced functional molecules and convenient defect healing techniques are to be exploited, in order to repair the high-density chalcogen vacancies in the channels. The importance of molecular engineering is also reflected by the requirement for degenerate doping at contact area. A long-term stable doping onto the ultrathin channels is the key for realizing practical high-performance devices. In the framework of holistic mobility engineering strategies, we expect that the electronic performance of the 2D vdW semiconductors will be further propelled to the level of their intrinsic behavior. In the regards, the molecular and interfacial chemistry is a key direction to be exploited for realizing the atomic electronics.

We then remark on the potential of 2D chalcogenides in the atomic electronics in terms of the electronic performance. In deeply downscaled bulk silicon FETs, the channels are heavily doped to reduce the depletion length and thus the channel mobility degrades accordingly due to the scattering from high-density dopants insides. The hole and electron mobility is only about 200--300\,\cmvs{}.\cite{Xu12IEDL_} In FETs made of ultrathin 2D channels, they operate in a fully depleted mode and only lightly doping is required. No considerable mobility degradation is expected to occur. From the point view of engineering, any 2D semiconductors could be electronically favorable in constructing FETs if the mobility is higher than or approaches the value of heavily doped silicon FETs. Recently a high hole mobility of 250\,\cmvs{} has been reached in monolayer \wse{} at room temperature,\cite{Fang12NL} indicating that the 2D semiconductors are electronically promising for next-generation microelectronics.

Apart from the applications in integrated circuits, the constant improvement on mobility would also benefit the 2D vdW semiconductors for other applications where the thickness of monolayer is not necessarily required, such as for the radio frequency circuits and the driving FET arrays in flat-panel displays.\cite{Kim12NC} For instance, from multilayer \ms{} flakes it is easy to achieve a high mobility of 100\,\cmvs{} that is generally higher than the prevailing amorphous silicon, InGaZnO,\cite{Nomura03S,Nomura04N,Fortunato12AM,Kamiya10STAM} and actively developed organic materials.

Furthermore, the elevated electronic performance would be favorable for broad potential applications in optoelectronics such as monolayer light emitting diodes, photodetectors and gas sensors. In general, higher mobility would allow for higher operating current and device performance. In particular if molecules with self-assembly ability on the vdW semiconductors can be developed, novel superlattices with periodic molecule/vdW material structure and designed functions would be possible. In this sense, the mobility engineering can extend the application fields of the 2D vdW semiconductors. With constant mobility and molecular engineering, we believe that the 2D vdW would generate numerous novel applications in near future.

In conclusion, we reviewed recent progress on the charge transport properties of FETs based on 2D chalcogenide semiconductors, in particular by unraveling the role of thickness on their carrier mobility. We discussed the physical origins and strategies devised for mobility engineering, with the ultimate goal of developing device with performance beyond the state-of-the-art. Specifically, various Coulomb impurities including gaseous adsorbates, dangling bonds/chemical groups of dielectric and other surficial residues are the main scattering centers. Contact quality also plays a  role as crucial as Coulomb impurities that affects mobility. Besides, vacancy healing could be used as an additional strategy to further improve device performance when both the contact and interface surroundings are optimized. The synergistic improvement of fundamental physico-chemical properties of 2D chalcogenides (chemical composition, spatial distribution and nature of structural and electronic defects, etc) and their interfacing with chemically optimized dielectric supports and functionalized electrodes (to suppress scattering centers, to tailor their environment, and to optimize contact resistance via lowering injection barrier), is the best solution to improves carrier mobility. We conclude that all the above adverse transport factors have to be optimized or suppressed in order to achieve technologically viable atomically thin body FETs and other novel (opto)electronic devices. The review sheds an in-depth light onto the charge transport behavior of the 2D semiconductors and would guide future performance optimization and device design.

\section*{Acknowledgments}
We acknowledge financial support from the European Commission through the Graphene Flagship (GA-604391), the FET project UPGRADE (No.\,309056), the Agence Nationale de la Recherche through the LabEx project Nanostructures in Interaction with their Environment (ANR-11-LABX-0058\_NIE), the Japan Society for the Promotion of Science (JSPS) through the Grant-in-Aid for Scientific Research (Kakenhi No.\,25107004), and the International Center for Frontier Research in Chemistry (icFRC).


\begin{mcitethebibliography}{299}
\providecommand*{\natexlab}[1]{#1}
\providecommand*{\mciteSetBstSublistMode}[1]{}
\providecommand*{\mciteSetBstMaxWidthForm}[2]{}
\providecommand*{\mciteBstWouldAddEndPuncttrue}
  {\def\EndOfBibitem{\unskip.}}
\providecommand*{\mciteBstWouldAddEndPunctfalse}
  {\let\EndOfBibitem\relax}
\providecommand*{\mciteSetBstMidEndSepPunct}[3]{}
\providecommand*{\mciteSetBstSublistLabelBeginEnd}[3]{}
\providecommand*{\EndOfBibitem}{}
\mciteSetBstSublistMode{f}
\mciteSetBstMaxWidthForm{subitem}
{(\emph{\alph{mcitesubitemcount}})}
\mciteSetBstSublistLabelBeginEnd{\mcitemaxwidthsubitemform\space}
{\relax}{\relax}

\bibitem[Whittingham(1978)]{Whitt78PiSSC_Review:chalcogenide}
M.~Whittingham, \emph{Prog. Solid State Chem.}, 1978, \textbf{12}, 41--99\relax
\mciteBstWouldAddEndPuncttrue
\mciteSetBstMidEndSepPunct{\mcitedefaultmidpunct}
{\mcitedefaultendpunct}{\mcitedefaultseppunct}\relax
\EndOfBibitem
\bibitem[Whittingham(2004)]{Whitt04CR_Review}
M.~Whittingham, \emph{Chem. Rev.}, 2004, \textbf{104}, 4271--4301\relax
\mciteBstWouldAddEndPuncttrue
\mciteSetBstMidEndSepPunct{\mcitedefaultmidpunct}
{\mcitedefaultendpunct}{\mcitedefaultseppunct}\relax
\EndOfBibitem
\bibitem[Novoselov \emph{et~al.}(2004)Novoselov, Geim, Morozov, Jiang, Zhang,
  Dubonos, Grigorieva, and Firsov]{Novoselov04S}
K.~Novoselov, A.~Geim, S.~Morozov, D.~Jiang, Y.~Zhang, S.~Dubonos,
  I.~Grigorieva and A.~Firsov, \emph{Science}, 2004, \textbf{306},
  666--669\relax
\mciteBstWouldAddEndPuncttrue
\mciteSetBstMidEndSepPunct{\mcitedefaultmidpunct}
{\mcitedefaultendpunct}{\mcitedefaultseppunct}\relax
\EndOfBibitem
\bibitem[Novoselov \emph{et~al.}(2005)Novoselov, Geim, Morozov, Jiang,
  Katsnelson, Grigorieva, Dubonos, and Firsov]{Novoselov05N}
K.~Novoselov, A.~Geim, S.~Morozov, D.~Jiang, M.~Katsnelson, I.~Grigorieva,
  S.~Dubonos and A.~Firsov, \emph{Nature}, 2005, \textbf{438}, 197--200\relax
\mciteBstWouldAddEndPuncttrue
\mciteSetBstMidEndSepPunct{\mcitedefaultmidpunct}
{\mcitedefaultendpunct}{\mcitedefaultseppunct}\relax
\EndOfBibitem
\bibitem[Zhang \emph{et~al.}(2005)Zhang, Tan, Stormer, and Kim]{Zhang05N}
Y.~Zhang, Y.~Tan, H.~Stormer and P.~Kim, \emph{Nature}, 2005, \textbf{438},
  201--204\relax
\mciteBstWouldAddEndPuncttrue
\mciteSetBstMidEndSepPunct{\mcitedefaultmidpunct}
{\mcitedefaultendpunct}{\mcitedefaultseppunct}\relax
\EndOfBibitem
\bibitem[Geim and Novoselov(2007)]{Geim07NM}
A.~K. Geim and K.~S. Novoselov, \emph{Nat. Mater.}, 2007, \textbf{6},
  183--191\relax
\mciteBstWouldAddEndPuncttrue
\mciteSetBstMidEndSepPunct{\mcitedefaultmidpunct}
{\mcitedefaultendpunct}{\mcitedefaultseppunct}\relax
\EndOfBibitem
\bibitem[Geim(2009)]{Geim09S}
A.~K. Geim, \emph{Science}, 2009, \textbf{324}, 1530--1534\relax
\mciteBstWouldAddEndPuncttrue
\mciteSetBstMidEndSepPunct{\mcitedefaultmidpunct}
{\mcitedefaultendpunct}{\mcitedefaultseppunct}\relax
\EndOfBibitem
\bibitem[Wilson and Yoffe(1969)]{Wilson69AP}
J.~A. Wilson and A.~D. Yoffe, \emph{Adv. Phys.}, 1969, \textbf{18},
  193--335\relax
\mciteBstWouldAddEndPuncttrue
\mciteSetBstMidEndSepPunct{\mcitedefaultmidpunct}
{\mcitedefaultendpunct}{\mcitedefaultseppunct}\relax
\EndOfBibitem
\bibitem[Yoffe(2002)]{Yoffe02AP_Review}
A.~Yoffe, \emph{Adv. Phys.}, 2002, \textbf{51}, 799--890\relax
\mciteBstWouldAddEndPuncttrue
\mciteSetBstMidEndSepPunct{\mcitedefaultmidpunct}
{\mcitedefaultendpunct}{\mcitedefaultseppunct}\relax
\EndOfBibitem
\bibitem[Mas-Balleste \emph{et~al.}(2011)Mas-Balleste, Gomez-Navarro,
  Gomez-Herrero, and Zamora]{Mas-B11N_Review}
R.~Mas-Balleste, C.~Gomez-Navarro, J.~Gomez-Herrero and F.~Zamora,
  \emph{Nanoscale}, 2011, \textbf{3}, 20--30\relax
\mciteBstWouldAddEndPuncttrue
\mciteSetBstMidEndSepPunct{\mcitedefaultmidpunct}
{\mcitedefaultendpunct}{\mcitedefaultseppunct}\relax
\EndOfBibitem
\bibitem[Neto and Novoselov(2011)]{Neto11RPP_Review}
A.~H.~C. Neto and K.~Novoselov, \emph{Rep. Prog. Phys.}, 2011, \textbf{74},
  082501\relax
\mciteBstWouldAddEndPuncttrue
\mciteSetBstMidEndSepPunct{\mcitedefaultmidpunct}
{\mcitedefaultendpunct}{\mcitedefaultseppunct}\relax
\EndOfBibitem
\bibitem[Hewak(2011)]{Hewak11NP_Review}
D.~Hewak, \emph{Nat. Photonics}, 2011, \textbf{5}, 474\relax
\mciteBstWouldAddEndPuncttrue
\mciteSetBstMidEndSepPunct{\mcitedefaultmidpunct}
{\mcitedefaultendpunct}{\mcitedefaultseppunct}\relax
\EndOfBibitem
\bibitem[Wang \emph{et~al.}(2012)Wang, Kalantar-Zadeh, Kis, Coleman, and
  Strano]{Wang12NN}
Q.~H. Wang, K.~Kalantar-Zadeh, A.~Kis, J.~N. Coleman and M.~S. Strano,
  \emph{Nat. Nanotechnol.}, 2012, \textbf{7}, 699--712\relax
\mciteBstWouldAddEndPuncttrue
\mciteSetBstMidEndSepPunct{\mcitedefaultmidpunct}
{\mcitedefaultendpunct}{\mcitedefaultseppunct}\relax
\EndOfBibitem
\bibitem[Butler \emph{et~al.}(2013)Butler, Hollen, Cao, Cui, Gupta,
  Guti\'{e}rrez, Heinz, Hong, Huang, Ismach, Johnston-Halperin, Kuno,
  Plashnitsa, Robinson, Ruoff, Salahuddin, Shan, Shi, Spencer, Terrones, Windl,
  and Goldberger]{Butle13AN_Review:2D}
S.~Z. Butler, S.~M. Hollen, L.~Cao, Y.~Cui, J.~A. Gupta, H.~R. Guti\'{e}rrez,
  T.~F. Heinz, S.~S. Hong, J.~Huang, A.~F. Ismach, E.~Johnston-Halperin,
  M.~Kuno, V.~V. Plashnitsa, R.~D. Robinson, R.~S. Ruoff, S.~Salahuddin,
  J.~Shan, L.~Shi, M.~G. Spencer, M.~Terrones, W.~Windl and J.~E. Goldberger,
  \emph{ACS Nano}, 2013, \textbf{7}, 2898--2926\relax
\mciteBstWouldAddEndPuncttrue
\mciteSetBstMidEndSepPunct{\mcitedefaultmidpunct}
{\mcitedefaultendpunct}{\mcitedefaultseppunct}\relax
\EndOfBibitem
\bibitem[Xu \emph{et~al.}(2013)Xu, Liang, Shi, and Chen]{Xu13CR}
M.~Xu, T.~Liang, M.~Shi and H.~Chen, \emph{Chem. Rev.}, 2013, \textbf{113},
  3766--3798\relax
\mciteBstWouldAddEndPuncttrue
\mciteSetBstMidEndSepPunct{\mcitedefaultmidpunct}
{\mcitedefaultendpunct}{\mcitedefaultseppunct}\relax
\EndOfBibitem
\bibitem[Chhowalla \emph{et~al.}(2013)Chhowalla, Shin, Eda, Li, Loh, and
  Zhang]{Chhowalla13NC}
M.~Chhowalla, H.~S. Shin, G.~Eda, L.-J. Li, K.~P. Loh and H.~Zhang, \emph{Nat.
  Chem.}, 2013, \textbf{5}, 263--275\relax
\mciteBstWouldAddEndPuncttrue
\mciteSetBstMidEndSepPunct{\mcitedefaultmidpunct}
{\mcitedefaultendpunct}{\mcitedefaultseppunct}\relax
\EndOfBibitem
\bibitem[Pu \emph{et~al.}(2014)Pu, Li, and Takenobu]{Pu14PCCP_Review:flexible}
J.~Pu, L.-J. Li and T.~Takenobu, \emph{Phys. Chem. Chem. Phys.}, 2014,
  \textbf{16}, 14996--15006\relax
\mciteBstWouldAddEndPuncttrue
\mciteSetBstMidEndSepPunct{\mcitedefaultmidpunct}
{\mcitedefaultendpunct}{\mcitedefaultseppunct}\relax
\EndOfBibitem
\bibitem[Kaul(2014)]{Kaul14JMR_Review:2D}
A.~B. Kaul, \emph{J. Mater. Res.}, 2014, \textbf{29}, 348--361\relax
\mciteBstWouldAddEndPuncttrue
\mciteSetBstMidEndSepPunct{\mcitedefaultmidpunct}
{\mcitedefaultendpunct}{\mcitedefaultseppunct}\relax
\EndOfBibitem
\bibitem[Fiori \emph{et~al.}(2014)Fiori, Bonaccorso, Iannaccone, Palacios,
  Neumaier, Seabaugh, Banerjee, and Colombo]{Fiori14NN_review:2D_electronics}
G.~Fiori, F.~Bonaccorso, G.~Iannaccone, T.~Palacios, D.~Neumaier, A.~Seabaugh,
  S.~K. Banerjee and L.~Colombo, \emph{Nat. Nanotechnol.}, 2014, \textbf{9},
  768--779\relax
\mciteBstWouldAddEndPuncttrue
\mciteSetBstMidEndSepPunct{\mcitedefaultmidpunct}
{\mcitedefaultendpunct}{\mcitedefaultseppunct}\relax
\EndOfBibitem
\bibitem[Pumera \emph{et~al.}(2014)Pumera, Sofer, and
  Ambrosi]{Pumer14JMCA_Review:chalcogenide}
M.~Pumera, Z.~Sofer and A.~Ambrosi, \emph{J. Mater. Chem. A}, 2014, \textbf{2},
  8981--8987\relax
\mciteBstWouldAddEndPuncttrue
\mciteSetBstMidEndSepPunct{\mcitedefaultmidpunct}
{\mcitedefaultendpunct}{\mcitedefaultseppunct}\relax
\EndOfBibitem
\bibitem[Jariwala \emph{et~al.}(2014)Jariwala, Sangwan, Lauhon, Marks, and
  Hersam]{Jariwala14AN}
D.~Jariwala, V.~K. Sangwan, L.~J. Lauhon, T.~J. Marks and M.~C. Hersam,
  \emph{ACS Nano}, 2014, \textbf{8}, 1102--1120\relax
\mciteBstWouldAddEndPuncttrue
\mciteSetBstMidEndSepPunct{\mcitedefaultmidpunct}
{\mcitedefaultendpunct}{\mcitedefaultseppunct}\relax
\EndOfBibitem
\bibitem[Ganatra and Zhang(2014)]{Ganat14An_Review:MoS2}
R.~Ganatra and Q.~Zhang, \emph{ACS Nano}, 2014, \textbf{8}, 4074--4099\relax
\mciteBstWouldAddEndPuncttrue
\mciteSetBstMidEndSepPunct{\mcitedefaultmidpunct}
{\mcitedefaultendpunct}{\mcitedefaultseppunct}\relax
\EndOfBibitem
\bibitem[Schwierz \emph{et~al.}(2015)Schwierz, Pezoldt, and Granzner]{Schwi15N}
F.~Schwierz, J.~Pezoldt and R.~Granzner, \emph{Nanoscale}, 2015, \textbf{7},
  8261--8283\relax
\mciteBstWouldAddEndPuncttrue
\mciteSetBstMidEndSepPunct{\mcitedefaultmidpunct}
{\mcitedefaultendpunct}{\mcitedefaultseppunct}\relax
\EndOfBibitem
\bibitem[Duan \emph{et~al.}(2015)Duan, Wang, Pan, Yu, and Duan]{Duan15CSR}
X.~Duan, C.~Wang, A.~Pan, R.~Yu and X.~Duan, \emph{Chem. Soc. Rev.}, 2015,
  10.1039/C5CS00507H\relax
\mciteBstWouldAddEndPuncttrue
\mciteSetBstMidEndSepPunct{\mcitedefaultmidpunct}
{\mcitedefaultendpunct}{\mcitedefaultseppunct}\relax
\EndOfBibitem
\bibitem[Najmaei \emph{et~al.}(2015)Najmaei, Yuan, Zhang, Ajayan, and
  Lou]{Najma15ACR_review:defect}
S.~Najmaei, J.~Yuan, J.~Zhang, P.~Ajayan and J.~Lou, \emph{Acc. Chem. Res.},
  2015, \textbf{48}, 31--40\relax
\mciteBstWouldAddEndPuncttrue
\mciteSetBstMidEndSepPunct{\mcitedefaultmidpunct}
{\mcitedefaultendpunct}{\mcitedefaultseppunct}\relax
\EndOfBibitem
\bibitem[Lembke \emph{et~al.}(2015)Lembke, Bertolazzi, and
  Kis]{Lembk15ACR_Review:2Delectronics}
D.~Lembke, S.~Bertolazzi and A.~Kis, \emph{Acc. Chem. Res.}, 2015, \textbf{48},
  100--110\relax
\mciteBstWouldAddEndPuncttrue
\mciteSetBstMidEndSepPunct{\mcitedefaultmidpunct}
{\mcitedefaultendpunct}{\mcitedefaultseppunct}\relax
\EndOfBibitem
\bibitem[Shi \emph{et~al.}(2015)Shi, Li, and Li]{Shi15CSR_Review:synthesis}
Y.~Shi, H.~Li and L.-J. Li, \emph{Chem. Soc. Rev.}, 2015, \textbf{44},
  2744--2756\relax
\mciteBstWouldAddEndPuncttrue
\mciteSetBstMidEndSepPunct{\mcitedefaultmidpunct}
{\mcitedefaultendpunct}{\mcitedefaultseppunct}\relax
\EndOfBibitem
\bibitem[Ji \emph{et~al.}(2015)Ji, Zhang, Zhang, and Liu]{Ji15CSR_}
Q.~Ji, Y.~Zhang, Y.~Zhang and Z.~Liu, \emph{Chem. Soc. Rev.}, 2015,
  \textbf{44}, 2587--2602\relax
\mciteBstWouldAddEndPuncttrue
\mciteSetBstMidEndSepPunct{\mcitedefaultmidpunct}
{\mcitedefaultendpunct}{\mcitedefaultseppunct}\relax
\EndOfBibitem
\bibitem[Wang \emph{et~al.}(2015)Wang, Yuan, Sae~Hong, Li, and Cui]{Wang15CSR_}
H.~Wang, H.~Yuan, S.~Sae~Hong, Y.~Li and Y.~Cui, \emph{Chem. Soc. Rev.}, 2015,
  \textbf{44}, 2664--2680\relax
\mciteBstWouldAddEndPuncttrue
\mciteSetBstMidEndSepPunct{\mcitedefaultmidpunct}
{\mcitedefaultendpunct}{\mcitedefaultseppunct}\relax
\EndOfBibitem
\bibitem[Wang \emph{et~al.}(2015)Wang, Wang, Wang, Wang, Yin, Xu, Huang, and
  He]{Wang15N}
F.~Wang, Z.~X. Wang, Q.~S. Wang, F.~M. Wang, L.~Yin, K.~Xu, Y.~Huang and J.~He,
  \emph{Nanotechnol.}, 2015, \textbf{26}, 292001\relax
\mciteBstWouldAddEndPuncttrue
\mciteSetBstMidEndSepPunct{\mcitedefaultmidpunct}
{\mcitedefaultendpunct}{\mcitedefaultseppunct}\relax
\EndOfBibitem
\bibitem[Ferrari \emph{et~al.}(2015)Ferrari, Bonaccorso, Fal{'}ko, Novoselov,
  Roche, Boggild, Borini, Koppens, Palermo, Pugno, Garrido, Sordan, Bianco,
  Ballerini, Prato, Lidorikis, Kivioja, Marinelli, Ryhanen, Morpurgo, Coleman,
  Nicolosi, Colombo, Fert, Garcia-Hernandez, Bachtold, Schneider, Guinea,
  Dekker, Barbone, Sun, Galiotis, Grigorenko, Konstantatos, Kis, Katsnelson,
  Vandersypen, Loiseau, Morandi, Neumaier, Treossi, Pellegrini, Polini,
  Tredicucci, Williams, Hee~Hong, Ahn, Min~Kim, Zirath, van Wees, van~der Zant,
  Occhipinti, Di~Matteo, Kinloch, Seyller, Quesnel, Feng, Teo, Rupesinghe,
  Hakonen, Neil, Tannock, Lofwander, and Kinaret]{Ferra15N_Review:2D:roadmap}
A.~C. Ferrari, F.~Bonaccorso, V.~Fal{'}ko, K.~S. Novoselov, S.~Roche,
  P.~Boggild, S.~Borini, F.~H.~L. Koppens, V.~Palermo, N.~Pugno, J.~A. Garrido,
  R.~Sordan, A.~Bianco, L.~Ballerini, M.~Prato, E.~Lidorikis, J.~Kivioja,
  C.~Marinelli, T.~Ryhanen, A.~Morpurgo, J.~N. Coleman, V.~Nicolosi,
  L.~Colombo, A.~Fert, M.~Garcia-Hernandez, A.~Bachtold, G.~F. Schneider,
  F.~Guinea, C.~Dekker, M.~Barbone, Z.~Sun, C.~Galiotis, A.~N. Grigorenko,
  G.~Konstantatos, A.~Kis, M.~Katsnelson, L.~Vandersypen, A.~Loiseau,
  V.~Morandi, D.~Neumaier, E.~Treossi, V.~Pellegrini, M.~Polini, A.~Tredicucci,
  G.~M. Williams, B.~Hee~Hong, J.-H. Ahn, J.~Min~Kim, H.~Zirath, B.~J. van
  Wees, H.~van~der Zant, L.~Occhipinti, A.~Di~Matteo, I.~A. Kinloch,
  T.~Seyller, E.~Quesnel, X.~Feng, K.~Teo, N.~Rupesinghe, P.~Hakonen, S.~R.~T.
  Neil, Q.~Tannock, T.~Lofwander and J.~Kinaret, \emph{Nanoscale}, 2015,
  \textbf{7}, 4598--4810\relax
\mciteBstWouldAddEndPuncttrue
\mciteSetBstMidEndSepPunct{\mcitedefaultmidpunct}
{\mcitedefaultendpunct}{\mcitedefaultseppunct}\relax
\EndOfBibitem
\bibitem[Schmidt \emph{et~al.}(2015)Schmidt, Giustiniano, and Eda]{Schmi15CSR}
H.~Schmidt, F.~Giustiniano and G.~Eda, \emph{Chem. Soc. Rev.}, 2015,
  \textbf{44}, 7715--7736\relax
\mciteBstWouldAddEndPuncttrue
\mciteSetBstMidEndSepPunct{\mcitedefaultmidpunct}
{\mcitedefaultendpunct}{\mcitedefaultseppunct}\relax
\EndOfBibitem
\bibitem[Peng \emph{et~al.}(2015)Peng, Ang, and Loh]{Peng15NT_Review:PV:MoS}
B.~Peng, P.~K. Ang and K.~P. Loh, \emph{Nano Today}, 2015, \textbf{10},
  128--137\relax
\mciteBstWouldAddEndPuncttrue
\mciteSetBstMidEndSepPunct{\mcitedefaultmidpunct}
{\mcitedefaultendpunct}{\mcitedefaultseppunct}\relax
\EndOfBibitem
\bibitem[Zeng and Cui(2015)]{Zeng15CSR}
H.~Zeng and X.~Cui, \emph{Chem. Soc. Rev.}, 2015, \textbf{44}, 2629--2642\relax
\mciteBstWouldAddEndPuncttrue
\mciteSetBstMidEndSepPunct{\mcitedefaultmidpunct}
{\mcitedefaultendpunct}{\mcitedefaultseppunct}\relax
\EndOfBibitem
\bibitem[Zhang \emph{et~al.}(2015)Zhang, Qiao, Shi, Wu, Jiang, and
  Tan]{Zhang15CSR_Review:PhononRaman}
X.~Zhang, X.-F. Qiao, W.~Shi, J.-B. Wu, D.-S. Jiang and P.-H. Tan, \emph{Chem.
  Soc. Rev.}, 2015, \textbf{44}, 2757--2785\relax
\mciteBstWouldAddEndPuncttrue
\mciteSetBstMidEndSepPunct{\mcitedefaultmidpunct}
{\mcitedefaultendpunct}{\mcitedefaultseppunct}\relax
\EndOfBibitem
\bibitem[Buscema \emph{et~al.}(2015)Buscema, Island, Groenendijk, Blanter,
  Steele, van~der Zant, and Castellanos-Gomez]{Busce15CSR_Review:Photo}
M.~Buscema, J.~O. Island, D.~J. Groenendijk, S.~I. Blanter, G.~A. Steele,
  H.~S.~J. van~der Zant and A.~Castellanos-Gomez, \emph{Chem. Soc. Rev.}, 2015,
  \textbf{44}, 3691--3718\relax
\mciteBstWouldAddEndPuncttrue
\mciteSetBstMidEndSepPunct{\mcitedefaultmidpunct}
{\mcitedefaultendpunct}{\mcitedefaultseppunct}\relax
\EndOfBibitem
\bibitem[Al-Alamy and Balchin(1977)]{Al-Al77JCG_synthesis:SnS:SnSe:TaS}
F.~Al-Alamy and A.~Balchin, \emph{J. Cryst. Growth}, 1977, \textbf{38},
  221--232\relax
\mciteBstWouldAddEndPuncttrue
\mciteSetBstMidEndSepPunct{\mcitedefaultmidpunct}
{\mcitedefaultendpunct}{\mcitedefaultseppunct}\relax
\EndOfBibitem
\bibitem[Feldman \emph{et~al.}(1995)Feldman, Wasserman, Srolovitz, and
  Tenne]{Feldman95S}
Y.~Feldman, E.~Wasserman, D.~J. Srolovitz and R.~Tenne, \emph{Science}, 1995,
  \textbf{267}, 222--225\relax
\mciteBstWouldAddEndPuncttrue
\mciteSetBstMidEndSepPunct{\mcitedefaultmidpunct}
{\mcitedefaultendpunct}{\mcitedefaultseppunct}\relax
\EndOfBibitem
\bibitem[Jen-La~Plante \emph{et~al.}(2010)Jen-La~Plante, Zeid, Yang, and
  Mokari]{Jen-LaPlante10JMC}
I.~Jen-La~Plante, T.~W. Zeid, P.~Yang and T.~Mokari, \emph{J. Mater. Chem.},
  2010, \textbf{20}, 6612--6617\relax
\mciteBstWouldAddEndPuncttrue
\mciteSetBstMidEndSepPunct{\mcitedefaultmidpunct}
{\mcitedefaultendpunct}{\mcitedefaultseppunct}\relax
\EndOfBibitem
\bibitem[Kim \emph{et~al.}(2011)Kim, Sun, Lu, Cheng, Zhu, Le, Rahman, and
  Bartels]{Kim11L}
D.~Kim, D.~Sun, W.~Lu, Z.~Cheng, Y.~Zhu, D.~Le, T.~S. Rahman and L.~Bartels,
  \emph{Langmuir}, 2011, \textbf{27}, 11650--11653\relax
\mciteBstWouldAddEndPuncttrue
\mciteSetBstMidEndSepPunct{\mcitedefaultmidpunct}
{\mcitedefaultendpunct}{\mcitedefaultseppunct}\relax
\EndOfBibitem
\bibitem[Jin \emph{et~al.}(2011)Jin, Cheng, Li, Cao, Li, Wang, Liu, and
  Zhao]{Jin11SSS}
H.~Jin, D.~Cheng, J.~Li, X.~Cao, B.~Li, X.~Wang, X.~Liu and X.~Zhao,
  \emph{Solid State Sci.}, 2011, \textbf{13}, 1166--1171\relax
\mciteBstWouldAddEndPuncttrue
\mciteSetBstMidEndSepPunct{\mcitedefaultmidpunct}
{\mcitedefaultendpunct}{\mcitedefaultseppunct}\relax
\EndOfBibitem
\bibitem[Liu \emph{et~al.}(2012)Liu, Zhang, Lee, Lin, Chang, Su, Chang, Li,
  Shi, Zhang, Lai, and Li]{Liu12NL}
K.-K. Liu, W.~Zhang, Y.-H. Lee, Y.-C. Lin, M.-T. Chang, C.~Su, C.-S. Chang,
  H.~Li, Y.~Shi, H.~Zhang, C.-S. Lai and L.-J. Li, \emph{Nano Lett.}, 2012,
  \textbf{12}, 1538--1544\relax
\mciteBstWouldAddEndPuncttrue
\mciteSetBstMidEndSepPunct{\mcitedefaultmidpunct}
{\mcitedefaultendpunct}{\mcitedefaultseppunct}\relax
\EndOfBibitem
\bibitem[Shi \emph{et~al.}(2012)Shi, Zhou, Lu, Fang, Lee, Hsu, Kim, Kim, Yang,
  Li, Idrobo, and Kong]{Shi12NL}
Y.~Shi, W.~Zhou, A.-Y. Lu, W.~Fang, Y.-H. Lee, A.~L. Hsu, S.~M. Kim, K.~K. Kim,
  H.~Y. Yang, L.-J. Li, J.-C. Idrobo and J.~Kong, \emph{Nano Lett.}, 2012,
  \textbf{12}, 2784--2791\relax
\mciteBstWouldAddEndPuncttrue
\mciteSetBstMidEndSepPunct{\mcitedefaultmidpunct}
{\mcitedefaultendpunct}{\mcitedefaultseppunct}\relax
\EndOfBibitem
\bibitem[Balendhran \emph{et~al.}(2012)Balendhran, Ou, Bhaskaran, Sriram,
  Ippolito, Vasic, Kats, Bhargava, Zhuiykov, and Kalantar-zadeh]{Balendhran12N}
S.~Balendhran, J.~Z. Ou, M.~Bhaskaran, S.~Sriram, S.~Ippolito, Z.~Vasic,
  E.~Kats, S.~Bhargava, S.~Zhuiykov and K.~Kalantar-zadeh, \emph{Nanoscale},
  2012, \textbf{4}, 461--466\relax
\mciteBstWouldAddEndPuncttrue
\mciteSetBstMidEndSepPunct{\mcitedefaultmidpunct}
{\mcitedefaultendpunct}{\mcitedefaultseppunct}\relax
\EndOfBibitem
\bibitem[Ding \emph{et~al.}(2012)Ding, Zhang, Chen, and Lou]{Ding12N}
S.~Ding, D.~Zhang, J.~S. Chen and X.~W.~D. Lou, \emph{Nanoscale}, 2012,
  \textbf{4}, 95--98\relax
\mciteBstWouldAddEndPuncttrue
\mciteSetBstMidEndSepPunct{\mcitedefaultmidpunct}
{\mcitedefaultendpunct}{\mcitedefaultseppunct}\relax
\EndOfBibitem
\bibitem[Zhan \emph{et~al.}(2012)Zhan, Liu, Najmaei, Ajayan, and Lou]{Zhan12S}
Y.~Zhan, Z.~Liu, S.~Najmaei, P.~M. Ajayan and J.~Lou, \emph{Small}, 2012,
  \textbf{8}, 966--971\relax
\mciteBstWouldAddEndPuncttrue
\mciteSetBstMidEndSepPunct{\mcitedefaultmidpunct}
{\mcitedefaultendpunct}{\mcitedefaultseppunct}\relax
\EndOfBibitem
\bibitem[Lee \emph{et~al.}(2012)Lee, Zhang, Zhang, Chang, Lin, Chang, Yu, Wang,
  Chang, Li, and Lin]{Lee12AM}
Y.-H. Lee, X.-Q. Zhang, W.~Zhang, M.-T. Chang, C.-T. Lin, K.-D. Chang, Y.-C.
  Yu, J.~T.-W. Wang, C.-S. Chang, L.-J. Li and T.-W. Lin, \emph{Adv. Mater.},
  2012, \textbf{24}, 2320--2325\relax
\mciteBstWouldAddEndPuncttrue
\mciteSetBstMidEndSepPunct{\mcitedefaultmidpunct}
{\mcitedefaultendpunct}{\mcitedefaultseppunct}\relax
\EndOfBibitem
\bibitem[Wang \emph{et~al.}(2013)Wang, Feng, Wu, and Jiao]{Wang13JACS}
X.~Wang, H.~Feng, Y.~Wu and L.~Jiao, \emph{J. Am. Chem. Soc.}, 2013,
  \textbf{135}, 5304--5307\relax
\mciteBstWouldAddEndPuncttrue
\mciteSetBstMidEndSepPunct{\mcitedefaultmidpunct}
{\mcitedefaultendpunct}{\mcitedefaultseppunct}\relax
\EndOfBibitem
\bibitem[Wu \emph{et~al.}(2013)Wu, De, Chang, Wang, Peng, Bao, and
  Pei]{Wu13APL_synthesis:CVDMSmu17}
W.~Wu, D.~De, S.-C. Chang, Y.~Wang, H.~Peng, J.~Bao and S.-S. Pei, \emph{Appl.
  Phys. Lett.}, 2013, \textbf{102}, 142106\relax
\mciteBstWouldAddEndPuncttrue
\mciteSetBstMidEndSepPunct{\mcitedefaultmidpunct}
{\mcitedefaultendpunct}{\mcitedefaultseppunct}\relax
\EndOfBibitem
\bibitem[Lei \emph{et~al.}(2013)Lei, Ge, Liu, Najmaei, Shi, You, Lou, Vajtai,
  and Ajayan]{Lei13NL_}
S.~Lei, L.~Ge, Z.~Liu, S.~Najmaei, G.~Shi, G.~You, J.~Lou, R.~Vajtai and P.~M.
  Ajayan, \emph{Nano Lett.}, 2013, \textbf{13}, 2777--2781\relax
\mciteBstWouldAddEndPuncttrue
\mciteSetBstMidEndSepPunct{\mcitedefaultmidpunct}
{\mcitedefaultendpunct}{\mcitedefaultseppunct}\relax
\EndOfBibitem
\bibitem[Wu \emph{et~al.}(2013)Wu, Huang, Aivazian, Ross, Cobden, and
  Xu]{Wu13AN_synthesis:nodefect}
S.~Wu, C.~Huang, G.~Aivazian, J.~S. Ross, D.~H. Cobden and X.~Xu, \emph{ACS
  Nano}, 2013, \textbf{7}, 2768--2772\relax
\mciteBstWouldAddEndPuncttrue
\mciteSetBstMidEndSepPunct{\mcitedefaultmidpunct}
{\mcitedefaultendpunct}{\mcitedefaultseppunct}\relax
\EndOfBibitem
\bibitem[Lee \emph{et~al.}(2013)Lee, Yu, Wang, Fang, Ling, Shi, Lin, Huang,
  Chang, Chang, Dresselhaus, Palacios, Li, and
  Kong]{Lee13NL_Synthesis:Seed:PTAS}
Y.-H. Lee, L.~Yu, H.~Wang, W.~Fang, X.~Ling, Y.~Shi, C.-T. Lin, J.-K. Huang,
  M.-T. Chang, C.-S. Chang, M.~Dresselhaus, T.~Palacios, L.-J. Li and J.~Kong,
  \emph{Nano Lett.}, 2013, \textbf{13}, 1852--1857\relax
\mciteBstWouldAddEndPuncttrue
\mciteSetBstMidEndSepPunct{\mcitedefaultmidpunct}
{\mcitedefaultendpunct}{\mcitedefaultseppunct}\relax
\EndOfBibitem
\bibitem[Huang \emph{et~al.}(2013)Huang, Yu, Li, and
  Cao]{Huang13JPCC_Synthesis:SnSe2}
L.~Huang, Y.~Yu, C.~Li and L.~Cao, \emph{J. Phys. Chem. C}, 2013, \textbf{117},
  6469--6475\relax
\mciteBstWouldAddEndPuncttrue
\mciteSetBstMidEndSepPunct{\mcitedefaultmidpunct}
{\mcitedefaultendpunct}{\mcitedefaultseppunct}\relax
\EndOfBibitem
\bibitem[Yu \emph{et~al.}(2013)Yu, Li, Liu, Su, Zhang, and
  Cao]{Yu13SR_synthesis:MoS2}
Y.~Yu, C.~Li, Y.~Liu, L.~Su, Y.~Zhang and L.~Cao, \emph{Sci. Rep.}, 2013,
  \textbf{3}, 1866\relax
\mciteBstWouldAddEndPuncttrue
\mciteSetBstMidEndSepPunct{\mcitedefaultmidpunct}
{\mcitedefaultendpunct}{\mcitedefaultseppunct}\relax
\EndOfBibitem
\bibitem[Lu \emph{et~al.}(2014)Lu, Utama, Lin, Gong, Zhang, Zhao, Pantelides,
  Wang, Dong, Liu, Zhou, and Xiong]{Lu14NL}
X.~Lu, M.~I.~B. Utama, J.~Lin, X.~Gong, J.~Zhang, Y.~Zhao, S.~T. Pantelides,
  J.~Wang, Z.~Dong, Z.~Liu, W.~Zhou and Q.~Xiong, \emph{Nano Lett.}, 2014,
  \textbf{14}, 2419--25\relax
\mciteBstWouldAddEndPuncttrue
\mciteSetBstMidEndSepPunct{\mcitedefaultmidpunct}
{\mcitedefaultendpunct}{\mcitedefaultseppunct}\relax
\EndOfBibitem
\bibitem[Ma \emph{et~al.}(2014)Ma, Isarraraz, Wang, Preciado, Klee, Bobek,
  Yamaguchi, Li, Odenthal, Nguyen, Barroso, Sun, von Son~Palacio, Gomez,
  Nguyen, Le, Pawin, Mann, Heinz, Rahman, and
  Bartels]{Ma14An_Synthesis:Isodoping}
Q.~Ma, M.~Isarraraz, C.~S. Wang, E.~Preciado, V.~Klee, S.~Bobek, K.~Yamaguchi,
  E.~Li, P.~M. Odenthal, A.~Nguyen, D.~Barroso, D.~Sun, G.~von Son~Palacio,
  M.~Gomez, A.~Nguyen, D.~Le, G.~Pawin, J.~Mann, T.~F. Heinz, T.~S. Rahman and
  L.~Bartels, \emph{ACS Nano}, 2014, \textbf{8}, 4672--7\relax
\mciteBstWouldAddEndPuncttrue
\mciteSetBstMidEndSepPunct{\mcitedefaultmidpunct}
{\mcitedefaultendpunct}{\mcitedefaultseppunct}\relax
\EndOfBibitem
\bibitem[Mann \emph{et~al.}(2014)Mann, Ma, Odenthal, Isarraraz, Le, Preciado,
  Barroso, Yamaguchi, von Son~Palacio, Nguyen, Tran, Wurch, Nguyen, Klee,
  Bobek, Sun, Heinz, Rahman, Kawakami, and
  Bartels]{Mann14AM_Synthesis:Isodoping}
J.~Mann, Q.~Ma, P.~M. Odenthal, M.~Isarraraz, D.~Le, E.~Preciado, D.~Barroso,
  K.~Yamaguchi, G.~von Son~Palacio, A.~Nguyen, T.~Tran, M.~Wurch, A.~Nguyen,
  V.~Klee, S.~Bobek, D.~Sun, T.~F. Heinz, T.~S. Rahman, R.~Kawakami and
  L.~Bartels, \emph{Adv. Mater.}, 2014, \textbf{26}, 1399--1404\relax
\mciteBstWouldAddEndPuncttrue
\mciteSetBstMidEndSepPunct{\mcitedefaultmidpunct}
{\mcitedefaultendpunct}{\mcitedefaultseppunct}\relax
\EndOfBibitem
\bibitem[Li \emph{et~al.}(2014)Li, Duan, Wu, Zhuang, Zhou, Zhang, Zhu, Hu, Ren,
  Guo, Ma, Fan, Wang, Xu, Pan, and Duan]{Li14JACS_Synthesis:Isodoping}
H.~Li, X.~Duan, X.~Wu, X.~Zhuang, H.~Zhou, Q.~Zhang, X.~Zhu, W.~Hu, P.~Ren,
  P.~Guo, L.~Ma, X.~Fan, X.~Wang, J.~Xu, A.~Pan and X.~Duan, \emph{J. Am. Chem.
  Soc.}, 2014, \textbf{136}, 3756--3759\relax
\mciteBstWouldAddEndPuncttrue
\mciteSetBstMidEndSepPunct{\mcitedefaultmidpunct}
{\mcitedefaultendpunct}{\mcitedefaultseppunct}\relax
\EndOfBibitem
\bibitem[Huang \emph{et~al.}(2014)Huang, Pu, Hsu, Chiu, Juang, Chang, Chang,
  Iwasa, Takenobu, and Li]{Huang14AN_Synthesis:WSe:inverter}
J.-K. Huang, J.~Pu, C.-L. Hsu, M.-H. Chiu, Z.-Y. Juang, Y.-H. Chang, W.-H.
  Chang, Y.~Iwasa, T.~Takenobu and L.-J. Li, \emph{ACS Nano}, 2014, \textbf{8},
  923--930\relax
\mciteBstWouldAddEndPuncttrue
\mciteSetBstMidEndSepPunct{\mcitedefaultmidpunct}
{\mcitedefaultendpunct}{\mcitedefaultseppunct}\relax
\EndOfBibitem
\bibitem[George \emph{et~al.}(2014)George, Mutlu, Ionescu, Wu, Jeong, Bay,
  Chai, Mkhoyan, Ozkan, and Ozkan]{Georg14AFM_syntesis:MoS:wafescale}
A.~S. George, Z.~Mutlu, R.~Ionescu, R.~J. Wu, J.~S. Jeong, H.~H. Bay, Y.~Chai,
  K.~A. Mkhoyan, M.~Ozkan and C.~S. Ozkan, \emph{Adv. Funct. Mater.}, 2014,
  \textbf{24}, 7461--7466\relax
\mciteBstWouldAddEndPuncttrue
\mciteSetBstMidEndSepPunct{\mcitedefaultmidpunct}
{\mcitedefaultendpunct}{\mcitedefaultseppunct}\relax
\EndOfBibitem
\bibitem[Wang \emph{et~al.}(2014)Wang, Gong, Shi, Chow, Keyshar, Ye, Vajtai,
  Lou, Liu, Ringe, Tay, and Ajayan]{Wang14An_syntesis:MoSe}
X.~Wang, Y.~Gong, G.~Shi, W.~L. Chow, K.~Keyshar, G.~Ye, R.~Vajtai, J.~Lou,
  Z.~Liu, E.~Ringe, B.~K. Tay and P.~M. Ajayan, \emph{ACS Nano}, 2014,
  \textbf{8}, 5125--5131\relax
\mciteBstWouldAddEndPuncttrue
\mciteSetBstMidEndSepPunct{\mcitedefaultmidpunct}
{\mcitedefaultendpunct}{\mcitedefaultseppunct}\relax
\EndOfBibitem
\bibitem[Lin \emph{et~al.}(2014)Lin, Lu, Perea-Lopez, Li, Lin, Peng, Lee, Sun,
  Calderin, Browning, Bresnehan, Kim, Mayer, Terrones, and
  Robinson]{Lin14AN_synthesis}
Y.-C. Lin, N.~Lu, N.~Perea-Lopez, J.~Li, Z.~Lin, X.~Peng, C.~H. Lee, C.~Sun,
  L.~Calderin, P.~N. Browning, M.~S. Bresnehan, M.~J. Kim, T.~S. Mayer,
  M.~Terrones and J.~A. Robinson, \emph{ACS Nano}, 2014, \textbf{8},
  3715--3723\relax
\mciteBstWouldAddEndPuncttrue
\mciteSetBstMidEndSepPunct{\mcitedefaultmidpunct}
{\mcitedefaultendpunct}{\mcitedefaultseppunct}\relax
\EndOfBibitem
\bibitem[Orofeo \emph{et~al.}(2014)Orofeo, Suzuki, Sekine, and
  Hibino]{Orofe14APL_synthesis}
C.~M. Orofeo, S.~Suzuki, Y.~Sekine and H.~Hibino, \emph{Appl. Phys. Lett.},
  2014, \textbf{105}, 083112\relax
\mciteBstWouldAddEndPuncttrue
\mciteSetBstMidEndSepPunct{\mcitedefaultmidpunct}
{\mcitedefaultendpunct}{\mcitedefaultseppunct}\relax
\EndOfBibitem
\bibitem[Gong \emph{et~al.}(2014)Gong, Liu, Lupini, Shi, Lin, Najmaei, Lin,
  Elias, Berkdemir, You, Terrones, Terrones, Vajtai, Pantelides, Pennycook,
  Lou, Zhou, and Ajayan]{Gong14NL_synthesis}
Y.~Gong, Z.~Liu, A.~R. Lupini, G.~Shi, J.~Lin, S.~Najmaei, Z.~Lin, A.~L. Elias,
  A.~Berkdemir, G.~You, H.~Terrones, M.~Terrones, R.~Vajtai, S.~T. Pantelides,
  S.~J. Pennycook, J.~Lou, W.~Zhou and P.~M. Ajayan, \emph{Nano Lett.}, 2014,
  \textbf{14}, 442--449\relax
\mciteBstWouldAddEndPuncttrue
\mciteSetBstMidEndSepPunct{\mcitedefaultmidpunct}
{\mcitedefaultendpunct}{\mcitedefaultseppunct}\relax
\EndOfBibitem
\bibitem[Huang \emph{et~al.}(2014)Huang, Al-Saab, Wang, Ou, Walker, Wang,
  Gholipour, Simpson, and Hewak]{Huang14N_synthesis}
C.-C. Huang, F.~Al-Saab, Y.~Wang, J.-Y. Ou, J.~C. Walker, S.~Wang,
  B.~Gholipour, R.~E. Simpson and D.~W. Hewak, \emph{Nanoscale}, 2014,
  \textbf{6}, 12792--12797\relax
\mciteBstWouldAddEndPuncttrue
\mciteSetBstMidEndSepPunct{\mcitedefaultmidpunct}
{\mcitedefaultendpunct}{\mcitedefaultseppunct}\relax
\EndOfBibitem
\bibitem[Zhang \emph{et~al.}(2014)Zhang, Yu, Chen, Tian, Liu, Cheng, Xie, Yang,
  Yang, Bai, Shi, and Zhang]{Zhang14AN_synthesis:MoS}
J.~Zhang, H.~Yu, W.~Chen, X.~Tian, D.~Liu, M.~Cheng, G.~Xie, W.~Yang, R.~Yang,
  X.~Bai, D.~Shi and G.~Zhang, \emph{ACS Nano}, 2014, \textbf{8},
  6024--6030\relax
\mciteBstWouldAddEndPuncttrue
\mciteSetBstMidEndSepPunct{\mcitedefaultmidpunct}
{\mcitedefaultendpunct}{\mcitedefaultseppunct}\relax
\EndOfBibitem
\bibitem[Ling \emph{et~al.}(2014)Ling, Lee, Lin, Fang, Yu, Dresselhaus, and
  Kong]{Ling14NL_Synthesis:MoS:seed}
X.~Ling, Y.-H. Lee, Y.~Lin, W.~Fang, L.~Yu, M.~S. Dresselhaus and J.~Kong,
  \emph{Nano Lett.}, 2014, \textbf{14}, 464--472\relax
\mciteBstWouldAddEndPuncttrue
\mciteSetBstMidEndSepPunct{\mcitedefaultmidpunct}
{\mcitedefaultendpunct}{\mcitedefaultseppunct}\relax
\EndOfBibitem
\bibitem[Lee \emph{et~al.}(2014)Lee, Lee, Bark, Oh, Ryu, Lee, Kim, Cho, Ahn,
  and Lee]{Lee14N_synthesis:MoS:waferscale}
Y.~Lee, J.~Lee, H.~Bark, I.-K. Oh, G.~H. Ryu, Z.~Lee, H.~Kim, J.~H. Cho, J.-H.
  Ahn and C.~Lee, \emph{Nanoscale}, 2014, \textbf{6}, 2821--2826\relax
\mciteBstWouldAddEndPuncttrue
\mciteSetBstMidEndSepPunct{\mcitedefaultmidpunct}
{\mcitedefaultendpunct}{\mcitedefaultseppunct}\relax
\EndOfBibitem
\bibitem[Tarasov \emph{et~al.}(2014)Tarasov, Campbell, Tsai, Hesabi, Feirer,
  Graham, Ready, and Vogel]{Taras14AFM_Synthesis:MoS2_3L:wafer}
A.~Tarasov, P.~M. Campbell, M.-Y. Tsai, Z.~R. Hesabi, J.~Feirer, S.~Graham,
  W.~J. Ready and E.~M. Vogel, \emph{Adv. Funct. Mater.}, 2014, \textbf{24},
  6389--6400\relax
\mciteBstWouldAddEndPuncttrue
\mciteSetBstMidEndSepPunct{\mcitedefaultmidpunct}
{\mcitedefaultendpunct}{\mcitedefaultseppunct}\relax
\EndOfBibitem
\bibitem[Zhang \emph{et~al.}(2014)Zhang, Wu, Zhu, Dumcenco, Hong, Mao, Deng,
  Chen, Yang, Jin, Chaki, Huang, Zhang, and Xie]{Zhang14AN_synthesis:MoWSe}
M.~Zhang, J.~Wu, Y.~Zhu, D.~O. Dumcenco, J.~Hong, N.~Mao, S.~Deng, Y.~Chen,
  Y.~Yang, C.~Jin, S.~H. Chaki, Y.-S. Huang, J.~Zhang and L.~Xie, \emph{ACS
  Nano}, 2014, \textbf{8}, 7130--7137\relax
\mciteBstWouldAddEndPuncttrue
\mciteSetBstMidEndSepPunct{\mcitedefaultmidpunct}
{\mcitedefaultendpunct}{\mcitedefaultseppunct}\relax
\EndOfBibitem
\bibitem[Han \emph{et~al.}(2015)Han, Kybert, Naylor, Lee, Ping, Park, Kang,
  Lee, Lee, Agarwal, and Johnson]{Han15NC_}
G.~H. Han, N.~J. Kybert, C.~H. Naylor, B.~S. Lee, J.~Ping, J.~H. Park, J.~Kang,
  S.~Y. Lee, Y.~H. Lee, R.~Agarwal and A.~T.~C. Johnson, \emph{Nat. Commun.},
  2015, \textbf{6}, 6128\relax
\mciteBstWouldAddEndPuncttrue
\mciteSetBstMidEndSepPunct{\mcitedefaultmidpunct}
{\mcitedefaultendpunct}{\mcitedefaultseppunct}\relax
\EndOfBibitem
\bibitem[Yun \emph{et~al.}(2015)Yun, Chae, Kim, Park, Park, Han, Lee, Kim, Oh,
  Seok, Jeong, Kim, and Lee]{Yun15An_syntesis:MoS:1L}
S.~J. Yun, S.~H. Chae, H.~Kim, J.~C. Park, J.-H. Park, G.~H. Han, J.~S. Lee,
  S.~M. Kim, H.~M. Oh, J.~Seok, M.~S. Jeong, K.~K. Kim and Y.~H. Lee, \emph{ACS
  Nano}, 2015, \textbf{9}, 5510--5519\relax
\mciteBstWouldAddEndPuncttrue
\mciteSetBstMidEndSepPunct{\mcitedefaultmidpunct}
{\mcitedefaultendpunct}{\mcitedefaultseppunct}\relax
\EndOfBibitem
\bibitem[Pisoni \emph{et~al.}(2015)Pisoni, Jacimovic, Barisic, Walter, Nafradi,
  Bugnon, Magrez, Berger, Revay, and Forro]{Pison15JPCC_synthesis:growthagent}
A.~Pisoni, J.~Jacimovic, O.~S. Barisic, A.~Walter, B.~Nafradi, P.~Bugnon,
  A.~Magrez, H.~Berger, Z.~Revay and L.~Forro, \emph{J. Phys. Chem. C}, 2015,
  \textbf{119}, 3918--3922\relax
\mciteBstWouldAddEndPuncttrue
\mciteSetBstMidEndSepPunct{\mcitedefaultmidpunct}
{\mcitedefaultendpunct}{\mcitedefaultseppunct}\relax
\EndOfBibitem
\bibitem[Jeon \emph{et~al.}(2015)Jeon, Jang, Jeon, Yoo, Jang, Park, and
  Lee]{Jeon15N_synthesis:MoS}
J.~Jeon, S.~K. Jang, S.~M. Jeon, G.~Yoo, Y.~H. Jang, J.-H. Park and S.~Lee,
  \emph{Nanoscale}, 2015, \textbf{7}, 1688--1695\relax
\mciteBstWouldAddEndPuncttrue
\mciteSetBstMidEndSepPunct{\mcitedefaultmidpunct}
{\mcitedefaultendpunct}{\mcitedefaultseppunct}\relax
\EndOfBibitem
\bibitem[Tao \emph{et~al.}(2015)Tao, Chai, Lu, Wong, Wong, Pan, Xiong, Chi, and
  Wang]{Tao15N_synthesis:MoS:sputtering}
J.~Tao, J.~Chai, X.~Lu, L.~M. Wong, T.~I. Wong, J.~Pan, Q.~Xiong, D.~Chi and
  S.~Wang, \emph{Nanoscale}, 2015, \textbf{7}, 2497--2503\relax
\mciteBstWouldAddEndPuncttrue
\mciteSetBstMidEndSepPunct{\mcitedefaultmidpunct}
{\mcitedefaultendpunct}{\mcitedefaultseppunct}\relax
\EndOfBibitem
\bibitem[Park \emph{et~al.}(2015)Park, Choudhary, Smith, Lee, Kim, and
  Choi]{Park15APL_synthesis:MoS:thickness}
J.~Park, N.~Choudhary, J.~Smith, G.~Lee, M.~Kim and W.~Choi, \emph{Appl. Phys.
  Lett.}, 2015, \textbf{106}, 012104\relax
\mciteBstWouldAddEndPuncttrue
\mciteSetBstMidEndSepPunct{\mcitedefaultmidpunct}
{\mcitedefaultendpunct}{\mcitedefaultseppunct}\relax
\EndOfBibitem
\bibitem[Kumar \emph{et~al.}(2015)Kumar, Dhar, Choudhury, Shivashankar, and
  Raghavan]{Kumar15N_synthesis:MoS:thickness}
V.~K. Kumar, S.~Dhar, T.~H. Choudhury, S.~A. Shivashankar and S.~Raghavan,
  \emph{Nanoscale}, 2015, \textbf{7}, 7802--7810\relax
\mciteBstWouldAddEndPuncttrue
\mciteSetBstMidEndSepPunct{\mcitedefaultmidpunct}
{\mcitedefaultendpunct}{\mcitedefaultseppunct}\relax
\EndOfBibitem
\bibitem[Kang \emph{et~al.}(2015)Kang, Xie, Huang, Han, Huang, Mak, Kim,
  Muller, and Park]{Kang15N_synthesis:MoS:waferscale}
K.~Kang, S.~Xie, L.~Huang, Y.~Han, P.~Y. Huang, K.~F. Mak, C.-J. Kim, D.~Muller
  and J.~Park, \emph{Nature}, 2015, \textbf{520}, 656--660\relax
\mciteBstWouldAddEndPuncttrue
\mciteSetBstMidEndSepPunct{\mcitedefaultmidpunct}
{\mcitedefaultendpunct}{\mcitedefaultseppunct}\relax
\EndOfBibitem
\bibitem[Britnell \emph{et~al.}(2012)Britnell, Gorbachev, Jalil, Belle,
  Schedin, Mishchenko, Georgiou, Katsnelson, Eaves, Morozov, Peres, Leist,
  Geim, Novoselov, and Ponomarenko]{britnell12s}
L.~Britnell, R.~V. Gorbachev, R.~Jalil, B.~D. Belle, F.~Schedin, A.~Mishchenko,
  T.~Georgiou, M.~I. Katsnelson, L.~Eaves, S.~V. Morozov, N.~M.~R. Peres,
  J.~Leist, A.~K. Geim, K.~S. Novoselov and L.~A. Ponomarenko, \emph{Science},
  2012, \textbf{335}, 947--950\relax
\mciteBstWouldAddEndPuncttrue
\mciteSetBstMidEndSepPunct{\mcitedefaultmidpunct}
{\mcitedefaultendpunct}{\mcitedefaultseppunct}\relax
\EndOfBibitem
\bibitem[Georgiou \emph{et~al.}(2013)Georgiou, Jalil, Belle, Britnell,
  Gorbachev, Morozov, Kim, Gholinia, Haigh, Makarovsky, Eaves, Ponomarenko,
  Geim, Novoselov, and Mishchenko]{georgiou13nn}
T.~Georgiou, R.~Jalil, B.~D. Belle, L.~Britnell, R.~V. Gorbachev, S.~V.
  Morozov, Y.-J. Kim, A.~Gholinia, S.~J. Haigh, O.~Makarovsky, L.~Eaves, L.~A.
  Ponomarenko, A.~K. Geim, K.~S. Novoselov and A.~Mishchenko, \emph{Nat.
  Nanotechnol.}, 2013, \textbf{8}, 100--103\relax
\mciteBstWouldAddEndPuncttrue
\mciteSetBstMidEndSepPunct{\mcitedefaultmidpunct}
{\mcitedefaultendpunct}{\mcitedefaultseppunct}\relax
\EndOfBibitem
\bibitem[Radisavljevic \emph{et~al.}(2011)Radisavljevic, Radenovic, Brivio,
  Giacometti, and Kis]{Radisavljevic11NN}
B.~Radisavljevic, A.~Radenovic, J.~Brivio, V.~Giacometti and A.~Kis, \emph{Nat.
  Nanotechnol.}, 2011, \textbf{6}, 147--150\relax
\mciteBstWouldAddEndPuncttrue
\mciteSetBstMidEndSepPunct{\mcitedefaultmidpunct}
{\mcitedefaultendpunct}{\mcitedefaultseppunct}\relax
\EndOfBibitem
\bibitem[Radisavljevic and Kis(2013)]{Radisavljevic13NM}
B.~Radisavljevic and A.~Kis, \emph{Nat. Mater.}, 2013, \textbf{12},
  815--820\relax
\mciteBstWouldAddEndPuncttrue
\mciteSetBstMidEndSepPunct{\mcitedefaultmidpunct}
{\mcitedefaultendpunct}{\mcitedefaultseppunct}\relax
\EndOfBibitem
\bibitem[Geim and Grigorieva(2013)]{Geim13N}
A.~K. Geim and I.~V. Grigorieva, \emph{Nature}, 2013, \textbf{499},
  419--425\relax
\mciteBstWouldAddEndPuncttrue
\mciteSetBstMidEndSepPunct{\mcitedefaultmidpunct}
{\mcitedefaultendpunct}{\mcitedefaultseppunct}\relax
\EndOfBibitem
\bibitem[Cao \emph{et~al.}(2012)Cao, Wang, Han, Ye, Zhu, Shi, Niu, Tan, Wang,
  Liu, and Feng]{Cao12NC}
T.~Cao, G.~Wang, W.~Han, H.~Ye, C.~Zhu, J.~Shi, Q.~Niu, P.~Tan, E.~Wang, B.~Liu
  and J.~Feng, \emph{Nat. Commun.}, 2012, \textbf{3}, 887\relax
\mciteBstWouldAddEndPuncttrue
\mciteSetBstMidEndSepPunct{\mcitedefaultmidpunct}
{\mcitedefaultendpunct}{\mcitedefaultseppunct}\relax
\EndOfBibitem
\bibitem[Gong \emph{et~al.}(2013)Gong, Liu, Yu, Xiao, Cui, Xu, and
  Yao]{Gong13NC}
Z.~Gong, G.-B. Liu, H.~Yu, D.~Xiao, X.~Cui, X.~Xu and W.~Yao, \emph{Nat.
  Commun.}, 2013, \textbf{4}, 2053\relax
\mciteBstWouldAddEndPuncttrue
\mciteSetBstMidEndSepPunct{\mcitedefaultmidpunct}
{\mcitedefaultendpunct}{\mcitedefaultseppunct}\relax
\EndOfBibitem
\bibitem[Jones \emph{et~al.}(2013)Jones, Yu, Ghimire, Wu, Aivazian, Ross, Zhao,
  Yan, Mandrus, Xiao, Yao, and Xu]{Jones13NN}
A.~M. Jones, H.~Yu, N.~J. Ghimire, S.~Wu, G.~Aivazian, J.~S. Ross, B.~Zhao,
  J.~Yan, D.~G. Mandrus, D.~Xiao, W.~Yao and X.~Xu, \emph{Nat. Nanotechnol.},
  2013, \textbf{8}, 634--638\relax
\mciteBstWouldAddEndPuncttrue
\mciteSetBstMidEndSepPunct{\mcitedefaultmidpunct}
{\mcitedefaultendpunct}{\mcitedefaultseppunct}\relax
\EndOfBibitem
\bibitem[Mak \emph{et~al.}(2012)Mak, He, Shan, and Heinz]{Mak12NN}
K.~F. Mak, K.~He, J.~Shan and T.~F. Heinz, \emph{Nat. Nanotechnol.}, 2012,
  \textbf{7}, 494--498\relax
\mciteBstWouldAddEndPuncttrue
\mciteSetBstMidEndSepPunct{\mcitedefaultmidpunct}
{\mcitedefaultendpunct}{\mcitedefaultseppunct}\relax
\EndOfBibitem
\bibitem[Suzuki \emph{et~al.}(2014)Suzuki, Sakano, Zhang, Akashi, Morikawa,
  Harasawa, Yaji, Kuroda, Miyamoto, Okuda, Ishizaka, Arita, and
  Iwasa]{Suzuki14NNb}
R.~Suzuki, M.~Sakano, Y.~J. Zhang, R.~Akashi, D.~Morikawa, A.~Harasawa,
  K.~Yaji, K.~Kuroda, K.~Miyamoto, T.~Okuda, K.~Ishizaka, R.~Arita and
  Y.~Iwasa, \emph{Nat. Nanotechnol.}, 2014, \textbf{9}, 611--617\relax
\mciteBstWouldAddEndPuncttrue
\mciteSetBstMidEndSepPunct{\mcitedefaultmidpunct}
{\mcitedefaultendpunct}{\mcitedefaultseppunct}\relax
\EndOfBibitem
\bibitem[Zeng \emph{et~al.}(2012)Zeng, Dai, Yao, Xiao, and Cui]{Zeng12NN}
H.~Zeng, J.~Dai, W.~Yao, D.~Xiao and X.~Cui, \emph{Nat. Nanotechnol.}, 2012,
  \textbf{7}, 490--493\relax
\mciteBstWouldAddEndPuncttrue
\mciteSetBstMidEndSepPunct{\mcitedefaultmidpunct}
{\mcitedefaultendpunct}{\mcitedefaultseppunct}\relax
\EndOfBibitem
\bibitem[Akinwande \emph{et~al.}(2014)Akinwande, Petrone, and Hone]{Akinw14NC}
D.~Akinwande, N.~Petrone and J.~Hone, \emph{Nat. Commun.}, 2014, \textbf{5},
  5678\relax
\mciteBstWouldAddEndPuncttrue
\mciteSetBstMidEndSepPunct{\mcitedefaultmidpunct}
{\mcitedefaultendpunct}{\mcitedefaultseppunct}\relax
\EndOfBibitem
\bibitem[Yu \emph{et~al.}(2013)Yu, Liu, Zhou, Yin, Li, Huang, and
  Duan]{Yu13NN_Photocurernt:G/MoS2/G}
W.~J. Yu, Y.~Liu, H.~Zhou, A.~Yin, Z.~Li, Y.~Huang and X.~Duan, \emph{Nat.
  Nanotechnol.}, 2013, \textbf{8}, 952--958\relax
\mciteBstWouldAddEndPuncttrue
\mciteSetBstMidEndSepPunct{\mcitedefaultmidpunct}
{\mcitedefaultendpunct}{\mcitedefaultseppunct}\relax
\EndOfBibitem
\bibitem[Duan \emph{et~al.}(2014)Duan, Wang, Shaw, Cheng, Chen, Li, Wu, Tang,
  Zhang, Pan, Jiang, Yu, Huang, and Duan]{Duan14NN_PV:WSe_MoS}
X.~Duan, C.~Wang, J.~C. Shaw, R.~Cheng, Y.~Chen, H.~Li, X.~Wu, Y.~Tang,
  Q.~Zhang, A.~Pan, J.~Jiang, R.~Yu, Y.~Huang and X.~Duan, \emph{Nat.
  Nanotechnol.}, 2014, \textbf{9}, 1024--1030\relax
\mciteBstWouldAddEndPuncttrue
\mciteSetBstMidEndSepPunct{\mcitedefaultmidpunct}
{\mcitedefaultendpunct}{\mcitedefaultseppunct}\relax
\EndOfBibitem
\bibitem[Lee \emph{et~al.}(2012)Lee, Min, Chang, Park, Nam, Kim, Kim, Ryu, and
  Im]{Lee12NL_Photodetector}
H.~S. Lee, S.-W. Min, Y.-G. Chang, M.~K. Park, T.~Nam, H.~Kim, J.~H. Kim,
  S.~Ryu and S.~Im, \emph{Nano Lett.}, 2012, \textbf{12}, 3695--3700\relax
\mciteBstWouldAddEndPuncttrue
\mciteSetBstMidEndSepPunct{\mcitedefaultmidpunct}
{\mcitedefaultendpunct}{\mcitedefaultseppunct}\relax
\EndOfBibitem
\bibitem[Lopez-Sanchez \emph{et~al.}(2013)Lopez-Sanchez, Lembke, Kayci,
  Radenovic, and Kis]{Lopez13NN_Photodetector}
O.~Lopez-Sanchez, D.~Lembke, M.~Kayci, A.~Radenovic and A.~Kis, \emph{Nat.
  Nanotechnol.}, 2013, \textbf{8}, 497--501\relax
\mciteBstWouldAddEndPuncttrue
\mciteSetBstMidEndSepPunct{\mcitedefaultmidpunct}
{\mcitedefaultendpunct}{\mcitedefaultseppunct}\relax
\EndOfBibitem
\bibitem[Jariwala \emph{et~al.}(2013)Jariwala, Sangwan, Wu, Prabhumirashi,
  Geier, Marks, Lauhon, and Hersam]{Jariw13PNAS}
D.~Jariwala, V.~K. Sangwan, C.-C. Wu, P.~L. Prabhumirashi, M.~L. Geier, T.~J.
  Marks, L.~J. Lauhon and M.~C. Hersam, \emph{Proc. Natl. Acad. Sci.}, 2013,
  \textbf{110}, 18076--18080\relax
\mciteBstWouldAddEndPuncttrue
\mciteSetBstMidEndSepPunct{\mcitedefaultmidpunct}
{\mcitedefaultendpunct}{\mcitedefaultseppunct}\relax
\EndOfBibitem
\bibitem[Britnell \emph{et~al.}(2013)Britnell, Ribeiro, Eckmann, Jalil, Belle,
  Mishchenko, Kim, Gorbachev, Georgiou, Morozov, Grigorenko, Geim, Casiraghi,
  Neto, and Novoselov]{Britnell13S}
L.~Britnell, R.~M. Ribeiro, A.~Eckmann, R.~Jalil, B.~D. Belle, A.~Mishchenko,
  Y.-J. Kim, R.~V. Gorbachev, T.~Georgiou, S.~V. Morozov, A.~N. Grigorenko,
  A.~K. Geim, C.~Casiraghi, A.~H.~C. Neto and K.~S. Novoselov, \emph{Science},
  2013, \textbf{340}, 1311--1314\relax
\mciteBstWouldAddEndPuncttrue
\mciteSetBstMidEndSepPunct{\mcitedefaultmidpunct}
{\mcitedefaultendpunct}{\mcitedefaultseppunct}\relax
\EndOfBibitem
\bibitem[Bernardi \emph{et~al.}(2013)Bernardi, Palummo, and
  Grossman]{Bernardi13NL}
M.~Bernardi, M.~Palummo and J.~C. Grossman, \emph{Nano Lett.}, 2013,
  \textbf{13}, 3664--3670\relax
\mciteBstWouldAddEndPuncttrue
\mciteSetBstMidEndSepPunct{\mcitedefaultmidpunct}
{\mcitedefaultendpunct}{\mcitedefaultseppunct}\relax
\EndOfBibitem
\bibitem[Xia \emph{et~al.}(2014)Xia, Wang, Xiao, Dubey, and
  Ramasubramaniam]{Xia14NP_Review:Photodetector}
F.~Xia, H.~Wang, D.~Xiao, M.~Dubey and A.~Ramasubramaniam, \emph{Nat.
  Photonics}, 2014, \textbf{8}, 899--907\relax
\mciteBstWouldAddEndPuncttrue
\mciteSetBstMidEndSepPunct{\mcitedefaultmidpunct}
{\mcitedefaultendpunct}{\mcitedefaultseppunct}\relax
\EndOfBibitem
\bibitem[Koppens \emph{et~al.}(2014)Koppens, Mueller, Avouris, Ferrari,
  Vitiello, and Polini]{Koppe14NN_Photodetector:Review_2D}
F.~H.~L. Koppens, T.~Mueller, P.~Avouris, A.~C. Ferrari, M.~S. Vitiello and
  M.~Polini, \emph{Nat. Nanotechnol.}, 2014, \textbf{9}, 780--793\relax
\mciteBstWouldAddEndPuncttrue
\mciteSetBstMidEndSepPunct{\mcitedefaultmidpunct}
{\mcitedefaultendpunct}{\mcitedefaultseppunct}\relax
\EndOfBibitem
\bibitem[Yan \emph{et~al.}(1992)Yan, Ourmazd, and Lee]{Yan92EDITo_Scalingrule}
R.-H. Yan, A.~Ourmazd and K.~Lee, \emph{IEEE Trans. Electron Devices}, 1992,
  \textbf{39}, 1704--1710\relax
\mciteBstWouldAddEndPuncttrue
\mciteSetBstMidEndSepPunct{\mcitedefaultmidpunct}
{\mcitedefaultendpunct}{\mcitedefaultseppunct}\relax
\EndOfBibitem
\bibitem[Chang \emph{et~al.}(2003)Chang, Choi, Ha, Ranade, Xiong, Bokor, Hu,
  and King]{Chang03PI}
L.~L. Chang, Y.~K. Choi, D.~W. Ha, P.~Ranade, S.~Y. Xiong, J.~Bokor, C.~M. Hu
  and T.~J. King, \emph{Proc. IEEE}, 2003, \textbf{91}, 1860--1873\relax
\mciteBstWouldAddEndPuncttrue
\mciteSetBstMidEndSepPunct{\mcitedefaultmidpunct}
{\mcitedefaultendpunct}{\mcitedefaultseppunct}\relax
\EndOfBibitem
\bibitem[Ieong \emph{et~al.}(2004)Ieong, Doris, Kedzierski, Rim, and
  Yang]{Ieong04S}
M.~Ieong, B.~Doris, J.~Kedzierski, K.~Rim and M.~Yang, \emph{Science}, 2004,
  \textbf{306}, 2057--2060\relax
\mciteBstWouldAddEndPuncttrue
\mciteSetBstMidEndSepPunct{\mcitedefaultmidpunct}
{\mcitedefaultendpunct}{\mcitedefaultseppunct}\relax
\EndOfBibitem
\bibitem[Vogel(2007)]{Vogel07NN}
E.~M. Vogel, \emph{Nat. Nanotechnol.}, 2007, \textbf{2}, 25--32\relax
\mciteBstWouldAddEndPuncttrue
\mciteSetBstMidEndSepPunct{\mcitedefaultmidpunct}
{\mcitedefaultendpunct}{\mcitedefaultseppunct}\relax
\EndOfBibitem
\bibitem[ITRS2013(2013)]{ITRS2013}
ITRS2013, \emph{The International Technology Roadmap for Semiconductors},
  Semiconductor industry association technical report, 2013\relax
\mciteBstWouldAddEndPuncttrue
\mciteSetBstMidEndSepPunct{\mcitedefaultmidpunct}
{\mcitedefaultendpunct}{\mcitedefaultseppunct}\relax
\EndOfBibitem
\bibitem[Schwierz(2010)]{schwierz10nn}
F.~Schwierz, \emph{Nat. Nanotechnol.}, 2010, \textbf{5}, 487--496\relax
\mciteBstWouldAddEndPuncttrue
\mciteSetBstMidEndSepPunct{\mcitedefaultmidpunct}
{\mcitedefaultendpunct}{\mcitedefaultseppunct}\relax
\EndOfBibitem
\bibitem[Wang \emph{et~al.}(2012)Wang, Yu, Lee, Shi, Hsu, Chin, Li, Dubey,
  Kong, and Palacios]{Wang12NL_Logic:MoS2}
H.~Wang, L.~Yu, Y.-H. Lee, Y.~Shi, A.~Hsu, M.~L. Chin, L.-J. Li, M.~Dubey,
  J.~Kong and T.~Palacios, \emph{Nano Lett.}, 2012, \textbf{12},
  4674--4680\relax
\mciteBstWouldAddEndPuncttrue
\mciteSetBstMidEndSepPunct{\mcitedefaultmidpunct}
{\mcitedefaultendpunct}{\mcitedefaultseppunct}\relax
\EndOfBibitem
\bibitem[Gomez \emph{et~al.}(2007)Gomez, Aberg, and Hoyt]{Gomez07IEDL}
L.~Gomez, I.~Aberg and J.~L. Hoyt, \emph{IEEE Electron Device Lett.}, 2007,
  \textbf{28}, 285--287\relax
\mciteBstWouldAddEndPuncttrue
\mciteSetBstMidEndSepPunct{\mcitedefaultmidpunct}
{\mcitedefaultendpunct}{\mcitedefaultseppunct}\relax
\EndOfBibitem
\bibitem[Sakaki \emph{et~al.}(1987)Sakaki, Noda, Hirakawa, Tanaka, and
  Matsusue]{Sakaki87APL}
H.~Sakaki, T.~Noda, K.~Hirakawa, M.~Tanaka and T.~Matsusue, \emph{Appl. Phys.
  Lett.}, 1987, \textbf{51}, 1934--1936\relax
\mciteBstWouldAddEndPuncttrue
\mciteSetBstMidEndSepPunct{\mcitedefaultmidpunct}
{\mcitedefaultendpunct}{\mcitedefaultseppunct}\relax
\EndOfBibitem
\bibitem[Gold(1987)]{Gold87PRB}
A.~Gold, \emph{Phys. Rev. B}, 1987, \textbf{35}, 723--733\relax
\mciteBstWouldAddEndPuncttrue
\mciteSetBstMidEndSepPunct{\mcitedefaultmidpunct}
{\mcitedefaultendpunct}{\mcitedefaultseppunct}\relax
\EndOfBibitem
\bibitem[Tao \emph{et~al.}(2015)Tao, Cinquanta, Chiappe, Grazianetti,
  Fanciulli, Dubey, Molle, and Akinwande]{Tao15NN_FET:2Dsilicene}
L.~Tao, E.~Cinquanta, D.~Chiappe, C.~Grazianetti, M.~Fanciulli, M.~Dubey,
  A.~Molle and D.~Akinwande, \emph{Nat. Nanotechnol.}, 2015, \textbf{10},
  227--231\relax
\mciteBstWouldAddEndPuncttrue
\mciteSetBstMidEndSepPunct{\mcitedefaultmidpunct}
{\mcitedefaultendpunct}{\mcitedefaultseppunct}\relax
\EndOfBibitem
\bibitem[Novoselov \emph{et~al.}(2005)Novoselov, Jiang, Schedin, Booth,
  Khotkevich, Morozov, and Geim]{Novos05PNAS_}
K.~S. Novoselov, D.~Jiang, F.~Schedin, T.~J. Booth, V.~V. Khotkevich, S.~V.
  Morozov and A.~K. Geim, \emph{Proc. Natl. Acad. Sci.}, 2005, \textbf{102},
  10451--10453\relax
\mciteBstWouldAddEndPuncttrue
\mciteSetBstMidEndSepPunct{\mcitedefaultmidpunct}
{\mcitedefaultendpunct}{\mcitedefaultseppunct}\relax
\EndOfBibitem
\bibitem[Ayari \emph{et~al.}(2007)Ayari, Cobas, Ogundadegbe, and
  Fuhrer]{Ayari07JAP_}
A.~Ayari, E.~Cobas, O.~Ogundadegbe and M.~S. Fuhrer, \emph{J. App. Phys.},
  2007, \textbf{101}, 014507\relax
\mciteBstWouldAddEndPuncttrue
\mciteSetBstMidEndSepPunct{\mcitedefaultmidpunct}
{\mcitedefaultendpunct}{\mcitedefaultseppunct}\relax
\EndOfBibitem
\bibitem[Sordan \emph{et~al.}(2009)Sordan, Traversi, and Russo]{sordan09apl}
R.~Sordan, F.~Traversi and V.~Russo, \emph{Appl. Phys. Lett.}, 2009,
  \textbf{94}, 073305\relax
\mciteBstWouldAddEndPuncttrue
\mciteSetBstMidEndSepPunct{\mcitedefaultmidpunct}
{\mcitedefaultendpunct}{\mcitedefaultseppunct}\relax
\EndOfBibitem
\bibitem[Traversi \emph{et~al.}(2009)Traversi, Russo, and
  Sordan]{traversi09apl}
F.~Traversi, V.~Russo and R.~Sordan, \emph{Appl. Phys. Lett.}, 2009,
  \textbf{94}, 223312\relax
\mciteBstWouldAddEndPuncttrue
\mciteSetBstMidEndSepPunct{\mcitedefaultmidpunct}
{\mcitedefaultendpunct}{\mcitedefaultseppunct}\relax
\EndOfBibitem
\bibitem[Li \emph{et~al.}(2010)Li, Miyazaki, Kumatani, Kanda, and
  Tsukagoshi]{li10nl}
S.-L. Li, H.~Miyazaki, A.~Kumatani, A.~Kanda and K.~Tsukagoshi, \emph{Nano
  Lett.}, 2010, \textbf{10}, 2357--2362\relax
\mciteBstWouldAddEndPuncttrue
\mciteSetBstMidEndSepPunct{\mcitedefaultmidpunct}
{\mcitedefaultendpunct}{\mcitedefaultseppunct}\relax
\EndOfBibitem
\bibitem[Harada \emph{et~al.}(2010)Harada, Yagi, Sato, and
  Yokoyama]{harada10apl}
N.~Harada, K.~Yagi, S.~Sato and N.~Yokoyama, \emph{Appl. Phys. Lett.}, 2010,
  \textbf{96}, 012102\relax
\mciteBstWouldAddEndPuncttrue
\mciteSetBstMidEndSepPunct{\mcitedefaultmidpunct}
{\mcitedefaultendpunct}{\mcitedefaultseppunct}\relax
\EndOfBibitem
\bibitem[Li \emph{et~al.}(2011)Li, Miyazaki, Hiura, Liu, and
  Tsukagoshi]{li11an}
S.-L. Li, H.~Miyazaki, H.~Hiura, C.~Liu and K.~Tsukagoshi, \emph{ACS Nano},
  2011, \textbf{5}, 500--506\relax
\mciteBstWouldAddEndPuncttrue
\mciteSetBstMidEndSepPunct{\mcitedefaultmidpunct}
{\mcitedefaultendpunct}{\mcitedefaultseppunct}\relax
\EndOfBibitem
\bibitem[Rizzi \emph{et~al.}(2012)Rizzi, Bianchi, Behnam, Carrion, Guerriero,
  Polloni, Pop, and Sordan]{rizzi12nl}
L.~G. Rizzi, M.~Bianchi, A.~Behnam, E.~Carrion, E.~Guerriero, L.~Polloni,
  E.~Pop and R.~Sordan, \emph{Nano Lett.}, 2012, \textbf{12}, 3948--3953\relax
\mciteBstWouldAddEndPuncttrue
\mciteSetBstMidEndSepPunct{\mcitedefaultmidpunct}
{\mcitedefaultendpunct}{\mcitedefaultseppunct}\relax
\EndOfBibitem
\bibitem[Guerriero \emph{et~al.}(2013)Guerriero, Polloni, Bianchi, Behnam,
  Carrion, Rizzi, Pop, and Sordan]{guerriero13an}
E.~Guerriero, L.~Polloni, M.~Bianchi, A.~Behnam, E.~Carrion, L.~G. Rizzi,
  E.~Pop and R.~Sordan, \emph{ACS Nano}, 2013, \textbf{7}, 5588--5594\relax
\mciteBstWouldAddEndPuncttrue
\mciteSetBstMidEndSepPunct{\mcitedefaultmidpunct}
{\mcitedefaultendpunct}{\mcitedefaultseppunct}\relax
\EndOfBibitem
\bibitem[Li \emph{et~al.}(2011)Li, Miyazaki, Lee, Liu, Kanda, and
  Tsukagoshi]{li11s}
S.-L. Li, H.~Miyazaki, M.~V. Lee, C.~Liu, A.~Kanda and K.~Tsukagoshi,
  \emph{Small}, 2011, \textbf{7}, 1552--1556\relax
\mciteBstWouldAddEndPuncttrue
\mciteSetBstMidEndSepPunct{\mcitedefaultmidpunct}
{\mcitedefaultendpunct}{\mcitedefaultseppunct}\relax
\EndOfBibitem
\bibitem[Miyazaki \emph{et~al.}(2010)Miyazaki, Li, Kanda, and
  Tsukagoshi]{Miyazaki10SST}
H.~Miyazaki, S.~Li, A.~Kanda and K.~Tsukagoshi, \emph{Semicond. Sci. Technol.},
  2010, \textbf{25}, 034008\relax
\mciteBstWouldAddEndPuncttrue
\mciteSetBstMidEndSepPunct{\mcitedefaultmidpunct}
{\mcitedefaultendpunct}{\mcitedefaultseppunct}\relax
\EndOfBibitem
\bibitem[Nakaharai \emph{et~al.}(2013)Nakaharai, Iijima, Ogawa, Suzuki, Li,
  Tsukagoshi, Sato, and Yokoyama]{Nakaharai13AN}
S.~Nakaharai, T.~Iijima, S.~Ogawa, S.~Suzuki, S.-L. Li, K.~Tsukagoshi, S.~Sato
  and N.~Yokoyama, \emph{ACS Nano}, 2013, \textbf{7}, 5694--5700\relax
\mciteBstWouldAddEndPuncttrue
\mciteSetBstMidEndSepPunct{\mcitedefaultmidpunct}
{\mcitedefaultendpunct}{\mcitedefaultseppunct}\relax
\EndOfBibitem
\bibitem[Aparecido-Ferreira \emph{et~al.}(2012)Aparecido-Ferreira, Miyazaki,
  Li, Komatsu, Nakaharai, and Tsukagoshi]{Aparecido-Ferreira12N}
A.~Aparecido-Ferreira, H.~Miyazaki, S.-L. Li, K.~Komatsu, S.~Nakaharai and
  K.~Tsukagoshi, \emph{Nanoscale}, 2012, \textbf{4}, 7842--7846\relax
\mciteBstWouldAddEndPuncttrue
\mciteSetBstMidEndSepPunct{\mcitedefaultmidpunct}
{\mcitedefaultendpunct}{\mcitedefaultseppunct}\relax
\EndOfBibitem
\bibitem[Miyazaki \emph{et~al.}(2012)Miyazaki, Li, Nakaharai, and
  Tsukagoshi]{Miyazaki12APL}
H.~Miyazaki, S.-L. Li, S.~Nakaharai and K.~Tsukagoshi, \emph{Appl. Phys.
  Lett.}, 2012, \textbf{100}, 163115\relax
\mciteBstWouldAddEndPuncttrue
\mciteSetBstMidEndSepPunct{\mcitedefaultmidpunct}
{\mcitedefaultendpunct}{\mcitedefaultseppunct}\relax
\EndOfBibitem
\bibitem[Nakaharai \emph{et~al.}(2014)Nakaharai, Iijima, Ogawa, Li, Tsukagoshi,
  Sato, and Yokoyama]{Nakah14ITN_}
S.~Nakaharai, T.~Iijima, S.~Ogawa, S.~L. Li, K.~Tsukagoshi, S.~Sato and
  N.~Yokoyama, \emph{IEEE Trans. Nanotechnol.}, 2014, \textbf{13},
  1039--1043\relax
\mciteBstWouldAddEndPuncttrue
\mciteSetBstMidEndSepPunct{\mcitedefaultmidpunct}
{\mcitedefaultendpunct}{\mcitedefaultseppunct}\relax
\EndOfBibitem
\bibitem[Nakaharai \emph{et~al.}(2012)Nakaharai, Iijima, Ogawa, Miyazaki, Li,
  Tsukagoshi, Sato, and Yokoyama]{Nakaharai12APE}
S.~Nakaharai, T.~Iijima, S.~Ogawa, H.~Miyazaki, S.~Li, K.~Tsukagoshi, S.~Sato
  and N.~Yokoyama, \emph{Appl. Phys. Express}, 2012, \textbf{5}, 015101\relax
\mciteBstWouldAddEndPuncttrue
\mciteSetBstMidEndSepPunct{\mcitedefaultmidpunct}
{\mcitedefaultendpunct}{\mcitedefaultseppunct}\relax
\EndOfBibitem
\bibitem[Fuhrer and Hone(2013)]{Fuhrer13NN}
M.~S. Fuhrer and J.~Hone, \emph{Nat. Nanotechnol.}, 2013, \textbf{8},
  146--147\relax
\mciteBstWouldAddEndPuncttrue
\mciteSetBstMidEndSepPunct{\mcitedefaultmidpunct}
{\mcitedefaultendpunct}{\mcitedefaultseppunct}\relax
\EndOfBibitem
\bibitem[Fivaz and Mooser(1967)]{Fivaz67PR}
R.~Fivaz and E.~Mooser, \emph{Phys. Rev.}, 1967, \textbf{163}, 743--755\relax
\mciteBstWouldAddEndPuncttrue
\mciteSetBstMidEndSepPunct{\mcitedefaultmidpunct}
{\mcitedefaultendpunct}{\mcitedefaultseppunct}\relax
\EndOfBibitem
\bibitem[Kaasbjerg \emph{et~al.}(2012)Kaasbjerg, Thygesen, and
  Jacobsen]{Kaasbjerg12PRB}
K.~Kaasbjerg, K.~S. Thygesen and K.~W. Jacobsen, \emph{Phys. Rev. B}, 2012,
  \textbf{85}, 115317\relax
\mciteBstWouldAddEndPuncttrue
\mciteSetBstMidEndSepPunct{\mcitedefaultmidpunct}
{\mcitedefaultendpunct}{\mcitedefaultseppunct}\relax
\EndOfBibitem
\bibitem[Kaasbjerg \emph{et~al.}(2013)Kaasbjerg, Thygesen, and
  Jauho]{Kaasb13PRB_Transport:Phonon}
K.~Kaasbjerg, K.~S. Thygesen and A.~P. Jauho, \emph{Phys. Rev. B}, 2013,
  \textbf{87}, 235312\relax
\mciteBstWouldAddEndPuncttrue
\mciteSetBstMidEndSepPunct{\mcitedefaultmidpunct}
{\mcitedefaultendpunct}{\mcitedefaultseppunct}\relax
\EndOfBibitem
\bibitem[Kim \emph{et~al.}(2012)Kim, Konar, Hwang, Lee, Lee, Yang, Jung, Kim,
  Yoo, Choi, Jin, Lee, Jena, Choi, and Kim]{Kim12NC}
S.~Kim, A.~Konar, W.-S. Hwang, J.~H. Lee, J.~Lee, J.~Yang, C.~Jung, H.~Kim,
  J.-B. Yoo, J.-Y. Choi, Y.~W. Jin, S.~Y. Lee, D.~Jena, W.~Choi and K.~Kim,
  \emph{Nat. Commun.}, 2012, \textbf{3}, 1011\relax
\mciteBstWouldAddEndPuncttrue
\mciteSetBstMidEndSepPunct{\mcitedefaultmidpunct}
{\mcitedefaultendpunct}{\mcitedefaultseppunct}\relax
\EndOfBibitem
\bibitem[Li \emph{et~al.}(2013)Li, Wakabayashi, Xu, Nakaharai, Komatsu, Li,
  Lin, Aparecido-Ferreira, and Tsukagoshi]{Li13NL}
S.-L. Li, K.~Wakabayashi, Y.~Xu, S.~Nakaharai, K.~Komatsu, W.-W. Li, Y.-F. Lin,
  A.~Aparecido-Ferreira and K.~Tsukagoshi, \emph{Nano Lett.}, 2013,
  \textbf{13}, 3546--3552\relax
\mciteBstWouldAddEndPuncttrue
\mciteSetBstMidEndSepPunct{\mcitedefaultmidpunct}
{\mcitedefaultendpunct}{\mcitedefaultseppunct}\relax
\EndOfBibitem
\bibitem[Ma and Jena(2014)]{Ma14PRX}
N.~Ma and D.~Jena, \emph{Phys. Rev. X}, 2014, \textbf{4}, 011043\relax
\mciteBstWouldAddEndPuncttrue
\mciteSetBstMidEndSepPunct{\mcitedefaultmidpunct}
{\mcitedefaultendpunct}{\mcitedefaultseppunct}\relax
\EndOfBibitem
\bibitem[Bouroushian(2010)]{bouro10_book:chalcogenide}
M.~Bouroushian, \emph{Electrochemistry of metal chalcogenides}, Springer,
  Berlin Heidelberg, 2010\relax
\mciteBstWouldAddEndPuncttrue
\mciteSetBstMidEndSepPunct{\mcitedefaultmidpunct}
{\mcitedefaultendpunct}{\mcitedefaultseppunct}\relax
\EndOfBibitem
\bibitem[Li \emph{et~al.}(2012)Li, Miyazaki, Song, Kuramochi, Nakaharai, and
  Tsukagoshi]{Li12AN}
S.-L. Li, H.~Miyazaki, H.~Song, H.~Kuramochi, S.~Nakaharai and K.~Tsukagoshi,
  \emph{ACS Nano}, 2012, \textbf{6}, 7381--7388\relax
\mciteBstWouldAddEndPuncttrue
\mciteSetBstMidEndSepPunct{\mcitedefaultmidpunct}
{\mcitedefaultendpunct}{\mcitedefaultseppunct}\relax
\EndOfBibitem
\bibitem[Lin \emph{et~al.}(2014)Lin, Dumcenco, Huang, and
  Suenaga]{Lin14NN_1Tphase:TEM}
Y.-C. Lin, D.~O. Dumcenco, Y.-S. Huang and K.~Suenaga, \emph{Nat.
  Nanotechnol.}, 2014, \textbf{9}, 391--396\relax
\mciteBstWouldAddEndPuncttrue
\mciteSetBstMidEndSepPunct{\mcitedefaultmidpunct}
{\mcitedefaultendpunct}{\mcitedefaultseppunct}\relax
\EndOfBibitem
\bibitem[Cho \emph{et~al.}(2015)Cho, Kim, Kim, Zhao, Seok, Keum, Baik, Choe,
  Chang, Suenaga, Kim, Lee, and Yang]{Cho15S}
S.~Cho, S.~Kim, J.~H. Kim, J.~Zhao, J.~Seok, D.~H. Keum, J.~Baik, D.-H. Choe,
  K.~J. Chang, K.~Suenaga, S.~W. Kim, Y.~H. Lee and H.~Yang, \emph{Science},
  2015, \textbf{349}, 625\relax
\mciteBstWouldAddEndPuncttrue
\mciteSetBstMidEndSepPunct{\mcitedefaultmidpunct}
{\mcitedefaultendpunct}{\mcitedefaultseppunct}\relax
\EndOfBibitem
\bibitem[Kappera \emph{et~al.}(2014)Kappera, Voiry, Yalcin, Branch, Gupta,
  Mohite, and Chhowalla]{Kappera14NM}
R.~Kappera, D.~Voiry, S.~E. Yalcin, B.~Branch, G.~Gupta, A.~D. Mohite and
  M.~Chhowalla, \emph{Nat. Mater.}, 2014, \textbf{13}, 1128--1134\relax
\mciteBstWouldAddEndPuncttrue
\mciteSetBstMidEndSepPunct{\mcitedefaultmidpunct}
{\mcitedefaultendpunct}{\mcitedefaultseppunct}\relax
\EndOfBibitem
\bibitem[Kappera \emph{et~al.}(2014)Kappera, Voiry, Yalcin, Jen, Acerce,
  Torrel, Branch, Lei, Chen, Najmaei, Lou, Ajayan, Gupta, Mohite, and
  Chhowalla]{Kappe14AM_FET:MoS2:1Tcontact}
R.~Kappera, D.~Voiry, S.~E. Yalcin, W.~Jen, M.~Acerce, S.~Torrel, B.~Branch,
  S.~Lei, W.~Chen, S.~Najmaei, J.~Lou, P.~M. Ajayan, G.~Gupta, A.~D. Mohite and
  M.~Chhowalla, \emph{APL Mater.}, 2014, \textbf{2}, 092516\relax
\mciteBstWouldAddEndPuncttrue
\mciteSetBstMidEndSepPunct{\mcitedefaultmidpunct}
{\mcitedefaultendpunct}{\mcitedefaultseppunct}\relax
\EndOfBibitem
\bibitem[Voiry \emph{et~al.}(2015)Voiry, Mohite, and Chhowalla]{Voiry15CSR}
D.~Voiry, A.~Mohite and M.~Chhowalla, \emph{Chem. Soc. Rev.}, 2015,
  \textbf{44}, 2702--2712\relax
\mciteBstWouldAddEndPuncttrue
\mciteSetBstMidEndSepPunct{\mcitedefaultmidpunct}
{\mcitedefaultendpunct}{\mcitedefaultseppunct}\relax
\EndOfBibitem
\bibitem[Molina-Sanchez and Wirtz(2011)]{Molina-Sanchez11PRB}
A.~Molina-Sanchez and L.~Wirtz, \emph{Phys. Rev. B}, 2011, \textbf{84},
  155413\relax
\mciteBstWouldAddEndPuncttrue
\mciteSetBstMidEndSepPunct{\mcitedefaultmidpunct}
{\mcitedefaultendpunct}{\mcitedefaultseppunct}\relax
\EndOfBibitem
\bibitem[Verble and Wieting(1970)]{Verble70PRL}
J.~L. Verble and T.~J. Wieting, \emph{Phys. Rev. Lett.}, 1970, \textbf{25},
  362--365\relax
\mciteBstWouldAddEndPuncttrue
\mciteSetBstMidEndSepPunct{\mcitedefaultmidpunct}
{\mcitedefaultendpunct}{\mcitedefaultseppunct}\relax
\EndOfBibitem
\bibitem[Wieting and Verble(1971)]{Wieting71PRB}
T.~J. Wieting and J.~L. Verble, \emph{Phys. Rev. B}, 1971, \textbf{3},
  4286--4292\relax
\mciteBstWouldAddEndPuncttrue
\mciteSetBstMidEndSepPunct{\mcitedefaultmidpunct}
{\mcitedefaultendpunct}{\mcitedefaultseppunct}\relax
\EndOfBibitem
\bibitem[Agnihotr and Sehgal(1972)]{Agnihotr72PM}
O.~P. Agnihotr and H.~K. Sehgal, \emph{Philos. Mag.}, 1972, \textbf{26},
  753\relax
\mciteBstWouldAddEndPuncttrue
\mciteSetBstMidEndSepPunct{\mcitedefaultmidpunct}
{\mcitedefaultendpunct}{\mcitedefaultseppunct}\relax
\EndOfBibitem
\bibitem[Agnihotr \emph{et~al.}(1973)Agnihotr, Sehgal, and Garg]{Agnihotr73SSC}
O.~P. Agnihotr, H.~K. Sehgal and A.~K. Garg, \emph{Solid State Commun.}, 1973,
  \textbf{12}, 135--138\relax
\mciteBstWouldAddEndPuncttrue
\mciteSetBstMidEndSepPunct{\mcitedefaultmidpunct}
{\mcitedefaultendpunct}{\mcitedefaultseppunct}\relax
\EndOfBibitem
\bibitem[Sekine \emph{et~al.}(1980)Sekine, Izumi, Nakashizu, Uchinokura, and
  Matsuura]{Sekine80JPSJ}
T.~Sekine, M.~Izumi, T.~Nakashizu, K.~Uchinokura and E.~Matsuura, \emph{J.
  Phys. Soc. Jpn.}, 1980, \textbf{49}, 1069--1077\relax
\mciteBstWouldAddEndPuncttrue
\mciteSetBstMidEndSepPunct{\mcitedefaultmidpunct}
{\mcitedefaultendpunct}{\mcitedefaultseppunct}\relax
\EndOfBibitem
\bibitem[Sugai and Ueda(1982)]{Sugai82PRB}
S.~Sugai and T.~Ueda, \emph{Phys. Rev. B}, 1982, \textbf{26}, 6554--6558\relax
\mciteBstWouldAddEndPuncttrue
\mciteSetBstMidEndSepPunct{\mcitedefaultmidpunct}
{\mcitedefaultendpunct}{\mcitedefaultseppunct}\relax
\EndOfBibitem
\bibitem[Sekine \emph{et~al.}(1984)Sekine, Uchinokura, Nakashizu, Matsuura, and
  Yoshizaki]{Sekine84JPSJ}
T.~Sekine, K.~Uchinokura, T.~Nakashizu, E.~Matsuura and R.~Yoshizaki, \emph{J.
  Phys. Soc. Jpn.}, 1984, \textbf{53}, 811--818\relax
\mciteBstWouldAddEndPuncttrue
\mciteSetBstMidEndSepPunct{\mcitedefaultmidpunct}
{\mcitedefaultendpunct}{\mcitedefaultseppunct}\relax
\EndOfBibitem
\bibitem[Wieting \emph{et~al.}(1980)Wieting, Grisel, and Levy]{Wieting80PB}
T.~Wieting, A.~Grisel and F.~Levy, \emph{Phys. B}, 1980, \textbf{99},
  337--342\relax
\mciteBstWouldAddEndPuncttrue
\mciteSetBstMidEndSepPunct{\mcitedefaultmidpunct}
{\mcitedefaultendpunct}{\mcitedefaultseppunct}\relax
\EndOfBibitem
\bibitem[Sekine \emph{et~al.}(1980)Sekine, Nakashizu, Toyoda, Uchinokura, and
  Matsuura]{SEKINE80SSC}
T.~Sekine, T.~Nakashizu, K.~Toyoda, K.~Uchinokura and E.~Matsuura, \emph{Solid
  State Commun.}, 1980, \textbf{35}, 371--373\relax
\mciteBstWouldAddEndPuncttrue
\mciteSetBstMidEndSepPunct{\mcitedefaultmidpunct}
{\mcitedefaultendpunct}{\mcitedefaultseppunct}\relax
\EndOfBibitem
\bibitem[Sourisseau \emph{et~al.}(1991)Sourisseau, Cruege, Fouassier, and
  Alba]{SOURISSEAU91CP}
C.~Sourisseau, F.~Cruege, M.~Fouassier and M.~Alba, \emph{Chem. Phys.}, 1991,
  \textbf{150}, 281--293\relax
\mciteBstWouldAddEndPuncttrue
\mciteSetBstMidEndSepPunct{\mcitedefaultmidpunct}
{\mcitedefaultendpunct}{\mcitedefaultseppunct}\relax
\EndOfBibitem
\bibitem[Frey \emph{et~al.}(1998)Frey, Tenne, Matthews, Dresselhaus, and
  Dresselhaus]{Frey98JMR}
G.~Frey, R.~Tenne, M.~Matthews, M.~Dresselhaus and G.~Dresselhaus, \emph{J.
  Mater. Res.}, 1998, \textbf{13}, 2412--2417\relax
\mciteBstWouldAddEndPuncttrue
\mciteSetBstMidEndSepPunct{\mcitedefaultmidpunct}
{\mcitedefaultendpunct}{\mcitedefaultseppunct}\relax
\EndOfBibitem
\bibitem[Lee \emph{et~al.}(2010)Lee, Yan, Brus, Heinz, Hone, and
  Ryu]{Lee10AN_Raman:MoS2}
C.~Lee, H.~Yan, L.~E. Brus, T.~F. Heinz, J.~Hone and S.~Ryu, \emph{ACS Nano},
  2010, \textbf{4}, 2695--2700\relax
\mciteBstWouldAddEndPuncttrue
\mciteSetBstMidEndSepPunct{\mcitedefaultmidpunct}
{\mcitedefaultendpunct}{\mcitedefaultseppunct}\relax
\EndOfBibitem
\bibitem[Terrones \emph{et~al.}(2014)Terrones, Corro, Feng, Poumirol, Rhodes,
  Smirnov, Pradhan, Lin, Nguyen, Elias, Mallouk, Balicas, Pimenta, and
  Terrones]{Terro14SR_Raman:MX2:newMode}
H.~Terrones, E.~D. Corro, S.~Feng, J.~M. Poumirol, D.~Rhodes, D.~Smirnov, N.~R.
  Pradhan, Z.~Lin, M.~A.~T. Nguyen, A.~L. Elias, T.~E. Mallouk, L.~Balicas,
  M.~A. Pimenta and M.~Terrones, \emph{Sci. Rep.}, 2014, \textbf{4}, 4215\relax
\mciteBstWouldAddEndPuncttrue
\mciteSetBstMidEndSepPunct{\mcitedefaultmidpunct}
{\mcitedefaultendpunct}{\mcitedefaultseppunct}\relax
\EndOfBibitem
\bibitem[Yamamoto \emph{et~al.}(2014)Yamamoto, Wang, Ni, Lin, Li, Aikawa, Jian,
  Ueno, Wakabayashi, and Tsukagoshi]{Yamam14AN_Raman:MoTe:NewMode}
M.~Yamamoto, S.~T. Wang, M.~Ni, Y.-F. Lin, S.-L. Li, S.~Aikawa, W.-B. Jian,
  K.~Ueno, K.~Wakabayashi and K.~Tsukagoshi, \emph{ACS Nano}, 2014, \textbf{8},
  3895--3903\relax
\mciteBstWouldAddEndPuncttrue
\mciteSetBstMidEndSepPunct{\mcitedefaultmidpunct}
{\mcitedefaultendpunct}{\mcitedefaultseppunct}\relax
\EndOfBibitem
\bibitem[Jim\'enez~Sandoval \emph{et~al.}(1991)Jim\'enez~Sandoval, Yang,
  Frindt, and Irwin]{Jim91PRB_Phonon:mode:1LMoS2}
S.~Jim\'enez~Sandoval, D.~Yang, R.~F. Frindt and J.~C. Irwin, \emph{Phys. Rev.
  B}, 1991, \textbf{44}, 3955--3962\relax
\mciteBstWouldAddEndPuncttrue
\mciteSetBstMidEndSepPunct{\mcitedefaultmidpunct}
{\mcitedefaultendpunct}{\mcitedefaultseppunct}\relax
\EndOfBibitem
\bibitem[Ataca \emph{et~al.}(2011)Ataca, Topsakal, Akturk, and
  Ciraci]{Ataca11JPCCb}
C.~Ataca, M.~Topsakal, E.~Akturk and S.~Ciraci, \emph{J. Phys. Chem. C}, 2011,
  \textbf{115}, 16354--16361\relax
\mciteBstWouldAddEndPuncttrue
\mciteSetBstMidEndSepPunct{\mcitedefaultmidpunct}
{\mcitedefaultendpunct}{\mcitedefaultseppunct}\relax
\EndOfBibitem
\bibitem[Ribeiro-Soares \emph{et~al.}(2014)Ribeiro-Soares, Almeida, Barros,
  Araujo, Dresselhaus, Can\ifmmode~\mbox{\c{c}}\else \c{c}\fi{}ado, and
  Jorio]{Ribei14PRB_Phonon:mode}
J.~Ribeiro-Soares, R.~M. Almeida, E.~B. Barros, P.~T. Araujo, M.~S.
  Dresselhaus, L.~G. Can\ifmmode~\mbox{\c{c}}\else \c{c}\fi{}ado and A.~Jorio,
  \emph{Phys. Rev. B}, 2014, \textbf{90}, 115438\relax
\mciteBstWouldAddEndPuncttrue
\mciteSetBstMidEndSepPunct{\mcitedefaultmidpunct}
{\mcitedefaultendpunct}{\mcitedefaultseppunct}\relax
\EndOfBibitem
\bibitem[Zhao \emph{et~al.}(2013)Zhao, Luo, Li, Zhang, Araujo, Gan, Wu, Zhang,
  Quek, Dresselhaus, and Xiong]{Zhao13NLb}
Y.~Zhao, X.~Luo, H.~Li, J.~Zhang, P.~T. Araujo, C.~K. Gan, J.~Wu, H.~Zhang,
  S.~Y. Quek, M.~S. Dresselhaus and Q.~Xiong, \emph{Nano Lett.}, 2013,
  \textbf{13}, 1007--1015\relax
\mciteBstWouldAddEndPuncttrue
\mciteSetBstMidEndSepPunct{\mcitedefaultmidpunct}
{\mcitedefaultendpunct}{\mcitedefaultseppunct}\relax
\EndOfBibitem
\bibitem[Kuc \emph{et~al.}(2011)Kuc, Zibouche, and Heine]{Kuc11PRB}
A.~Kuc, N.~Zibouche and T.~Heine, \emph{Phys. Rev. B}, 2011, \textbf{83},
  245213\relax
\mciteBstWouldAddEndPuncttrue
\mciteSetBstMidEndSepPunct{\mcitedefaultmidpunct}
{\mcitedefaultendpunct}{\mcitedefaultseppunct}\relax
\EndOfBibitem
\bibitem[Yun \emph{et~al.}(2012)Yun, Han, Hong, Kim, and Lee]{Yun12PRB}
W.~S. Yun, S.~W. Han, S.~C. Hong, I.~G. Kim and J.~D. Lee, \emph{Phys. Rev. B},
  2012, \textbf{85}, 033305\relax
\mciteBstWouldAddEndPuncttrue
\mciteSetBstMidEndSepPunct{\mcitedefaultmidpunct}
{\mcitedefaultendpunct}{\mcitedefaultseppunct}\relax
\EndOfBibitem
\bibitem[Kumar and Ahluwalia(2012)]{Kumar12PB_Dielectric:MX2}
A.~Kumar and P.~K. Ahluwalia, \emph{Phys. B}, 2012, \textbf{407},
  4627--4634\relax
\mciteBstWouldAddEndPuncttrue
\mciteSetBstMidEndSepPunct{\mcitedefaultmidpunct}
{\mcitedefaultendpunct}{\mcitedefaultseppunct}\relax
\EndOfBibitem
\bibitem[Yoon \emph{et~al.}(2011)Yoon, Ganapathi, and Salahuddin]{Yoon11NL_}
Y.~Yoon, K.~Ganapathi and S.~Salahuddin, \emph{Nano Lett.}, 2011, \textbf{11},
  3768--3773\relax
\mciteBstWouldAddEndPuncttrue
\mciteSetBstMidEndSepPunct{\mcitedefaultmidpunct}
{\mcitedefaultendpunct}{\mcitedefaultseppunct}\relax
\EndOfBibitem
\bibitem[Beal and Hughes(1979)]{Beal79JPCSSP}
A.~R. Beal and H.~P. Hughes, \emph{J. Phys. C: Solid State Phys.}, 1979,
  \textbf{12}, 881\relax
\mciteBstWouldAddEndPuncttrue
\mciteSetBstMidEndSepPunct{\mcitedefaultmidpunct}
{\mcitedefaultendpunct}{\mcitedefaultseppunct}\relax
\EndOfBibitem
\bibitem[Yim \emph{et~al.}(2014)Yim, O'Brien, McEvoy, Winters, Mirza, Lunney,
  and Duesberg]{Yim14APL_Dielectric:MoS2:thickness}
C.~Yim, M.~O'Brien, N.~McEvoy, S.~Winters, I.~Mirza, J.~G. Lunney and G.~S.
  Duesberg, \emph{Appl. Phys. Lett.}, 2014, \textbf{104}, 103114\relax
\mciteBstWouldAddEndPuncttrue
\mciteSetBstMidEndSepPunct{\mcitedefaultmidpunct}
{\mcitedefaultendpunct}{\mcitedefaultseppunct}\relax
\EndOfBibitem
\bibitem[Zhang \emph{et~al.}(2015)Zhang, Ma, Wan, Rong, Xie, Wang, and
  Dai]{Zhang15SR_Dielectric:MoS2}
H.~Zhang, Y.~Ma, Y.~Wan, X.~Rong, Z.~Xie, W.~Wang and L.~Dai, \emph{Sci. Rep.},
  2015, \textbf{5}, 8440\relax
\mciteBstWouldAddEndPuncttrue
\mciteSetBstMidEndSepPunct{\mcitedefaultmidpunct}
{\mcitedefaultendpunct}{\mcitedefaultseppunct}\relax
\EndOfBibitem
\bibitem[Mukherjee \emph{et~al.}(2015)Mukherjee, Tseng, Gunlycke, Amara, Eda,
  and Simsek]{Mukhe15OME_Dielectric:MoS2_1L}
B.~Mukherjee, F.~Tseng, D.~Gunlycke, K.~K. Amara, G.~Eda and E.~Simsek,
  \emph{Optical Materials Express}, 2015, \textbf{5}, 447--455\relax
\mciteBstWouldAddEndPuncttrue
\mciteSetBstMidEndSepPunct{\mcitedefaultmidpunct}
{\mcitedefaultendpunct}{\mcitedefaultseppunct}\relax
\EndOfBibitem
\bibitem[Liu \emph{et~al.}(2014)Liu, Shen, Su, Hsu, Li, and
  Li]{Liu14APL_Dielectric:MX2}
H.-L. Liu, C.-C. Shen, S.-H. Su, C.-L. Hsu, M.-Y. Li and L.-J. Li, \emph{Appl.
  Phys. Lett.}, 2014, \textbf{105}, 201905\relax
\mciteBstWouldAddEndPuncttrue
\mciteSetBstMidEndSepPunct{\mcitedefaultmidpunct}
{\mcitedefaultendpunct}{\mcitedefaultseppunct}\relax
\EndOfBibitem
\bibitem[Li \emph{et~al.}(2014)Li, Chernikov, Zhang, Rigosi, Hill, van~der
  Zande, Chenet, Shih, Hone, and Heinz]{Li14PRB_Dielectric:MX2}
Y.~Li, A.~Chernikov, X.~Zhang, A.~Rigosi, H.~M. Hill, A.~M. van~der Zande,
  D.~A. Chenet, E.-M. Shih, J.~Hone and T.~F. Heinz, \emph{Phys. Rev. B}, 2014,
  \textbf{90}, 205422\relax
\mciteBstWouldAddEndPuncttrue
\mciteSetBstMidEndSepPunct{\mcitedefaultmidpunct}
{\mcitedefaultendpunct}{\mcitedefaultseppunct}\relax
\EndOfBibitem
\bibitem[Late \emph{et~al.}(2012)Late, Liu, Matte, Dravid, and
  Rao]{Late12AN_FET:absorbate:MoS2}
D.~J. Late, B.~Liu, H.~S. S.~R. Matte, V.~P. Dravid and C.~N.~R. Rao, \emph{ACS
  Nano}, 2012, \textbf{6}, 5635--5641\relax
\mciteBstWouldAddEndPuncttrue
\mciteSetBstMidEndSepPunct{\mcitedefaultmidpunct}
{\mcitedefaultendpunct}{\mcitedefaultseppunct}\relax
\EndOfBibitem
\bibitem[Sangwan \emph{et~al.}(2013)Sangwan, Arnold, Jariwala, Marks, Lauhon,
  and Hersam]{Sangw13NL_noise}
V.~K. Sangwan, H.~N. Arnold, D.~Jariwala, T.~J. Marks, L.~J. Lauhon and M.~C.
  Hersam, \emph{Nano Lett.}, 2013, \textbf{13}, 4351--4355\relax
\mciteBstWouldAddEndPuncttrue
\mciteSetBstMidEndSepPunct{\mcitedefaultmidpunct}
{\mcitedefaultendpunct}{\mcitedefaultseppunct}\relax
\EndOfBibitem
\bibitem[Lembke \emph{et~al.}(2015)Lembke, Allain, and Kis]{Lembk15N_}
D.~Lembke, A.~Allain and A.~Kis, \emph{Nanoscale}, 2015, \textbf{7},
  6255--6260\relax
\mciteBstWouldAddEndPuncttrue
\mciteSetBstMidEndSepPunct{\mcitedefaultmidpunct}
{\mcitedefaultendpunct}{\mcitedefaultseppunct}\relax
\EndOfBibitem
\bibitem[Baugher \emph{et~al.}(2013)Baugher, Churchill, Yang, and
  Jarillo-Herrero]{Baugher13NL}
B.~W.~H. Baugher, H.~O.~H. Churchill, Y.~Yang and P.~Jarillo-Herrero,
  \emph{Nano Lett.}, 2013, \textbf{13}, 4212--4216\relax
\mciteBstWouldAddEndPuncttrue
\mciteSetBstMidEndSepPunct{\mcitedefaultmidpunct}
{\mcitedefaultendpunct}{\mcitedefaultseppunct}\relax
\EndOfBibitem
\bibitem[Chu \emph{et~al.}(2014)Chu, Schmidt, Pu, Wang, \"{O}zyilmaz, Takenobu,
  and Eda]{Chu14SR_FET:MoS2:IonicGated}
L.~Chu, H.~Schmidt, J.~Pu, S.~Wang, B.~\"{O}zyilmaz, T.~Takenobu and G.~Eda,
  \emph{Sci. Rep.}, 2014, \textbf{4}, 7293\relax
\mciteBstWouldAddEndPuncttrue
\mciteSetBstMidEndSepPunct{\mcitedefaultmidpunct}
{\mcitedefaultendpunct}{\mcitedefaultseppunct}\relax
\EndOfBibitem
\bibitem[Kang \emph{et~al.}(2014)Kang, Liu, and
  Banerjee]{Kang14APL_FET:MoS2:MoContact}
J.~Kang, W.~Liu and K.~Banerjee, \emph{Appl. Phys. Lett.}, 2014, \textbf{104},
  093106\relax
\mciteBstWouldAddEndPuncttrue
\mciteSetBstMidEndSepPunct{\mcitedefaultmidpunct}
{\mcitedefaultendpunct}{\mcitedefaultseppunct}\relax
\EndOfBibitem
\bibitem[Lee \emph{et~al.}(2013)Lee, Yu, Cui, Petrone, Lee, Choi, Lee, Lee,
  Yoo, Watanabe, Taniguchi, Nuckolls, Kim, and
  Hone]{Lee13AN_BN:u45_3LMoS_thickness}
G.-H. Lee, Y.-J. Yu, X.~Cui, N.~Petrone, C.-H. Lee, M.~S. Choi, D.-Y. Lee,
  C.~Lee, W.~J. Yoo, K.~Watanabe, T.~Taniguchi, C.~Nuckolls, P.~Kim and
  J.~Hone, \emph{ACS Nano}, 2013, \textbf{7}, 7931--7936\relax
\mciteBstWouldAddEndPuncttrue
\mciteSetBstMidEndSepPunct{\mcitedefaultmidpunct}
{\mcitedefaultendpunct}{\mcitedefaultseppunct}\relax
\EndOfBibitem
\bibitem[Chan \emph{et~al.}(2013)Chan, Komatsu, Li, Xu, Darmawan, Kuramochi,
  Nakaharai, Aparecido-Ferreira, Watanabe, Taniguchi, and Tsukagoshi]{Chan13N}
M.~Y. Chan, K.~Komatsu, S.-L. Li, Y.~Xu, P.~Darmawan, H.~Kuramochi,
  S.~Nakaharai, A.~Aparecido-Ferreira, K.~Watanabe, T.~Taniguchi and
  K.~Tsukagoshi, \emph{Nanoscale}, 2013, \textbf{5}, 9572--9576\relax
\mciteBstWouldAddEndPuncttrue
\mciteSetBstMidEndSepPunct{\mcitedefaultmidpunct}
{\mcitedefaultendpunct}{\mcitedefaultseppunct}\relax
\EndOfBibitem
\bibitem[Jin \emph{et~al.}(2013)Jin, Kang, Kim, Lee, and
  Lee]{Jin13JAP_SuspendedMoS2_mu1}
T.~Jin, J.~Kang, E.~S. Kim, S.~Lee and C.~Lee, \emph{J. Appl. Phys.}, 2013,
  \textbf{114}, 164509\relax
\mciteBstWouldAddEndPuncttrue
\mciteSetBstMidEndSepPunct{\mcitedefaultmidpunct}
{\mcitedefaultendpunct}{\mcitedefaultseppunct}\relax
\EndOfBibitem
\bibitem[Klots \emph{et~al.}(2014)Klots, Newaz, Wang, Prasai, Krzyzanowska,
  Lin, Caudel, Ghimire, Yan, Ivanov, Velizhanin, Burger, Mandrus, Tolk,
  Pantelides, and Bolotin]{Klots14SR_SuspendedMoS2:Photocurrent}
A.~R. Klots, A.~K.~M. Newaz, B.~Wang, D.~Prasai, H.~Krzyzanowska, J.~Lin,
  D.~Caudel, N.~J. Ghimire, J.~Yan, B.~L. Ivanov, K.~A. Velizhanin, A.~Burger,
  D.~G. Mandrus, N.~H. Tolk, S.~T. Pantelides and K.~I. Bolotin, \emph{Sci.
  Rep.}, 2014, \textbf{4}, 6608\relax
\mciteBstWouldAddEndPuncttrue
\mciteSetBstMidEndSepPunct{\mcitedefaultmidpunct}
{\mcitedefaultendpunct}{\mcitedefaultseppunct}\relax
\EndOfBibitem
\bibitem[Guo \emph{et~al.}(2014)Guo, Han, Li, Xiang, Wei, Gao, and
  Chen]{Guo14AN}
Y.~Guo, Y.~Han, J.~Li, A.~Xiang, X.~Wei, S.~Gao and Q.~Chen, \emph{ACS Nano},
  2014, \textbf{8}, 7771--7779\relax
\mciteBstWouldAddEndPuncttrue
\mciteSetBstMidEndSepPunct{\mcitedefaultmidpunct}
{\mcitedefaultendpunct}{\mcitedefaultseppunct}\relax
\EndOfBibitem
\bibitem[Qiu \emph{et~al.}(2012)Qiu, Pan, Yao, Li, Shi, and
  Wang]{Qiu12APL_Transport:2L}
H.~Qiu, L.~Pan, Z.~Yao, J.~Li, Y.~Shi and X.~Wang, \emph{Appl. Phys. Lett.},
  2012, \textbf{100}, 123104\relax
\mciteBstWouldAddEndPuncttrue
\mciteSetBstMidEndSepPunct{\mcitedefaultmidpunct}
{\mcitedefaultendpunct}{\mcitedefaultseppunct}\relax
\EndOfBibitem
\bibitem[Das \emph{et~al.}(2013)Das, Chen, Penumatcha, and
  Appenzeller]{Das13NLb}
S.~Das, H.-Y. Chen, A.~V. Penumatcha and J.~Appenzeller, \emph{Nano Lett.},
  2013, \textbf{13}, 100--105\relax
\mciteBstWouldAddEndPuncttrue
\mciteSetBstMidEndSepPunct{\mcitedefaultmidpunct}
{\mcitedefaultendpunct}{\mcitedefaultseppunct}\relax
\EndOfBibitem
\bibitem[Perera \emph{et~al.}(2013)Perera, Lin, Chuang, Chamlagain, Wang, Tan,
  Cheng, Tom\'{a}nek, and Zhou]{Perer13AN_}
M.~M. Perera, M.-W. Lin, H.-J. Chuang, B.~P. Chamlagain, C.~Wang, X.~Tan,
  M.~M.-C. Cheng, D.~Tom\'{a}nek and Z.~Zhou, \emph{ACS Nano}, 2013,
  \textbf{7}, 4449--4458\relax
\mciteBstWouldAddEndPuncttrue
\mciteSetBstMidEndSepPunct{\mcitedefaultmidpunct}
{\mcitedefaultendpunct}{\mcitedefaultseppunct}\relax
\EndOfBibitem
\bibitem[Fang \emph{et~al.}(2013)Fang, Tosun, Seol, Chang, Takei, Guo, and
  Javey]{Fang13NL}
H.~Fang, M.~Tosun, G.~Seol, T.~C. Chang, K.~Takei, J.~Guo and A.~Javey,
  \emph{Nano Lett.}, 2013, \textbf{13}, 1991--1995\relax
\mciteBstWouldAddEndPuncttrue
\mciteSetBstMidEndSepPunct{\mcitedefaultmidpunct}
{\mcitedefaultendpunct}{\mcitedefaultseppunct}\relax
\EndOfBibitem
\bibitem[Wang \emph{et~al.}(2014)Wang, Liu, Tang, Jin, Zhao, and
  Xiu]{Wang14SR_Rc:MoS2:Permalloy}
W.~Wang, Y.~Liu, L.~Tang, Y.~Jin, T.~Zhao and F.~Xiu, \emph{Sci. Rep.}, 2014,
  \textbf{4}, 6928\relax
\mciteBstWouldAddEndPuncttrue
\mciteSetBstMidEndSepPunct{\mcitedefaultmidpunct}
{\mcitedefaultendpunct}{\mcitedefaultseppunct}\relax
\EndOfBibitem
\bibitem[Pu \emph{et~al.}(2012)Pu, Yomogida, Liu, Li, Iwasa, and
  Takenobu]{Pu12NL_}
J.~Pu, Y.~Yomogida, K.-K. Liu, L.-J. Li, Y.~Iwasa and T.~Takenobu, \emph{Nano
  Lett.}, 2012, \textbf{12}, 4013--4017\relax
\mciteBstWouldAddEndPuncttrue
\mciteSetBstMidEndSepPunct{\mcitedefaultmidpunct}
{\mcitedefaultendpunct}{\mcitedefaultseppunct}\relax
\EndOfBibitem
\bibitem[Zou \emph{et~al.}(2014)Zou, Wang, Chiu, Wu, Xiao, Jiang, Wu, Mai,
  Chen, Li, Ho, and Liao]{Zou14AM}
X.~Zou, J.~Wang, C.-H. Chiu, Y.~Wu, X.~Xiao, C.~Jiang, W.-W. Wu, L.~Mai,
  T.~Chen, J.~Li, J.~C. Ho and L.~Liao, \emph{Adv. Mater.}, 2014, \textbf{26},
  6255--6261\relax
\mciteBstWouldAddEndPuncttrue
\mciteSetBstMidEndSepPunct{\mcitedefaultmidpunct}
{\mcitedefaultendpunct}{\mcitedefaultseppunct}\relax
\EndOfBibitem
\bibitem[Neal \emph{et~al.}(2013)Neal, Liu, Gu, and Ye]{Neal13AN_FET:MoS2}
A.~T. Neal, H.~Liu, J.~Gu and P.~D. Ye, \emph{ACS Nano}, 2013, \textbf{7},
  7077--7082\relax
\mciteBstWouldAddEndPuncttrue
\mciteSetBstMidEndSepPunct{\mcitedefaultmidpunct}
{\mcitedefaultendpunct}{\mcitedefaultseppunct}\relax
\EndOfBibitem
\bibitem[Liu \emph{et~al.}(2012)Liu, Neal, and Ye]{Liu12AN}
H.~Liu, A.~T. Neal and P.~D. Ye, \emph{ACS Nano}, 2012, \textbf{6},
  8563--8569\relax
\mciteBstWouldAddEndPuncttrue
\mciteSetBstMidEndSepPunct{\mcitedefaultmidpunct}
{\mcitedefaultendpunct}{\mcitedefaultseppunct}\relax
\EndOfBibitem
\bibitem[Roy \emph{et~al.}(2014)Roy, Tosun, Kang, Sachid, Desai, Hettick, Hu,
  and Javey]{Roy14AN}
T.~Roy, M.~Tosun, J.~S. Kang, A.~B. Sachid, S.~B. Desai, M.~Hettick, C.~C. Hu
  and A.~Javey, \emph{ACS Nano}, 2014, \textbf{8}, 6259--6264\relax
\mciteBstWouldAddEndPuncttrue
\mciteSetBstMidEndSepPunct{\mcitedefaultmidpunct}
{\mcitedefaultendpunct}{\mcitedefaultseppunct}\relax
\EndOfBibitem
\bibitem[Zhang \emph{et~al.}(2012)Zhang, Ye, Matsuhashi, and Iwasa]{Zhang12NL_}
Y.~Zhang, J.~T. Ye, Y.~Matsuhashi and Y.~Iwasa, \emph{Nano Lett.}, 2012,
  \textbf{12}, 1136--1140\relax
\mciteBstWouldAddEndPuncttrue
\mciteSetBstMidEndSepPunct{\mcitedefaultmidpunct}
{\mcitedefaultendpunct}{\mcitedefaultseppunct}\relax
\EndOfBibitem
\bibitem[Na \emph{et~al.}(2014)Na, Joo, Shin, Huh, Kim, Piao, Jin, Jang, Choi,
  Shim, and Kim]{Na14N_noise}
J.~Na, M.-K. Joo, M.~Shin, J.~Huh, J.-S. Kim, M.~Piao, J.-E. Jin, H.-K. Jang,
  H.~J. Choi, J.~H. Shim and G.-T. Kim, \emph{Nanoscale}, 2014, \textbf{6},
  433--441\relax
\mciteBstWouldAddEndPuncttrue
\mciteSetBstMidEndSepPunct{\mcitedefaultmidpunct}
{\mcitedefaultendpunct}{\mcitedefaultseppunct}\relax
\EndOfBibitem
\bibitem[Wang \emph{et~al.}(2015)Wang, Stepanov, Gray, and
  Lau]{Wang15N_FET:MoS:suspended}
F.~Wang, P.~Stepanov, M.~Gray and C.~N. Lau, \emph{Nanotechnol.}, 2015,
  \textbf{26}, 105709--105709\relax
\mciteBstWouldAddEndPuncttrue
\mciteSetBstMidEndSepPunct{\mcitedefaultmidpunct}
{\mcitedefaultendpunct}{\mcitedefaultseppunct}\relax
\EndOfBibitem
\bibitem[Pradhan \emph{et~al.}(2013)Pradhan, Rhodes, Zhang, Talapatra,
  Terrones, Ajayan, and Balicas]{Pradhan13APL}
N.~R. Pradhan, D.~Rhodes, Q.~Zhang, S.~Talapatra, M.~Terrones, P.~M. Ajayan and
  L.~Balicas, \emph{Appl. Phys. Lett.}, 2013, \textbf{102}, 123105\relax
\mciteBstWouldAddEndPuncttrue
\mciteSetBstMidEndSepPunct{\mcitedefaultmidpunct}
{\mcitedefaultendpunct}{\mcitedefaultseppunct}\relax
\EndOfBibitem
\bibitem[Dankert \emph{et~al.}(2014)Dankert, Langouche, Kamalakar, and
  Dash]{Dankert14AN}
A.~Dankert, L.~Langouche, M.~V. Kamalakar and S.~P. Dash, \emph{ACS Nano},
  2014, \textbf{8}, 476--482\relax
\mciteBstWouldAddEndPuncttrue
\mciteSetBstMidEndSepPunct{\mcitedefaultmidpunct}
{\mcitedefaultendpunct}{\mcitedefaultseppunct}\relax
\EndOfBibitem
\bibitem[Sanne \emph{et~al.}(2015)Sanne, Ghosh, Rai, Movva, Sharma, Rao,
  Mathew, and Banerjee]{Sanne15APL_FET:MoS:Si3N4}
A.~Sanne, R.~Ghosh, A.~Rai, H.~C.~P. Movva, A.~Sharma, R.~Rao, L.~Mathew and
  S.~K. Banerjee, \emph{Appl. Phys. Lett.}, 2015, \textbf{106}, 062101\relax
\mciteBstWouldAddEndPuncttrue
\mciteSetBstMidEndSepPunct{\mcitedefaultmidpunct}
{\mcitedefaultendpunct}{\mcitedefaultseppunct}\relax
\EndOfBibitem
\bibitem[Najmaei \emph{et~al.}(2014)Najmaei, Zou, Er, Li, Jin, Gao, Zhang,
  Park, Ge, Lei, Kono, Shenoy, Yakobson, George, Ajayan, and
  Lou]{Najma14NL_Interface:SAM_mu13}
S.~Najmaei, X.~Zou, D.~Er, J.~Li, Z.~Jin, W.~Gao, Q.~Zhang, S.~Park, L.~Ge,
  S.~Lei, J.~Kono, V.~B. Shenoy, B.~I. Yakobson, A.~George, P.~M. Ajayan and
  J.~Lou, \emph{Nano Lett.}, 2014, \textbf{14}, 1354--1361\relax
\mciteBstWouldAddEndPuncttrue
\mciteSetBstMidEndSepPunct{\mcitedefaultmidpunct}
{\mcitedefaultendpunct}{\mcitedefaultseppunct}\relax
\EndOfBibitem
\bibitem[Liu \emph{et~al.}(2013)Liu, Si, Najmaei, Neal, Du, Ajayan, Lou, and
  Ye]{Liu13NL_}
H.~Liu, M.~Si, S.~Najmaei, A.~T. Neal, Y.~Du, P.~M. Ajayan, J.~Lou and P.~D.
  Ye, \emph{Nano Lett.}, 2013, \textbf{13}, 2640--2646\relax
\mciteBstWouldAddEndPuncttrue
\mciteSetBstMidEndSepPunct{\mcitedefaultmidpunct}
{\mcitedefaultendpunct}{\mcitedefaultseppunct}\relax
\EndOfBibitem
\bibitem[Zhu \emph{et~al.}(2014)Zhu, Low, Lee, Wang, Farmer, Kong, Xia, and
  Avouris]{Zhu14NC_trspt:bandtail}
W.~Zhu, T.~Low, Y.-H. Lee, H.~Wang, D.~B. Farmer, J.~Kong, F.~Xia and
  P.~Avouris, \emph{Nat. Commun.}, 2014, \textbf{5}, 3087\relax
\mciteBstWouldAddEndPuncttrue
\mciteSetBstMidEndSepPunct{\mcitedefaultmidpunct}
{\mcitedefaultendpunct}{\mcitedefaultseppunct}\relax
\EndOfBibitem
\bibitem[Yang \emph{et~al.}(2014)Yang, Wang, and
  Feng]{Yang14N_FET:MoS:thickness}
R.~Yang, Z.~Wang and P.~X. .~L. Feng, \emph{Nanoscale}, 2014, \textbf{6},
  12383--12390\relax
\mciteBstWouldAddEndPuncttrue
\mciteSetBstMidEndSepPunct{\mcitedefaultmidpunct}
{\mcitedefaultendpunct}{\mcitedefaultseppunct}\relax
\EndOfBibitem
\bibitem[Larentis \emph{et~al.}(2012)Larentis, Fallahazad, and
  Tutuc]{Laren12APL_FET:MoSe}
S.~Larentis, B.~Fallahazad and E.~Tutuc, \emph{Appl. Phys. Lett.}, 2012,
  \textbf{101}, 223104\relax
\mciteBstWouldAddEndPuncttrue
\mciteSetBstMidEndSepPunct{\mcitedefaultmidpunct}
{\mcitedefaultendpunct}{\mcitedefaultseppunct}\relax
\EndOfBibitem
\bibitem[Pradhan \emph{et~al.}(2014)Pradhan, Rhodes, Xin, Memaran, Bhaskaran,
  Siddiq, Hill, Ajayan, and Balicas]{Pradh14AN_FET:MoS}
N.~R. Pradhan, D.~Rhodes, Y.~Xin, S.~Memaran, L.~Bhaskaran, M.~Siddiq, S.~Hill,
  P.~M. Ajayan and L.~Balicas, \emph{ACS Nano}, 2014, \textbf{8},
  7923--7929\relax
\mciteBstWouldAddEndPuncttrue
\mciteSetBstMidEndSepPunct{\mcitedefaultmidpunct}
{\mcitedefaultendpunct}{\mcitedefaultseppunct}\relax
\EndOfBibitem
\bibitem[Chamlagain \emph{et~al.}(2014)Chamlagain, Li, Ghimire, Chuang, Perera,
  Tu, Xu, Pan, Xaio, Yan, Mandrus, and Zhou]{Chamlagain14AN}
B.~Chamlagain, Q.~Li, N.~J. Ghimire, H.-J. Chuang, M.~M. Perera, H.~Tu, Y.~Xu,
  M.~Pan, D.~Xaio, J.~Yan, D.~Mandrus and Z.~Zhou, \emph{ACS Nano}, 2014,
  \textbf{8}, 5079--5088\relax
\mciteBstWouldAddEndPuncttrue
\mciteSetBstMidEndSepPunct{\mcitedefaultmidpunct}
{\mcitedefaultendpunct}{\mcitedefaultseppunct}\relax
\EndOfBibitem
\bibitem[Abderrahmane \emph{et~al.}(2014)Abderrahmane, Ko, Thu, Ishizawa,
  Takamura, and Sandhu]{Abder14N_FET:MoSe}
A.~Abderrahmane, P.~J. Ko, T.~V. Thu, S.~Ishizawa, T.~Takamura and A.~Sandhu,
  \emph{Nanotechnol.}, 2014, \textbf{25}, 365202\relax
\mciteBstWouldAddEndPuncttrue
\mciteSetBstMidEndSepPunct{\mcitedefaultmidpunct}
{\mcitedefaultendpunct}{\mcitedefaultseppunct}\relax
\EndOfBibitem
\bibitem[Pradhan \emph{et~al.}(2014)Pradhan, Rhodes, Feng, Xin, Memaran, Moon,
  Terrones, Terrones, and Balicas]{Pradh14AN_FET:MoTe}
N.~R. Pradhan, D.~Rhodes, S.~Feng, Y.~Xin, S.~Memaran, B.-H. Moon, H.~Terrones,
  M.~Terrones and L.~Balicas, \emph{ACS Nano}, 2014, \textbf{8},
  5911--5920\relax
\mciteBstWouldAddEndPuncttrue
\mciteSetBstMidEndSepPunct{\mcitedefaultmidpunct}
{\mcitedefaultendpunct}{\mcitedefaultseppunct}\relax
\EndOfBibitem
\bibitem[Lin \emph{et~al.}(2014)Lin, Xu, Wang, Li, Yamamoto,
  Aparecido-Ferreira, Li, Sun, Nakaharai, Jian, Ueno, and
  Tsukagoshi]{Lin14AM_FET:MoTe}
Y.-F. Lin, Y.~Xu, S.-T. Wang, S.-L. Li, M.~Yamamoto, A.~Aparecido-Ferreira,
  W.~Li, H.~Sun, S.~Nakaharai, W.-B. Jian, K.~Ueno and K.~Tsukagoshi,
  \emph{Adv. Mater.}, 2014, \textbf{26}, 3263--3269\relax
\mciteBstWouldAddEndPuncttrue
\mciteSetBstMidEndSepPunct{\mcitedefaultmidpunct}
{\mcitedefaultendpunct}{\mcitedefaultseppunct}\relax
\EndOfBibitem
\bibitem[Lezama \emph{et~al.}(2014)Lezama, Ubaldini, Longobardi, Giannini,
  Renner, Kuzmenko, and Morpurgo]{Lezam142M_FET:MoTe}
I.~G. Lezama, A.~Ubaldini, M.~Longobardi, E.~Giannini, C.~Renner, A.~B.
  Kuzmenko and A.~F. Morpurgo, \emph{2D Materials}, 2014, \textbf{1},
  021002\relax
\mciteBstWouldAddEndPuncttrue
\mciteSetBstMidEndSepPunct{\mcitedefaultmidpunct}
{\mcitedefaultendpunct}{\mcitedefaultseppunct}\relax
\EndOfBibitem
\bibitem[Fathipour \emph{et~al.}(2014)Fathipour, Ma, Hwang, Protasenko,
  Vishwanath, Xing, Xu, Jena, Appenzeller, and
  Seabaugh]{Fathi14APL_MoTe:Pdetector}
S.~Fathipour, N.~Ma, W.~S. Hwang, V.~Protasenko, S.~Vishwanath, H.~G. Xing,
  H.~Xu, D.~Jena, J.~Appenzeller and A.~Seabaugh, \emph{Appl. Phys. Lett.},
  2014, \textbf{105}, 192101\relax
\mciteBstWouldAddEndPuncttrue
\mciteSetBstMidEndSepPunct{\mcitedefaultmidpunct}
{\mcitedefaultendpunct}{\mcitedefaultseppunct}\relax
\EndOfBibitem
\bibitem[Ovchinnikov \emph{et~al.}(2014)Ovchinnikov, Allain, Huang, Dumcenco,
  and Kis]{Ovchi14AN_FET:WS}
D.~Ovchinnikov, A.~Allain, Y.-S. Huang, D.~Dumcenco and A.~Kis, \emph{ACS
  Nano}, 2014, \textbf{8}, 8174--8181\relax
\mciteBstWouldAddEndPuncttrue
\mciteSetBstMidEndSepPunct{\mcitedefaultmidpunct}
{\mcitedefaultendpunct}{\mcitedefaultseppunct}\relax
\EndOfBibitem
\bibitem[Jo \emph{et~al.}(2014)Jo, Ubrig, Berger, Kuzmenko, and
  Morpurgo]{Jo14NL_Opto-LED:WS2}
S.~Jo, N.~Ubrig, H.~Berger, A.~B. Kuzmenko and A.~F. Morpurgo, \emph{Nano
  Lett.}, 2014, \textbf{14}, 2019--2025\relax
\mciteBstWouldAddEndPuncttrue
\mciteSetBstMidEndSepPunct{\mcitedefaultmidpunct}
{\mcitedefaultendpunct}{\mcitedefaultseppunct}\relax
\EndOfBibitem
\bibitem[Withers \emph{et~al.}(2014)Withers, Bointon, Hudson, Craciun, and
  Russo]{Withe14SR_FET:WS:BN}
F.~Withers, T.~H. Bointon, D.~C. Hudson, M.~F. Craciun and S.~Russo, \emph{Sci.
  Rep.}, 2014, \textbf{4}, 4967\relax
\mciteBstWouldAddEndPuncttrue
\mciteSetBstMidEndSepPunct{\mcitedefaultmidpunct}
{\mcitedefaultendpunct}{\mcitedefaultseppunct}\relax
\EndOfBibitem
\bibitem[Liu \emph{et~al.}(2014)Liu, Hu, Yue, Della~Fera, Ling, Mao, and
  Wei]{Liu14AN_FET:WS:mu230}
X.~Liu, J.~Hu, C.~Yue, N.~Della~Fera, Y.~Ling, Z.~Mao and J.~Wei, \emph{ACS
  Nano}, 2014, \textbf{8}, 10396--10402\relax
\mciteBstWouldAddEndPuncttrue
\mciteSetBstMidEndSepPunct{\mcitedefaultmidpunct}
{\mcitedefaultendpunct}{\mcitedefaultseppunct}\relax
\EndOfBibitem
\bibitem[Braga \emph{et~al.}(2012)Braga, Guti\'{e}rrez~Lezama, Berger, and
  Morpurgo]{Braga12NL}
D.~Braga, I.~Guti\'{e}rrez~Lezama, H.~Berger and A.~F. Morpurgo, \emph{Nano
  Lett.}, 2012, \textbf{12}, 5218--5223\relax
\mciteBstWouldAddEndPuncttrue
\mciteSetBstMidEndSepPunct{\mcitedefaultmidpunct}
{\mcitedefaultendpunct}{\mcitedefaultseppunct}\relax
\EndOfBibitem
\bibitem[Allain and Kis(2014)]{Allai14AN_FET:WSe}
A.~Allain and A.~Kis, \emph{ACS Nano}, 2014, \textbf{8}, 7180--7185\relax
\mciteBstWouldAddEndPuncttrue
\mciteSetBstMidEndSepPunct{\mcitedefaultmidpunct}
{\mcitedefaultendpunct}{\mcitedefaultseppunct}\relax
\EndOfBibitem
\bibitem[Das \emph{et~al.}(2014)Das, Gulotty, Sumant, and Roelofs]{Das14NL}
S.~Das, R.~Gulotty, A.~V. Sumant and A.~Roelofs, \emph{Nano Lett.}, 2014,
  \textbf{14}, 2861--2866\relax
\mciteBstWouldAddEndPuncttrue
\mciteSetBstMidEndSepPunct{\mcitedefaultmidpunct}
{\mcitedefaultendpunct}{\mcitedefaultseppunct}\relax
\EndOfBibitem
\bibitem[Wang \emph{et~al.}(2015)Wang, Yang, Chen, Watanabe, Taniguchi,
  Churchill, and Jarillo-Herrero]{Wang15NL_FET:WSe_BN}
J.~I.-J. Wang, Y.~Yang, Y.-A. Chen, K.~Watanabe, T.~Taniguchi, H.~O.~H.
  Churchill and P.~Jarillo-Herrero, \emph{Nano Lett.}, 2015, \textbf{15},
  1898--1903\relax
\mciteBstWouldAddEndPuncttrue
\mciteSetBstMidEndSepPunct{\mcitedefaultmidpunct}
{\mcitedefaultendpunct}{\mcitedefaultseppunct}\relax
\EndOfBibitem
\bibitem[Chuang \emph{et~al.}(2014)Chuang, Tan, Ghimire, Perera, Chamlagain,
  Cheng, Yan, Mandrus, Tom\'{a}nek, and Zhou]{Chuan14NL_Rc:WSe2:IonicLiquid}
H.-J. Chuang, X.~Tan, N.~J. Ghimire, M.~M. Perera, B.~Chamlagain, M.~M.-C.
  Cheng, J.~Yan, D.~Mandrus, D.~Tom\'{a}nek and Z.~Zhou, \emph{Nano Lett.},
  2014, \textbf{14}, 3594--3601\relax
\mciteBstWouldAddEndPuncttrue
\mciteSetBstMidEndSepPunct{\mcitedefaultmidpunct}
{\mcitedefaultendpunct}{\mcitedefaultseppunct}\relax
\EndOfBibitem
\bibitem[Pradhan \emph{et~al.}(2015)Pradhan, Rhodes, Memaran, Poumirol,
  Smirnov, Talapatra, Feng, Perea-Lopez, Elias, Terrones, Ajayan, and
  Balicas]{Pradh15SR_FET:WSe}
N.~R. Pradhan, D.~Rhodes, S.~Memaran, J.~M. Poumirol, D.~Smirnov, S.~Talapatra,
  S.~Feng, N.~Perea-Lopez, A.~L. Elias, M.~Terrones, P.~M. Ajayan and
  L.~Balicas, \emph{Sci. Rep.}, 2015, \textbf{5}, 8979\relax
\mciteBstWouldAddEndPuncttrue
\mciteSetBstMidEndSepPunct{\mcitedefaultmidpunct}
{\mcitedefaultendpunct}{\mcitedefaultseppunct}\relax
\EndOfBibitem
\bibitem[Song \emph{et~al.}(2013)Song, Li, Gao, Xu, Ueno, Cheng, and
  K.Tsukagoshi]{Song13N}
H.~S. Song, S.~L. Li, L.~Gao, Y.~Xu, K.~Ueno, Y.~B. Cheng and K.Tsukagoshi,
  \emph{Nanoscale}, 2013, \textbf{5}, 9666--9670\relax
\mciteBstWouldAddEndPuncttrue
\mciteSetBstMidEndSepPunct{\mcitedefaultmidpunct}
{\mcitedefaultendpunct}{\mcitedefaultseppunct}\relax
\EndOfBibitem
\bibitem[De \emph{et~al.}(2013)De, Manongdo, See, Zhang, Guloy, and
  Peng]{De13N_FET:SnS:mu0.8}
D.~De, J.~Manongdo, S.~See, V.~Zhang, A.~Guloy and H.~Peng,
  \emph{Nanotechnol.}, 2013, \textbf{24}, 025202\relax
\mciteBstWouldAddEndPuncttrue
\mciteSetBstMidEndSepPunct{\mcitedefaultmidpunct}
{\mcitedefaultendpunct}{\mcitedefaultseppunct}\relax
\EndOfBibitem
\bibitem[Jariwala \emph{et~al.}(2013)Jariwala, Sangwan, Late, Johns, Dravid,
  Marks, Lauhon, and Hersam]{Jariwala13APL}
D.~Jariwala, V.~K. Sangwan, D.~J. Late, J.~E. Johns, V.~P. Dravid, T.~J. Marks,
  L.~J. Lauhon and M.~C. Hersam, \emph{Appl. Phys. Lett.}, 2013, \textbf{102},
  173107\relax
\mciteBstWouldAddEndPuncttrue
\mciteSetBstMidEndSepPunct{\mcitedefaultmidpunct}
{\mcitedefaultendpunct}{\mcitedefaultseppunct}\relax
\EndOfBibitem
\bibitem[Yu \emph{et~al.}(2014)Yu, Pan, Shen, Wang, Ong, Xu, Xin, Pan, Wang,
  Sun, Wang, Zhang, Zhang, Shi, and Wang]{Yu14NC_MoS2_mu80}
Z.~Yu, Y.~Pan, Y.~Shen, Z.~Wang, Z.-Y. Ong, T.~Xu, R.~Xin, L.~Pan, B.~Wang,
  L.~Sun, J.~Wang, G.~Zhang, Y.~W. Zhang, Y.~Shi and X.~Wang, \emph{Nat.
  Commun.}, 2014, \textbf{5}, 5290\relax
\mciteBstWouldAddEndPuncttrue
\mciteSetBstMidEndSepPunct{\mcitedefaultmidpunct}
{\mcitedefaultendpunct}{\mcitedefaultseppunct}\relax
\EndOfBibitem
\bibitem[Cui \emph{et~al.}(2015)Cui, Lee, Kim, Arefe, Huang, Lee, Chenet,
  Zhang, Wang, Ye, Pizzocchero, Jessen, Watanabe, Taniguchi, Muller, Low, Kim,
  and Hone]{Cui15NN_FET:G/MoS/BN_mu34k}
X.~Cui, G.-H. Lee, Y.~D. Kim, G.~Arefe, P.~Y. Huang, C.-H. Lee, D.~A. Chenet,
  X.~Zhang, L.~Wang, F.~Ye, F.~Pizzocchero, B.~S. Jessen, K.~Watanabe,
  T.~Taniguchi, D.~A. Muller, T.~Low, P.~Kim and J.~Hone, \emph{Nat.
  Nanotechnol.}, 2015, \textbf{10}, 534--540\relax
\mciteBstWouldAddEndPuncttrue
\mciteSetBstMidEndSepPunct{\mcitedefaultmidpunct}
{\mcitedefaultendpunct}{\mcitedefaultseppunct}\relax
\EndOfBibitem
\bibitem[Ye \emph{et~al.}(2012)Ye, Zhang, Akashi, Bahramy, Arita, and
  Iwasa]{Ye12S_IonicLiquid:SC}
J.~T. Ye, Y.~J. Zhang, R.~Akashi, M.~S. Bahramy, R.~Arita and Y.~Iwasa,
  \emph{Science}, 2012, \textbf{338}, 1193--1196\relax
\mciteBstWouldAddEndPuncttrue
\mciteSetBstMidEndSepPunct{\mcitedefaultmidpunct}
{\mcitedefaultendpunct}{\mcitedefaultseppunct}\relax
\EndOfBibitem
\bibitem[Schmidt \emph{et~al.}(2014)Schmidt, Wang, Chu, Toh, Kumar, Zhao, Neto,
  Martin, Adam, Oezyilmaz, and Eda]{Schmidt14NL}
H.~Schmidt, S.~Wang, L.~Chu, M.~Toh, R.~Kumar, W.~Zhao, A.~H.~C. Neto,
  J.~Martin, S.~Adam, B.~Oezyilmaz and G.~Eda, \emph{Nano Lett.}, 2014,
  \textbf{14}, 1909--1913\relax
\mciteBstWouldAddEndPuncttrue
\mciteSetBstMidEndSepPunct{\mcitedefaultmidpunct}
{\mcitedefaultendpunct}{\mcitedefaultseppunct}\relax
\EndOfBibitem
\bibitem[Ghatak \emph{et~al.}(2011)Ghatak, Pal, and Ghosh]{Ghatak11AN}
S.~Ghatak, A.~N. Pal and A.~Ghosh, \emph{ACS Nano}, 2011, \textbf{5},
  7707--7712\relax
\mciteBstWouldAddEndPuncttrue
\mciteSetBstMidEndSepPunct{\mcitedefaultmidpunct}
{\mcitedefaultendpunct}{\mcitedefaultseppunct}\relax
\EndOfBibitem
\bibitem[Qiu \emph{et~al.}(2013)Qiu, Xu, Wang, Ren, Nan, Ni, Chen, Yuan, Miao,
  Song, Long, Shi, Sun, Wang, and Wang]{Qiu13NC}
H.~Qiu, T.~Xu, Z.~Wang, W.~Ren, H.~Nan, Z.~Ni, Q.~Chen, S.~Yuan, F.~Miao,
  F.~Song, G.~Long, Y.~Shi, L.~Sun, J.~Wang and X.~Wang, \emph{Nat. Commun.},
  2013, \textbf{4}, 2642\relax
\mciteBstWouldAddEndPuncttrue
\mciteSetBstMidEndSepPunct{\mcitedefaultmidpunct}
{\mcitedefaultendpunct}{\mcitedefaultseppunct}\relax
\EndOfBibitem
\bibitem[Li and Tsukagoshi(2015)]{Li15JPSJ}
S.-L. Li and K.~Tsukagoshi, \emph{J. Phys. Soc. Jpn.}, 2015, \textbf{84},
  121011\relax
\mciteBstWouldAddEndPuncttrue
\mciteSetBstMidEndSepPunct{\mcitedefaultmidpunct}
{\mcitedefaultendpunct}{\mcitedefaultseppunct}\relax
\EndOfBibitem
\bibitem[Li \emph{et~al.}(2014)Li, Komatsu, Nakaharai, Lin, Yamamoto, Duan, and
  Tsukagoshi]{Li14AN_MoS2:Contact}
S.-L. Li, K.~Komatsu, S.~Nakaharai, Y.-F. Lin, M.~Yamamoto, X.~Duan and
  K.~Tsukagoshi, \emph{ACS Nano}, 2014, \textbf{8}, 12836--12842\relax
\mciteBstWouldAddEndPuncttrue
\mciteSetBstMidEndSepPunct{\mcitedefaultmidpunct}
{\mcitedefaultendpunct}{\mcitedefaultseppunct}\relax
\EndOfBibitem
\bibitem[Sze and Ng(2007)]{Sze07_textbook}
S.~M. Sze and K.~K. Ng, \emph{Physics of Semiconductor Devices}, John Wiley \&
  Sons, New Jersey, 3rd edn, 2007\relax
\mciteBstWouldAddEndPuncttrue
\mciteSetBstMidEndSepPunct{\mcitedefaultmidpunct}
{\mcitedefaultendpunct}{\mcitedefaultseppunct}\relax
\EndOfBibitem
\bibitem[Heinze \emph{et~al.}(2002)Heinze, Tersoff, Martel, Derycke,
  Appenzeller, and Avouris]{Heinze02PRL}
S.~Heinze, J.~Tersoff, R.~Martel, V.~Derycke, J.~Appenzeller and P.~Avouris,
  \emph{Phys. Rev. Lett.}, 2002, \textbf{89}, 106801\relax
\mciteBstWouldAddEndPuncttrue
\mciteSetBstMidEndSepPunct{\mcitedefaultmidpunct}
{\mcitedefaultendpunct}{\mcitedefaultseppunct}\relax
\EndOfBibitem
\bibitem[Liu \emph{et~al.}(2014)Liu, Si, Deng, Neal, Du, Najmaei, Ajayan, Lou,
  and Ye]{Liu14AN}
H.~Liu, M.~Si, Y.~Deng, A.~T. Neal, Y.~Du, S.~Najmaei, P.~M. Ajayan, J.~Lou and
  P.~D. Ye, \emph{ACS Nano}, 2014, \textbf{8}, 1031--1038\relax
\mciteBstWouldAddEndPuncttrue
\mciteSetBstMidEndSepPunct{\mcitedefaultmidpunct}
{\mcitedefaultendpunct}{\mcitedefaultseppunct}\relax
\EndOfBibitem
\bibitem[Chen \emph{et~al.}(2005)Chen, Appenzeller, Knoch, Lin, and
  Avouris]{chen05nl}
Z.~Chen, J.~Appenzeller, J.~Knoch, Y.~Lin and P.~Avouris, \emph{Nano Lett.},
  2005, \textbf{5}, 1497--1502\relax
\mciteBstWouldAddEndPuncttrue
\mciteSetBstMidEndSepPunct{\mcitedefaultmidpunct}
{\mcitedefaultendpunct}{\mcitedefaultseppunct}\relax
\EndOfBibitem
\bibitem[Murrmann and Widmann(1969)]{Murrm69ITED_Rc:CurrentCrowding}
H.~Murrmann and D.~Widmann, \emph{IEEE Trans. Electron Devices}, 1969,
  \textbf{16}, 1022--1024\relax
\mciteBstWouldAddEndPuncttrue
\mciteSetBstMidEndSepPunct{\mcitedefaultmidpunct}
{\mcitedefaultendpunct}{\mcitedefaultseppunct}\relax
\EndOfBibitem
\bibitem[Li(2006)]{Li06_transport:introductory}
S.~S. Li, \emph{Semiconductor Physical Electronics}, Springer, New York,
  2006\relax
\mciteBstWouldAddEndPuncttrue
\mciteSetBstMidEndSepPunct{\mcitedefaultmidpunct}
{\mcitedefaultendpunct}{\mcitedefaultseppunct}\relax
\EndOfBibitem
\bibitem[Ridley(1997)]{ridley1997electrons}
B.~K. Ridley, \emph{Electrons and Phonons in Semiconductor Multilayers},
  Cambridge University Press, 1997\relax
\mciteBstWouldAddEndPuncttrue
\mciteSetBstMidEndSepPunct{\mcitedefaultmidpunct}
{\mcitedefaultendpunct}{\mcitedefaultseppunct}\relax
\EndOfBibitem
\bibitem[Lundstrom(2000)]{Lundstrom00}
M.~Lundstrom, \emph{Fundamentals of Carrier Transport}, Cambridge University
  Press, 2nd edn, 2000\relax
\mciteBstWouldAddEndPuncttrue
\mciteSetBstMidEndSepPunct{\mcitedefaultmidpunct}
{\mcitedefaultendpunct}{\mcitedefaultseppunct}\relax
\EndOfBibitem
\bibitem[Ando \emph{et~al.}(1982)Ando, Fowler, and Stern]{Ando82RMP}
T.~Ando, A.~B. Fowler and F.~Stern, \emph{Rev. Mod. Phys.}, 1982, \textbf{54},
  437--672\relax
\mciteBstWouldAddEndPuncttrue
\mciteSetBstMidEndSepPunct{\mcitedefaultmidpunct}
{\mcitedefaultendpunct}{\mcitedefaultseppunct}\relax
\EndOfBibitem
\bibitem[Ando(2006)]{Ando06JPSJ}
T.~Ando, \emph{J. Phys. Soc. Jpn.}, 2006, \textbf{75}, 074716\relax
\mciteBstWouldAddEndPuncttrue
\mciteSetBstMidEndSepPunct{\mcitedefaultmidpunct}
{\mcitedefaultendpunct}{\mcitedefaultseppunct}\relax
\EndOfBibitem
\bibitem[Das~Sarma \emph{et~al.}(2011)Das~Sarma, Adam, Hwang, and
  Rossi]{DasSarma11RMP}
S.~Das~Sarma, S.~Adam, E.~H. Hwang and E.~Rossi, \emph{Rev. Mod. Phys.}, 2011,
  \textbf{83}, 407--470\relax
\mciteBstWouldAddEndPuncttrue
\mciteSetBstMidEndSepPunct{\mcitedefaultmidpunct}
{\mcitedefaultendpunct}{\mcitedefaultseppunct}\relax
\EndOfBibitem
\bibitem[Jena and Konar(2007)]{jena07prl}
D.~Jena and A.~Konar, \emph{Phys. Rev. Lett.}, 2007, \textbf{98}, 136805\relax
\mciteBstWouldAddEndPuncttrue
\mciteSetBstMidEndSepPunct{\mcitedefaultmidpunct}
{\mcitedefaultendpunct}{\mcitedefaultseppunct}\relax
\EndOfBibitem
\bibitem[Gelmont \emph{et~al.}(1995)Gelmont, Shur, and Stroscio]{Gelmont95JAP}
B.~L. Gelmont, M.~Shur and M.~Stroscio, \emph{J. Appl. Phys.}, 1995,
  \textbf{77}, 657--660\relax
\mciteBstWouldAddEndPuncttrue
\mciteSetBstMidEndSepPunct{\mcitedefaultmidpunct}
{\mcitedefaultendpunct}{\mcitedefaultseppunct}\relax
\EndOfBibitem
\bibitem[Moore and Ferry(1980)]{moore80jap}
B.~T. Moore and D.~K. Ferry, \emph{J. Appl. Phys.}, 1980, \textbf{51},
  2603--2605\relax
\mciteBstWouldAddEndPuncttrue
\mciteSetBstMidEndSepPunct{\mcitedefaultmidpunct}
{\mcitedefaultendpunct}{\mcitedefaultseppunct}\relax
\EndOfBibitem
\bibitem[Fischetti \emph{et~al.}(2001)Fischetti, Neumayer, and
  Cartier]{fischetti01jap}
M.~V. Fischetti, D.~A. Neumayer and E.~A. Cartier, \emph{J. Appl. Phys.}, 2001,
  \textbf{90}, 4587--4608\relax
\mciteBstWouldAddEndPuncttrue
\mciteSetBstMidEndSepPunct{\mcitedefaultmidpunct}
{\mcitedefaultendpunct}{\mcitedefaultseppunct}\relax
\EndOfBibitem
\bibitem[Stassen \emph{et~al.}(2004)Stassen, de~Boer, Iosad, and
  Morpurgo]{stass04apl_}
A.~F. Stassen, R.~W.~I. de~Boer, N.~N. Iosad and A.~F. Morpurgo, \emph{Appl.
  Phys. Lett.}, 2004, \textbf{85}, 3899--3901\relax
\mciteBstWouldAddEndPuncttrue
\mciteSetBstMidEndSepPunct{\mcitedefaultmidpunct}
{\mcitedefaultendpunct}{\mcitedefaultseppunct}\relax
\EndOfBibitem
\bibitem[Veres \emph{et~al.}(2003)Veres, Ogier, Leeming, Cupertino, and
  Mohialdin~Khaffaf]{veres03afm_}
J.~Veres, S.~Ogier, S.~Leeming, D.~Cupertino and S.~Mohialdin~Khaffaf,
  \emph{Adv. Funct. Mater.}, 2003, \textbf{13}, 199--204\relax
\mciteBstWouldAddEndPuncttrue
\mciteSetBstMidEndSepPunct{\mcitedefaultmidpunct}
{\mcitedefaultendpunct}{\mcitedefaultseppunct}\relax
\EndOfBibitem
\bibitem[Hulea \emph{et~al.}(2006)Hulea, Fratini, Xie, Mulder, Iossad,
  Rastelli, Ciuchi, and Morpurgo]{hulea06nm_rip}
I.~N. Hulea, S.~Fratini, H.~Xie, C.~L. Mulder, N.~N. Iossad, G.~Rastelli,
  S.~Ciuchi and A.~F. Morpurgo, \emph{Nat. Mater.}, 2006, \textbf{5},
  982--986\relax
\mciteBstWouldAddEndPuncttrue
\mciteSetBstMidEndSepPunct{\mcitedefaultmidpunct}
{\mcitedefaultendpunct}{\mcitedefaultseppunct}\relax
\EndOfBibitem
\bibitem[Chen \emph{et~al.}(2008)Chen, Jang, Xiao, Ishigami, and
  Fuhrer]{chen08nn}
J.-H. Chen, C.~Jang, S.~Xiao, M.~Ishigami and M.~S. Fuhrer, \emph{Nat.
  Nanotechnol.}, 2008, \textbf{3}, 206--209\relax
\mciteBstWouldAddEndPuncttrue
\mciteSetBstMidEndSepPunct{\mcitedefaultmidpunct}
{\mcitedefaultendpunct}{\mcitedefaultseppunct}\relax
\EndOfBibitem
\bibitem[Fratini and Guinea(2008)]{fratini08prb}
S.~Fratini and F.~Guinea, \emph{Phys. Rev. B}, 2008, \textbf{77}, 195415\relax
\mciteBstWouldAddEndPuncttrue
\mciteSetBstMidEndSepPunct{\mcitedefaultmidpunct}
{\mcitedefaultendpunct}{\mcitedefaultseppunct}\relax
\EndOfBibitem
\bibitem[Sabio \emph{et~al.}(2008)Sabio, Seoanez, Fratini, Guinea, Castro~Neto,
  and Sols]{sabio08prb}
J.~Sabio, C.~Seoanez, S.~Fratini, F.~Guinea, A.~H. Castro~Neto and F.~Sols,
  \emph{Phys. Rev. B}, 2008, \textbf{77}, 195409\relax
\mciteBstWouldAddEndPuncttrue
\mciteSetBstMidEndSepPunct{\mcitedefaultmidpunct}
{\mcitedefaultendpunct}{\mcitedefaultseppunct}\relax
\EndOfBibitem
\bibitem[Konar \emph{et~al.}(2010)Konar, Fang, and Jena]{konar10prb}
A.~Konar, T.~Fang and D.~Jena, \emph{Phys. Rev. B}, 2010, \textbf{82},
  115452\relax
\mciteBstWouldAddEndPuncttrue
\mciteSetBstMidEndSepPunct{\mcitedefaultmidpunct}
{\mcitedefaultendpunct}{\mcitedefaultseppunct}\relax
\EndOfBibitem
\bibitem[Zou \emph{et~al.}(2010)Zou, Hong, Keefer, and Zhu]{zou10prl}
K.~Zou, X.~Hong, D.~Keefer and J.~Zhu, \emph{Phys. Rev. Lett.}, 2010,
  \textbf{105}, 126601\relax
\mciteBstWouldAddEndPuncttrue
\mciteSetBstMidEndSepPunct{\mcitedefaultmidpunct}
{\mcitedefaultendpunct}{\mcitedefaultseppunct}\relax
\EndOfBibitem
\bibitem[DaSilva \emph{et~al.}(2010)DaSilva, Zou, Jain, and Zhu]{dasilva10prl}
A.~M. DaSilva, K.~Zou, J.~K. Jain and J.~Zhu, \emph{Phys. Rev. Lett.}, 2010,
  \textbf{104}, 236601\relax
\mciteBstWouldAddEndPuncttrue
\mciteSetBstMidEndSepPunct{\mcitedefaultmidpunct}
{\mcitedefaultendpunct}{\mcitedefaultseppunct}\relax
\EndOfBibitem
\bibitem[Caulfield and Fisher(1997)]{Caulf97JPCM_defect:vacancy}
J.~C. Caulfield and A.~J. Fisher, \emph{J. Phys.: Condens. Matter}, 1997,
  \textbf{9}, 3671--3686\relax
\mciteBstWouldAddEndPuncttrue
\mciteSetBstMidEndSepPunct{\mcitedefaultmidpunct}
{\mcitedefaultendpunct}{\mcitedefaultseppunct}\relax
\EndOfBibitem
\bibitem[Feng \emph{et~al.}(2014)Feng, Su, Chen, and
  Liu]{Feng14MCP_defect:vacancy}
L.-P. Feng, J.~Su, S.~Chen and Z.-T. Liu, \emph{Mater. Chem. Phys.}, 2014,
  \textbf{148}, 5--9\relax
\mciteBstWouldAddEndPuncttrue
\mciteSetBstMidEndSepPunct{\mcitedefaultmidpunct}
{\mcitedefaultendpunct}{\mcitedefaultseppunct}\relax
\EndOfBibitem
\bibitem[Feng \emph{et~al.}(2014)Feng, Su, and Liu]{Feng14JAC_defect:vacancy}
L.-P. Feng, J.~Su and Z.-T. Liu, \emph{J. Alloys Compd.}, 2014, \textbf{613},
  122--127\relax
\mciteBstWouldAddEndPuncttrue
\mciteSetBstMidEndSepPunct{\mcitedefaultmidpunct}
{\mcitedefaultendpunct}{\mcitedefaultseppunct}\relax
\EndOfBibitem
\bibitem[Gan and Zhao(2014)]{Gan14PLA_defect:vacancy}
Y.~Gan and H.~Zhao, \emph{Phys. Lett. A}, 2014, \textbf{378}, 2910--2914\relax
\mciteBstWouldAddEndPuncttrue
\mciteSetBstMidEndSepPunct{\mcitedefaultmidpunct}
{\mcitedefaultendpunct}{\mcitedefaultseppunct}\relax
\EndOfBibitem
\bibitem[Heine(2015)]{Heine15ACR_defect:e-structure}
T.~Heine, \emph{Acc. Chem. Res.}, 2015, \textbf{48}, 65--72\relax
\mciteBstWouldAddEndPuncttrue
\mciteSetBstMidEndSepPunct{\mcitedefaultmidpunct}
{\mcitedefaultendpunct}{\mcitedefaultseppunct}\relax
\EndOfBibitem
\bibitem[Santosh \emph{et~al.}(2014)Santosh, Longo, Addou, Wallace, and
  Cho]{Santo14N_}
K.~C. Santosh, R.~C. Longo, R.~Addou, R.~M. Wallace and K.~Cho,
  \emph{Nanotechnol.}, 2014, \textbf{25}, 375703\relax
\mciteBstWouldAddEndPuncttrue
\mciteSetBstMidEndSepPunct{\mcitedefaultmidpunct}
{\mcitedefaultendpunct}{\mcitedefaultseppunct}\relax
\EndOfBibitem
\bibitem[Spirko \emph{et~al.}(2004)Spirko, Neiman, Oelker, and
  Klier]{Spirk04SS_}
J.~A. Spirko, M.~L. Neiman, A.~M. Oelker and K.~Klier, \emph{Surf. Sci.}, 2004,
  \textbf{572}, 191--205\relax
\mciteBstWouldAddEndPuncttrue
\mciteSetBstMidEndSepPunct{\mcitedefaultmidpunct}
{\mcitedefaultendpunct}{\mcitedefaultseppunct}\relax
\EndOfBibitem
\bibitem[Tianmin \emph{et~al.}(2015)Tianmin, Shengbao, Yuming, Jiajia, Haiqing,
  and Zhiyong]{Tianm15RMMaE_defect:vacancy}
L.~Tianmin, W.~Shengbao, Z.~Yuming, L.~Jiajia, J.~Haiqing and Z.~Zhiyong,
  \emph{Rare Metal Materials and Engineering}, 2015, \textbf{44},
  608--611\relax
\mciteBstWouldAddEndPuncttrue
\mciteSetBstMidEndSepPunct{\mcitedefaultmidpunct}
{\mcitedefaultendpunct}{\mcitedefaultseppunct}\relax
\EndOfBibitem
\bibitem[Komsa \emph{et~al.}(2012)Komsa, Kotakoski, Kurasch, Lehtinen, Kaiser,
  and Krasheninnikov]{Komsa12PRL_defect:e-structureTEM}
H.-P. Komsa, J.~Kotakoski, S.~Kurasch, O.~Lehtinen, U.~Kaiser and A.~V.
  Krasheninnikov, \emph{Phys. Rev. Lett.}, 2012, \textbf{109}, 035503\relax
\mciteBstWouldAddEndPuncttrue
\mciteSetBstMidEndSepPunct{\mcitedefaultmidpunct}
{\mcitedefaultendpunct}{\mcitedefaultseppunct}\relax
\EndOfBibitem
\bibitem[Moses \emph{et~al.}(2007)Moses, Hinnemann, Topsoe, and
  Norskov]{Moses07JC_defect:vacancy}
P.~G. Moses, B.~Hinnemann, H.~Topsoe and J.~K. Norskov, \emph{J. Catal.}, 2007,
  \textbf{248}, 188--203\relax
\mciteBstWouldAddEndPuncttrue
\mciteSetBstMidEndSepPunct{\mcitedefaultmidpunct}
{\mcitedefaultendpunct}{\mcitedefaultseppunct}\relax
\EndOfBibitem
\bibitem[Paul and Payen(2003)]{Paul03JPCB_defect:vacancy}
J.~F. Paul and E.~Payen, \emph{J. Phys. Chem. B}, 2003, \textbf{107},
  4057--4064\relax
\mciteBstWouldAddEndPuncttrue
\mciteSetBstMidEndSepPunct{\mcitedefaultmidpunct}
{\mcitedefaultendpunct}{\mcitedefaultseppunct}\relax
\EndOfBibitem
\bibitem[Morales-Guio and Hu(2014)]{Moral14ACR_Energy:H2}
C.~G. Morales-Guio and X.~Hu, \emph{Acc. Chem. Res.}, 2014, \textbf{47},
  2671--81\relax
\mciteBstWouldAddEndPuncttrue
\mciteSetBstMidEndSepPunct{\mcitedefaultmidpunct}
{\mcitedefaultendpunct}{\mcitedefaultseppunct}\relax
\EndOfBibitem
\bibitem[Kibsgaard \emph{et~al.}(2012)Kibsgaard, Chen, Reinecke, and
  Jaramillo]{Kibsg12NM_Energy:H2}
J.~Kibsgaard, Z.~Chen, B.~N. Reinecke and T.~F. Jaramillo, \emph{Nat. Mater.},
  2012, \textbf{11}, 963--969\relax
\mciteBstWouldAddEndPuncttrue
\mciteSetBstMidEndSepPunct{\mcitedefaultmidpunct}
{\mcitedefaultendpunct}{\mcitedefaultseppunct}\relax
\EndOfBibitem
\bibitem[Voiry \emph{et~al.}(2013)Voiry, Yamaguchi, Li, Silva, Alves, Fujita,
  Chen, Asefa, Shenoy, Eda, , and Chhowalla]{Voiry13NM_Energy:H2}
D.~Voiry, H.~Yamaguchi, J.~Li, R.~Silva, D.~C.~B. Alves, T.~Fujita, M.~Chen,
  T.~Asefa, V.~B. Shenoy, G.~Eda,  and M.~Chhowalla, \emph{Nat. Mater.}, 2013,
  \textbf{12}, 850--855\relax
\mciteBstWouldAddEndPuncttrue
\mciteSetBstMidEndSepPunct{\mcitedefaultmidpunct}
{\mcitedefaultendpunct}{\mcitedefaultseppunct}\relax
\EndOfBibitem
\bibitem[Karunadasa \emph{et~al.}(2012)Karunadasa, Montalvo, Sun, Majda, Long,
  and Chang]{Karun12S_Energy:H2}
H.~I. Karunadasa, E.~Montalvo, Y.~Sun, M.~Majda, J.~R. Long and C.~J. Chang,
  \emph{Science}, 2012, \textbf{335}, 698--702\relax
\mciteBstWouldAddEndPuncttrue
\mciteSetBstMidEndSepPunct{\mcitedefaultmidpunct}
{\mcitedefaultendpunct}{\mcitedefaultseppunct}\relax
\EndOfBibitem
\bibitem[Komsa and Krasheninnikov(2015)]{Komsa15PRB_Vacancy:MoS}
H.-P. Komsa and A.~V. Krasheninnikov, \emph{Phys. Rev. B}, 2015, \textbf{91},
  125304\relax
\mciteBstWouldAddEndPuncttrue
\mciteSetBstMidEndSepPunct{\mcitedefaultmidpunct}
{\mcitedefaultendpunct}{\mcitedefaultseppunct}\relax
\EndOfBibitem
\bibitem[van~der Zande \emph{et~al.}(2013)van~der Zande, Huang, Chenet,
  Berkelbach, You, Lee, Heinz, Reichman, Muller, and Hone]{Zande13NM}
A.~M. van~der Zande, P.~Y. Huang, D.~A. Chenet, T.~C. Berkelbach, Y.~You, G.-H.
  Lee, T.~F. Heinz, D.~R. Reichman, D.~A. Muller and J.~C. Hone, \emph{Nat.
  Mater.}, 2013, \textbf{12}, 554--561\relax
\mciteBstWouldAddEndPuncttrue
\mciteSetBstMidEndSepPunct{\mcitedefaultmidpunct}
{\mcitedefaultendpunct}{\mcitedefaultseppunct}\relax
\EndOfBibitem
\bibitem[Hong \emph{et~al.}(2015)Hong, Hu, Probert, Li, Lv, Yang, Gu, Mao,
  Feng, Xie, Zhang, Wu, Zhang, Jin, Ji, Zhang, Yuan, and
  Zhang]{Hong15NC_defect:vacancy}
J.~Hong, Z.~Hu, M.~Probert, K.~Li, D.~Lv, X.~Yang, L.~Gu, N.~Mao, Q.~Feng,
  L.~Xie, J.~Zhang, D.~Wu, Z.~Zhang, C.~Jin, W.~Ji, X.~Zhang, J.~Yuan and
  Z.~Zhang, \emph{Nat. Commun.}, 2015, \textbf{6}, 6293\relax
\mciteBstWouldAddEndPuncttrue
\mciteSetBstMidEndSepPunct{\mcitedefaultmidpunct}
{\mcitedefaultendpunct}{\mcitedefaultseppunct}\relax
\EndOfBibitem
\bibitem[Ravich \emph{et~al.}(1971)Ravich, Efimova, and
  Tamarchenko]{Ravic71pssb_defect:PbTe}
Y.~I. Ravich, B.~A. Efimova and V.~I. Tamarchenko, \emph{Phys. Stat. Sol. b},
  1971, \textbf{43}, 11--33\relax
\mciteBstWouldAddEndPuncttrue
\mciteSetBstMidEndSepPunct{\mcitedefaultmidpunct}
{\mcitedefaultendpunct}{\mcitedefaultseppunct}\relax
\EndOfBibitem
\bibitem[Zayachuk(1997)]{Zayac97S_defect:PbTe}
D.~Zayachuk, \emph{Semiconductors}, 1997, \textbf{31}, 173--176\relax
\mciteBstWouldAddEndPuncttrue
\mciteSetBstMidEndSepPunct{\mcitedefaultmidpunct}
{\mcitedefaultendpunct}{\mcitedefaultseppunct}\relax
\EndOfBibitem
\bibitem[Hwang \emph{et~al.}(2007)Hwang, Adam, and Das~Sarma]{Hwang07PRLb}
E.~H. Hwang, S.~Adam and S.~Das~Sarma, \emph{Phys. Rev. Lett.}, 2007,
  \textbf{98}, 186806\relax
\mciteBstWouldAddEndPuncttrue
\mciteSetBstMidEndSepPunct{\mcitedefaultmidpunct}
{\mcitedefaultendpunct}{\mcitedefaultseppunct}\relax
\EndOfBibitem
\bibitem[Hwang and Das~Sarma(2008)]{Hwang08PRB}
E.~H. Hwang and S.~Das~Sarma, \emph{Phys. Rev. B}, 2008, \textbf{77},
  195412\relax
\mciteBstWouldAddEndPuncttrue
\mciteSetBstMidEndSepPunct{\mcitedefaultmidpunct}
{\mcitedefaultendpunct}{\mcitedefaultseppunct}\relax
\EndOfBibitem
\bibitem[Najmaei \emph{et~al.}(2013)Najmaei, Liu, Zhou, Zou, Shi, Lei,
  Yakobson, Idrobo, Ajayan, and Lou]{Najmaei13NM}
S.~Najmaei, Z.~Liu, W.~Zhou, X.~Zou, G.~Shi, S.~Lei, B.~I. Yakobson, J.-C.
  Idrobo, P.~M. Ajayan and J.~Lou, \emph{Nat. Mater.}, 2013, \textbf{12},
  754--759\relax
\mciteBstWouldAddEndPuncttrue
\mciteSetBstMidEndSepPunct{\mcitedefaultmidpunct}
{\mcitedefaultendpunct}{\mcitedefaultseppunct}\relax
\EndOfBibitem
\bibitem[Zhang \emph{et~al.}(2013)Zhang, Zhang, Ji, Ju, Yuan, Shi, Gao, Ma,
  Liu, Chen, Song, Hwang, Cui, and Liu]{Zhang13AN_defect:GB:WS}
Y.~Zhang, Y.~Zhang, Q.~Ji, J.~Ju, H.~Yuan, J.~Shi, T.~Gao, D.~Ma, M.~Liu,
  Y.~Chen, X.~Song, H.~Y. Hwang, Y.~Cui and Z.~Liu, \emph{ACS Nano}, 2013,
  \textbf{7}, 8963--8971\relax
\mciteBstWouldAddEndPuncttrue
\mciteSetBstMidEndSepPunct{\mcitedefaultmidpunct}
{\mcitedefaultendpunct}{\mcitedefaultseppunct}\relax
\EndOfBibitem
\bibitem[Zou \emph{et~al.}(2013)Zou, Liu, and Yakobson]{Zou13NL}
X.~Zou, Y.~Liu and B.~I. Yakobson, \emph{Nano Lett.}, 2013, \textbf{13},
  253--258\relax
\mciteBstWouldAddEndPuncttrue
\mciteSetBstMidEndSepPunct{\mcitedefaultmidpunct}
{\mcitedefaultendpunct}{\mcitedefaultseppunct}\relax
\EndOfBibitem
\bibitem[Liu \emph{et~al.}(2013)Liu, Kang, Sarkar, Khatami, Jena, and
  Banerjee]{Liu13NL}
W.~Liu, J.~Kang, D.~Sarkar, Y.~Khatami, D.~Jena and K.~Banerjee, \emph{Nano
  Lett.}, 2013, \textbf{13}, 1983--1990\relax
\mciteBstWouldAddEndPuncttrue
\mciteSetBstMidEndSepPunct{\mcitedefaultmidpunct}
{\mcitedefaultendpunct}{\mcitedefaultseppunct}\relax
\EndOfBibitem
\bibitem[Fang \emph{et~al.}(2012)Fang, Chuang, Chang, Takei, Takahashi, and
  Javey]{Fang12NL}
H.~Fang, S.~Chuang, T.~C. Chang, K.~Takei, T.~Takahashi and A.~Javey,
  \emph{Nano Lett.}, 2012, \textbf{12}, 3788--3792\relax
\mciteBstWouldAddEndPuncttrue
\mciteSetBstMidEndSepPunct{\mcitedefaultmidpunct}
{\mcitedefaultendpunct}{\mcitedefaultseppunct}\relax
\EndOfBibitem
\bibitem[Kwak \emph{et~al.}(2014)Kwak, Hwang, Calderon, Alsalman, Munoz,
  Schutter, and Spencer]{Kwak14NL_FET:MoS:Gcontact}
J.~Y. Kwak, J.~Hwang, B.~Calderon, H.~Alsalman, N.~Munoz, B.~Schutter and M.~G.
  Spencer, \emph{Nano Lett.}, 2014, \textbf{14}, 4511--4516\relax
\mciteBstWouldAddEndPuncttrue
\mciteSetBstMidEndSepPunct{\mcitedefaultmidpunct}
{\mcitedefaultendpunct}{\mcitedefaultseppunct}\relax
\EndOfBibitem
\bibitem[Liu \emph{et~al.}(2015)Liu, Wu, Cheng, Yang, Zhu, He, Ding, Li, Guo,
  Weiss, Huang, and Duan]{Liu15NL_FET:G/MoS/BN}
Y.~Liu, H.~Wu, H.-C. Cheng, S.~Yang, E.~Zhu, Q.~He, M.~Ding, D.~Li, J.~Guo,
  N.~O. Weiss, Y.~Huang and X.~Duan, \emph{Nano Lett.}, 2015, \textbf{15},
  3030--3034\relax
\mciteBstWouldAddEndPuncttrue
\mciteSetBstMidEndSepPunct{\mcitedefaultmidpunct}
{\mcitedefaultendpunct}{\mcitedefaultseppunct}\relax
\EndOfBibitem
\bibitem[Tongay \emph{et~al.}(2013)Tongay, Zhou, Ataca, Liu, Kang, Matthews,
  You, Li, Grossman, and Wu]{Tongay13NL}
S.~Tongay, J.~Zhou, C.~Ataca, J.~Liu, J.~S. Kang, T.~S. Matthews, L.~You,
  J.~Li, J.~C. Grossman and J.~Wu, \emph{Nano Lett.}, 2013, \textbf{13},
  2831--2836\relax
\mciteBstWouldAddEndPuncttrue
\mciteSetBstMidEndSepPunct{\mcitedefaultmidpunct}
{\mcitedefaultendpunct}{\mcitedefaultseppunct}\relax
\EndOfBibitem
\bibitem[Dolui \emph{et~al.}(2013)Dolui, Rungger, and
  Sanvito]{Dolui13PRB_CIsourcce:danglingbondandNa}
K.~Dolui, I.~Rungger and S.~Sanvito, \emph{Phys. Rev. B}, 2013, \textbf{87},
  165402\relax
\mciteBstWouldAddEndPuncttrue
\mciteSetBstMidEndSepPunct{\mcitedefaultmidpunct}
{\mcitedefaultendpunct}{\mcitedefaultseppunct}\relax
\EndOfBibitem
\bibitem[Bao \emph{et~al.}(2013)Bao, Cai, Kim, Sridhara, and Fuhrer]{Bao13APL}
W.~Bao, X.~Cai, D.~Kim, K.~Sridhara and M.~S. Fuhrer, \emph{Appl. Phys. Lett.},
  2013, \textbf{102}, 042104\relax
\mciteBstWouldAddEndPuncttrue
\mciteSetBstMidEndSepPunct{\mcitedefaultmidpunct}
{\mcitedefaultendpunct}{\mcitedefaultseppunct}\relax
\EndOfBibitem
\bibitem[Feng \emph{et~al.}(2014)Feng, Zheng, Cao, and Hu]{Feng14AM}
W.~Feng, W.~Zheng, W.~Cao and P.~Hu, \emph{Adv. Mater.}, 2014, \textbf{26},
  6587--6593\relax
\mciteBstWouldAddEndPuncttrue
\mciteSetBstMidEndSepPunct{\mcitedefaultmidpunct}
{\mcitedefaultendpunct}{\mcitedefaultseppunct}\relax
\EndOfBibitem
\bibitem[Dean \emph{et~al.}(2010)Dean, Young, Meric, Lee, Wang, Sorgenfrei,
  Watanabe, Taniguchi, Kim, Shepard, and Hone]{dean10nn}
C.~R. Dean, A.~F. Young, I.~Meric, C.~Lee, L.~Wang, S.~Sorgenfrei, K.~Watanabe,
  T.~Taniguchi, P.~Kim, K.~L. Shepard and J.~Hone, \emph{Nat. Nanotechnol.},
  2010, \textbf{5}, 722--726\relax
\mciteBstWouldAddEndPuncttrue
\mciteSetBstMidEndSepPunct{\mcitedefaultmidpunct}
{\mcitedefaultendpunct}{\mcitedefaultseppunct}\relax
\EndOfBibitem
\bibitem[Dean \emph{et~al.}(2011)Dean, Young, Cadden-Zimansky, Wang, Ren,
  Watanabe, Taniguchi, Kim, Hone, and Shepard]{dean11np}
C.~R. Dean, A.~F. Young, P.~Cadden-Zimansky, L.~Wang, H.~Ren, K.~Watanabe,
  T.~Taniguchi, P.~Kim, J.~Hone and K.~L. Shepard, \emph{Nat. Phys.}, 2011,
  \textbf{7}, 693--696\relax
\mciteBstWouldAddEndPuncttrue
\mciteSetBstMidEndSepPunct{\mcitedefaultmidpunct}
{\mcitedefaultendpunct}{\mcitedefaultseppunct}\relax
\EndOfBibitem
\bibitem[Xue \emph{et~al.}(2011)Xue, Sanchez-Yamagishi, Bulmash, Jacquod,
  Deshpande, Watanabe, Taniguchi, Jarillo-Herrero, and Leroy]{xue11nm}
J.~Xue, J.~Sanchez-Yamagishi, D.~Bulmash, P.~Jacquod, A.~Deshpande,
  K.~Watanabe, T.~Taniguchi, P.~Jarillo-Herrero and B.~J. Leroy, \emph{Nat.
  Mater.}, 2011, \textbf{10}, 282--285\relax
\mciteBstWouldAddEndPuncttrue
\mciteSetBstMidEndSepPunct{\mcitedefaultmidpunct}
{\mcitedefaultendpunct}{\mcitedefaultseppunct}\relax
\EndOfBibitem
\bibitem[Ponomarenko \emph{et~al.}(2009)Ponomarenko, Yang, Mohiuddin,
  Katsnelson, Novoselov, Morozov, Zhukov, Schedin, Hill, and
  Geim]{Ponomarenko09PRL}
L.~A. Ponomarenko, R.~Yang, T.~M. Mohiuddin, M.~I. Katsnelson, K.~S. Novoselov,
  S.~V. Morozov, A.~A. Zhukov, F.~Schedin, E.~W. Hill and A.~K. Geim,
  \emph{Phys. Rev. Lett.}, 2009, \textbf{102}, 206603\relax
\mciteBstWouldAddEndPuncttrue
\mciteSetBstMidEndSepPunct{\mcitedefaultmidpunct}
{\mcitedefaultendpunct}{\mcitedefaultseppunct}\relax
\EndOfBibitem
\bibitem[Wang \emph{et~al.}(2015)Wang, Li, Zou, Ho, Liao, Xiao, Jiang, Hu,
  Wang, and Li]{Wang15S}
J.-L. Wang, S.-L. Li, X.~Zou, J.~Ho, L.~Liao, X.~Xiao, C.~Jiang, W.~Hu, J.~Wang
  and J.-C. Li, \emph{Small}, 2015,  10.1002/smll.201501260\relax
\mciteBstWouldAddEndPuncttrue
\mciteSetBstMidEndSepPunct{\mcitedefaultmidpunct}
{\mcitedefaultendpunct}{\mcitedefaultseppunct}\relax
\EndOfBibitem
\bibitem[Lu \emph{et~al.}(2015)Lu, Carvalho, Chan, Liu, Liu, Tok, Loh,
  Castro~Neto, and Sow]{Lu15NL_defect:healing}
J.~Lu, A.~Carvalho, X.~K. Chan, H.~Liu, B.~Liu, E.~S. Tok, K.~P. Loh, A.~H.
  Castro~Neto and C.~H. Sow, \emph{Nano Lett.}, 2015, \textbf{15},
  3524--32\relax
\mciteBstWouldAddEndPuncttrue
\mciteSetBstMidEndSepPunct{\mcitedefaultmidpunct}
{\mcitedefaultendpunct}{\mcitedefaultseppunct}\relax
\EndOfBibitem
\bibitem[Liu \emph{et~al.}(2012)Liu, Gu, and Ye]{Liu12IEDL_b}
H.~Liu, J.~Gu and P.~D. Ye, \emph{IEEE Electron Device Lett.}, 2012,
  \textbf{33}, 1273--1275\relax
\mciteBstWouldAddEndPuncttrue
\mciteSetBstMidEndSepPunct{\mcitedefaultmidpunct}
{\mcitedefaultendpunct}{\mcitedefaultseppunct}\relax
\EndOfBibitem
\bibitem[Radisavljevic and Kis(2013)]{Radis13NN_CorrectionCg}
B.~Radisavljevic and A.~Kis, \emph{Nat. Nanotechnol.}, 2013, \textbf{8},
  147--148\relax
\mciteBstWouldAddEndPuncttrue
\mciteSetBstMidEndSepPunct{\mcitedefaultmidpunct}
{\mcitedefaultendpunct}{\mcitedefaultseppunct}\relax
\EndOfBibitem
\bibitem[Xu \emph{et~al.}(2012)Xu, Ho, Andrieu, Smith, Nguyen, Weber, Poiroux,
  Faynot, and Liu]{Xu12IEDL_}
N.~Xu, B.~Ho, F.~Andrieu, L.~Smith, B.-y. Nguyen, O.~Weber, T.~Poiroux,
  O.~Faynot and T.-J.~K. Liu, \emph{IEEE Electron Device Lett.}, 2012,
  \textbf{33}, 318--320\relax
\mciteBstWouldAddEndPuncttrue
\mciteSetBstMidEndSepPunct{\mcitedefaultmidpunct}
{\mcitedefaultendpunct}{\mcitedefaultseppunct}\relax
\EndOfBibitem
\bibitem[Nomura \emph{et~al.}(2003)Nomura, Ohta, Ueda, Kamiya, Hirano, and
  Hosono]{Nomura03S}
K.~Nomura, H.~Ohta, K.~Ueda, T.~Kamiya, M.~Hirano and H.~Hosono,
  \emph{Science}, 2003, \textbf{300}, 1269--1272\relax
\mciteBstWouldAddEndPuncttrue
\mciteSetBstMidEndSepPunct{\mcitedefaultmidpunct}
{\mcitedefaultendpunct}{\mcitedefaultseppunct}\relax
\EndOfBibitem
\bibitem[Nomura \emph{et~al.}(2004)Nomura, Ohta, Takagi, Kamiya, Hirano, and
  Hosono]{Nomura04N}
K.~Nomura, H.~Ohta, A.~Takagi, T.~Kamiya, M.~Hirano and H.~Hosono,
  \emph{Nature}, 2004, \textbf{432}, 488--492\relax
\mciteBstWouldAddEndPuncttrue
\mciteSetBstMidEndSepPunct{\mcitedefaultmidpunct}
{\mcitedefaultendpunct}{\mcitedefaultseppunct}\relax
\EndOfBibitem
\bibitem[Fortunato \emph{et~al.}(2012)Fortunato, Barquinha, and
  Martins]{Fortunato12AM}
E.~Fortunato, P.~Barquinha and R.~Martins, \emph{Adv. Mater.}, 2012,
  \textbf{24}, 2945--2986\relax
\mciteBstWouldAddEndPuncttrue
\mciteSetBstMidEndSepPunct{\mcitedefaultmidpunct}
{\mcitedefaultendpunct}{\mcitedefaultseppunct}\relax
\EndOfBibitem
\bibitem[Kamiya \emph{et~al.}(2010)Kamiya, Nomura, and Hosono]{Kamiya10STAM}
T.~Kamiya, K.~Nomura and H.~Hosono, \emph{Sci. Technol. Adv. Mater}, 2010,
  \textbf{11}, 044305\relax
\mciteBstWouldAddEndPuncttrue
\mciteSetBstMidEndSepPunct{\mcitedefaultmidpunct}
{\mcitedefaultendpunct}{\mcitedefaultseppunct}\relax
\EndOfBibitem
\end{mcitethebibliography}
\bibliographystyle{rsc} 

\providecommand*{\mcitethebibliography}{\thebibliography}
\csname @ifundefined\endcsname{endmcitethebibliography}
{\let\endmcitethebibliography\endthebibliography}{}

\vspace*{-12pt}
\onecolumn
\setstretch{1.05}
\small
\begin{longtable}[b!t]{llllllllllll}
\caption{Carrier mobility values of 2D chalcogenides with slight or no mobility engineering, roughly listed in ascending orders of preparation method (mechanically exfoliated or synthesized) and  channel thickness (numbers of layers NL or nanometers), and in a descending order of room temperature mobility.} \label{tbl:mobility_typical} \\
\hline
Channel & Channel  & Contact    &Thermal       & Dielectric \&   & Measurement  & $\mu$ near RT     &$\mu$ at LT   & $\gamma$ value  & Ref.\\  
material& thickness  & \& doping&annealing     &encapsulation    & pressure     &\,\cmvs{}        &\,\cmvs{}   & near RT         & no.\\  
\hline
\ms{}&	1L&	Au&	&	TG:Al$_2$O$_3$&	&	80 (?)&	&	&	\citenum{Lee12NL_Photodetector}\\
\ms{}&	1L&	Ti&	in. 0.7Pa 300\CT{} 1h&	TE:Si$_3$N$_4$&	humidity&	71.8 (?)&	&	&	\citenum{Late12AN_FET:absorbate:MoS2}\\
\ms{}&	1L&	Au&	&	BG:SiO$_2$&	&	64&	147 (6.5\,K)&	&	\citenum{Sangw13NL_noise}\\
\ms{}&	1L&	Au&	ex. Ar/H$_2$ 200\CT{} 2h&	BG:SiO$_2$&	$\sim$0.13mPa&	59$^\text{4W}$&	&	&	\citenum{Lembk15N_}\\
&	&	&	{+} in. 120\CT{} 12 h&	&	&	&	&	&	\citenum{Lembk15N_}\\
\ms{}&	1L&	Ti &	ex. Ar/H$_2$ 350\CT{} 3h&	BG:SiO$_2$&	$\sim$0.13mPa&	$\sim$20$^\text{Hall}$&	250 (4\,K)&	1.7&	\citenum{Baugher13NL}\\
\ms{}&	1L&	Au/IL&	ex. 200\CT{} 2h&	TG:IG&	vac.&	&	230 (10\,K)&	1.22&	\citenum{Chu14SR_FET:MoS2:IonicGated}\\
\ms{}&	1L&	Mo&	147\CT{} 2h&	BG:Al$_2$O$_3$&	&	11--13&	&	&	\citenum{Kang14APL_FET:MoS2:MoContact}\\
\ms{}&	1L&	Ti &	ex. vac. 200\CT{}&	\multicolumn{2}{l}{BG:SiO$_2$$\rightarrow$BN}&		\multicolumn{2}{l}{0.5$\rightarrow$7.6--12}&		&	\citenum{Lee13AN_BN:u45_3LMoS_thickness}\\
\ms{}&	1L&	Au&	ex. N$_2$ 250\CT{} 1h&	BG:SiO$_2$&	&	$\sim$1&	&	&	\citenum{Chan13N}\\
\ms{}&	1L&	Au&	ex. N$_2$ 250\CT{} 1h&	$\rightarrow$BG:BN&	&	$\rightarrow\sim$10&	&	&	\citenum{Chan13N}\\
\ms{}&	1L&	Ti&	&	BG:SiO$_2$&	vac., &	1.1--10&	&	&	\citenum{Late12AN_FET:absorbate:MoS2}\\
\ms{}&	1L&	Cr&	&	suspended&	$\sim$0.13mPa&	0.9&	&	&	\citenum{Jin13JAP_SuspendedMoS2_mu1}\\
\ms{}&	1L&	Cr&	&	BG:SiO$_2$&	$\sim$0.13mPa&	0.1&	&	&	\citenum{Jin13JAP_SuspendedMoS2_mu1}\\
\ms{}&	1L&	Cr&	&	suspended&	&	0.05&	&	&	\citenum{Klots14SR_SuspendedMoS2:Photocurrent}\\
\ms{}&	2L&	Ti &	in. 120\CT{} 20h&	BG:SiO$_2$&	$\sim$0.13mPa&	$\sim$80$^\text{Hall}$&	375 (3\,K)&	1.1&	\citenum{Baugher13NL}\\
\ms{}&	2L&	Au&	ex. 200\CT{} 2h&	TG:IG&	vac.&	&	450 (2\,K)&	1.9--2.9 &	\citenum{Chu14SR_FET:MoS2:IonicGated}\\
\ms{}&	2L&	Ti&	no&	BG:SiO$_2$&	vac.&	35&	&	&	\citenum{Guo14AN}\\
\ms{}&	2L&	Au&	ex. Ar/H$_2$ 200\CT{} 2h&	BG:SiO$_2$&	$\sim$0.13mPa&	33$^\text{4W}$&	&	&	\citenum{Lembk15N_}\\
&	&	&	{+} in. 120\CT{} 12 h&	&	&	&	&	&	\citenum{Lembk15N_}\\
\ms{}&	2L&	Au&	ex. N$_2$ 250\CT{} 1h&	BG:BN&	&	$\sim$27&	&	&	\citenum{Chan13N}\\
\ms{}&	2L&	Au&	&	TG:Al$_2$O$_3$&	&	27&	&	&	\citenum{Lee12NL_Photodetector}\\
\ms{}&	2L&	Ti &	ex. vac. 200\CT{}&	BG:SiO$_2$&	&	$\sim$7&	&	&	\citenum{Lee13AN_BN:u45_3LMoS_thickness}\\
\ms{}&	2L&	Ti &	ex. vac. 200\CT{}&	$\rightarrow$BG:BN&	&	$\rightarrow$24&	&	&	\citenum{Lee13AN_BN:u45_3LMoS_thickness}\\
\ms{}&	2L&	Mo&	147\CT{} 2h&	BG:Al$_2$O$_3$&	&	11--14&	&	&	\citenum{Kang14APL_FET:MoS2:MoContact}\\
\ms{}&	2L&	Au&	ex. N$_2$ 250\CT{} 1h&	BG:SiO$_2$&	&	$\sim$3.5&	&	&	\citenum{Chan13N}\\
\ms{}&	2L&	Ti&	in. vac. 77\CT{}&	BG:SiO$_2$&	$\sim$0.13mPa&	4&	&	&	\citenum{Qiu12APL_Transport:2L}\\
\ms{}&	2L&	Ti&	ex. Ar/H$_2$ 400\CT{}&	BG:SiO$_2$&	air&	0.12&	&	&	\citenum{Qiu12APL_Transport:2L}\\
\ms{}&	2--3L&	Au$\rightarrow$1T&	no&	BG:SiO$_2$&	air&	19$\rightarrow$46&	&	&	\citenum{Kappera14NM}\\
\ms{}&	2--3L&	Au$\rightarrow$1T&	no&	TG:HfO$_2$&	air&	3.5$\rightarrow$12.5&	&	&	\citenum{Kappera14NM}\\
\ms{}&	2\,nm&	Sc&	&	BG:SiO$_2$&	&	26&	&	&	\citenum{Das13NLb}\\
\ms{}&	3L&	Au&	ex. 200\CT{} 2h&	TG:IG&	vac.&	65--95&	820 (2\,K)&	1.9--2.4 &	\citenum{Chu14SR_FET:MoS2:IonicGated}\\
\ms{}&	3L&	Ti &	&	TG:IL&	$\sim$0.13mPa&	63&	&	&	\citenum{Perer13AN_}\\
\ms{}&	3L&	Ti &	ex. vac. 200\CT{}&	BG:SiO$_2$&	&	$\sim$9&	&	&	\citenum{Lee13AN_BN:u45_3LMoS_thickness}\\
\ms{}&	3L&	Ti &	ex. vac. 200\CT{}&	$\rightarrow$BG:BN&	&	$\rightarrow$45&	&	&	\citenum{Lee13AN_BN:u45_3LMoS_thickness}\\
\ms{}&	3L&	Au&	ex. Ar/H$_2$ 200\CT{} 2h&	BG:SiO$_2$&	$\sim$0.13mPa&	36$^\text{4W}$&	&	&	\citenum{Lembk15N_}\\
&	&	&	{+} in. 120\CT{} 12 h&	&	&	&	&	&	\citenum{Lembk15N_}\\
\ms{}&	3L&	Mo&	147\CT{} 2h&	BG:Al$_2$O$_3$&	&	$\sim$27&	&	&	\citenum{Kang14APL_FET:MoS2:MoContact}\\
\ms{}&	3L&	Ni&	&	TG:ZrO$_2$&	&	25&	&	&	\citenum{Fang13NL}\\
\ms{}&	3L&	Permalloy&	in. 87\CT{} 2h&	BG:SiO$_2$&	vac.&	$\sim$27 (200K)&	$\sim$54 (2\,K)&	$\sim$0.6&	\citenum{Wang14SR_Rc:MoS2:Permalloy}\\
\ms{}&	3L&	Ni&	&	TG:IG&	&	12&	&	&	\citenum{Pu12NL_}\\
\ms{}&	3L&	Au&	&	TG:Al$_2$O$_3$&	&	10&	&	&	\citenum{Lee12NL_Photodetector}\\
\ms{}&	3--5L&	Cr&	&	TG:Y$_2$O$_3$/HfO$_2$&	&	47.7$\pm$11.9&	&	&	\citenum{Zou14AM}\\
\ms{}&	3--5L&	Cr&	&	TG:Al$_2$O$_3$/HfO$_2$&	&	37.4$\pm$11.4&	&	&	\citenum{Zou14AM}\\
\ms{}&	3--5L&	Cr&	&	TG:MgO/HfO$_2$&	&	15.9$\pm$7.2&	&	&	\citenum{Zou14AM}\\
\ms{}&	4L&	Mo&	147\CT{} 2h&	BG:Al$_2$O$_3$&	&	22--26&	&	&	\citenum{Kang14APL_FET:MoS2:MoContact}\\
\ms{}&	4L&	Ti &	ex. vac. 200\CT{}&	BG:SiO$_2$&	&	$\sim$5&	&	&	\citenum{Lee13AN_BN:u45_3LMoS_thickness}\\
\ms{}&	4L&	Ni &	&	BG:SiO$_2$&	vac.&	&	310 (1K)&	&	\citenum{Neal13AN_FET:MoS2}\\
\ms{}&	5L&	Mo&	147\CT{} 2h&	BG:Al$_2$O$_3$&	&	25--26&	&	&	\citenum{Kang14APL_FET:MoS2:MoContact}\\
\ms{}&	5L&	Ti &	ex. vac. 200\CT{}&	BG:SiO$_2$&	&	$\sim$15&	&	&	\citenum{Lee13AN_BN:u45_3LMoS_thickness}\\
\ms{}&	5L&	Ti &	&	BG:SiO$_2$&	$\sim$0.13mPa&	$\sim$5 (295K)&	$\sim$0.3 (140\,K)&	&	\citenum{Perer13AN_}\\
\ms{}&	5L&	Ti &	&	TG:IL&	$\sim$0.13mPa&	$\sim$100 (180K)&	$\sim$220 (77\,K)&	1&	\citenum{Perer13AN_}\\
\ms{}&	6L&	Au&	no&	BG:SiO$_2$&	vac.&	49&	&	&	\citenum{Guo14AN}\\
\ms{}&	6L&	Ti&	no&	BG:SiO$_2$&	vac.&	42&	&	&	\citenum{Guo14AN}\\
\ms{}&	5\,nm/7L&	Ti &	&	BG:SiO$_2$&	$\sim$0.13mPa&	$\sim$75 (295K)&	$\sim$180 (140\,K)&	&	\citenum{Perer13AN_}\\
\ms{}&	5\,nm&	Ni or Au&	no&	BG:SiO$_2$&	&	28&	&	&	\citenum{Liu12AN}\\
\ms{}&	5--6\,nm&	Ni &	&	BG:SiO$_2$&	vac.&	24&	&	&	\citenum{Neal13AN_FET:MoS2}\\
\ms{}&	8nm&	Ti &	&	BG:SiO$_2$&	$\sim$0.13mPa&	$\sim$40 (300K)&	$\sim$390 (77\,K)&	1.7&	\citenum{Perer13AN_}\\
\ms{}&	8nm&	Ti &	&	TG:IL&	$\sim$0.13mPa&	$\sim$160 (100K)&	$\sim$390 (77\,K)&	1.2&	\citenum{Perer13AN_}\\
\ms{}&	$\sim$10\,nm&	Sc&	&	BG:SiO$_2$&	&	184&	&	&	\citenum{Das13NLb}\\
\ms{}&	$\sim$10\,nm&	Ti&	&	BG:SiO$_2$&	&	125&	&	&	\citenum{Das13NLb}\\
\ms{}&	$\sim$10\,nm&	Ni&	&	BG:SiO$_2$&	&	36&	&	&	\citenum{Das13NLb}\\
\ms{}&	$\sim$10\,nm&	Pt&	&	BG:SiO$_2$&	&	21&	&	&	\citenum{Das13NLb}\\
\ms{}&	10\,nm&	BLG&	in. 200\CT{} 3h&	TG:BN&	&	26, 33(no Rc)&	&	&	\citenum{Roy14AN}\\
\ms{}&	10\,nm&	Ti&	&	TG:IL&	&	44$^\text{Hall}$ (220\,K)&	&	&	\citenum{Zhang12NL_}\\
\ms{}:p&	&	&	&	&	&	86$^\text{Hall}$ (220\,K)&	&	&	\citenum{Zhang12NL_}\\
\ms{}&	11\,nm&	Ti&	&	BG:SiO$_2$&	&	8.4&	&	&	\citenum{Na14N_noise}\\
\ms{}&	11\,nm&	Ti&	&	TG:Al$_2$O$_3$&	&	9.8&	&	&	\citenum{Na14N_noise}\\
\ms{}&	1--17\,nm&	Cr&	as-fabricated&	suspended&	vac.&	0.01--46&	&	&	\citenum{Wang15N_FET:MoS:suspended}\\
\ms{}&	1--17\,nm&	Cr&	ex. Ar/H$_2$ 200\CT{} 1h&	suspended&	vac.&	0.5--105&	&	&	\citenum{Wang15N_FET:MoS:suspended}\\
\ms{}&	12\,nm&	Ti&	ex. Ar/H$_2$ 200\CT{} 2h&	BG:SiO$_2$&	vac.(PPMS)&	$\sim$150$^\text{4W}$&	&	&	\citenum{Pradhan13APL}\\
\ms{}&	12\,nm &	Ti&	ex. Ar/H$_2$ 200\CT{} 2h&	BG:SiO$_2$&	vac. (PPMS)&	91&	&	&	\citenum{Pradhan13APL}\\
\ms{}&	13\,nm&	\multicolumn{2}{l}{Co$\rightarrow$1nm\,TiO$_2$}	&	BG:SiO$_2$&	vac.&	12$\rightarrow$76&	&	&	\citenum{Dankert14AN}\\
\ms{}-M&	1L&	Ti/Au&	&	BG:SiO$_2$&	ambient$^\text{RT}$ &	3--37$^\text{all}$ &	90--110 (90\,K)&	1.6&	\citenum{Kang15N_synthesis:MoS:waferscale}\\
\ms{}-M&	1L&	Ti/Au&	&	BG:SiO$_2$&	vac.$^\text{Low T}$&	25--35$^\text{best}$&	 &	&	\citenum{Kang15N_synthesis:MoS:waferscale}\\
\ms{}-C&	1L&	Ag&	&	DE:Si$_3$N$_4$&	$\sim$0.13mPa&	24&	58 (77\,K)&	0.65&	\citenum{Sanne15APL_FET:MoS:Si3N4}\\
&	&	&	&	TG:HfO$_2$&	&	&	&	&	\citenum{Sanne15APL_FET:MoS:Si3N4}\\
\ms{}-C&	1L&	Ti&	&	BG:-SH&	<1.3mPa&	13&	&	&	\citenum{Najma14NL_Interface:SAM_mu13}\\
\ms{}-C&	1L&	Ti&	200\CT{}&	TG:Al$_2$O$_3$&	&	11$\pm$3&	&	&	\citenum{Liu13NL_}\\
\ms{}-S&	1L&	Cr&	&	BG:SiO$_2$&	&	7 (2--12)&	&	&	\citenum{Tao15N_synthesis:MoS:sputtering}\\
\ms{}-C&	1L&	Ti &	&	TG:HfO$_2$&	&	6, 30$^\text{band}$&	\multicolumn{2}{l}{1.3, 13$^\text{band}$ (50\,K)}&	 	\citenum{Zhu14NC_trspt:bandtail}\\
\ms{}-C&	1L&	Ti&	ex. Ar/H$_2$ 350\CT{} 2h&	BG:SiO$_2$&	&	3.6&	$\sim$3.6 (90\,K)&	&	\citenum{Jeon15N_synthesis:MoS}\\
\ms{}-C&	1L&	Ti&	&	BG:-NH2&	<1.3mPa&	3.6&	&	&	\citenum{Najma14NL_Interface:SAM_mu13}\\
\ms{}-C&	1L&	Ti&	&	BG:SiO$_2$&	<1.3mPa&	1.9&	&	&	\citenum{Najma14NL_Interface:SAM_mu13}\\
\ms{}-A&	1L&	Ti&	&	BG:SiO$_2$&	$\sim$1.3mPa&	1.2&	&	&	\citenum{Lee13NL_Synthesis:Seed:PTAS}\\
\ms{}-C&	2L&	Ti&	ex. Ar/H$_2$ 350\CT{} 2h&	BG:SiO$_2$&	&	8.2&	$\sim$8.2 (90\,K)&	&	\citenum{Jeon15N_synthesis:MoS}\\
\ms{}-C&	2--3L&	1T \ms{}&	no&	BG:SiO$_2$&	air&	24$\rightarrow$56&	&	&	\citenum{Kappe14AM_FET:MoS2:1Tcontact}\\
\ms{}-C&	3L&	Ti&	ex. Ar/H$_2$ 350\CT{} 2h&	BG:SiO$_2$&	&	15.6&	$\sim$15.6 (90\,K)&	&	\citenum{Jeon15N_synthesis:MoS}\\
\ms{}-C&	5.7\,nm&	Ti/Ni&	&	BG:SiO$_2$&	&	9.9&	&	&	\citenum{Yang14N_FET:MoS:thickness}\\
\ms{}-C&	70\,nm&	Ti/Ni&	&	BG:SiO$_2$&	&	42&	&	&	\citenum{Yang14N_FET:MoS:thickness}\\
\hline									
\mse{}-C&	1L&	Ti&	ex. vac. 120\CT{} 2h&	BG:SiO$_2$&	&	50&	&	&	\citenum{Wang14An_syntesis:MoSe}\\
\mse{}&	few nm&	Ni&	&	BG:SiO$_2$&	$\sim$13$\mu$Pa&	$\sim$40&	180 (78\,K)&	2.1&	\citenum{Laren12APL_FET:MoSe}\\
\mse{}&	7\,nm&	Ti&	ex. Ar/H$_2$ 250\CT{} 2h&	BG:SiO$_2$&	&	$\sim$200 (275\,K)&	$\sim$150 (h) &	&	\citenum{Pradh14AN_FET:MoS}\\
\mse{}&:p	7\,nm&	Ti&	+ in. 120\CT{} 24h&	BG:SiO$_2$&	&	$\sim$150 (275\,K)&	&	&	\citenum{Pradh14AN_FET:MoS}\\
\mse{}&	10--12\,nm&	Ti &	&	BG:SiO$_2$&	$\sim$0.13mPa&	$\sim$110&	500 (100\,K)&	1.2&	\citenum{Chamlagain14AN}\\
\mse{}&	20\,nm&	Ti$^\text{BC}$&	ex. N$_2$ 400\CT{} 2h&	BG:SiO$_2$&	&	&	$\sim$20&	&	\citenum{Abder14N_FET:MoSe}\\
\mse{}&	5--14\,nm&	Ti &	&	BG:SiO$_2$&	$\sim$0.13mPa&	$\sim$50&	600 (77\,K)&	1.7&	\citenum{Chamlagain14AN}\\
\hline									
\mte{}:p&	2L&	Ti&	ex. Ar/H$_2$ 250\CT{} 2h&	BG:SiO$_2$&	vac. (PPMS)&	11&	&	&	\citenum{Pradh14AN_FET:MoTe}\\
&	&	&	 in. 120\CT{} 24h&	&	&	&	&	&	\citenum{Pradh14AN_FET:MoTe}\\
\mte{}:p&	3L$\rightarrow$7L &	Ti&	ex. Ar/H$_2$ 250\CT{} 2h&	BG:SiO$_2$&	vac. (PPMS)&	20$\rightarrow$27&	&	&	\citenum{Pradh14AN_FET:MoTe}\\
&	&	&	+ in. 120\CT{} 24h&	&	&	&	&	&	\citenum{Pradh14AN_FET:MoTe}\\
\mte{}:p&	3L&	Ti&	&	BG:SiO$_2$&	vac.&	0.3&	&	&	\citenum{Lin14AM_FET:MoTe}\\
\mte{}&	3L&	Ti&	&	BG:SiO$_2$&	vac.&	0.03&	&	&	\citenum{Lin14AM_FET:MoTe}\\
\mte{} &	8\,nm&	Ti&	&	TG:IL&	vac.&	30 (270K)&	&	&	\citenum{Lezam142M_FET:MoTe}\\
\mte{}:p&	8\,nm&	Ti&	&	TG:IL&	vac.&	5 (270K)&	&	&	\citenum{Lezam142M_FET:MoTe}\\
\mte{}:p&	30L&	Ti&	ex. Ar/H$_2$ 300\CT{} 3h&	BG:SiO$_2$&	vac.&	6.4&	&	&	\citenum{Fathi14APL_MoTe:Pdetector}\\
\hline									
\ws{}&	1L&	Au&	ex. Ar/H$_2$ 200\CT{} 2h&	BG:SiO$_2$&	vac. (PPMS)&	$\sim$50$\pm$7&	140 (7\,K)&	0.73&	\citenum{Ovchi14AN_FET:WS}\\
&	&	&	+ in. 120\CT{} 24h&	&	&	&	&	&	\citenum{Ovchi14AN_FET:WS}\\
\ws{}&	1L&	Au&	ex. Ar/H$_2$ 200\CT{} 2h&	TG:IL&	<0.13mPa&	19 (240\,K)&	&	&	\citenum{Jo14NL_Opto-LED:WS2}\\
\ws{}:p&	1L&	Au&	ex. Ar/H$_2$ 200\CT{} 2h&	TG:IL&	<0.13mPa&	12 (240\,K)&	&	&	\citenum{Jo14NL_Opto-LED:WS2}\\
\ws{}&	1L&	Cr&	in. current&	BG:SiO$_2$&	vac.&	0.23&	&	&	\citenum{Withe14SR_FET:WS:BN}\\
\ws{}&	2L&	Au&	ex. Ar/H$_2$ 200\CT{} 2h&	BG:SiO$_2$&	vac. (PPMS)&	$\sim$30&	>300 (5\,K)&	1.75&	\citenum{Ovchi14AN_FET:WS}\\
&	&	&	+ in. 120\CT{} 24h&	&	&	&	&	&	\citenum{Ovchi14AN_FET:WS}\\
\ws{}&	2L&	Au&	ex. Ar/H$_2$ 200\CT{} 2h&	TG:IL&	<0.13mPa&	44 (230\,K) &	&	&	\citenum{Jo14NL_Opto-LED:WS2}\\
\ws{}:p&	2L&	Au&	ex. Ar/H$_2$ 200\CT{} 2h&	TG:IL&	<0.13mPa&	43 (230\,K)&	&	&	\citenum{Jo14NL_Opto-LED:WS2}\\
\ws{}&	4L&	Cr&	in. current&	BG:SiO$_2$&	vac.&	17&	&	&	\citenum{Withe14SR_FET:WS:BN}\\
\ws{}&	4L&	Cr&	in. current&	BG:BN&	vac.&	80&	&	&	\citenum{Withe14SR_FET:WS:BN}\\
\ws{}&	few nm&	Ti&	&	BG:SiO$_2$&	vac.&	16&	&	&	\citenum{Duan14NN_PV:WSe_MoS}\\
\ws{}&	$\sim$7\,nm&	Au&	ex. Ar/H$_2$ 200\CT{} 2h&	BG:SiO$_2$&	&	80&	250 ($\sim$3.5\,K)&	1.15&	\citenum{Liu14AN_FET:WS:mu230}\\
\ws{}:p&	20--60\,nm&	Ti&	&	TG:IL&	<0.13mPa&	60--100&	&	&	\citenum{Braga12NL}\\
\ws{}&	20--60\,nm&	Ti&	&	TG:IL&	<0.13mPa&	20--60&	&	&	\citenum{Braga12NL}\\
\ws{}-C&	1L&	Ti&	ex. N$_2$ 200\CT{} 5h&	BG:SiO$_2$&	vac.&	15--24&	&	&	\citenum{Yun15An_syntesis:MoS:1L}\\
\ws{}-M&	1L&	Ti/Au&	&	BG:SiO$_2$&	ambient&	$\sim$5$^\text{median}$&	&	&	\citenum{Kang15N_synthesis:MoS:waferscale}\\
\ws{}-M&	1L&	Ti/Au&	&	BG:SiO$_2$&	ambient&	18$^\text{best}$&	&	&	\citenum{Kang15N_synthesis:MoS:waferscale}\\
\ws{}-A&	1L&	Ti&	&	BG:SiO$_2$&	$\sim$1.3mPa&	0.01&	&	&	\citenum{Lee13NL_Synthesis:Seed:PTAS}\\
\ws{}-C&	7.5\,nm&	Au&	Ar/H$_2$ 200\CT{} 2h&	BG:SiO$_2$&	&	234&	&	&	\citenum{Liu14AN_FET:WS:mu230}\\
\ws{}-C&	8\,nm&	Au&	Ar/H$_2$ 200\CT{} 2h&	BG:SiO$_2$&	&	$\sim$250&	$\sim$70(3.5\,K)&	1.15&	\citenum{Liu14AN_FET:WS:mu230}\\
\hline									
\wse{}:p&	1L&	Ni&	&	TG:IG&	&	90&	&	&	\citenum{Huang14AN_Synthesis:WSe:inverter}\\
\wse{}&	1L&	Ni&	&	TG:IG&	&	7&	&	&	\citenum{Huang14AN_Synthesis:WSe:inverter}\\
\wse{}:p&	1L&	Pd&	in. vac. 80\CT{} &	TG:IL&	vac.&	180 (250K)&	255 (4\,K)&	&	\citenum{Allai14AN_FET:WSe}\\
\wse{}&	1L&	Pd&	in. vac. 80\CT{} &	TG:IL&	vac.&	30 (250K)&	160 (4\,K)&	&	\citenum{Allai14AN_FET:WSe}\\
\wse{}&	1L&	Pd&	in. vac. 80\CT{} &	TG:IL&	vac.&	30 (250K)&	100$^\text{Hall}$ (4\,K)&	&	\citenum{Allai14AN_FET:WSe}\\
\wse{}:p&	2L&	SLG&	&	TG:BN&	&	45&	&	&	\citenum{Das14NL}\\
\wse{}&	2L&	SLG&	&	TG:BN&	&	34&	&	&	\citenum{Das14NL}\\
\wse{}:p&	3L&	Cr/Pd/IL&	&	DE:BN&	vac.&	\multicolumn{2}{l}{>600$^\text{Hall}$ (220K)}&		&	\citenum{Wang15NL_FET:WSe_BN}\\
\wse{}&	3L&	Au&	&	TG:ZrO$_2$&	&	110&	&	&	\citenum{Fang13NL}\\
\wse{}&	few nm&	Au&	&	BG:SiO$_2$&	vac.&	82&	&	&	\citenum{Duan14NN_PV:WSe_MoS}\\
\wse{}&	6\,nm&	Gr./IL&	&	TE:BN&	$\sim$0.13mPa&	$\sim$200 (160\,K)&	$\sim$330 (77\,K)&	&	\citenum{Chuan14NL_Rc:WSe2:IonicLiquid}\\
\wse{}:p&	6\,nm&	Gr./IL&	&	TE:BN&	$\sim$0.13mPa&	$\sim$200 (160\,K)&	$\sim$270 (77\,K)&	&	\citenum{Chuan14NL_Rc:WSe2:IonicLiquid}\\
\wse{}&	7\,nm&	Gr./IL&	&	TE:Al$_2$O$_3$&	$\sim$0.13mPa&	&	$\sim$130 (77\,K)&	&	\citenum{Chuan14NL_Rc:WSe2:IonicLiquid}\\
\wse{}:p&	7\,nm&	Gr./IL&	&	TE:Al$_2$O$_3$&	$\sim$0.13mPa&	&	$\sim$57 (77\,K) &	&	\citenum{Chuan14NL_Rc:WSe2:IonicLiquid}\\
\wse{}&	8\,nm&	Gr./IL&	&	TG:IL&	$\sim$0.13mPa&	&	$\sim$250 (77\,K)&	&	\citenum{Chuan14NL_Rc:WSe2:IonicLiquid}\\
\wse{}:p&	8\,nm&	Gr./IL&	&	TG:IL&	$\sim$0.13mPa&	&	$\sim$110 (77\,K)&	&	\citenum{Chuan14NL_Rc:WSe2:IonicLiquid}\\
\wse{}:p&	8\,nm/12L&	Ti&	ex. Ar/H$_2$ 200\CT{} 2h&	BG:SiO$_2$&	vac. (PPMS)&	150&	550 (<50\,K)&	&	\citenum{Pradh15SR_FET:WSe}\\
\wse{}&	8\,nm/12L&	Ti&	ex. Ar/H$_2$ 200\CT{} 2h&	BG:SiO$_2$&	vac. (PPMS)&	300&	665 (100\,K)&	&	\citenum{Pradh15SR_FET:WSe}\\
\hline									
\sns{}&	1L&	Ti&	&	TG:Al$_2$O$_3$&	$\sim$1.3mPa&	50&	&	&	\citenum{Song13N}\\
\sns{}&	15\,nm&	Cr&	&	BG:SiO$_2$&	0.13--1.3mPa&	0.8&	0.1 (100\,K)&	&	\citenum{De13N_FET:SnS:mu0.8}\\
\hline
\multicolumn{10}{l}{Abbreviation and note.}\\
\multicolumn{10}{l}{C: Chemical Vapour Deposition (CVD),  A: Atmospheric Pressure CVD (APCVD), M: Metal-Organic CVD (MOCVD), S: sputtering}\\
\multicolumn{10}{l}{p: p-type conduction, NL: number of layers (N is an integer)}\\
\multicolumn{10}{l}{ex.: \textit{ex situ}, in.: \textit{in situ},  vac.:vacuum, PPMS: Physical Property Measurement System, RT: room temperature, LT: low temperature}\\
\multicolumn{10}{l}{BG: bottom gated, TG: top gated, DE: double-side encapsulated, BC: bottom contacted, IL: ionic liquid, IG: ionic gel, Gr.: graphene}\\
\multicolumn{10}{l}{+ (annealing): \textit{in situ} annealing is also used besides \textit{ex situ} annealing.}\\
\multicolumn{10}{l}{X$\rightarrow$Y: Experimental condition (or mobility) changes from X to Y.}\\
\multicolumn{10}{l}{?: Mobility value deserves to be checked due to top/bottom gate coupling.}\\
 \hline
\end{longtable}
\normalsize
\setstretch{1.125}

\end{document}